# North South University

Department of Electrical & Computer Engineering

# A Practical Framework for Storing and Searching Encrypted Data on Cloud Storage

A Thesis submitted in partial fulfillment of
the requirements for the award of the degree of

Master of Science in Computer Science and Engineering

Submitted by

**Md. Mazharul Islam**

ID: 1935366650

Under the Supervision of

**Dr. Rajesh Palit**

July, 2022

# DECLARATION

I hereby declare that the work presented in this report is the outcome of the design and development work performed by myself, Md. Mazharul Islam, under the supervision of Dr. Rajesh Palit, Professor, Department of Electrical & Computer Engineering, North South University, as a course work of CSE599 (Thesis). I also declare that no part of this report has been taken from other works without reference.

Signature

________________
Md. Mazharul Islam
ID 1935366650

# APPROVAL

This thesis report titled "**A Practical Framework for Storing and Searching Encrypted Data on Cloud Storage**" is submitted by MD. MAZHARUL ISLAM (ID: 1935366650) to the Department of Electrical & Computer Engineering (ECE), North South University, has been accepted as the final term thesis report.

Signatures

________________________________

Dr. Rajesh Palit
Professor
Department of Electrical and Computer Engineering
Supervisor

________________________________

Dr. Hafiz Abdur Rahman
Professor
Department of Electrical and Computer Engineering
Committee Member (Internal)

________________________________

Dr. Salekul Islam
Professor & Head
Department of Computer Science and Engineering
United International University
Committee Member (External)

________________________________

Dr. Mohammad Rezaul Bari
Associate Professor & Chair
Department of Electrical and Computer Engineering

# ACKNOWLEDGEMENTS

The title of my research is **"A Practical Framework for Storing and Searching Encrypted Data on Cloud Storage"**. Foremost, I want to offer this work to Almighty God for the wisdom, strength, and sound health that He has given me to complete the study. Then, I am thankful to many people in the entire journey of my research work who provided me with guidance and motivation to move forward when I lost my enthusiasm.

I like to acknowledge those significant contributions that make me capable of reaching the final stage. I would like to express my gratitude to my supervisor Dr. Rajesh Palit, Ph.D., for his valuable advice, conceptual ideas, consistent support, and constructive criticisms despite his hectic schedule. His guidance helped me in all the time of research and writing of this thesis. I acknowledge with thanks the member of the research committee for their thought-provoking ideas and guidance that lead to this accomplishment. I would like to thank the faculty and stuff members of the Department of Electrical and Computer Engineering (ECE) to support me to complete my degree.

Lastly and most importantly, I acknowledge the sacrifices and encouragements of my beloved family members, especially my parents, who have constantly been providing support and encouragement throughout my studies.

# ABSTRACT

Security has become a significant concern with the increased popularity of cloud storage services. It comes with the vulnerability of being accessed by third parties. Security is one of the major hurdles in the cloud server for the user when the user data that reside in local storage is outsourced to the cloud. It has given rise to security concerns involved in data confidentiality even after the deletion of data from cloud storage. This becomes a severe issue when the replicated copies remain hidden in the server even after the owner has deleted the original file. Thus, data must be encrypted before uploading to the public cloud to protect it from data security and unauthorized access. By encrypting the data, the third parties cannot access it easily. Though, it raises a serious problem when the encrypted data needs to be shared with more people than the data owner initially designated. Besides, searching over the encrypted files is important to find the data quickly. However, searching on encrypted data is a fundamental issue in cloud storage. The method of searching over encrypted data represents a significant challenge in the cloud.

As multiple copies of the same data are kept in the storage server for reliability, some copies remain in the cloud storage through a user deleting the data. To address this issue, a technique termed File Assured DEletion (FADE) is introduced that encrypts the data before storing it in cloud storage. It encrypts outsourced data files to guarantee their privacy and integrity of deletes files to make them unrecoverable to anyone (including those who manage the cloud storage) upon revocations of file access policies.

Since cloud servers are not within the trusted domain of users, encryption and access control are needed to protect the data. Encrypted data sharing among users is important because that allows users to access the selected subset of data from the cloud without revealing his/her access rights to the cloud server. Encrypted file sharing support was improved significantly by delivering complete control over file access to the owner.

Searchable encryption allows a cloud server to conduct a search over encrypted data on behalf of the data users without learning the underlying plaintexts. While many academic SE schemes show provable security, they usually expose some query information, making them less practical, weak in usability, and challenging to deploy. They do not provide effective data utilization for large dataset files in the cloud. Also, sharing encrypted data with other authorized users must provide each document's secret key. However, this way has many limitations due to the difficulty of key management and distribution.

To address the above issues, we introduce a secure and practical searchable encryption scheme with provable security strength for cloud applications that supports efficient search using an inverted index. It improves search efficiency and enhances the security strength of SE using symmetric cryptography combined with Identity-Based Encryption (IBE). The proposed scheme is more secure and efficient than the other proposed models.

We have designed the system using the existing cryptographic approaches, ensuring the search on encrypted data over the cloud. The primary focus of our proposed model is to ensure user privacy and security through a less computationally intensive, user-friendly system with a trusted third-party entity. To demonstrate our proposed model, we have implemented a web application called "CryptoSearch" as an overlay system on top of a well-known cloud storage domain. It exhibits secure search on encrypted data with no compromise to the user-friendliness and the scheme's functional performance in real-world applications.

# TABLE OF CONTENTS

# LIST OF FIGURES

# LIST OF TABLES

# LIST OF ACRONYMS

| Acronyms | Full Meaning |
|---|---|
| ABE | Attribute Based Encryption |
| ASE | Asymmetric Searchable Encryption |
| CP-ABE | Ciphertext-policy Attribute Based Encryption |
| CS | Cloud Server |
| CSP | Cloud Service Provider |
| DO | Data Owner |
| DU | Data User |
| EHR | Electronic Health Record |
| HE | Homomorphic Encryption |
| IaaS | Infrastructure-as-a-Service |
| IBE | Identity Based Encryption |
| KP-ABE | Key-policy Attribute Based Encryption |
| KSAC | Keyword Search with Access Control |
| PaaS | Platform-as-a-Service |
| PKSE | Public Key Searchable Encryption |
| PSE | Privacy Preserving Searchable Encryption |
| SaaS | Storage-as-a-Service |
| SE | Searchable Encryption |
| SFADE | Simplified File Assured DEletion |
| SSE | Symmetric Searchable Encryption |
| TTP | Trusted Third Party |

# 1. CHAPTER ONE: INTRODUCTION

## 1.1 Cloud Computing and Cloud Infrastructure

Nowadays, a new generation of technology is transforming the world of cloud computing. Cloud computing has recently emerged as a disruptive trend in IT industries and research communities. The latest paradigm plays a significant role in the IT industry and the government sector [1]. Today more and more users use internet-based services like data storage, processing, and services collectively known as cloud computing. Cloud computing has developed from various technologies such as autonomic computing, virtualization, grid computing, and other technologies. Secure storage is essential and important because it provides virtualized resources on the Internet [2].

Cloud computing provides the facility for various applications operating over thousands of computers and servers to access the services through the Internet concurrently. With the evolution of cloud computing, it has become easier for users to store, retrieve, and share their data. It provides flexibility to work from anywhere at any time [3]. It offers various benefits to users and service providers, which are shown in Figure 1.1[1].

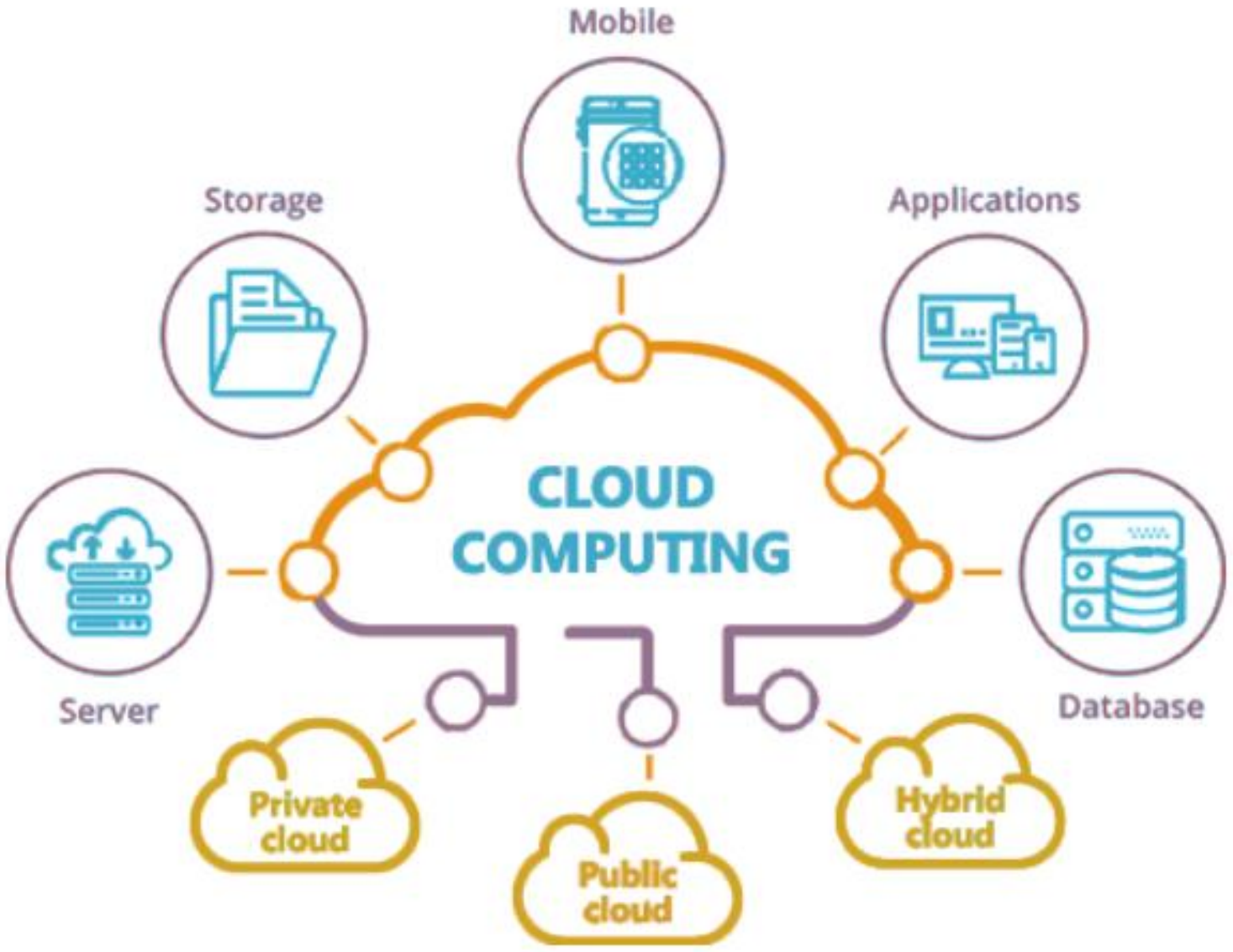

Figure 1.1 Cloud Computing Benefits

Cloud computing can be defined based on processing, storage resources, the service-oriented interface, the exploitation of virtualization techniques, etc. The National Institute of Standards and Technology (NIST) has given a complete reference definition. NIST defined "Cloud computing as a pay-per-use model for enabling available, convenient, on-demand network access to a shared pool of configurable computing resources (e.g., networks, servers, storage, applications, services) that can be rapidly provisioned and released with minimal management effort or service provider interaction” [4].

[1] Peters, L. (2019, February 22). *Cloud Computing Trends for 2019*. Networks Unlimited. https://www.networksunlimited.com/cloud-computing-trends-for-2019/

The benefits brought by this new computing model include but are not limited to relief of the burden for storage management, universal data access with independent geographical locations, and avoidance of capital expenditure on hardware, software, personnel maintenances, etc. Nowadays, more and more companies and individuals from a large number of big data applications have outsourced their data and deployed their services into cloud servers for easy data management, efficient data mining, and query processing tasks. It brings great convenience to consumers, where shared resources, data, and information are provided to computers on-demand. The consumer has to pay as per use. It offers organizations more excellent choice, agility, and flexibility while driving efficiency gains and lowering overall IT costs [5]. The emergence of cloud infrastructure has significantly reduced the costs of hardware and software resources in computing infrastructure.

According to a case study done by Yamasaki et al. (2015) [6] on IaaS and SaaS in Public Cloud, three kinds of services can be provided through cloud computing such as; Infrastructure-as-a-Service (IaaS), Platform-as-a-Service (PaaS), and Software-as-a-Service (SaaS) is shown in figure 1.2[2].

Infrastructure-as-a-Service (IaaS) offers the customer's ability to deliver computing, storage, networking, and other primary computer services through which the consumer can install and manage virtual machines, including operating systems and applications. The user does not handle the cloud infrastructure but may have control over operating systems, storage, and apps and may have little control over selected networking components. Example: Amazon (EC2) delivers physical and virtual services to customers, including consumer specifications, memory, OS, and storage. As a service provider, IaaS provides the virtual server with one or more central operating units running different options (IaaS). It is a centralized, fully automated package that owns and hosts a service provider and provides customers with computing services accompanied by storage and networking facilities on request.

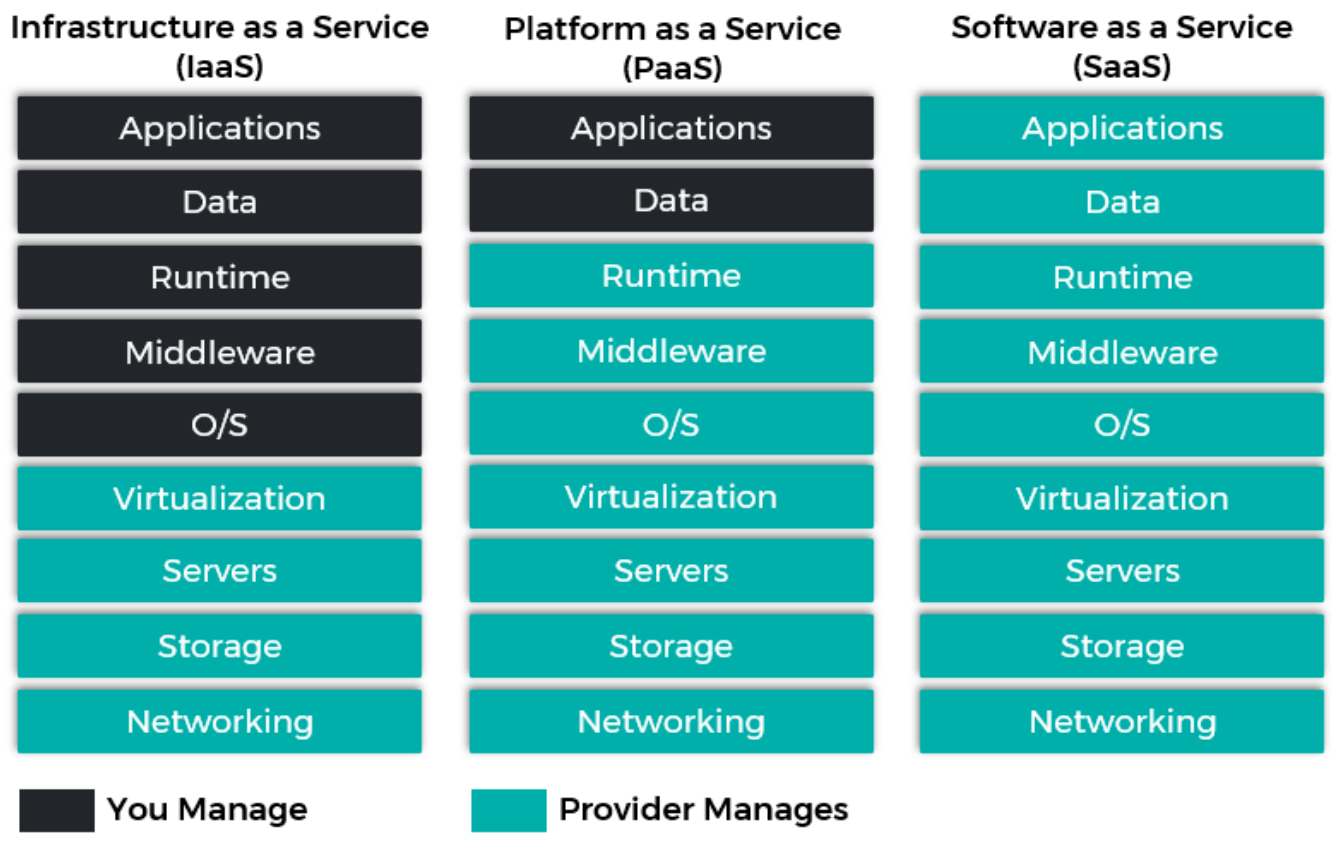

Figure 1.2 Cloud Infrastructure

[2] Nap, I. (2020, December 15). *What are the Differences Between IaaS, PaaS and SaaS?* INAP. https://www.inap.com/blog/iaas-paas-saas-differences/

Platform-as-a-Service (PaaS) provides an acceptable framework or medium for the developer to build applications and programs and distribute them on the network without getting the production environment installed or managed. To run available software or build and test the latest, PaaS allows clients to rent software-defined servers and attached resources. The client is not in charge of the cloud's hardware, such as servers, networks, storage, and OS. However, the client controls the applications and their configuration. Google application engine and Microsoft Azure are the most recent examples of PaaS. It is focused on the creation and usage of cloud software by deployers and developers. The multi-layer architecture is highly scalable, e.g., Salesforces.com and Azure. This model uses tools and libraries that act as the framework.

In addition to that, there is another kind of service that can be delivered over the cloud; known as Storage-as-a-Service. The opportunity to use apps of the provider that operate on a cloud platform is provided to the consumer. Applications can be accessed by a web application interface, such as a web browser (e.g., a web-based email) or the application interface, from separate client devices. With the potential exception of a small range of device setups for consumers, the customer does not handle or monitor the cloud infrastructure. It includes the network, servers, operating systems, storage, or even specific applications. SaaS primarily focuses on the end-user interface, as end users can use and manage this cloud-built software. Examples of SaaS are CRM, Google apps, Deskaway, and Wipro w-SaaS.

Cloud computing does not refer to a single thing; instead, it describes various individual components or services. The combination of different cloud computing models, referred to as Cloud stack, is shown in figure 1.3[3].

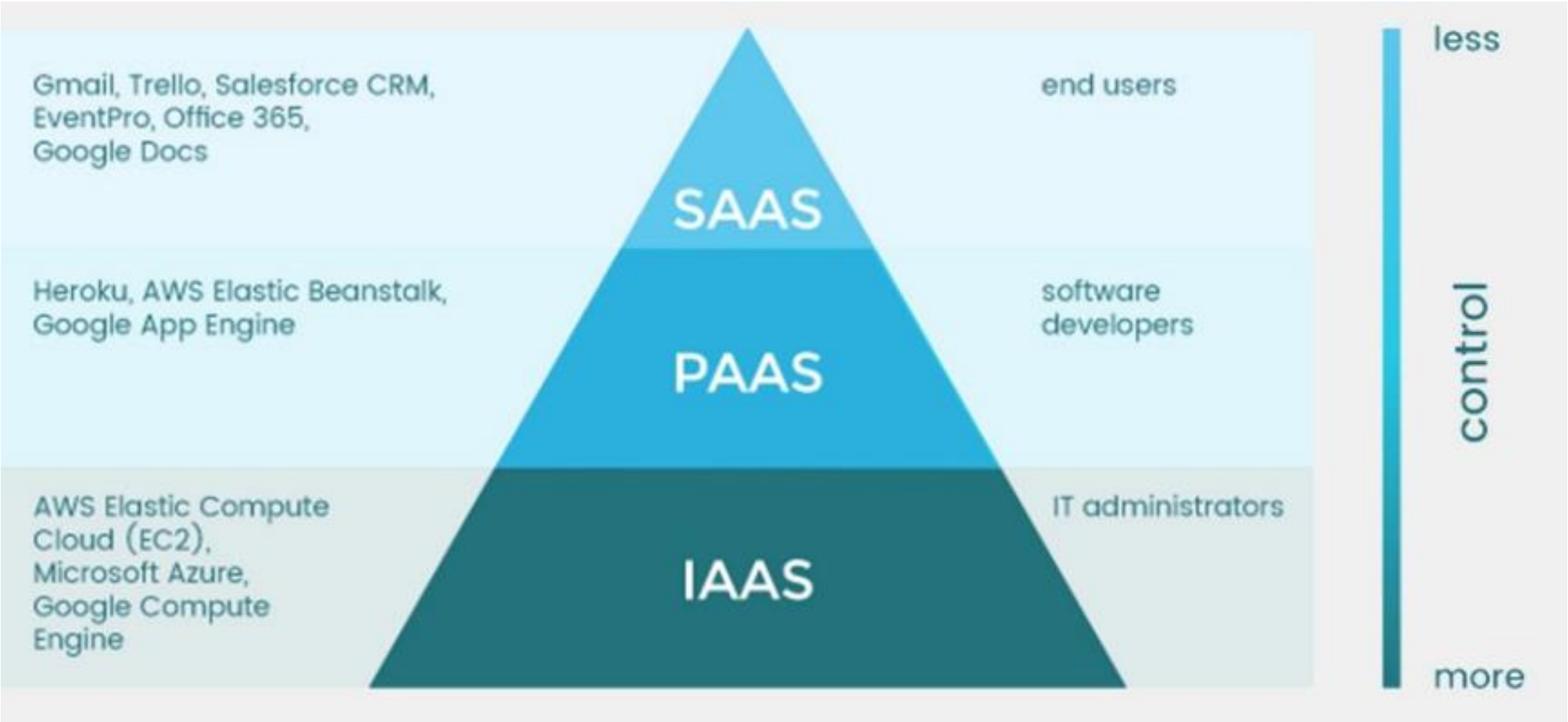

Figure 1.3 Cloud Stack

Cloud-based systems have more advantages than traditional systems. Users can store and maintain massive data quickly and enjoy high-quality data storage services formed by cloud computing [7].

[3] Plesky, E. (2019, October 25). *IaaS vs PaaS vs Saas – various cloud service models compared*. Plesk. https://www.plesk.com/blog/various/iaas-vs-paas-vs-saas-various-cloud-service-models-compared/

Cloud computing systems provide large-scale infrastructures for high-performance computing that are dynamically adapted to user and application needs.

Cloud storage services [8] are widely used in different applications due to their many advantages. Many services like emails, personal health records, documents, emails, and office applications like accounting, HR, purchase, and CRM are delivered from the cloud. Despite these benefits offered by the cloud, users believe that security is the major hurdle in the adoption of the cloud and, hence, resist outsourcing data in plain text to the cloud even if data privacy is provided by law. They feel that security breaches are more on the cloud than on traditional IT infrastructure. Various challenges associated with the cloud are security, malicious insiders, lack of resources, lack of expertise, compliance, lack of ability to manage multiple cloud services, managing costs, governance control, and performance, etc. Out of these, security is the prime concern in the cloud.

Many challenges remain in the security of cloud computing. Such major challenges are access and management, trust, business sustainability, and the development of scalable service. When users outsource their sensitive data to cloud storage, the transmitted data is vulnerable to intrusion [9] by illegal entities, especially under critical infrastructures. Meanwhile, the user lost their capabilities to control the data effectively. By accessing the data, the cloud server and the illegal user can try to acquire the information contained in the data, and the personal security problem of the user is faced with great challenges. Threats Cloud computing faces just as many security threats currently found in the existing computing platforms, networks, intranets, and enterprises. These threats and risk vulnerabilities come in various forms, as shown in figure 1.4[4].

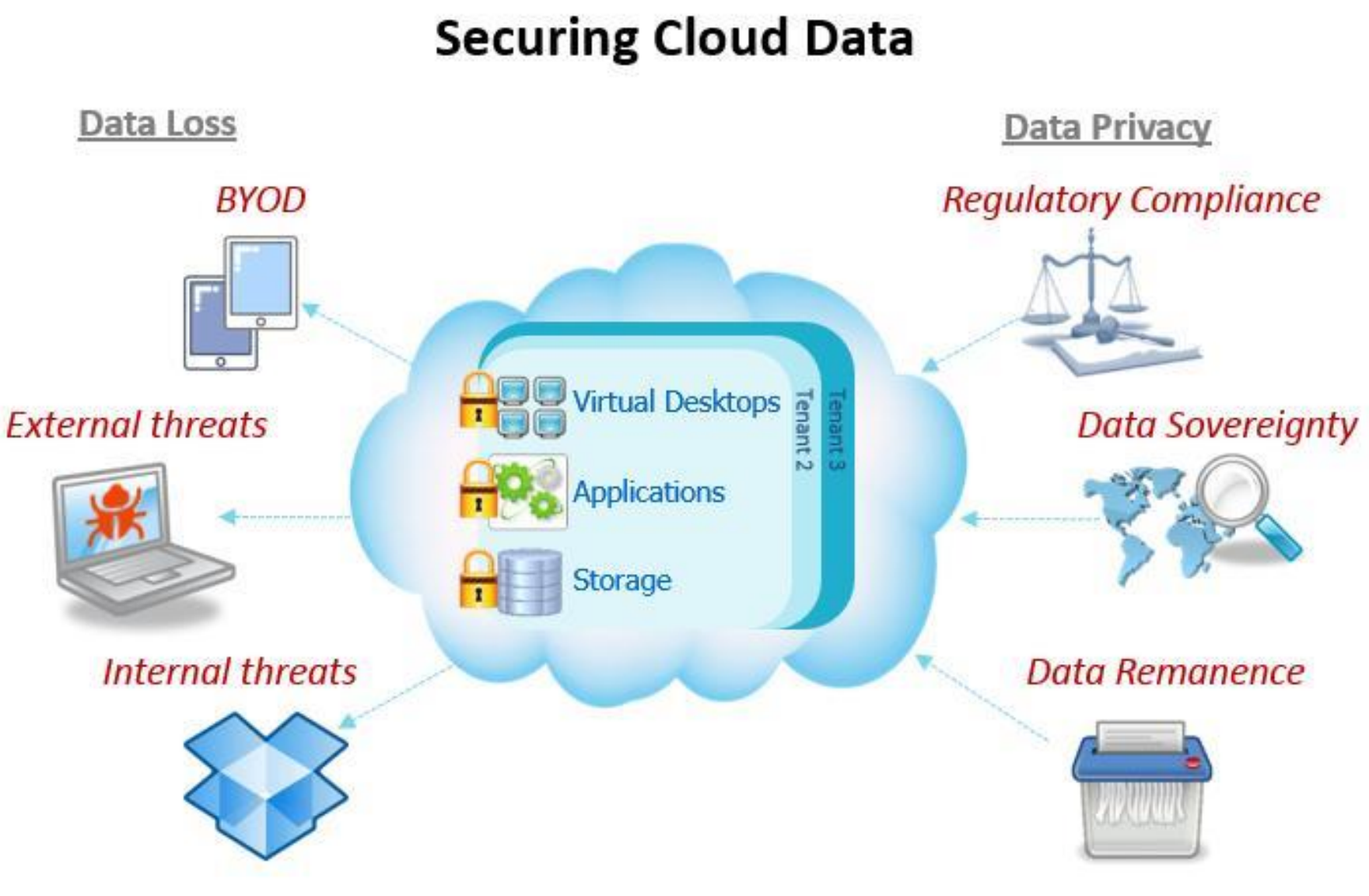

Figure 1.4 Cloud Computing Security Issues

Therefore, the security of cloud computing is far from satisfying. Cloud security has become a challenging problem for data of users and companies when they adopt cloud services. Cloud security can be classified into storage and computation security [10]. Storage security means the users' data privacy of online storage, such as avoiding data leakage, data integrity, assured data deletion, etc. Since cloud services data is stored online, data owners lose the ultimate control of their data. Then they cannot physically protect data from attackers' interception or manipulation. For the cloud computation aspect, when users delegate the cloud server to do some computation, such as keyword search or scientific computation, the cloud may not perform a secure and exact computation to save computing resources. Hence users also need to guarantee data privacy during cloud computation.

[4] Bharat, M & Nagageetha, M & Narendra, Gurram & Santosh Kumar, Gurram & Anusha, K & Dasore, Abhishek. (2018). Securing Data Storage in Cloud by Using Cryptography Conventions. International Journal of Pure and Applied Mathematics. 118.

## 1.2 Cloud Security Requirements

The security of cloud computing is more difficult to control because its environment becomes dynamic and demanding**.** The cloud infrastructure must be capable enough to implement the appropriate security measures at its premises. Although the services provided by the cloud are regularly being improved, there is still a great need to protect data stored in the cloud. Therefore, traditional security mechanisms are not sufficient and efficient because of the communication overhead and the heavy computations**.** The most important requirement is to build up trust between the user and the service provider.

For cloud data protection, the following security measures need to be implemented:

**Authentication**: This technique helps the communicating entities to prove their identity and assures authentic communication. This service also guarantees that no other unauthorized entity can masquerade itself as the authorized entity to take undue advantage of ongoing communication.

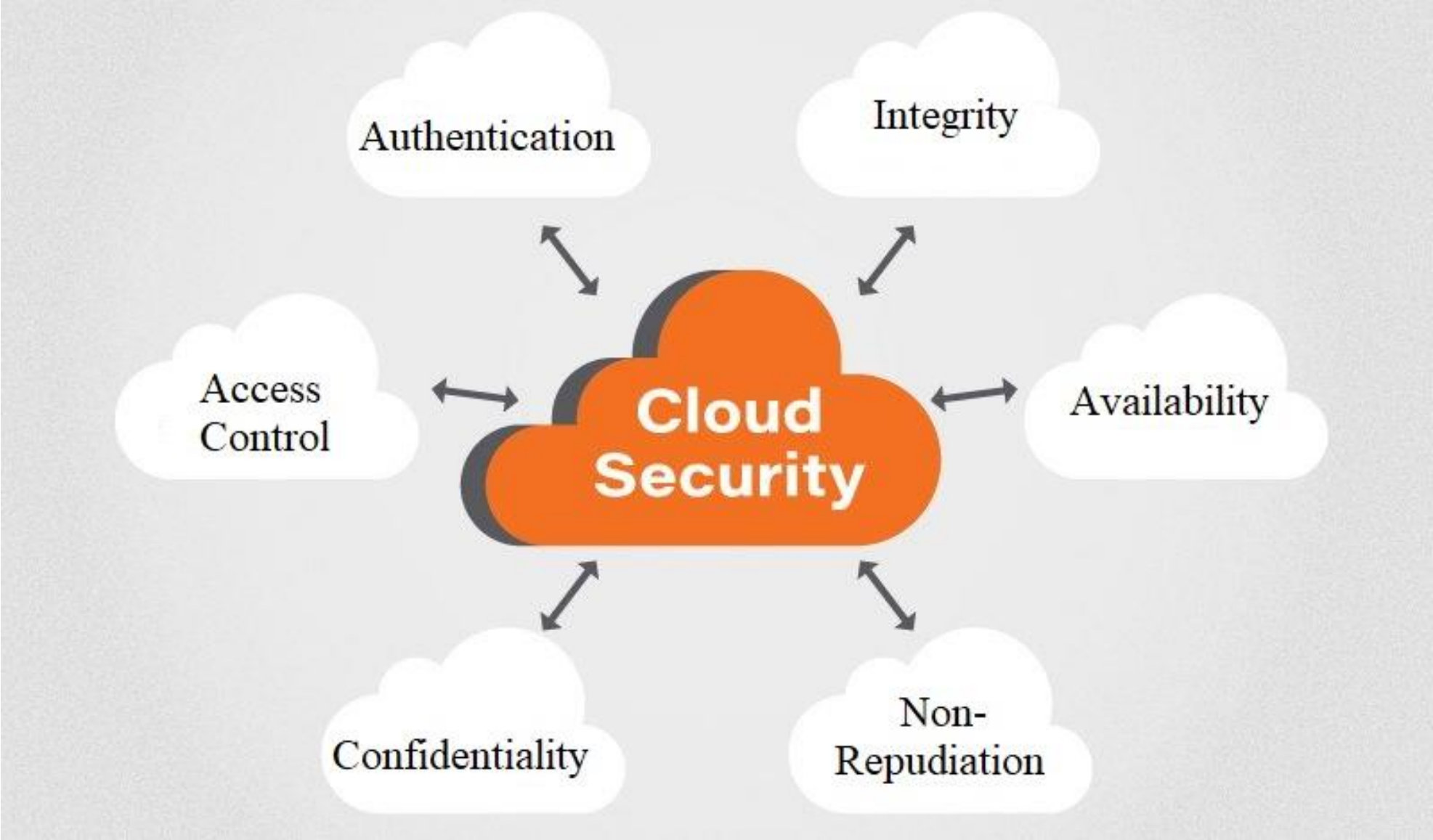

Figure 1.5 Cloud Security Requirements

**Access control**: Authentication and identification of the entity must be carried out to give access rights to the entity. It is the process of restricting access to systems and applications according to the level of security requirements.

**Confidentiality**: The attacker cannot look at the frequency, length, and other attributes of traffic flowing through the network. Unauthorized exposure of information must be protected to maintain the confidentiality of sensitive cloud data.

**Integrity**: The received data must be free from duplication, modification, and reordering. Only authorized users can make changes to it. This service assures the correctness and validity of data being transmitted through the network.

**Availability**: To maintain offsite backup regularly, Denial of Service attacks must prevent from the systems. This service assures that information is available to authorized users whenever required.

**Non-Repudiation**: This service proves that the authorized sender and receiver have sent and received the information, respectively. For this, accurate and traceable records must be maintained.

However, many more challenges remain in cloud computing security; such major challenges are access and management, trust, business sustainability, and the development of scalable service.

## 1.3 Cloud Storage

Cloud computing offers ubiquitous and on-demand access to flexible computation and storage resources as a new computing paradigm. With the rapid development of cloud computing, cloud storage has enabled the provision of high data availability, easy access to data, and reduced infrastructure costs from outsourcing data to remote servers [11]. Cloud computing and storage solutions provide users and enterprises with various capabilities to store and process their data. Nowadays, cloud computing enables an economic paradigm in which more and more sensitive information is being centralized for data service outsourcing, such as emails, personal health records, government documents, confidential business documents, etc. which is shown in figure 1.6[5].

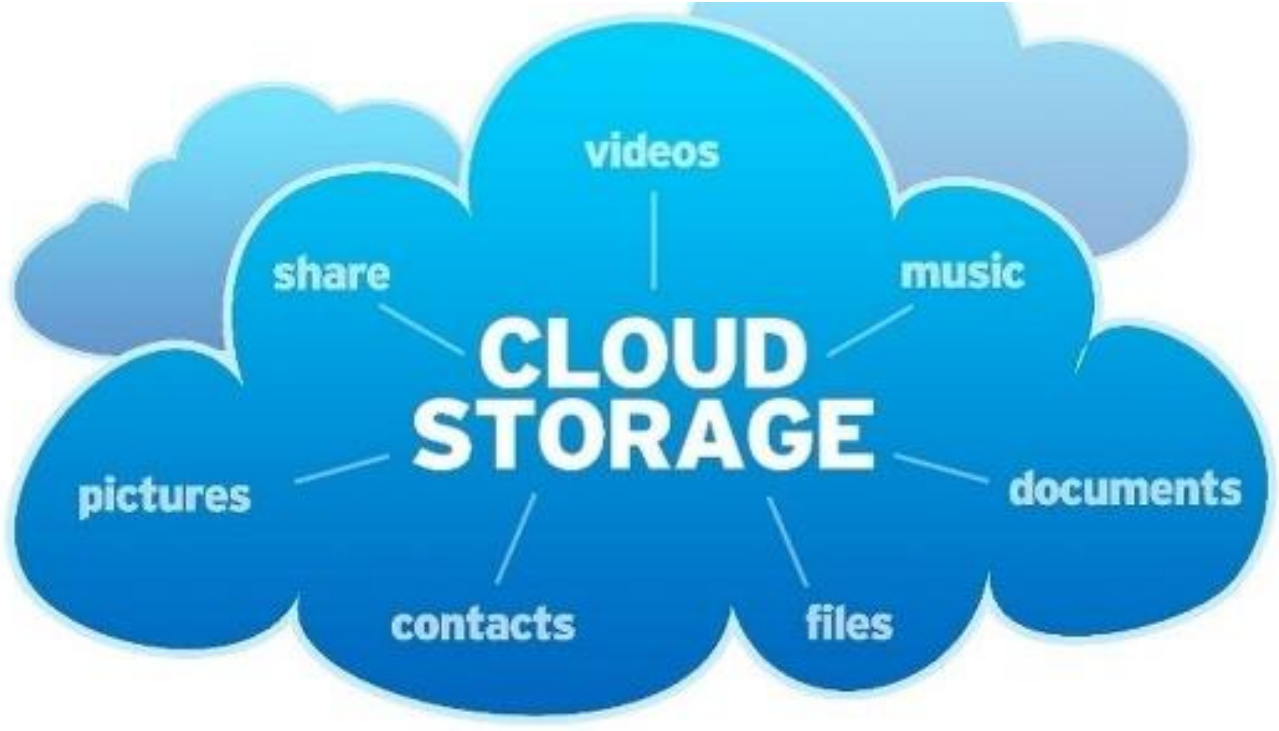

Figure 1.6 Cloud Storage

[5] Baer, T. (2019, January 17). *Finding the data buried in cloud storage*. ZDNet. https://www.zdnet.com/article/finding-the-data-buried-in-cloud-storage/

A cloud storage services provide transparency to the customers, which means the customers no need to know how the cloud service provider manages, implements, or operates their data within the cloud. It provides easy collaboration, remote access, scalability, universal access, economical storage, remote storage, and data management and maintenance.

A Cloud storage system is a service model in which data are maintained, managed, and backed up remotely on the cloud side, and meanwhile, data keeps available to the users over a network. Cloud storage provides benefits including easy access to data, less usage of physical storage devices, and reducing the infrastructure to maintain the data.) David Simms conducted a survey, and he found that almost 95% of people rely on the cloud for their storage [12]. The biggest advantage offered by the clouds is data outsourcing.

Nowadays, more and more companies and individuals from many big data applications have outsourced their data and deployed their services into cloud servers for easy data management, efficient data mining, and query processing tasks). For business users, rather than building their own data center, the company can leverage cloud storage service to store their data on the cloud storage server. Many users prefer cloud storage services to relieve the burden of maintenance costs and the overhead of storing data locally. cloud storage provides users with on-demand storage services, such as Dropbox, Amazon Simple Storage Services (S3), and Google Drive.) Moreover, cloud users can access their data anywhere and on any device with an Internet connection.

Although data outsourcing reduces computation and storage costs for users, the privacy of sensitive data is still a challenging issue (sensitive data include personal health records, confidential documents, private photographs, and any document which is personal to the user). Individuals and enterprises may be reluctant to outsource their sensitive data to an untrusted third-party cloud service provider with such suspicion. This information needs protection from different kinds of cyber threats. Users are using cheaper data storage and computation offered in the cloud environment. They also face many problems with reliability, privacy-preserving, and optimized searching of their outsourced data [14]. Outsourcing sensitive data into the cloud leads to privacy issues. This issue is because once the data are outsourced to the cloud server, the user loses physical control over the data. Most cloud servers are honest but curious, i.e., we can trust cloud service providers for their services. However, they may be interested in accessing our data.

Hence, it is now necessary to protect the privacy of sensitive data in the cloud. The most common solution to guarantee user privacy is the encryption of user data before outsourcing into the cloud. Data owners must encrypt his/her documents locally before uploading them to the public cloud storage because the documents almost contain sensitive information and must be kept secured from all except the authorized users. Also, searching on encrypted data is a very important issue in cloud storage. The method of searching the encrypted data represents a big challenge in cloud storage.

## 1.4 Searchable Encryption

Consider a search function on plaintexts. A user sends query keywords to the server in order to retrieve corresponding documents. After searching, the server will return the search results to the user. However, during the search process, both the knowledge of the contents stored on the server and the query keywords are exposed to the semi-trusted server. Because of this, the problem of encrypted search is now of interest to many sub-fields in computer science (e.g., databases, security, cryptography, privacy. Therefore, the main challenge in cloud computing is effective searching. Previously searching was suitable for only unencrypted data, which does not offer security over the cloud; hence encryption of data came into existence.

Searching capability is one of the most valuable and essential applications that many users want to perform on the data stored on a remote server. More specifically, the user should be able to search for a particular query and retrieve the relevant documents corresponding to the query from the cloud server. Suppose the cloud server that hosts the private data from multiple users is untrusted. In that case, the data will be completely exposed to attackers [14]. To protect data privacy and combat unsolicited access in the cloud and beyond, sensitive data, for example, e-mails, personal health records, photo albums, tax documents, financial transactions, and so on. A general approach to protecting data confidentiality is to encrypt data owners before outsourcing to the commercial, public cloud. Once the data have been encrypted, the server cannot reveal the data. However, this will cause a huge cost in terms of data usability. For example, the existing techniques on keyword-based information retrieval, which are widely used on plaintext data, cannot be directly applied to encrypted data. Therefore, it is necessary to search the encrypted data without breaking the confidentiality of the data.

With files encrypted on a remote server, retrieving files based on their content is difficult. However, when the user wants to search for related files containing a particular keyword, how to process and search the encrypted data becomes an intractable problem. In the past, there were two methods to solve it. A trivial solution is to download all encrypted data to the local and then decryption query, leading to a massive waste of bandwidth and computation resources. Also, this approach is inefficient as it incurs high communication and computation costs at the user's end.

For example, consider a user Alice who stores her work documents on an untrusted file system. Suppose Alice wishes to retrieve all documents containing the word “aardvark.” Since the document files are encrypted, and the server cannot be trusted with the document keys or their contents, Alice has to download all the document files, decrypt them, and then search the decrypted documents on her local machine; This naive solution is inefficient. This method needs to download many unneeded files, which wastes network overhead and requires much computational cost for decryption. This way is not feasible in practice. Downloading all the data from the cloud and decrypting it locally is impractical.

Another extreme method is the user sends the secret key to the cloud server to decrypt the query. However, the cloud server is not fully trusted.

Thus, achieving efficient data retrieval while ensuring data security becomes a challenging issue. The ideal solution lets the server search the encrypted documents and return only relevant ones while ensuring that it learns nothing about the keyword or document contents [15].

Searchable encryption is a cryptographic primitive developed for performing the keyword search over encrypted data. Searchable encryption allows a cloud server to conduct keyword searches over encrypted data on behalf of the data users without learning the underlying plaintexts [16]. With the searchable encryption technique, users can store and share their data in an encrypted format to improve data security, and other users can search in this secure encrypted data.

Searchable encryption is a multidisciplinary field (Figure 1.7[6]) based on Information Retrieval for searchable index generation, Algorithms for devising an efficient search scheme, and cryptography to maintain security and privacy requirements.

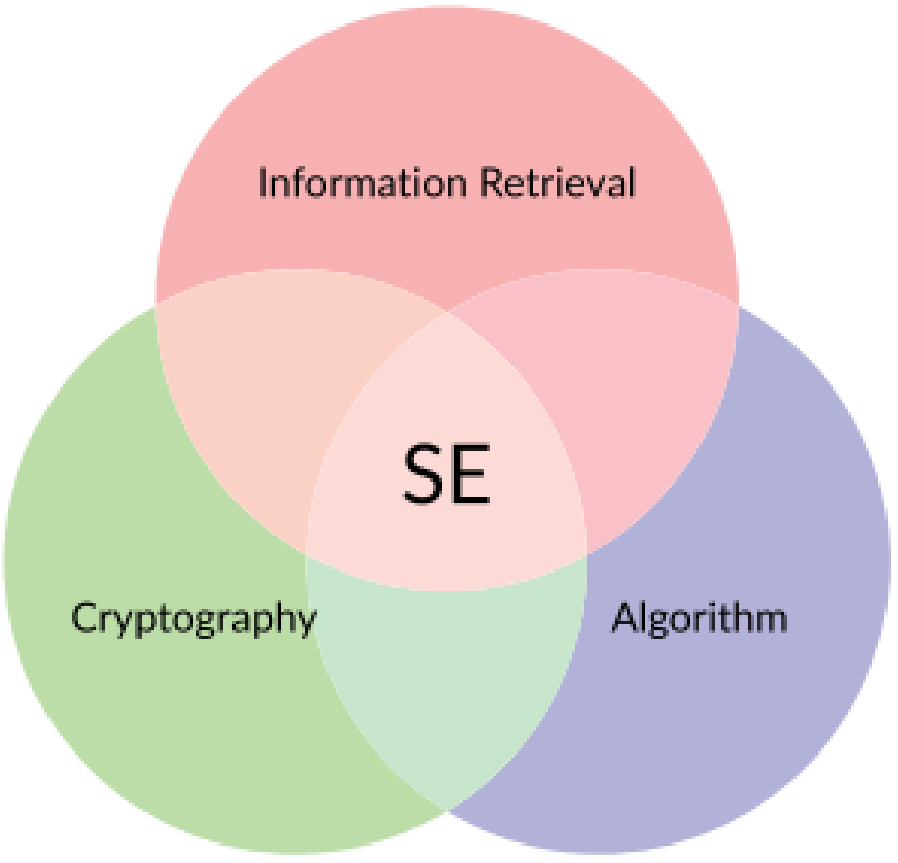

Figure 1.7 Searchable encryption: a multidisciplinary field

The first searchable encryption's basic idea is to encrypt each word in a text file individually. Then, as can easily be known, the search cost of the given basic searchable encryption would be very high. The subsequent research efforts are put into developing an index that can support more efficient keyword searches. The data owner encrypts the index and document collection in SE schemes before outsourcing it to the cloud server. The index contains the set of keywords from all the documents. The user generates an encrypted text (trapdoor) and sends it to the cloud server to perform a search. The cloud server returns the documents related to the generated trapdoor [17].

Many researchers have proposed a series of efficient search schemes over encrypted cloud data [61]. Different kinds of Searchable Mechanisms have been proposed based on different aspects till now. In chapter 2, the most popular searchable encryption proposed schemas are discussed for better understanding. The evolution of SE schemes highlighting various milestones achieved in this field in the previous years is presented in Table 1.

[6] Handa, R., Krishna, C.R. and Aggarwal, N., 2019. Searchable encryption: a survey on privacy-preserving search schemes on encrypted outsourced data. *Concurrency and Computation: Practice and Experience*, *31*(17), p.e5201.

**Searchable Encryption Timeline**

| | |
|---|---|
| 2000 | Symmetric Searchable Encryption |
| 2003 | Secure Indexes |
| 2004 | Public-key Encryption |
| 2008 | Multi-keyword Conjunctive Search |
| 2010 | Ranked Single-keyword search \| Fuzzy Search |
| 2011 | Ranked Multi-keyword search |
| 2013 | Synonym-based Searching |
| 2014 | Hardware-based SE solution |
| 2015 | Dynamic Multi-Keyword Ranked Search |
| 2016 | Fine-grained Search Using Attribute-Based Encryption |
| 2017 | Privacy-Preserving Multi-Keyword Top Similarity Search |
| 2018 | Semantic Aware Search |
| 2019 | Novel Multi-Keyword Search |
| 2020 | Practical Multi-keyword Ranked Search with Access Control |
| 2021 | Verifiable Boolean Search |

Table 1.1 Searchable Encryption Timeline

A searchable encryption service contains three entities: a data owner, a data user (data users), and the untrusted cloud. The data owner is like a user who outsourced the original data to a third-party cloud by renting the storage service from the cloud provider with both storage space and SE-based access service. Considering the privacy of the outsourced data, the data owner encrypts the data before uploading them to the cloud. The data user is the cloud service consumer, says Alice, who has been authorized to search and read the outsourced data in an encrypted format. She issues a keyword query with a generated search token to the cloud. The SE service of the cloud searches over the encrypted data with the search token and returns the matching result to the data user. Figure 1.8[7] illustrates the system model for searchable encryption service.

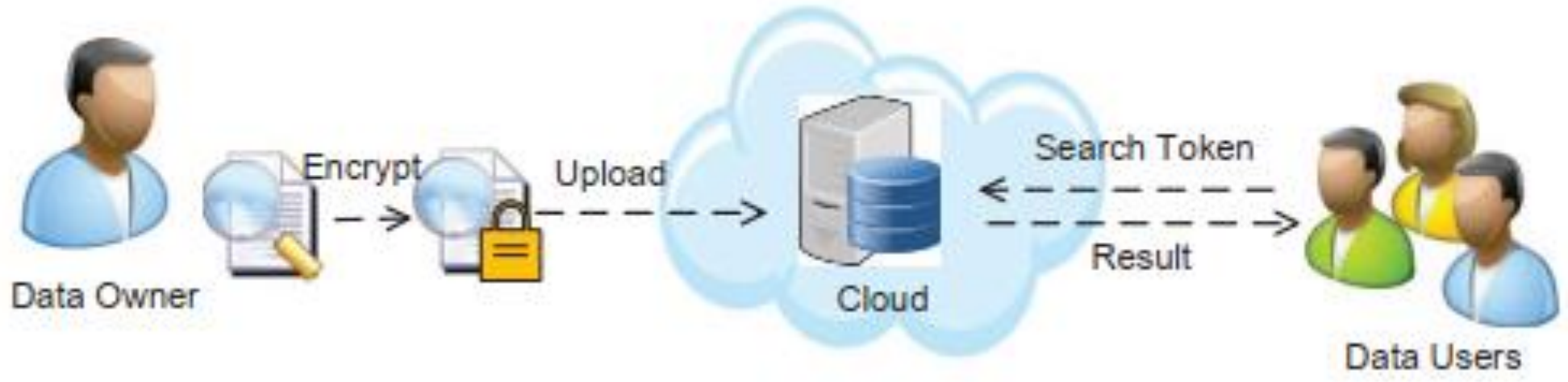

Figure 1.8    System Model of Searchable Encryption

Many existing SE schemes consider the cloud server to be honest but curious. The cloud server will faithfully perform the protocols but is sufficiently curious about the outsourced data and the queries sent by data users and attempt to deduce sensitive information from such information [18]. The honest but curious cloud server assumption is not reasonable in practice. In addition, the cloud server may only perform a fraction of the search operations and return forged results to save on computation and storage costs. Thus, practical SE schemes need to support verification of the returned results. The cloud server should be able to prove that the returned results are complete, correct, and sound, as explained below.

- **Completeness:** The cloud server returns a result set $R = \{R_1....R_n\}$for a given query *q*. For each document $D_i \in$ DB (*q*), if the encrypted form $R_i$ is in the set *R*, we say the result set is complete.
- **Correctness:** The cloud server returns a result set $R = \{R_1....R_n\}$ for a given query *q*. If the decrypted document Di has not been tampered with for each Ri R, we say that the document $R_i$ *(or* $D_i$*)* or the result set *R* is correct.
- **Soundness:** The cloud server returns a result set $R = \{R_1....R_n\}$for a given query *q*. For each $R_i \in R$, if the decrypted document $D_i$ satisfies the query *q*, we say the document $R_i$ *(or* $D_i$*)* or the result set R is sound.

## 1.5 Existing Models and Limitations

Privacy-preserving is one of the major hurdles in the cloud for the user, especially when the user data that reside in local storage is outsourced and shared over the cloud server. Most cloud service providers keep multiple copies of files to ensure reliability to their users. Nevertheless, this turns into a serious privacy issue when the replicated copies remain hidden in the server even after the owner has deleted the original file. So, Patwary and Palit (2019) [19] designed a system using the existing cryptographic approaches, ensuring the deletion of files by securing the file storing and sharing system for cloud infrastructure. The primary focus of their proposed model is to ensure user privacy and security through a user-friendly system that is less computationally intensive and does not rely upon any trusted entity. Deng et al. (2020) [20] introduced an Identity-Based Encryption Transformation for Flexible Sharing of Encrypted Data in Public Cloud, which permits the complicated certificate management in usual secure distributed systems. In both of the designed models, the searching on the securing file was missing, which is the basic criteria in the cloud storage to find the specific document.

[7] Renwick, S.L. and Martin, K.M., 2017. Practical architectures for deployment of searchable encryption in a cloud environment. *Cryptography*, *1*(3), p.19.

Song et al. (2000) [21] first proposed a searchable symmetric encryption scheme. The search cost of this scheme increases with the linear time of the ciphertext length, and the efficiency is low. After that, searchable encryption (SE) technology has been studied by many researchers. Li et al. (2017) [22] proposed achieving Secure and Efficient Dynamic Searchable Symmetric Encryption over medical cloud data which can achieve two important security features, i.e., forward privacy and backward privacy. However, their scheme is lacking in terms of storage, search, and updating complexity.

Chaudhari and Das (2019) [23] designed a searchable encryption scheme where the data owner predefines which attributes data users have will be given search permission with fine-grained access control. In this scheme, CSP first verifies the user's access right and then decides whether to return the search results to the user. Thus, the fine-grained search control can further protect the security of outsourced files. Unfortunately, their scheme only can search one keyword at one time, and the biggest disadvantage of such a scheme is that if a data user needs to search a file containing multiple keywords at the same time, he/she has to launch multiple searches for each keyword, which inevitably results in a lot of communication and computation cost.

Liu et al. (2021) [24] proposed an identity-based searchable encryption framework in cloud systems. They have introduced the Key Generator Center (KGC) and Identity-Certifying Authority (ICA) entities, who have the control of verifying certificates and partial secret keys to the data user and data owner. But this process incurs additional overhead. Specifically, the KGC can gain access to any user's secret key, and key escrow problems occur when the KGC is malicious. In addition, the users in these schemes require an additional generated certificate or a public–secret value pair to encrypt keywords or generate trapdoors. Consequently, users require more storage space to store certificates and keys. It becomes less convenient for the users in this scheme because the users have to use additional information instead of merely using their identities to encrypt keywords and generate trapdoors. Shen et al. (2016) [25] proposed Keyword Search with Access Control (KSAC) on encrypted data where it provides the search capability deviation but less efficient access policy update with compromising data privacy.

All the above three schemes are based on Identity-based Encryption (IBE) technology to realize a fine-grained search or file access control, i.e., only the attributes of data users match the access control policy of the data owner, data users can obtain the search results.

At present, most SE schemes adopt the inverted index, which is a matrix of files and keywords to represent their relevance. When a data user uploads a trapdoor, CSP matches it with each index, i.e., the cloud server needs to perform *n* search operations if the cloud has *n* outsourced files, which requires high computation cost. The first effort to build a secure inverted index comes from Curtmola et al. (2015) [26]. To extend the work, wang et al. (2015) [27] proposed an inverted index with strong privacy where the scheme preserves the high search efficiency inherited from the inverted index while lifting the one-time-only search limitation. The scheme lacks a probabilistic trapdoor generation algorithm, making it vulnerable to the search pattern. Buecenna et al. (2019) [28] proposed a secured inverted index with access control management using two key techniques, homomorphic encryption, and the dummy documents technique. Though their scheme can be compressed to the table of encrypted scores and the double score formula, the scheme can't perform much better results due to dual inverted index and large memory overhead.

There are two major limitations of the existing inverted index-based searchable encryption schemes.

- First of all, the keyword privacy is compromised once a keyword is searched. As a result, the index must be rebuilt for the keyword once it has been searched. Such a solution is counterproductive due to the high overhead suffered.
- Secondly, the existing inverted index-based searchable schemes do not support conjunctive multi-keyword search, which is the most common form of queries nowadays.

The inverted index can retrieve all files containing some keyword in one search. Unfortunately, all of the above schemes are only suitable for the single Data Owner scenario. When there are multiple Data Owners, the scheme needs to create an inverted index structure with different keys for each Data Owner, which is not conducive to multiple Data User searches.

Deriving from single keyword to multi-keyword search, Golle et al. (2004) [29] first designed a multi-keyword searchable encryption scheme, which achieves a breakthrough from single keyword to multiple keywords. Subsequently, Xia et al. (2016) [30] proposed an MSE scheme using the *k*-nearest neighbor (*k*-NN) technique; this scheme can not only satisfy multiple keywords search but also sort the search results by the elegance scores between file index and trapdoor. However, due to the indexes and trapdoors being generated using the matrix multiplication, the time cost of search increases linearly with the increase of matrix dimension. The above two schemes cause a huge challenge to data users and data owners' storage, communication, and computing ability, especially for mobile cloud users who have limited computing capability.

A major challenge exposed from the existing schemes is the difficulty of protecting users' query privacy. The challenge roots in two facts: 1) the existing solutions use a deterministic trapdoor generation function for queries; and 2) once a keyword is searched, the encrypted inverted list for this keyword is revealed to the cloud server. The existing solutions to the problem where we have also observed a challenging tradeoff between memory overhead, search capability, and security when storing data in encrypted form.

Nevertheless, all of the above SE schemes do not mention how to deal with the problem of keyword updates. In fact, few SE schemes now consider keyword updates. If a keyword goes wrong or does not match a file, the file will be deleted from the cloud. The whole file outsourcing needs to be redone again, which obviously is inefficient.

Also, most of the existing schemes have built the cloud service provider by the algorithms, where the encrypted document and index are outsourced to the cloud server. The cloud server firstly searches on the index and then returns the encrypted documents to the data owner or data user. But in a real-life scenario, the most commonly used cloud service provider such as Google Drive, Amazon Cloud, OneDrive, and DropBox has no index-based searchable mechanism so far. So, most of the existing proposed schemes are theoretical but need to be implemented based on real and daily based usage.

## 1.6 Our Proposed Model Contribution

As all these models for encrypted search over the cloud have their own potential and drawbacks, we have designed a secure system architecture that resolves most of the limitations of previously Proposed models keep the potential of their predecessors intact.

We have proposed a practical solution of searchable encryption with strong theoretical security guarantees that require only small amounts of overhead in terms of bandwidth and storage. The scheme supports keyword search, which enables the server to return only the encrypted data that satisfies an encrypted query without decrypting it. Our solution utilizes the notion of a keyword index, which the Data Owner creates. The keyword index associates each keyword with its associated files. All keyword searches by the Data User are based on this index; It is worth noting that in this framework, the Data User can have complete control over what words are keywords and which keywords are associated with which files. This power can be useful for many applications.

Our architecture makes use of an encrypted index which is generated prior to searching. We have developed a new inverted index-based scheme to obtain accurate search results quickly on the user side and protect sensitive data from Cloud Service providers and unauthorized entities.

In this work, we not only provide a complexity analysis of our algorithm but also perform a simulation study using a real-world dataset to evaluate the performance. We have chosen Google Drive cloud storage as a cloud server where the encrypted document and indexes are stored in our implementation. We have implemented a web application of our proposed scheme using React JS and MongoDB. For encrypting documents, we chose AES because it satisfies all of the security requirements that and it offers a variety of different modes for encryption and pseudorandom number generation. In order to share files, our model also relies upon the RSA algorithm. But our proposed model is designed in such a way that; RSA key generation is done only once for each user, and RSA decryption is also done only once per decryption throughout the lifecycle of a file which reduces the load from the CPU of the user device and increases the performance of our model significantly. In our implementation, we also have one secret key used to generate an inverted index.

Though our model relies on an additional server, privacy is not in the hands of the server administrator. It is because the server does not contain any information that can be used by anyone else other than the user. Our contributions are summarized as follows.

**Multi Data Owners:** Our system supports encrypted data search under scenarios in which multiple data owners encrypt all data records. By deploying an improved multi data owners identity-based encryption scheme, all data owners can distribute their search capability to the data user under different data owners without additional negotiations.

**Multi Data Users:** Due to the use of identity-based encryption, this work also satisfies multi-client requirements. Because all search capabilities are encrypted before being sent to the data user, only the allowed data user with the corresponding identity can obtain a valid searchable keyword. In fact, the search is controlled by providing different keyword capabilities for authorized data users.

**Search Efficiency:** We designed the search scheme to allow both single and multi-keyword queries and provide result similarity for effective data retrieval instead of returning undifferentiated results. Also, the procedure is simple and fast (More specifically, for a document of length n, the encryption and search algorithms only need O(n) stream cipher and block cipher operations); and is introduced almost no space and communication overhead.

**Strong Privacy:** Our system provides strong privacy guarantees. In contrast, the existing schemes are searched on the index to the cloud server so that there is a possibility to leak and reveal the keyword to the server. But in our proposed model, the search procedure of a keyword to the inverted index has happened locally through a local server. So, short of breaking the encryption, the cloud server has no way to learn anything about the content of keyword queries or the stored documents, except possibly the lengths of the documents.

**Inverted Index Structure:** A new model is developed to search for secure encryption over the index without learning about queried keywords from unauthorized entities. We take advantage of the inverted index structure to build a secure index that achieves greater flexibility and efficiency. The inverted index is one of the most popular and efficient index structures for encrypted search. Our scheme is built based on an inverted index structure, which is dynamic and easy to update.

**Non-interactive:** Our system also provides an efficient approach to enable noninteractive authorized search. Once the authority determines the set of authorized keywords for the data user, it only performs a one-time calculation to generate the encrypted keyword for the search for the data user. Moreover, the size of the encrypted keyword in our scheme is constant, regardless of the number of authorized keywords. Our system also allows for dynamic user enrollment since a user joining does not affect other users' settings.

To improve scalability, we conducted rigorous security and cost analysis. The implementation experiment showed that our scheme is suitable for practical usage and is more efficient than existing searchable encryption schemes. Therefore, the focus of this report is to explore the possibility of designing a scheme that simultaneously supports all these functions securely and efficiently.

## 1.7 Organization of the Report

The rest of the paper is organized as follows. We have an in-depth discussion regarding how all the operations in the searchable encryption scheme work from scratch to advance modules in Chapter 2. Chapter 3 discusses literature reviews on related works and previous research in encrypted searching. In Chapter 4, we discuss how our proposed model works, why it is more secure, and how it prevails over the shortcomings of its predecessors. We discuss the implementation details of our proposed model in Chapter 5. We conclude in Chapter 6, discuss the limitations of our scheme, and focus on future works.

# 2. CHAPTER TWO: BACKGROUND

## 2.1 Cryptography

Cryptography uses advanced mathematical principles to store and transmit data in a particular form so that only those for whom it is intended can read and process it [31]. Encryption is a key concept in cryptography – It is a process whereby a message is encoded in a format that cannot be read or understood by an eavesdropper. The technique is old and was first used by Caesar to encrypt his messages using the Caesar cipher. A plain text from a user can be encrypted to a ciphertext, then sent through a communication channel, and no eavesdropper can interfere with the plain text. The ciphertext is decrypted to the original plain text when it reaches the receiver end.

### Cryptography Terms

- **Encryption**: It is the process of locking up information using cryptography. Information that has been locked this way is encrypted.
- **Decryption**: The process of unlocking the encrypted information using cryptographic techniques.
- **Key**: A secret like a password used to encrypt and decrypt information. There are a few different types of keys used in cryptography.
- **Steganography** is the science of hiding information from people who would snoop on someone. The difference between steganography and encryption is that the would-be snoopers may not be able to tell there is any confidential information in the first place.

### 2.1.1 Symmetric Key Cryptography

This is the most straightforward kind of encryption that involves only one secret key to encrypt and decrypt information. Symmetric encryption is an old and best-known technique is shown in figure 2.1[8].

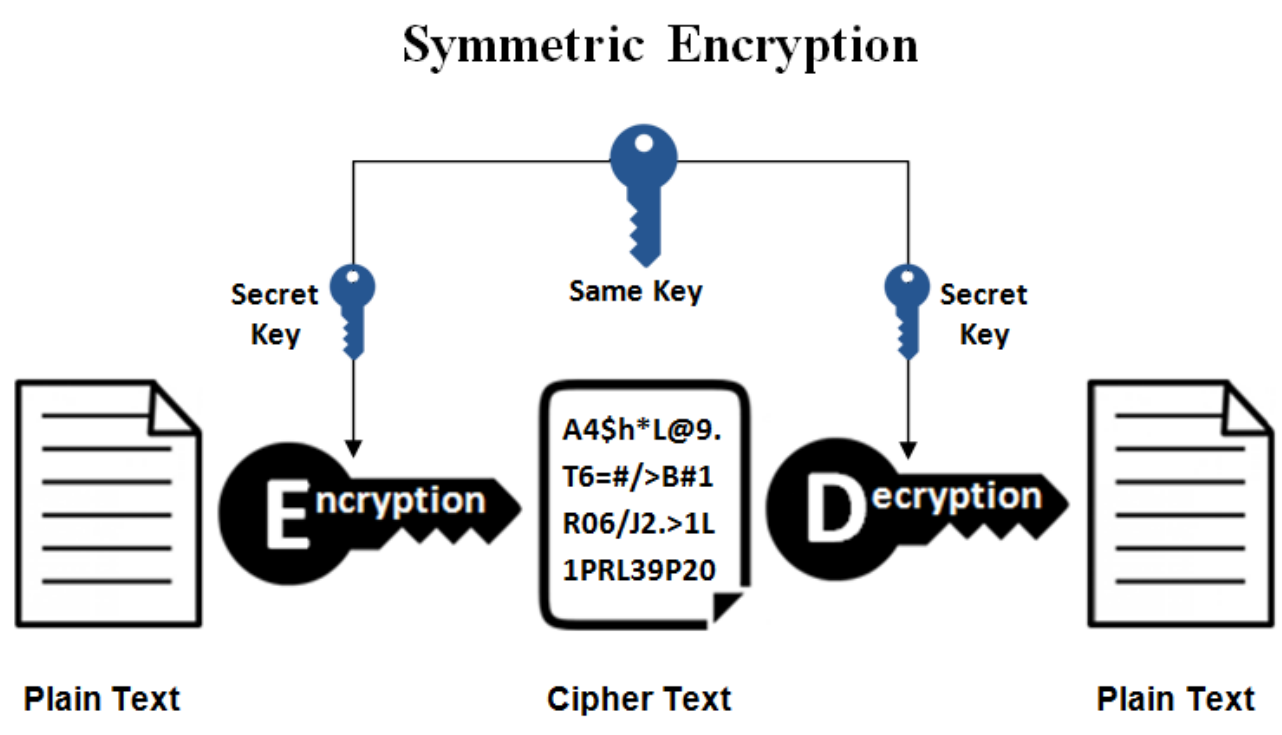

Figure 2.1 Symmetric Key Encryption

[8] 'Symmetric vs. Asymmetric Encryption – What Are Differences?' SSL2BUY, https://www.ssl2buy.com/wiki/symmetric-vs-asymmetric-encryption-what-are-differences. Accessed 23 May 2022

It uses a secret key that can either be a number, a word, or a string of random letters. It is blended with the plain text of a message to change the content in a particular way. The sender and the recipient should know the secret key that is used to encrypt and decrypt all the messages. Blowfish, AES, RC4, DES, RC5, and RC6 are examples of symmetric encryption. The most widely used symmetric algorithm is AES-128, AES-192, and AES-256.

The main disadvantage of symmetric key encryption is that all parties involved have to exchange the key used to encrypt the data before they can decrypt it.

### 2.1.2 Asymmetric Key Cryptography

Asymmetric encryption is also known as public-key cryptography, a relatively new method compared to symmetric encryption. Asymmetric encryption uses two keys to encrypt plain text. Secret keys are exchanged over the Internet or an extensive network is shown in figure 2.2[9].

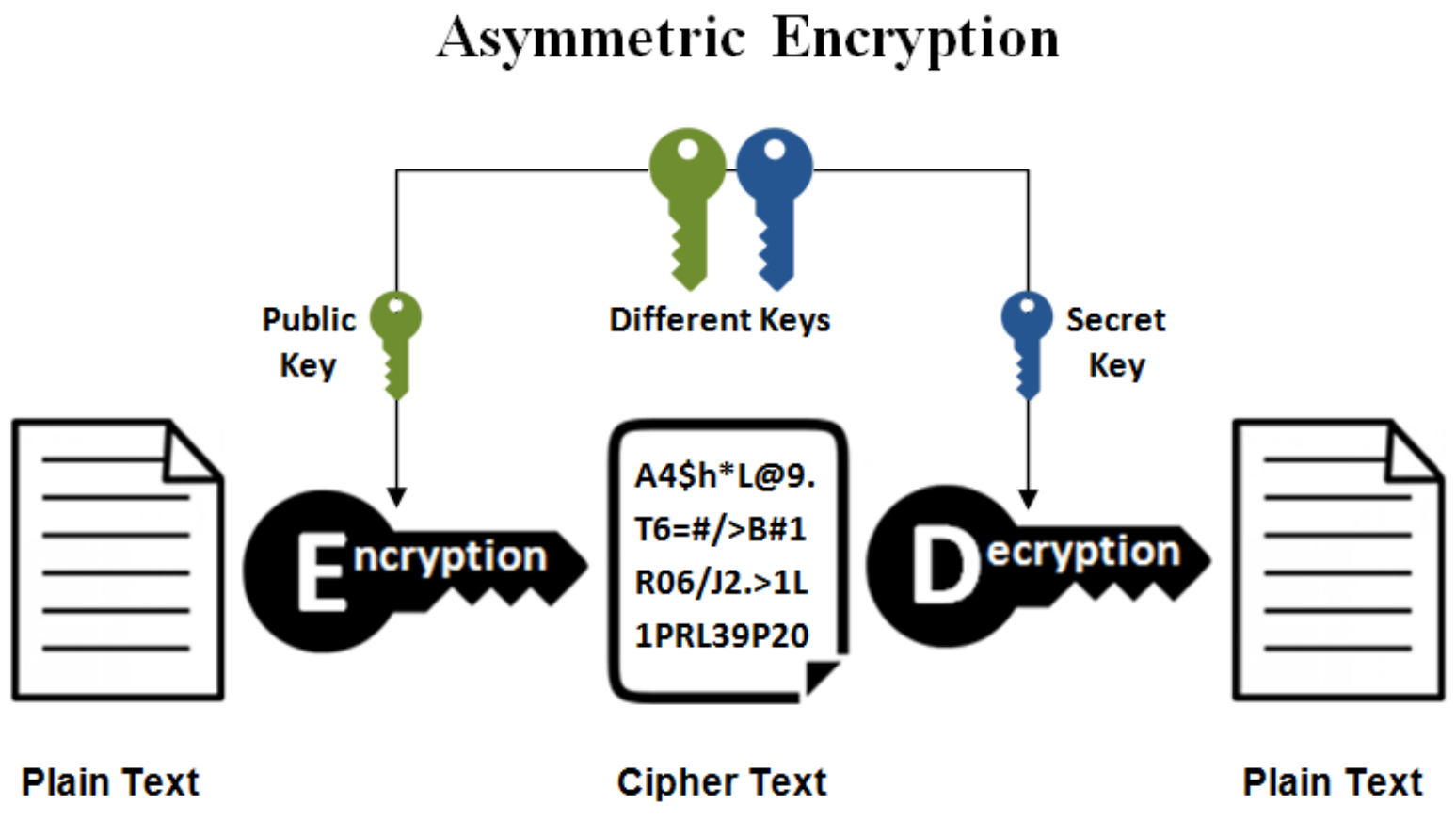

Figure 2.2 Asymmetric Key Encryption

It ensures that malicious persons do not misuse the keys. It is important to note that anyone with a secret key can decrypt the message, and this is why asymmetric encryption uses two related keys to boost security. A public key is made freely available to anyone who might want to send a message. The second private key is kept a secret so that the data user can only know.

An encrypted message using a public key can only be decrypted using a private key. At the same time, also, a message encrypted using a private key can be decrypted using a public key. Security of the public key is not required because it is publicly available and can be passed over the internet. Asymmetric key has far better power in ensuring the security of information transmitted during communication.

Asymmetric encryption is mainly used in communication channels, especially over the Internet. Popular asymmetric key encryption algorithm includes ElGamal, RSA, DSA, Elliptic curve techniques, and PKCS.

[9] 'Symmetric vs. Asymmetric Encryption – What Are Differences?' SSL2BUY, https://www.ssl2buy.com/wiki/symmetric-vs-asymmetric-encryption-what-are-differences. Accessed 23 May 2022

## 2.2 Data Encryption

Privacy-preserving is one of the major hurdles in the cloud for the user, especially when the user data that reside in local storage is outsourced and computed onto the cloud [32]. Firewalls could secure the sensitive data that a cloud service provider is holding, intrusion detection systems, and CSP has full control over the infrastructure of the cloud, including the lower level of system stack and system hardware. Although mitigate concerns are still taken, privacy breaches are likely to occur in the paradigm. In a few cases, the service provider is not fully trusted, but still, we need the service. Therefore, some methods should be empowered to protect the user data and user queries from an unauthorized persons in the cloud environment. Thus, data must be encrypted before sending data onto the cloud to protect it from data privacy and unsolicited access which is shown in figure 2.3[10].

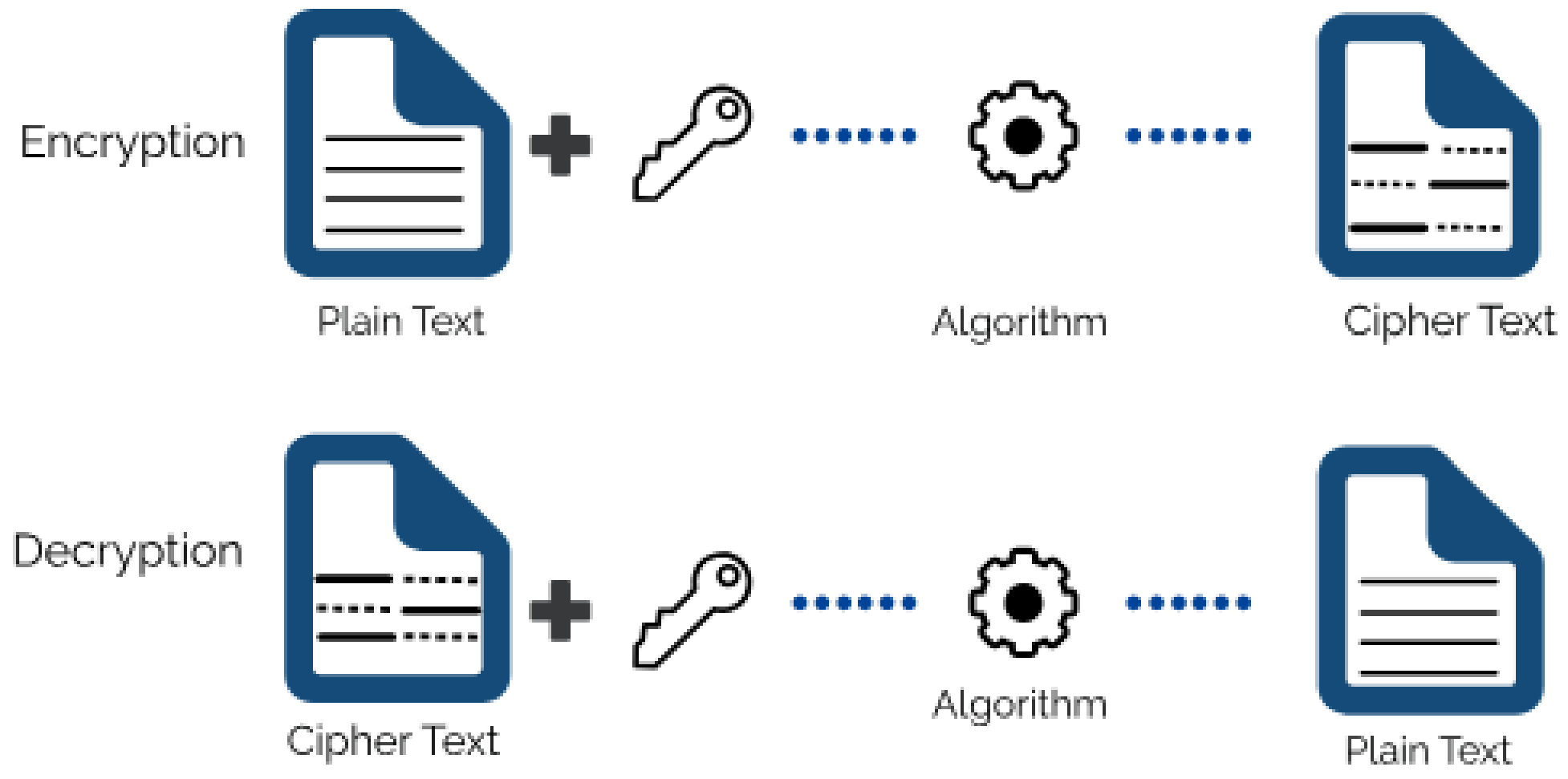

Figure 2.3 Data Encryption Process

The most straightforward method of addressing data privacy concerns is to encrypt data before uploading it to the cloud [33]. Data encryption is referred to as the most effective method for data protection. It refers to mathematical calculation and algorithmic schemes that transform plaintext into ciphertext, a non-readable form to unauthorized parties. Data needs to be encrypted before outsourcing to the cloud containing sensitive information like email, health records, financial transactions, government documents, etc. [34]. By encrypting the data, customers are assured that data confidentiality is preserved irrespective of the CSPs' actions.

### 2.2.1 File Assured Deletion

The dependency on cloud storage has been growing immensely with the use of public cloud storage facilities. However, this has given rise to security concerns involved in data confidentiality even after the deletion of data. As multiple copies of the same data are kept in the storage server for reliability, some copies of the data remain in the cloud storage through a user deletes the data. To address this issue, a technique termed File Assured DEletion (FADE) was proposed by Tang et al.

[10] Davis, R., 1978. The data encryption standard in perspective. *IEEE Communications Society Magazine*, *16*(6), pp.5-9.

(2012) [35] that encrypts the data before storing it in cloud storage. It is another approach to data encryption to ensure the security concern of deleted files. It is a practical, implementable, and readily deployable cloud storage system that protects deleted data with policy-based file assured deletion. FADE is built upon standard cryptographic techniques. It encrypts outsourced data files to guarantee their privacy and integrity and, most importantly, deletes files to make them unrecoverable to anyone (including those who manage the cloud storage) upon revocations of file access policies.

This technique involves the use of a third-party key manager to store the keys, which involves a relatively complex system architecture. Moreover, the requirement of simultaneous multiple network connections makes it suitable for corporate clients. Simplified File Assured DEletion (SFADE), which is easy to implement and friendlier to ordinary users. It introduced a new model which eradicates the reliance on the key manager system, and at the same time, it ensures assured deletion. The new system has not only been suggested theoretically, but we have also implemented a prototype to validate the model. This proposed work will benefit the cloud storage users by helping them maintain their own data confidentiality.

A more simplified model SFADE was proposed by Habib et al. (2013) [36] which is shown in figure 2.4[11]. It does not require any third-party key manager. A randomly generated key is used for encrypting files. That key is then encrypted by a secret key generated using the user's passphrase. In 2017, an extended version of SFADE named SFADE+ was proposed by Nusrat et al. (2017) [37], supporting file sharing. It uses the RSA algorithm to share files securely. The private key is used for file encryption. It is generated based on the user's passphrase and is not stored anywhere. The public key is used for decryption, and it is stored in a repository along with the location of the files to be stored.

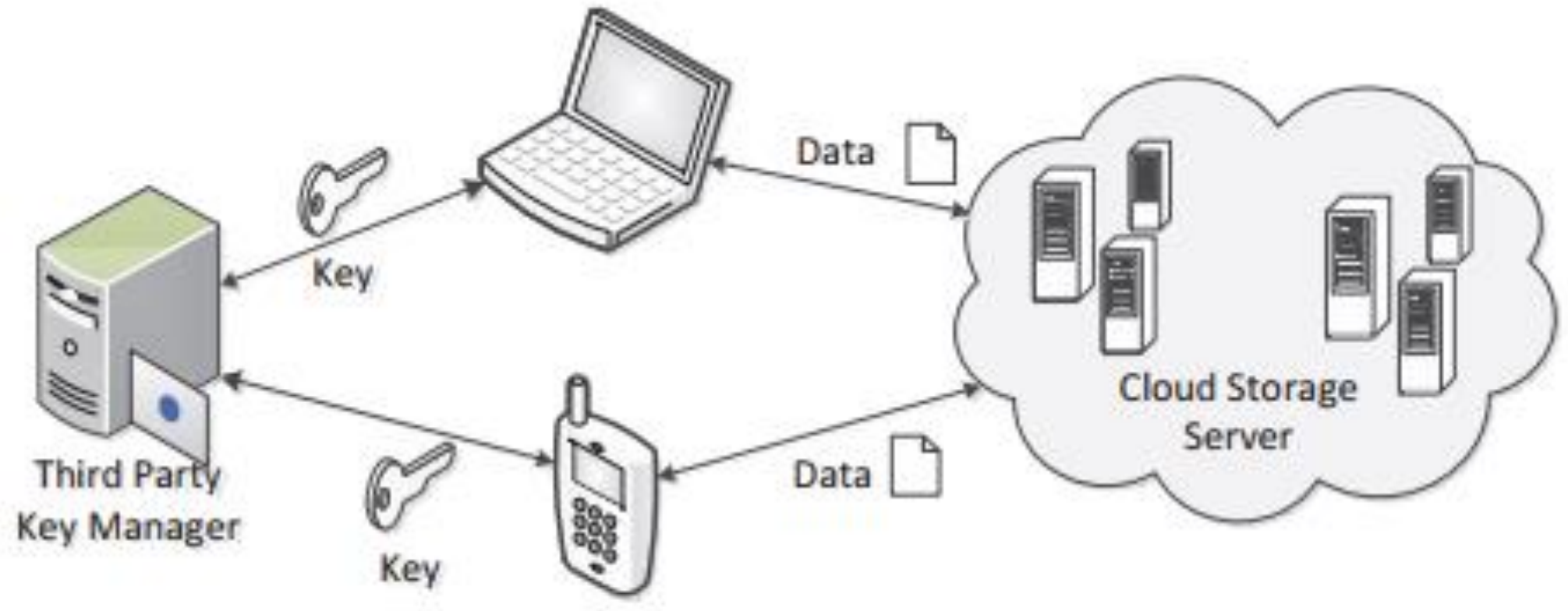

Figure 2.4 System Architecture of FADE

[11] Habib, A.B., Khanam, T. and Palit, R., 2013, August. Simplified file assured deletion (sfade)-a user friendly overlay approach for data security in cloud storage system. In *2013 International Conference on Advances in Computing, Communications and Informatics (ICACCI)* (pp. 1640-1644). IEEE.

### 2.2.2 Importance of Cloud Encryption

Cloud encryption can help organizations boost data privacy, enable remote work flexibility, and ensure regulatory compliance—issues that have come to the fore in the modern era. Let us look at some benefits of cloud encryption is shown in figure 2.5[12].

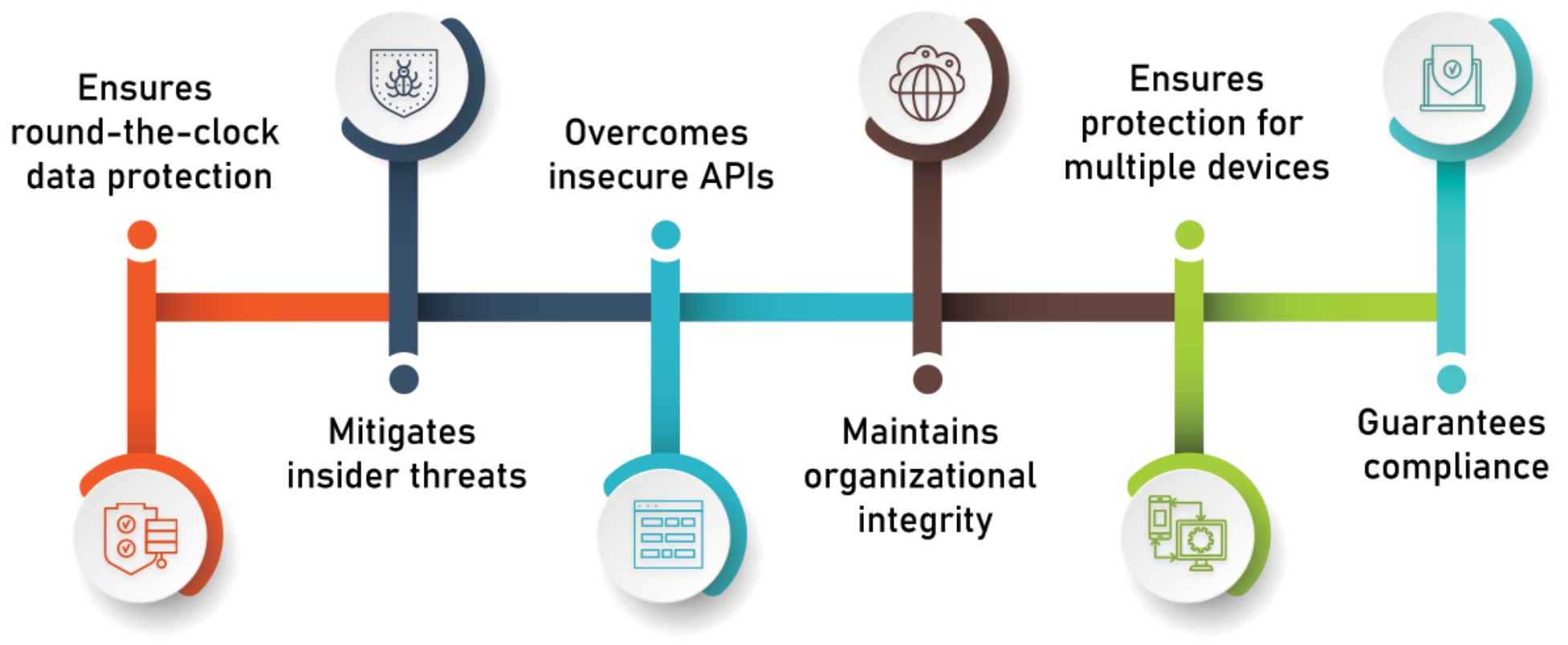

Figure 2.5 Importance of Cloud Encryption

**Ensures round-the-clock data protection**

Generally, enterprise data is exposed to the highest level of risk when undergoing a transfer or stored in a third-party environment, such as a cloud server. Cloud encryption ensures security for both data-at-rest and data-in-motion. As workflow structures become more flexible and employees stretch their shifts and shuffle their devices or locations, data must be subject to 24/7 protection. If not, there is a high chance that it will be accessed by unscrupulous elements looking to cause damage to the enterprise.

Cloud encryption solutions jump into action and protect data, whether stored or transferred. Regardless of the process that the data is going through, cloud encryption solutions prevent unauthorized access.

**Mitigates insider threat**

External elements are not the only risk factor for an organization's data security, especially during remote work when scrutiny is stretched thin and monitoring is not always efficient. Should they choose to do so, employees, business partners, and contractors with malicious intentions can wreak even more havoc than a cybercriminal that is not affiliated with the organization.

While cloud encryption is not a replacement for negligence or lack of training, it certainly helps shift control of the enterprise data to an experienced and trustworthy cloud service provider. This introduces a new layer of security and helps prevent employees from causing any damage to the company.

[12] Ashtari, H. (2021, September 1). *What Is Cloud Encryption? Definition, Importance, Methods, and Best Practices*. Spiceworks. https://www.spiceworks.com/tech/cloud/articles/what-is-cloud-encryption/

**Overcomes insecure APIs**

Organizations that operate in a cloud environment often rely on APIs to control various elements of their online infrastructure. APIs are built into mobile or web applications and can access cloud data externally (for clients and contractors) or internally (for employees). Whether external or internal, APIs with weak security protocols can introduce a cloud-based security risk, especially when data is being transferred. For instance, an insecure external API may serve as a gateway that offers unauthorized access to cybercriminals looking to steal data.

Cloud encryption services, especially comprehensive ones that encrypt data before the upload process begins, can help mitigate risks posed by insecure APIs and prevent sensitive data from falling into the wrong hands, even if a leak occurs.

**Maintains organizational integrity**

In recent years, it continues to increase the frequency of cyberattacks, especially in healthcare, banking & finance, education, and government sectors. This can be attributed to the transition toward storing data on the cloud rather than on local databases that can no longer be accessed by employees working remotely.

Cloud databases are connected using wired and wireless technologies and present a simple way to store large amounts of data, including employee, customer, sales, and financial records. However, the rise in the popularity of remote workplaces has given cybercriminals many more avenues to exploit the shortcomings of cloud computing platforms.

Unencrypted cloud data is susceptible to unauthorized access. Hackers cloak malicious packets as local traffic and introduce them into organizational cloud databases through illegal methods. Further, cybercriminals can benefit from modifying data to commit fraud. However, when cloud data is encrypted, stealing or modifying it becomes impossible.

**Ensures protection for multiple devices**

Gone are the days when employees had a dedicated endpoint to work on. Remote workers are using whatever device their company’s policy allows. However, some of these devices can be less secure than others. The transfer of information among devices introduces another layer of vulnerability, making encryption critical for protecting data across several devices.

Apart from stored data, comprehensive cloud encryption solutions can also help encrypt communication, passwords, and even web traffic—elements that are agreed upon as best practices for data security. No matter what data ends up compromised, whether due to a breach at the endpoint level or on the cloud provider’s side, bad actors would only obtain useless information without the correct decryption key.

**Guarantees compliance**

Remote work may be a potential regulatory compliance nightmare for organizations across verticals, especially with directives requiring companies to know exactly where their data is stored, how it is being transferred and processed, the parties that have access to it, and how it is secured.

At the provider level, regulations in some jurisdictions may require cloud solution vendors to hold specific compliance credentials and meet other cybersecurity requirements. Therefore, just one instance of careless transfer of sensitive data to or from the cloud or choosing a non-compliant provider may put the entire enterprise at risk of being pulled up for non-compliance and lead to severe financial and legal repercussions.

Cloud encryption is not only a secure solution for sharing and saving data. It can also be set to comply with the restrictions required within an organization as well as compliance with relevant regulatory bodies such as Federal Information Security Management Act (FISMA), Federal Information Processing Standards (FIPS), Health Insurance Portability and Accountability Act (HIPAA), and Payment Card Industry Data Security Standard (PCI/DSS).

## 2.3 Basic Framework of Searchable Encryption

The basic framework is provided for the reader to understand better the encryption and search process in a typical setting [38]. Let us assume a user, Alice, has a certain number of documents and wants to store them on an untrusted server, Bob. For example, Alice may be a mobile user with a low-bandwidth connection who wants to store her email messages on Bob securely. Since Bob is untrusted, Alice has to encrypt her documents and only store the resulting ciphertext on Bob. Alice uses a secret key, K, only known to her and not to Bob, for encryption.

After storing her documents, Alice wants to retrieve the documents containing the word W. In order to perform an efficient search for W, she encrypts W using the secret key K and sends the encrypted result along with a trapdoor for W to Bob. A trapdoor is a one-way function $f$: $X \to Y$ with the additional property that given some extra information (called trapdoor information, and in this case, it is provided by the secret key, K) it becomes feasible to find for any $y \in Im(f)$, an $x \in X$ such that $f(x) = y$ [39].

After determining the relevant encrypted documents, Alice decrypts using the same secret key, K. This scenario, along with the necessary steps discussed for encryption and query search, is depicted in Figure 2.6[13].

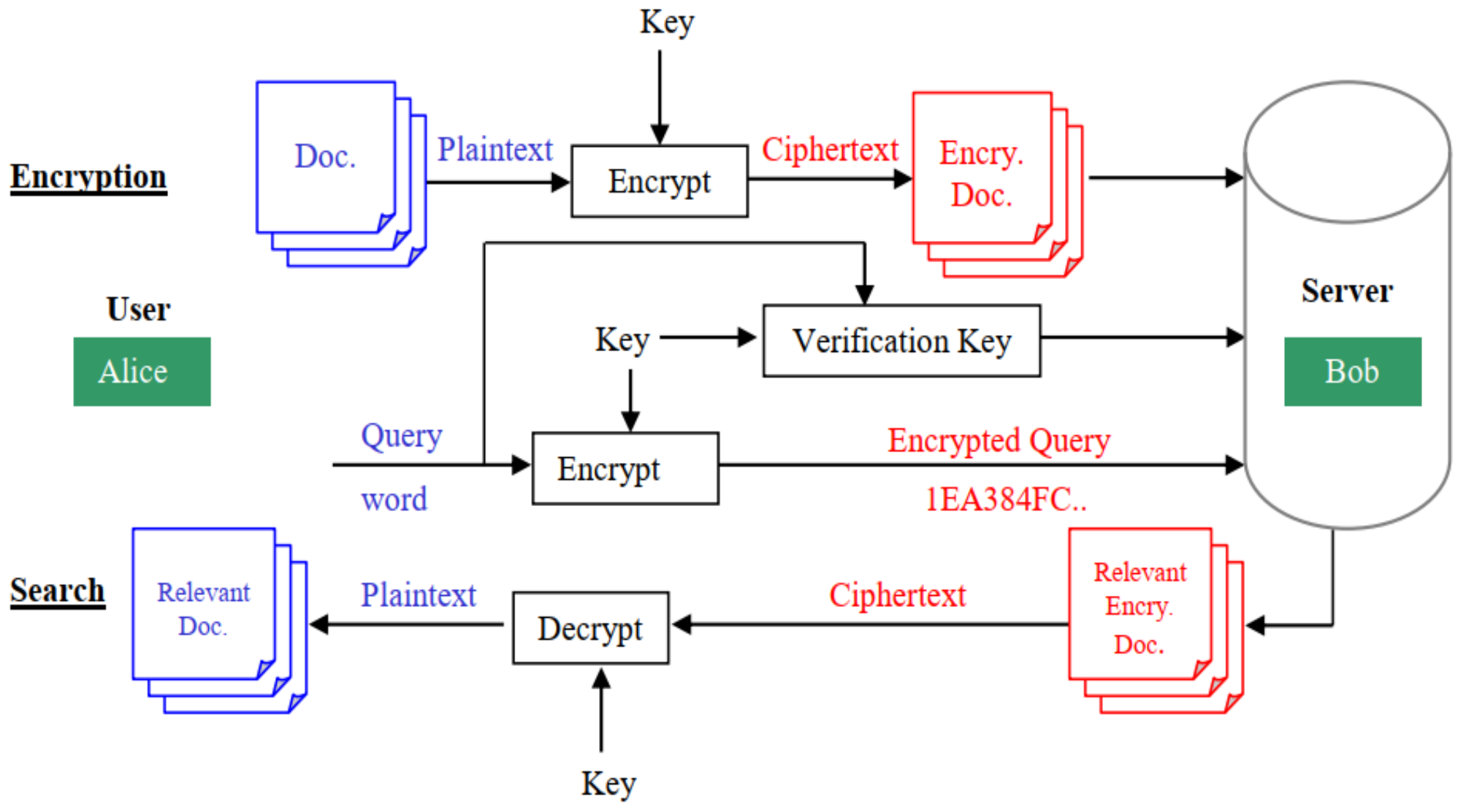

Figure 2.6 Basic Framework of Searchable Encryption

[13] Ucal, M., 2005. Searching on encrypted data. Department of Electrical and Computer Engineering University of Maryland College Park, MD August.

A general searchable encryption scheme consists of four algorithms:

**Setup($1^\lambda$) → *K***: The setup algorithm takes as input a security parameter $1^\lambda$, and then it Output the keys of the scheme.

**Enc(*K*,*D*) → (*I*,*C*):** The encryption algorithm takes as inputs the data files collection Along with the keys generated above, encrypts the data files and associated keywords index.

**Trapdoor(*K*,*w*) → $td_w$:** The trapdoor algorithm takes as inputs the target keyword and Secret key. It generates the trapdoor by encrypting keywords with the secret key.

**Test( $td_w$, *I*) → {0,1}:** The test algorithm takes as inputs the encrypted keyword index and The trapdoor of the target keyword, by computing it, returns 1 if successful; otherwise returns 0. Users and storage servers can apply the above four algorithms to complete the search.

## 2.4 Model of searchable encryption

A searchable encryption scheme includes three parties: a trusted data owner DO, a semi-trusted cloud server CS, and a collection of DU users authorized to search. The task for each party is as follows in figure 2.7[14].

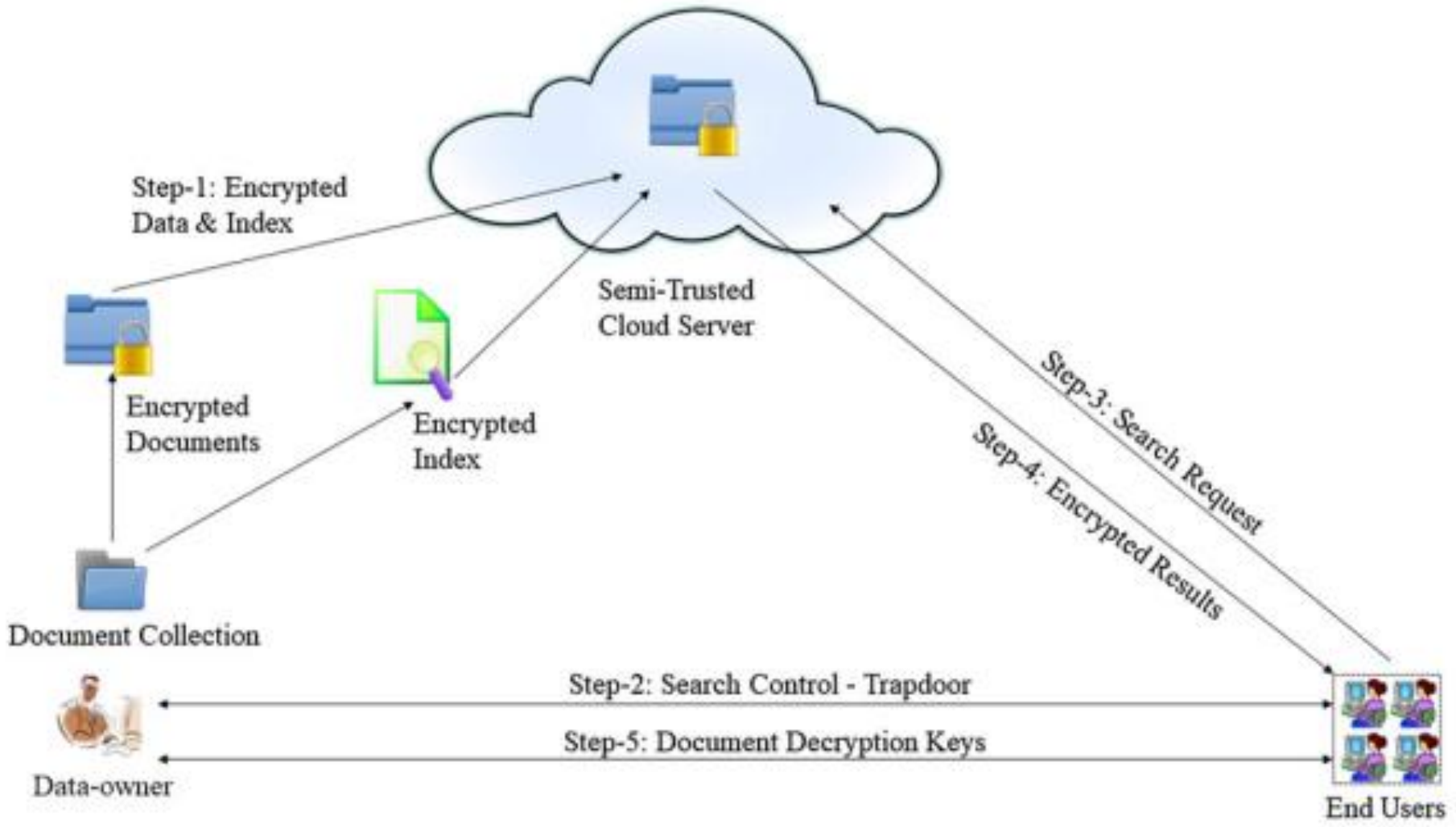

Figure 2.7 Model of Searchable Encryption

**Data Owner:** Data owner (individual or organization) is the owner of the data and is responsible for building the encrypted searchable index (if index-based searching is supported) after extracting the keywords of interest from the document collection. The encrypted documents and the searchable index are uploaded to a cloud server (refer to Step 1).

**Data User/End User:** A Data User wishes to access the data outsourced on the cloud. The data user may be a data owner (in the case of symmetric searchable encryption [SSE]) or a third party authorized to access the data (in the case of asymmetric searchable encryption [ASE]). The concept of sharing outsourced encrypted data, i.e., allowing users other than the DO to retrieve the documents based on the keywords of interest. Data users generate a query after obtaining a trapdoor from the DO (refer to Step 2) and share it with the CS (refer to Step 3).

[14] Wang, Y., Wang, J. and Chen, X., 2016. Secure searchable encryption: a survey. *Journal of communications and information networks*, *1*(4), pp.52-65.

**Cloud Server:** Cloud Server is an entity that provides a storage facility to the data owner and a search facility to the end-users. The cloud server searches for the desired documents by comparing the search query with the encrypted index or encrypted documents (refer to Step 4) and forwards the encrypted results to the end-user. Based on the retrieved results, the desired documents are downloaded by the end-user and decrypted after obtaining the decryption keys from the DO (refer to Step 5).

## 2.5 Architecture of Searchable Encryption

Every searchable encryption scheme consists of the following basic algorithms (other than document encryption and decryption) to support searching over outsourced data. These search schemes provide different query capabilities.

**1. KEYGEN:** The *keygen* algorithm is used to initialize the system. The secret keys are generated depending on the type of encryption method used for index generation and document encryption. If symmetric key cryptography is used for index generation, the *keygen* algorithm generates the set *SK* of symmetric keys. If public-key cryptography is used, it generates the public key (*PK*) and secret key (*SK*) parameters. In some of the SE schemes, these secret parameters are shared with the authorized end users so that they can generate the valid trapdoors corresponding to the keywords of interest, whereas other schemes require the Data Owner (DO) to generate queries on behalf of the end-users.

**2. BUILDINDEX:** The *buildindex* algorithm generates the encrypted searchable index corresponding to the documents in the collection. To generate the index, the data owner extracts the keywords from the documents in the collection and outsources the encrypted index (in some cases, the index includes the relevance score in the posting lists). Additionally, the documents are encrypted and outsourced to the cloud server.

**3. TRAPDOOR:** The *trapdoor* algorithm is used by the end-users to generate the search query (*Q*) based on the secret information obtained from the DO corresponding to the keywords of interest. In some cases, the query may be generated by the DO on behalf of the end-users.

**4. SEARCH:** Upon receiving the encrypted query (*Q*), the cloud server invokes the *search* algorithm, which compares the encrypted query with the entries in the index, thus generating the list of matching documents. Additionally, the retrieved documents can be ranked, i.e., organized according to the relevance of the search query if the underlying SE scheme provides the support. If the scheme supports document-based searching, then the search query is compared with the content of the outsourced documents.

## 2.6 Searchable Index

Given a data item, which can be of various forms (e.g., an email, a document, a video clip, etc.), the search can be done in two ways.

One is to perform a full-domain search. A search will sequentially go through every data item to test some criteria. For instance, a search can test whether some term appears more than a threshold time and another term does not appear in the content. The full-domain search takes linear operations to cover all data items. It is flexible because the criteria can be anything and be defined on the fly. The downside is its inefficiency when the data items are of big size. The other is index-based search or keyword-based search, where every data item is first characterized by a list of keywords used to build a search index for the data item. Later, when a search takes place, the criteria will be tested based on certain indexes instead of the contents of data items.

Searchable encryption is an essential technique that allows search directly over encrypted data. Similar to searching over plaintext documents, the typical approach of searchable encryption schemes is to build a secure index for the entire document set. Note that secure indexes help search on encrypted data where the encrypted documents and their indexes on the remote server are frequently updated. Then in the search stage, which is done on the cloud server-side, only the secure index is referenced.

Searchable encryption is used to execute keyword-based search directly on encrypted data and reduce the costly operation in cloud computing. Searchable encryption schemes usually build up an index for each keyword of interest and associate the index with the files that contain the keyword [40]. An effective keyword search can be realized by integrating the trapdoors of keywords within the index information. In contrast, both file content and keyword privacy are well-preserved. Keyword indexes search in constant time for documents containing specified keywords. The server performs the searching of keywords on the index without learning anything about the data.

Informally, a secure index allows users with an "encrypted text" for a word *x* to test the index only for *x*; The index reveals no information about its contents without valid encrypted text, and encrypted text can only be generated with a secret key.

Using an index is much faster in terms of search time if we have documents that are large in size. Also, performing a search with an encrypted index needs less sophisticated constructions than a non-keyword-based approach. However, index-based schemes suffer from insecure updates. For example, if the data owner changes the documents, he must update the index. If he adds a new document and does not update the list of pointers for one keyword entry, the cloud server would be able to error message that the keyword does not appear in the new document that was just added.

There are two basic ways to construct indexes in the literature, namely forward index and inverted index. Like plain-text retrieval, the index utilized in Searchable Encryption schemes can be forward or inverted. With this approach, the search can be done much more efficiently since it does not need to go through the contents of all data items. The downside is that search criteria may not be as flexible as in the case of full-domain search. The performance will depend on the selection of keywords and how the index is constructed and maintained.

### 2.6.1 Forward Index

Forward Index is an index constructed per document, i.e., it maps the document to the list of keywords it contains (Table 2.1). The use of a forward index was proposed by Goh [15] and used by other researchers as it required only one index entry to be updated on the addition (or deletion/update) of a document in the collection. However, the search time is linear, i.e., the time to search is proportional to the number of entries in the index. Generating a forward index is fast than compared to the inverted index. Adding a new document requires adding a new entry to the existing index. Hence, it requires no rebuilding of the index.

| Item Identifiers | Keywords |
|---|---|
| Item ID1 | Keyword1, keyword3, keyword5 |
| Item ID2 | keyword5 |
| Item ID3 | Keyword2, keyword4 |
| Item ID4 | Keyword1, Keyword2, keyword4 |
| Item ID5 | Keyword4, Keyword5, keyword6 |
| Item ID6 | Keyword2, keyword3, Keyword4, keyword5 |

Table 2.1 Forward Index format

If we consider documents such as $D_1 \ldots D_n$ with multiple keywords for each document where some documents have the same keyword, then the forward index can be built, which is shown in table 2.2. The forward index can be considered a 2-D array, each row consisting of a tuple. The array's length is equivalent to the number of keywords that can be searched in the system. The first entry of each tuple is a document list. In contrast, the second entry of each tuple consists of the keywords that the corresponding keyword contains in the document.

| Document | Keywords |
|---|---|
| $D_1$ | Antony, Brutus, Caesar, Cleopatra… |
| $D_2$ | Antony, Brutus, Caesar, Calpurnia… |
| $D_3$ | Mercy, Worser … |
| $D_4$ | Brutus, Caesar, Mercy… |
| … | … |
| $D_n$ | Antony, Caesar, Mercy… |

Table 2.2 Forward Index example

### 2.6.2 Inverted Index

The inverted index is one of the most popular data structures used in document retrieval systems. It is one of the most efficient searchable secure index structures and has been widely adopted in plaintext search [11].

Secure index-based searchable encryption enhances search scalability. It allows the encryption of each document to be independent of the searchable encryption scheme used. First, a search on the inverted index can be very efficient, especially for a large dataset. Search is directly pointed to the related documents, i.e., the inverted lists that match the query keyword. Secondly, the inverted index has been widely adopted in large dataset search schemes. Index generation uses a hash and a mapping function. It is linear in the upper bound of the number of distinct keywords. The search is linear in the number of indexes.

Therefore, secure inverted indexes can be applied incrementally over the already built inverted indexes. An inverted index is an index data structure storing a mapping from content, such as Keywords to a set of documents [41]. The purpose of an inverted index is to allow fast, full Text searches. It is the most popular data structure used in document retrieval systems, They are used, for example, on a large scale in search engines.

Actually, many search engines incorporate an inverted index when evaluating a search query to quickly locate documents containing the words in a query and then rank these documents by relevance. Because the inverted index stores a list of documents containing each word, the search engine can use direct access to find the documents associated with each word in the query to retrieve the matching documents quickly.

The inverted index is a fundamental technique to support keyword search. It maps the keywords to the list of documents containing them (Table 2.3). We thus follow the notion of the inverted index to have our design for similarity search. The inverted index can be considered a 2-D array, each row consisting of a tuple. The array's length is equivalent to the number of keywords that can be searched in the system. The first entry of each tuple is a keyword. In contrast, the second entry of each tuple consists of the file IDs that the corresponding files contain the keyword. Using an inverted index requires searching for a term once in the index and retrieving all the documents appearing in the corresponding posting list as a match.

| Keywords | Item Identifiers |
|---|---|
| Keyword1 | Item ID1, Item ID5 |
| Keyword2 | Item ID3, Item ID4, Item ID5, Item ID7 |
| Keyword3 | Item ID1, Item ID4, Item ID7 |
| Keyword4 | Item ID3, Item ID5, Item ID6, Item ID7 |
| Keyword5 | Item ID1, Item ID2, Item ID6, Item ID7 |
| Keyword6 | Item ID6 |

Table 2.3 Inverted Index format

If we consider documents such as $D_1 \ldots D_n$ with multiple keywords for each document where some documents have the same keyword, the inverted index can be built, shown in table 2.4.

| Keywords | Documents |
|---|---|
| Antony | $D_1, D_2, D_6 \ldots$ |
| Brutus | $D_1, D_2, D_4 \ldots$ |
| Caesar | $D_1, D_2, D_3, D_5 \ldots$ |
| Calpurnia | $D_2, \ldots$ |
| Mercy | $D_1, D_3, D_4, D_6$ |
| … | … |
| Worser | $D_2, D_4, D_6 \ldots$ |

Table 2.4 Inverted Index example

As we are using the inverted index method in our proposed model, here we are discussing how this works in our scheme. The inverted index–based of unencrypted data structure is shown in table 2.5 1. Let $D = \{D_1, D_2, \ldots, D_m\}$ be the set of documents to be stored in an untrusted cloud server, where $|D| = m$ is the total number of documents. Each document $D_i$ contains a set of keywords. For simplicity, we are using a single keyword, but we can use multiple keywords according to each document by comma (,) for each keyword. Let $W = \{w_1, w_2, \ldots, w_n\}$ be the set of keywords in $D$, where $|W| = n$ is the total number of keywords. $Di \in [n] \subseteq D$ denotes the set of documents containing the keyword $w_i$.

The keywords and the sets of documents are respectively encrypted by secret key *SK* and corresponding random keys $R_{1, \ldots,} R_n$ and are stored in the cloud. Note that secret key *SK* and random keys $R_{1, \ldots,} R_n$ are different encryption schemes: SK is a searchable encryption scheme supporting binary search proposed in this article, and E can be any secure symmetric encryption scheme, such as Advanced Encryption Standard (AES). For example, if we are considered an unencrypted inverted index where the plaintext keywords "Headache", "Diabetes", "Cold", and "Allergies" are selected from the document ID $D_{1, \ldots,} D_{10}$ randomly, which is listed in table 2.5. Now, to build an encrypted inverted index, we could use the secret key *SK* for encrypted the plaintext keywords, and for document encryption, we could use random keys $R_{1, \ldots,} R_n$ accordingly with document ID with a reference number which is listed in table 2.6. After the encryption, the inverted index will be stored in the cloud service provider.

| Plaintext Keyword | Document ID |
|---|---|
| Headache | $D_3, D_5, D_7, D_{10}$ |
| Diabetes | $D_2, D_3$ |
| Cold | $D_5, D_7$ |
| … | … |
| Allergies | $D_2, D_3, D_6$ |

Table 2.5 Plaintext Inverted Index

| Encrypted Keyword | Encrypted Document ID |
|---|---|
| 874f718fb3df | $\mathbf{R_3}(D_3), \mathbf{R_5}(D_5), \mathbf{R_7}(D_7), \mathbf{R_{10}}(D_{10})$ |
| 659689db1c6 | $\mathbf{R_2}(D_2), \mathbf{R_3}(D_3)$ |
| 26b94b7c17a | $\mathbf{R_5}(D_5), \mathbf{R_7}(D_7)$ |
| … | …. |
| 9619da7d95c | $\mathbf{R_2}(D_2), \mathbf{R_3}(D_3), \mathbf{R_6}(D_6)$ |

Table 2.6 Encrypted Inverted Index

Now, if the data owner/data user wants to search to get the encrypted document, first, he needs to download the whole encrypted inverted index locally. The data owner/user will search on the encrypted index by using the secret key *SK* with plaintext keyword, i.e., "Cold". After generating the encrypted keyword, he will search on the encrypted inverted index, and the index will return if there is any match with the encrypted keyword. It will return all the documents based on the keyword match, i.e., $D_5$, $D_7$, to the user. As the document is in encrypted form with corresponding random keys $R_5$ and $R_7$, it will return the File ID, which was stored in the cloud server. This way, the inverted index is used in our proposed model, which is shown in table 2.7. The time complexity of such a search method is linear in efficiently and efficiently optimized number of keywords.

| Keyword Matching | Document ID | File ID |
|---|---|---|
| Cold : 26b94b7c17a | $\mathbf{R_5}(D_5)$ | 1IVBMUp9xNYJTa4MFKFiY1bTvR9ihOqf7 |
| Cold : 26b94b7c17a | $\mathbf{R_7}(D_7)$ | 1ezIJxI_NmmzuvtRmwgPM-Xyeplg12Gkwg |

Table 2.7 Encrypted Result

The most significant advantage of using the inverted index is the search efficiency, especially for a large volume of documents. The search operation is performed on a much smaller document set consisting of the inverted lists that match the query keywords. However, securing an inverted index and its associated search schemes is not a trivial task. Also, the update is complicated as an addition (or deletion) of a document requires rebuilding the entire index.

### 2.6.3 Performance and Properties

In the following, we state the performance and properties of the inverted index approach:

- **Main Feature:** An inverted index in the form of (key, value), where the key is an element representing a keyword w, such as an encrypted keyword (or keyed hash value), and value is a list of elements representing the set of messages that contain w.

- **Storage:** Schemes under this construction require similar storage space as schemes with direct index since encrypted messages and masked index are stored. Different schemes may have different storage efficiency due to how an index is designed.

- **Search:** Search time is sublinear (and, in fact, optimal in many cases) compared to linear search time. This is because searching for a keyword immediately returns the list of document identifiers matching the keyword. Thus, keyword matching efficiency is equivalent to the O(1) efficiency of a hash table (dictionary) and O(r) on retrieval of the r matching documents or document identifiers. This is the main advantage of the inverted index under this approach.

- **Characteristics (static or dynamic):** Static index is a method of indexing. A fixed number of keywords is allocated along with documents to store the records. However, creating a dynamic scheme is generally not as straightforward. Suppose one is to add/remove a document through the index. In that case, each of the keyword tokens must be linearly scanned through to add/remove a document entry. This is the main disadvantage of this approach.

- **Security (leakage):** Schemes of this type have a similar leakage profile with constructions using the direct index, in which they also leak keyword distributions and access patterns. Depending on the underlying design of the index, certain schemes leak a total number of keywords, while specific schemes do not.

## 2.7 Encryption Techniques

Due to encrypted data, computation and searching of data is an important task because traditional plaintext searching techniques can not directly be applied to encrypted cloud data. However, encryption for outsourced data can protect privacy against unauthorized behaviors and protect the data confidentiality in the cloud; it also complicates data utilization, such as searching for encrypted data, in a complicated issue [42]. Many searching techniques are available for searching for encrypted cloud data from cloud servers. Multiple encryption techniques have been invented to search on encrypt the data. Some of the most popular encryption techniques are discussed below.

Encrypted search has become an important problem in security. This is due to a combination of three things: (1) search is now the primary way we access our data; (2) we are outsourcing more and more of our data to third parties; and (3) we trust these third parties less and less.

Several searchable encryption schemes have been proposed later based on Song et al. [21] scheme. Figure 2.8 shows the taxonomy of searchable encryption. In this section, we discussed in detail each of the encryption techniques. Further, we examined different schemes in each of the encryption techniques; most of these schemes were published between 2005 and 2021 and divided based on keyword search functionality.

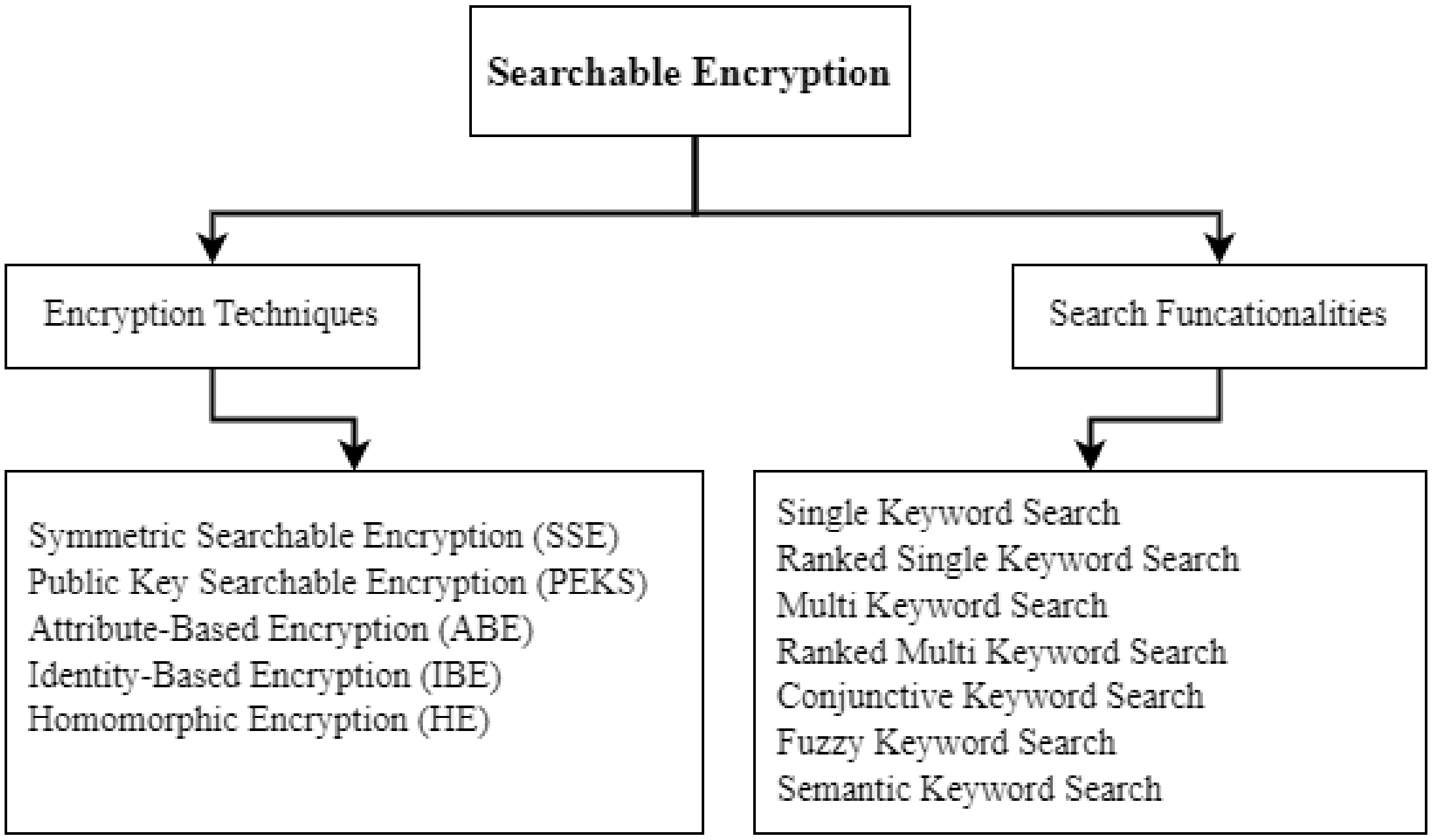

Figure 2.8 Taxonomy of Searchable Encryption

From the perspective of cryptography, SE technology mainly includes two types such as searchable encryption (SSE) and another is public-key encryption with keyword search (PEKS). SSE is related to the private key primitive. It allows only the private key holder to produce ciphertexts and create trapdoors for search. In contrast, PEKS, on the other hand, is related to the public key primitive. It enables several users who know the public key to produce ciphertexts but only allows the private key holder to create trapdoors for search. More details are discussed below.

### 2.7.1 Symmetric Searchable Encryption (SSE)

In searching on private-key-encrypted data, the user himself encrypts the data to organize it arbitrarily (before encryption) and includes additional data structures to allow for efficient access to relevant data. The data and the additional data structures can then be encrypted and stored on the server. Only someone with a private key can access it. In this setting, the initial work for the user (i.e., for preprocessing the data) is at least as extensive as the data. However, subsequent work (i.e., for accessing the data) is minimal relative to the data size for both the user and the server.

The SSE consists of a data sender, data receiver, and cloud service provider. A data sender encrypts their documents, indexes them with the private key, and uploads them to the remote service provider. A data receiver with a corresponding private key can perform the search operation. He generates the trapdoor he wants to search the keyword with the private key and sends it to the server. After receiving the trapdoor, the service provider enables the test of whether a given ciphertext contains the search keyword without knowing the plaintext message of the encrypted data and the keyword. Then, the service provider returns the query results to the data receiver. Finally, the receiver can decrypt the ciphertext which the server sends.

Symmetric searchable encryption (SSE), introduced by Song et al. [21], is one solution to the aforementioned problem. In SSE, a data owner (DO) can generate encrypted keywords for each encrypted file by using a symmetric key shared with the data user (DU) before their data are uploaded to the cloud systems. Subsequently, the DU can generate a trapdoor for specified keywords and submit them to the cloud systems to search for encrypted files related to these keywords. Because of these properties, SSE is well suited to cloud computing, and various SSE approaches have been proposed. However, SSE is restricted to the key sharing problem of symmetric cryptosystems. Specifically, the DO and DU must agree on a shared key before encrypting keywords and generating trapdoors. The model of the SSE system is shown in figure 2.9[15].

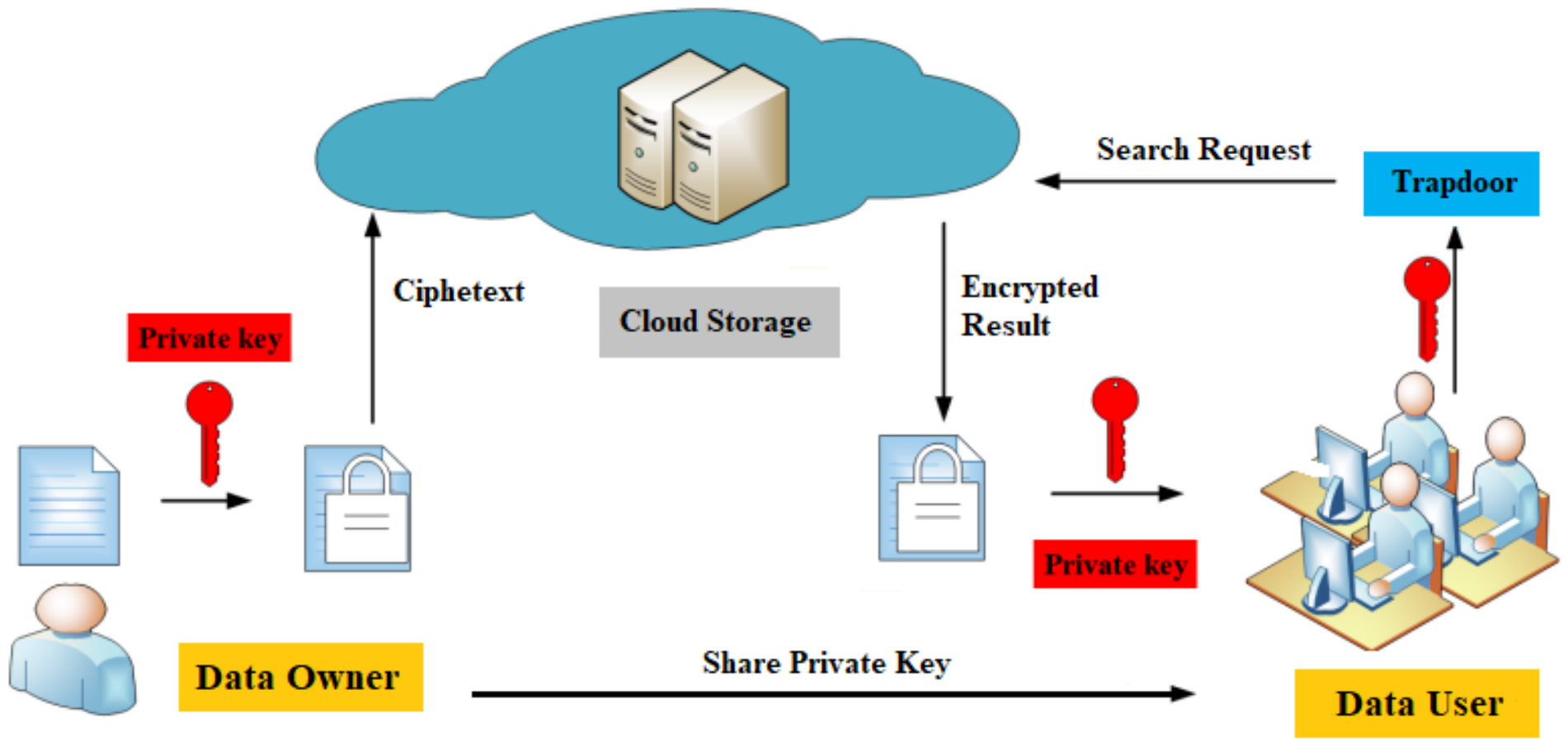

Figure 2.9 Model of Searchable Symmetric Encryption (SSE) System

Searchable symmetric encryption (SSE) is a widely popular cryptographic technique that supports search functionality over encrypted data on the cloud. However, most existing SSE schemes leak the search pattern despite their usefulness. An adversary can tell whether two queries are for the same keyword [43]. In recent years, it has been shown that search pattern leakage can be exploited to launch attacks to compromise the confidentiality of the client's queried keywords.

[15] Zhou, Y., Li, N., Tian, Y., An, D. and Wang, L., 2020. Public key encryption with keyword search in cloud: a survey. *Entropy*, *22*(4), p.421.

### 2.7.2 Public-key Encryption with Keyword Search (PEKS)

Except for symmetric searchable encryption, another important work on encrypted data search is the public key encryption scheme with keyword search (PEKS), which was first proposed by Boneh et al. (2004) [44]. In the public-key setting, anyone with a public key can write to the data stored on the server. However, only authorized users with a private key can search. Public key solutions are usually very computationally expensive, however. Furthermore, the keyword privacy could not be protected in the public-key setting. The server could encrypt any keyword with the public key and then use the received trapdoor to evaluate this ciphertext.

The PEKS consists of a data sender, data receiver, and cloud service provider. A data sender encrypts their documents, indexes them with the public key, and uploads them to the remote service provider. A data receiver with a corresponding private key can perform the search operation. He generates the trapdoor he wants to search the keyword with the private key and sends it to the server. After receiving the trapdoor, the service provider enables the test of whether a given ciphertext contains the search keyword without knowing the plaintext message of the encrypted data and the keyword. Then, the service provider returns the query results to the data receiver. Finally, the receiver can decrypt the ciphertext which the server sends.

However, Public key solutions are usually very computationally expensive. Furthermore, the keyword privacy could not be protected in the public-key setting. The server could encrypt any keyword with the public key and then use the received trapdoor to evaluate this ciphertext. A secure channel is necessary to transmit the key in this scenario. Also, the salient problem of PEKS is the offline keyword guessing attack. PEKS is more suitable for use in some insecure networks. It does not require the encryption party and the decryption party to negotiate the key in advance. Figure 2.10[16] shows the model of the PEKS system.

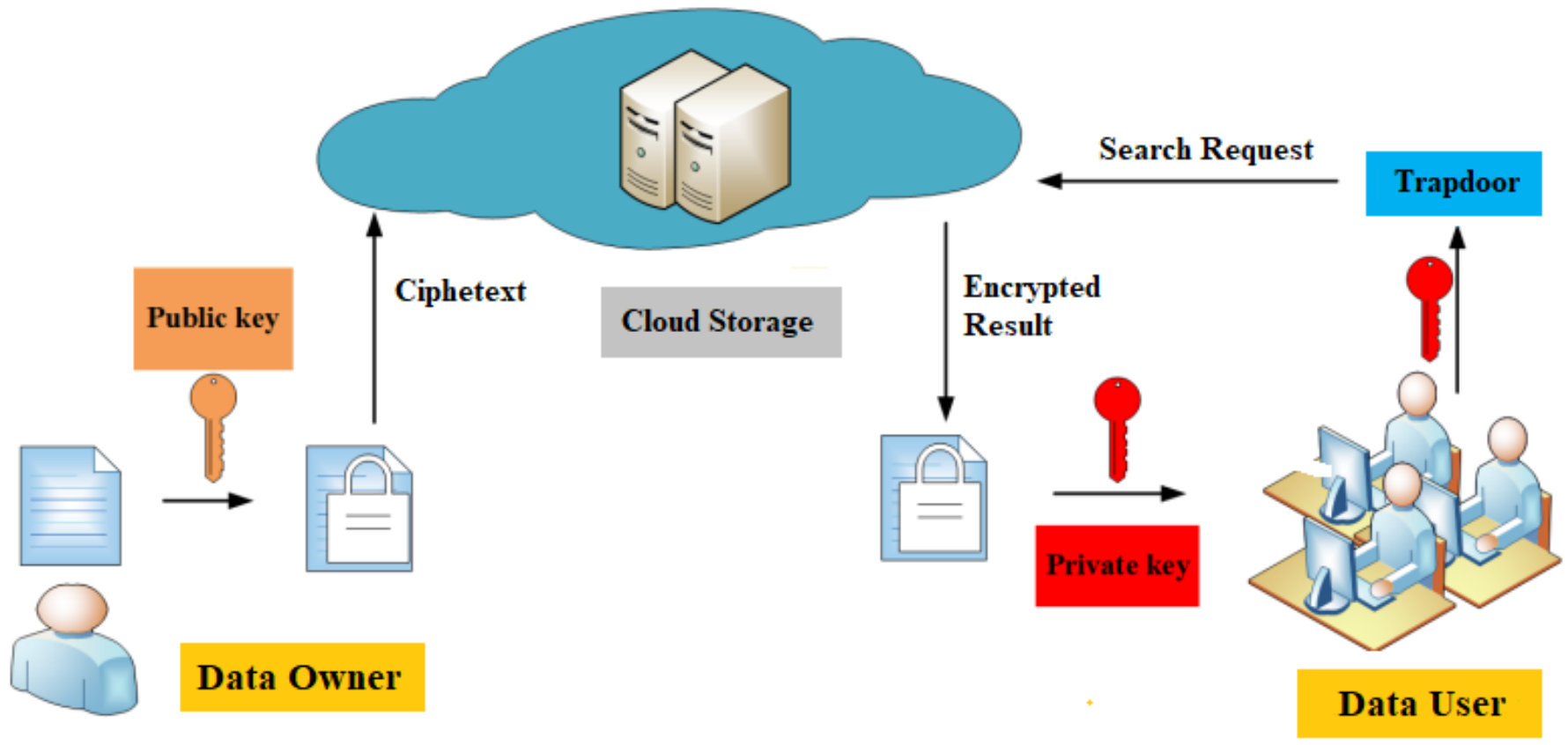

Figure 2.10 Model of public-key encryption with keyword search (PEKS) system

[16] Zhou, Y., Li, N., Tian, Y., An, D. and Wang, L., 2020. Public key encryption with keyword search in cloud: a survey. *Entropy*, *22*(4), p.421.

### 2.7.3 Attribute-Based Encryption (ABE)

Attribute-based encryption (ABE) has been widely applied in searchable encryption in recent years. Sahai and Waters (2005) [45] first defined the concept of attribute-based encryption as an extension of identity-based encryption. It regards identity as a series of attribute sets; attributes are the information elements of the user. When a data user's attributes satisfy the access policy formulated by the data owner, the user could decrypt the ciphertext. Attribute-based encryption (ABE) protocol provides fine-grained access control for encrypted data based on the client's attributes. Such protocol allows the client whose attributes satisfy the access policy to decrypt the encrypted messages under specific policies.

In an ABE system, a user's keys and ciphertexts are labeled with sets of descriptive attributes, and a particular key can decrypt a particular ciphertext only if there is a match between the attributes of the ciphertext and the user's key [46]. In ABSE, a trusted third-party authority (TA) generates the private keys, and data owners encrypt the documents under defined access policies, such that the users decrypt the documents only when the attributes are matched with the access policy

Attribute-based Searchable Encryption (ABSE) integrates an access policy with the encrypted documents to control the user access. The access policy contains information about all the users who can access the document. Suppose user attributes (e.g., name, department, course, gender, etc.) are not satisfied with the defined access policy. In that case, the user is not authorized to access the documents.

Classical attribute-based encryption systems can be divided into two categories: Ciphertext-policy (CP) ABE and Key-policy (KP) ABE.

**Ciphertext-policy attribute-based encryption (CP-ABE)**

The CP-ABE was first proposed by Bethencourt et al. (2007) [46], who use system attributes to describe the client's credentials; moreover, private keys are computed concerning a set of attributes and the access policy determines which clients can decrypt the encrypted data. Ciphertext-policy attribute-based encryption (CP-ABE), where the private key is generated with an attribute set and the ciphertext is related to an access policy is shown in figure 2.11[17].

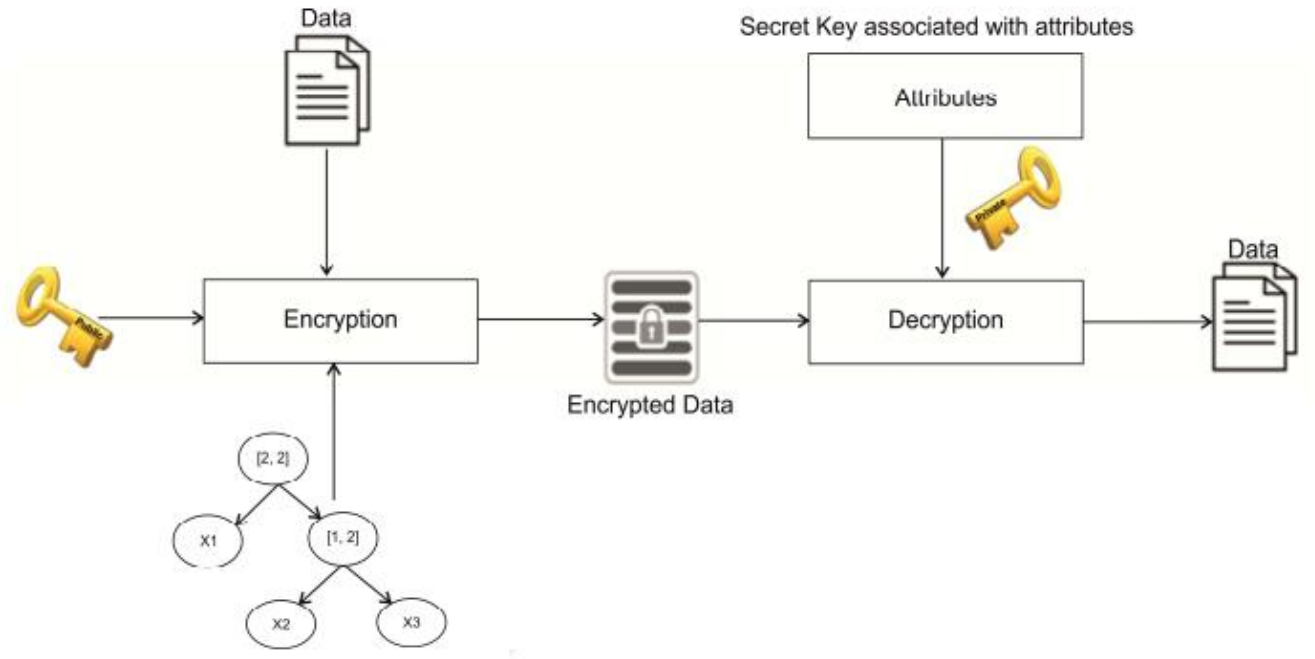

Figure 2.11 Cipher Policy Attribute-Based Encryption

[17] Kumar, P. and Alphonse, P.J.A., 2018. Attribute based encryption in cloud computing: A survey, gap analysis, and future directions. *Journal of Network and Computer Applications*, *108*, pp.37-52.

Unfortunately, one of the primary efficiency defects in existing CP-ABE schemes is that encryption or decryption operations involve time-consuming pairing operations. The number of other operations increases with the complexity of the access policy [47]. Hence, it is critical to essentially eliminate computational costs for resource-limited end-users in a fine-grained keyword search system.

**Key-policy attribute-based encryption (KP-ABE)**

In KP-ABE, the role of access policy and attributes are converted, and the access policy is used to design the client's private key. The ciphertext is generated for a set of attributes. The first KP-ABE is suggested by Goyal et al. [48]. Key-policy attribute-based encryption (KP-ABE), where the private key of a user is related to an access policy and the ciphertext corresponds to an attribute set. In the cryptosystem, ciphertexts are labeled with sets of attributes, and private keys are associated with access structures that control which ciphertexts a user can decrypt is shown in figure 2.12[18].

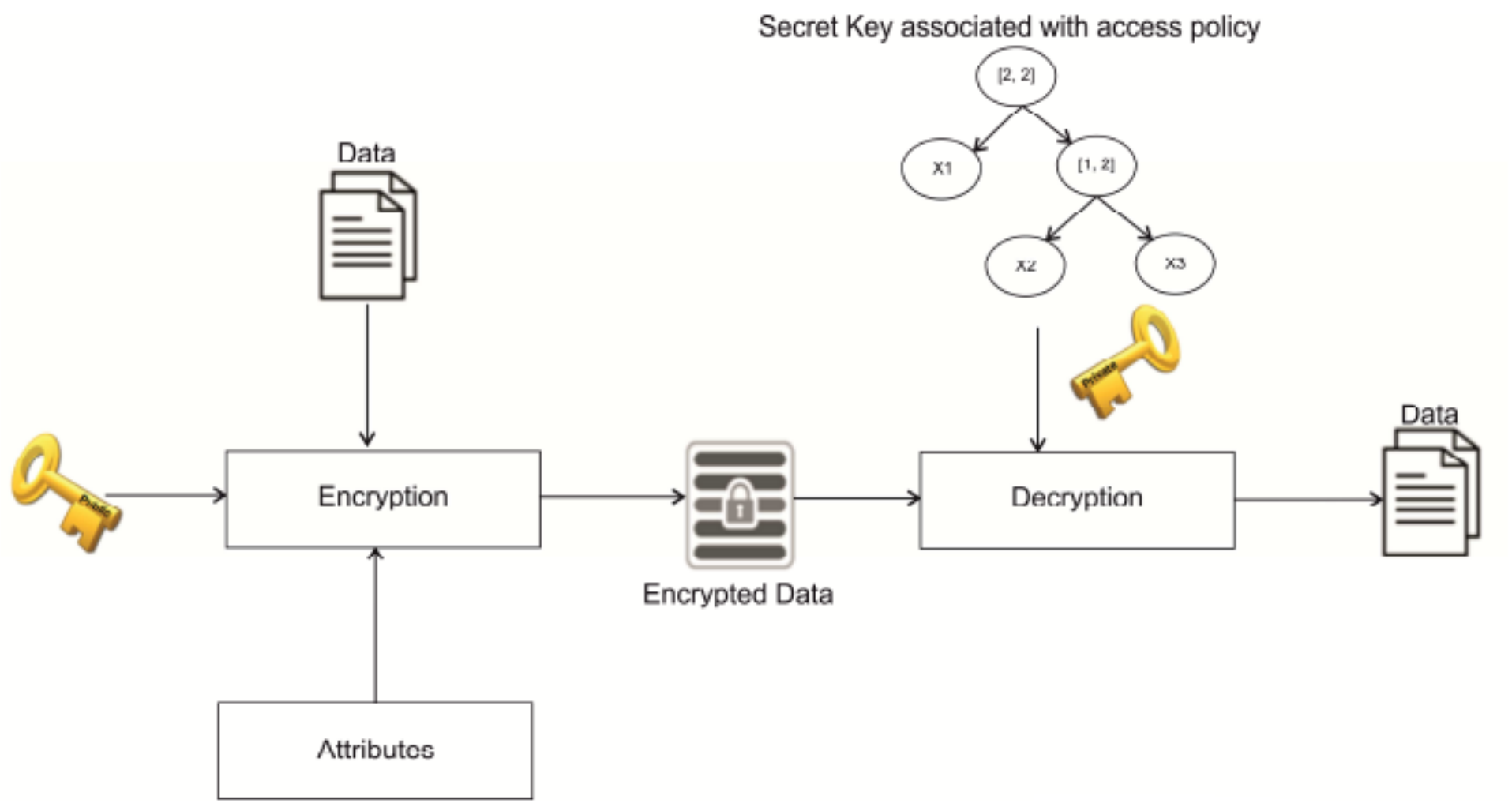

Figure 2.12 Key Policy Attribute-Based Encryption

Both of these schemes play an important role in secure system construction, and since the construction of the first attribute-based encryption scheme, many works have been performed [49], [50], [51] to improve the efficiency, practicality, and expressiveness of earlier efforts.

### 2.7.4 Identity Based Encryption (IBE)

When the encrypted data needs to be shared with more people beyond those initially designated by the data owner, it raises a serious problem. Identity-based encryption (IBE) was firstly proposed by Shamir [52] in 1984, which simplifies the management of public keys and certificates in traditional public-key encryption based on PKI to address this problem. Identity-based encryption (IBE) is a form of public-key cryptography. A third-party server uses a simple identifier, such as an email address, IP address, telephone number, ID, etc., to generate a public key that can be used for encrypting and decrypting electronic messages. The private key generation (PKG) can generate the private key according to the user's authentication and request. The data encrypted by one user can be correctly decrypted by all the authorized users in the system.

[18] Kumar, P. and Alphonse, P.J.A., 2018. Attribute based encryption in cloud computing: A survey, gap analysis, and future directions. *Journal of Network and Computer Applications*, *108*, pp.37-52.

In IBE, data users are identified and authorized for data access based on their recognizable identities, which avoids complicated certificate management in usual secure distributed systems. More importantly, IBE provides a mechanism that generates an IBE ciphertext so that a new group of users not specified during the IBE encryption could access the underlying data

Suppose Alice wants to send a message to Bob; the communication step of the two parties using identity-based encryption is shown in Figure 2.13[19]. In an IBE scheme, a data sender uploads ciphertexts to a service provider. The receiver contacts KGC using its identity to get a corresponding private key and generates a search trapdoor using its private key to send the server provider. Finally, the server provider conducts a keyword search.

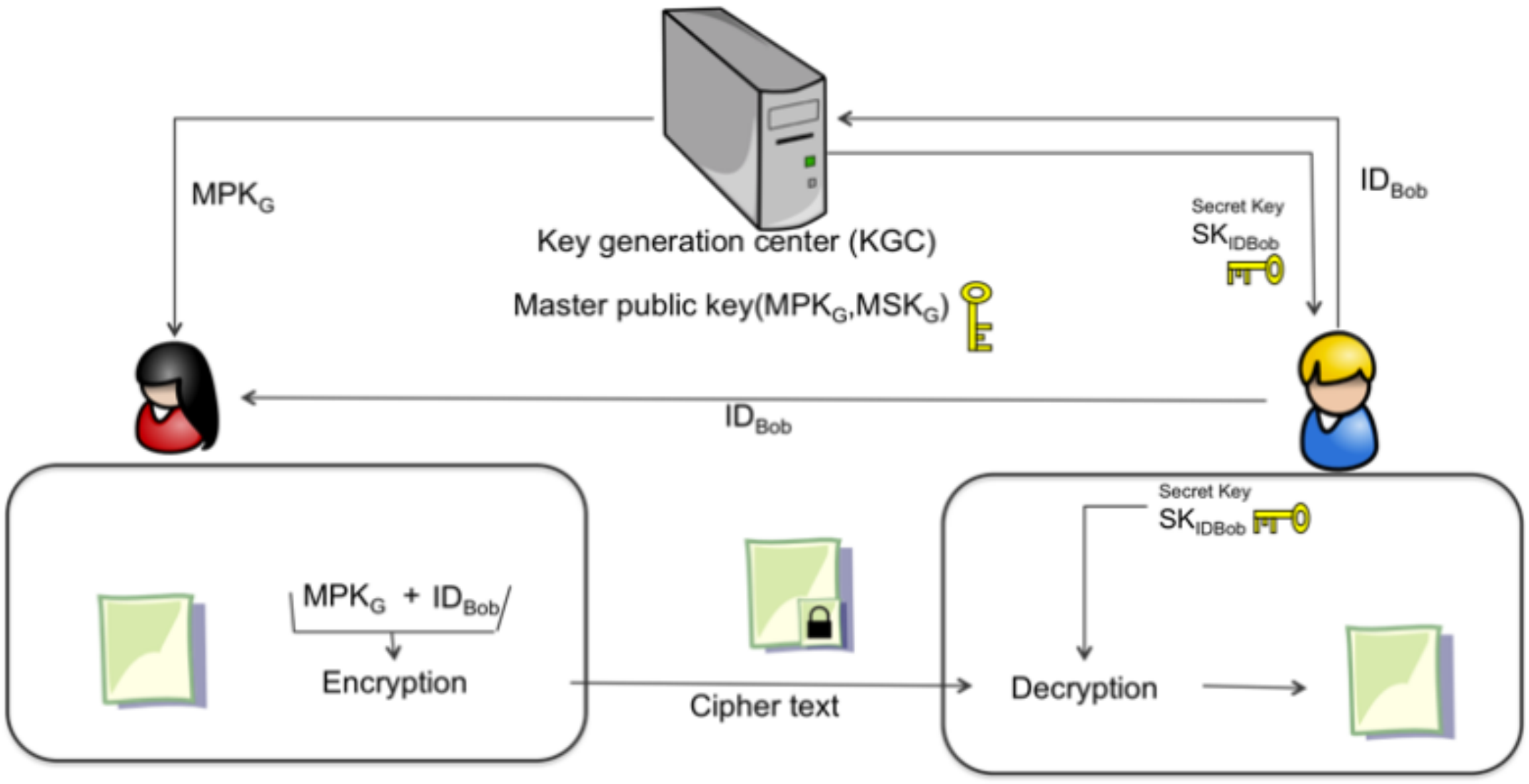

Figure 2.13 Identity-based encryption (IBE) system

Boneh et al. [44] provided the first PEKS scheme based on IBE, in which the keyword acted as the identity. Emura et al. [53] proposed a PEKS scheme supporting keyword revocable based on partially-anonymous identity-based encryption. In their scheme, a keyword trapdoor is generated even if the keyword is revoked, which can resist the security risks caused by the keyword trapdoor. Recently, Wang et al. [54] proposed a secure channel-free identity-based searchable encryption scheme in a peer-to-peer group, which allowed multiple users to share in a peer-to-peer group and search the private data in the cloud.

[19] Naima, Meddah & Toumanari, Ahmed. (2016). Reinforce cloud computing access control with key policy attribute-based anonymous proxy reencryption. International Journal of Cloud Computing. 5. 187. 10.1504/IJCC.2016.080044.

### 2.7.5 Homomorphic Encryption (HE)

Homomorphic Encryption (HE) is a particular type of encryption that allows one to perform an algebraic operation on ciphertexts without decrypting them. This makes HE an exciting tool for searching over encrypted data since meaningful computation can be executed on the encrypted data is shown in figure 2.14[20]. The first apparition of fully homomorphic encryption was in 1978 with Rivest et al. under privacy homomorphism [55]. However, privacy homomorphism stayed a conjecture without an effective solution until 2009. In his thesis, Craig Gentry presented the first semantic secure privacy homomorphism under the name of a fully homomorphic encryption scheme [56]. The scheme uses a bootstrapping theorem to reduce the noise generated after processing ciphertexts. The cleartext space of Gentry's scheme is the binary field, such as space that allows us to evaluate any circuit on encrypted data using the homomorphic capacities of the scheme. The pairing-based HE scheme proposed by Boneh et al. [57] can perform an arbitrary number of additions and one multiplication.

Fully homomorphic encryption is a powerful tool that can allow searching over encrypted data and realizing PIR protocols [58]. It is generally believed that FHE can solve the problem of querying encrypted data since any meaningful computation can be performed on the encrypted data. However, one issue with FHE is the performance since current schemes are computationally expensive and have a high storage overhead.

For some applications, so-called somewhat homomorphic encryption schemes [59] can be used. These schemes are more efficient than FHE but allow only a certain amount of additions and multiplications.

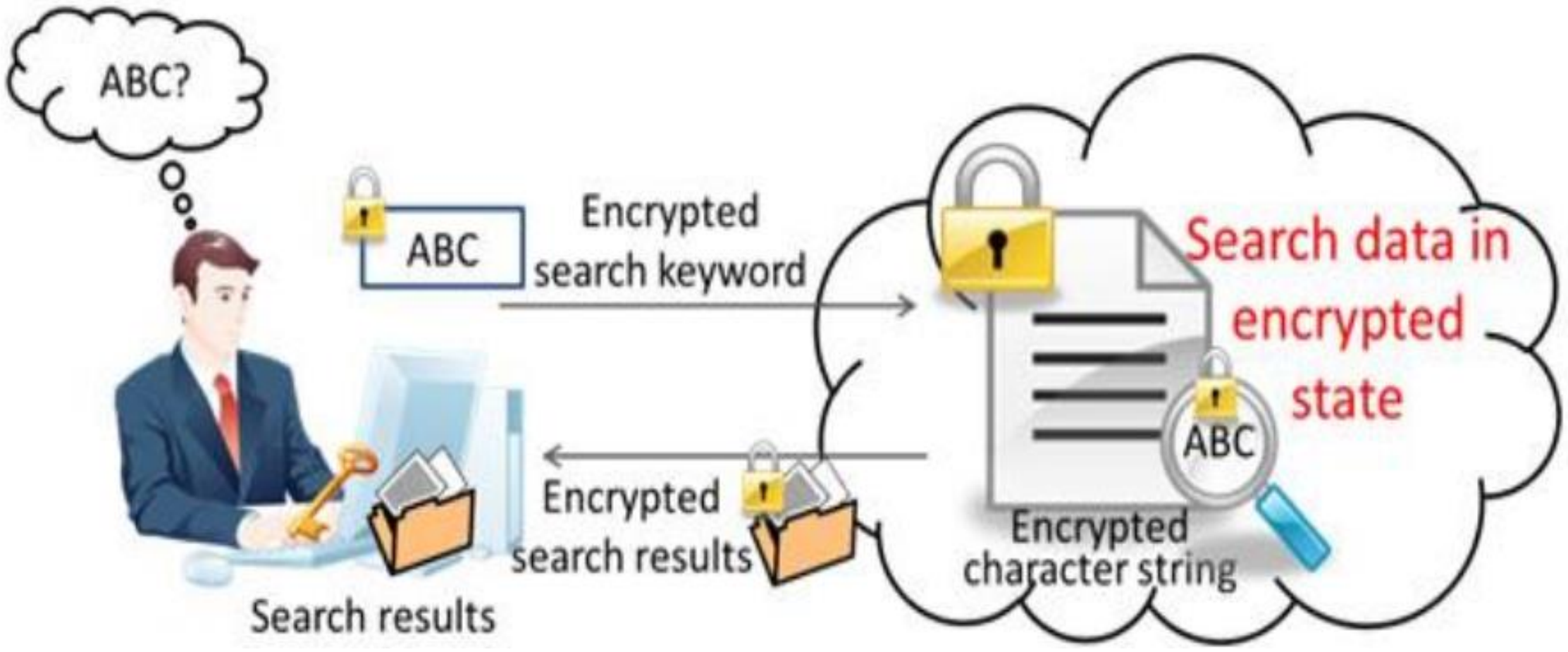

Figure 2.14 Homomorphic Encryption (HE) System

The primary issue when using somewhat or fully HE for querying encrypted data in different applications is that the resulting schemes require a linear search time in the length of the dataset. This is too slow for many mission-critical applications in practice.

---

[20] Fontaine, C. and Galand, F., 2007. A survey of homomorphic encryption for nonspecialists. *EURASIP Journal on Information Security*, *2007*, pp.1-10.

## 2.8 Search Functionalities

Various encryption algorithms are used to protect information from unauthorized disclosure, maintain data confidentiality, and secure storage. However, searching over encrypted data was difficult to attain. Therefore, keyword-based searching has been introduced where the desired file is retrieved when searched for a particular keyword. Keyword-based search is one of the most popular techniques to search documents on encrypted cloud data. Keyword search techniques are widely used in plain-text scenarios, and the user is allowed to retrieve select files from the storage space [60].

The flowchart of a typical keyword-based scheme is provided in Figure 2.15. Generally, this scheme has a small memory overhead, and they are faster when compared to non-keyword-based schemes. The search time is in the order of the number of keywords. On the other hand, having a keyword list limits the search capability of the user. It requires updates when a document or a keyword is changed, added, or deleted on the remote server.

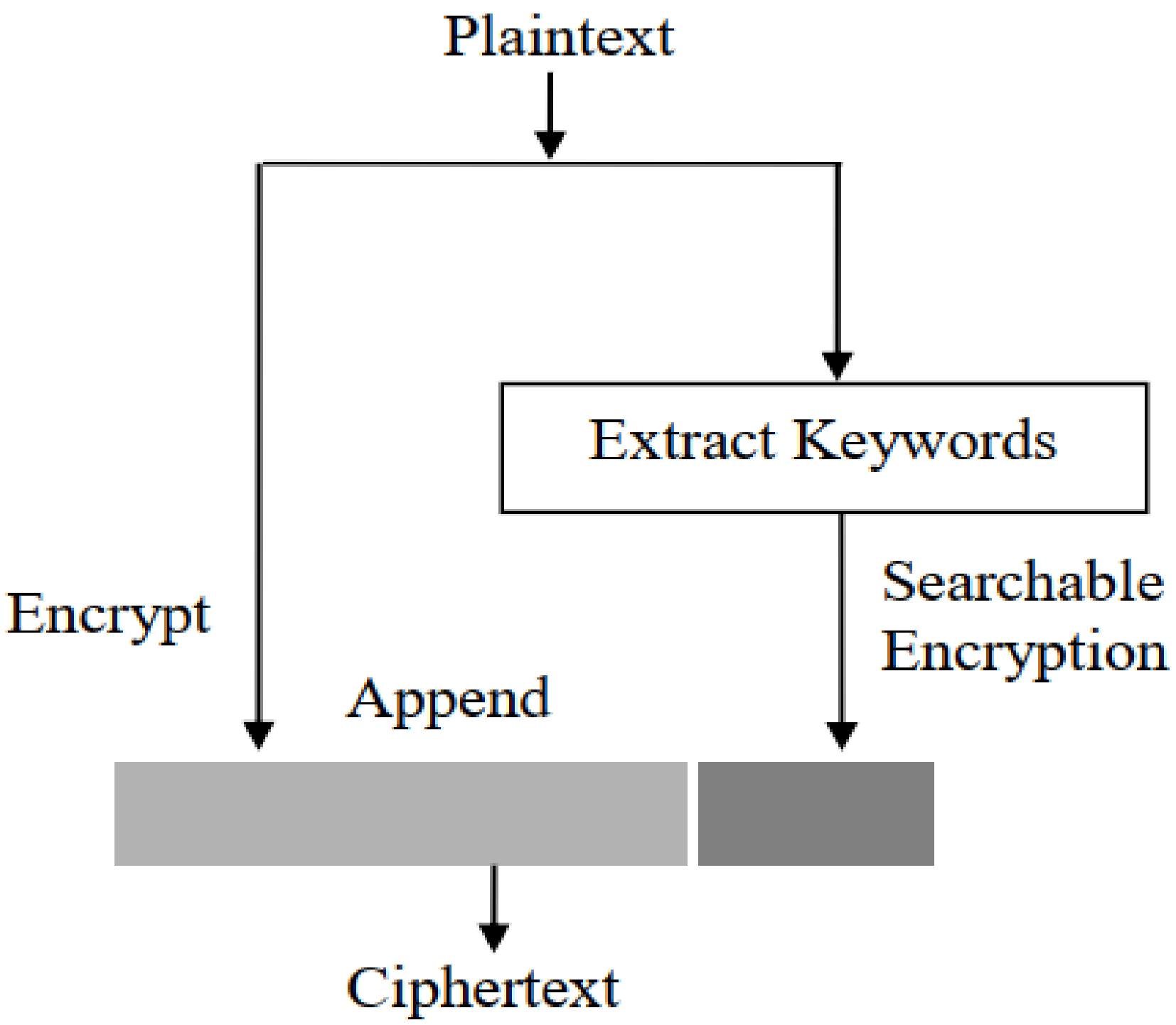

Figure 2.15 Keyword Based Scheme

Keyword-based searchable encryption is the mechanism by which data are encrypted to allow a user with a “trapdoor” for a keyword to efficiently retrieve some of the encrypted data containing the specific keyword over a remote server [61]. The scheme for searchable keyword encryption is considered a crucial building block that solves the security problems of privacy and data confidentiality in many settings, such as outsourced database systems and mail servers, files, etc.

### 2.8.1 Single Keyword Searchable Encryption

Single-keyword search allows users to search over encrypted data stored in the cloud server with only one keyword. The user query must contain precisely one keyword to generate the trapdoor. Then CS returns the result related to that keyword. There have been many works proposed based on a single-keyword searching mechanism. Some of these schemes are as follows.

Traditional single keyword searchable encryption schemes usually build an encrypted searchable index. Its content is hidden from the server unless it is given appropriate trapdoors generated via secret key(s). It was first studied by Song et al. [21] in the symmetric key setting, and improvements and advanced security definitions are given in Goh [15] and Curtmola et al. [26].

To improve the search efficiency, Chang and Mitzenmacher (2005) [62] proposed a scheme to retrieve the files by searching with the indexed keywords efficiently. In this scheme, they have used pseudo-random bits to mask the index keywords and sent them to the server. However, the entire database must be searched for a specific query, which increases computation overhead and also requires additional storage overhead. Further, Chase and Kamara [63] proposed schemes for performing queries on structured data. In this, search queries are applied to the labeled data. While performing encryption, the data becomes labeled data by padding the data elements to be of the same length. However, only one label is assigned to the entire database. Thus, it requires padding the entire result set to achieve security. But Cash et al. (2013) [64] used one label for each document that the keyword contains, i.e., If the keyword contains *k* documents, then it uses *k* different labels. It avoids making use of extra padding and also enables the parallel search.

All of the schemes discussed above are relevant to retrieving the search results based on an exact match. There is no advantage to typo mistakes and other kinds of minor mistakes made by the user.

Data owner outsources their *n* data files to the cloud. Before outsourcing the files to the cloud server, they encrypt their file for security purposes. Even though they are encrypted, they can also retain their ability to search for practical data usage. Before outsourcing their information, create an index *i* using a set of *k* distinct keywords extracted from the file collection *D*, and store both the index *I* and the encrypted file collection *D* on the cloud server. After this, on receiving search query *Tw* from the data user, the server is actually responsible for finding the index *i* and provides results without revealing the actual content of sensitive data. For data users, To search for a keyword *w* authorization is performed with trapdoor generation *Tw* and submitted to the cloud server.

However, a single-keyword search is used to search on the cloud server with only one keyword in the query. However, it is not suitable for real-time applications. It is because searching with only one keyword may not identify the required file accurately. So, the multi-keyword search was introduced to improve search efficiency and get accurate results.

### 2.8.2 Ranked Single Keyword Search

Ranked single keyword search generated the results rank-wise instead of just providing matched results. It can enable the quick search of the most relevant data. Sending back only the top-k most relevant documents can effectively decrease network traffic. In this technique, single random keywords are input to the cloud server.

Wang et al. (2010) [65] presented a secure ranked single keyword search over encrypted cloud data. They build a public-key searchable encryption scheme based on the inverted index. Their scheme served the high search efficiency inherited from the inverted index. It features a probabilistic trapdoor generation algorithm and protects the search pattern. To meet more robust security requirements, they strengthened their scheme with an efficient oblivious transfer protocol that hides the access pattern from the cloud. However, their performance and energy consumption would be a problem since their algorithm was complicated and needed many computing resources. The simulation results demonstrated that their scheme was less suitable for practical usage with moderate overhead. However, in their work, the terms are closely related to the files, which could lead to a potential information leak

Fu et al. [66] proposed an efficient solution for supporting ranked keyword search problems. In this technique, the single random keyword is the input to the cloud server, and the cloud server generates the most related file that matches the input keyword. Ranked keyword search generates the results rank-wise instead of just providing matched results. It will reduce the cost of searching and provide the most related results to improve the user experience.

Ranked search greatly enhances system usability by returning the matching files in a ranked order regarding certain relevance criteria (e.g., keyword frequency), thus making one step closer to the practical deployment of privacy-preserving data hosting services in Cloud Computing.

| Word | $w_i$ | | | | |
|---|---|---|---|---|---|
| File ID | $F_{i_1}$ | $F_{i_2}$ | $F_{i_3}$ | $\cdots$ | $F_{i_{N_i}}$ |
| Relevance Score | 6.52 | 2.29 | 13.42 | 4.76 | 13.80 |

Table 2.8 An example posting list of inverted indexes

A ranking function calculates relevance scores of matching files to a given search request. The most widely used statistical measurement for evaluating relevance scores in the information retrieval community uses the TF × IDF rule. The TF (term frequency) is simply the number of times a given term or keyword (we will use them interchangeably hereafter) appears within a file (to measure the importance of the term within the particular file), and IDF (inverse document frequency) is obtained by dividing the number of files in the whole collection by the number of files containing the term (to measure the overall importance of the term within the whole collection).

Ranked search greatly enhances system usability by returning the matching files in a ranked order regarding relevance criteria (e.g., Keyword frequency), so achieving the privacy-preserving data hosting service in the ranked keyword search method protects the relevance score of keywords from leaking the information about a keyword that integrate the new crypto primitive order-preserving symmetric encryption and appropriately modify it to protect the sensitive weight information. This technique provides some functionality.

1. It provides an effective protocol, which fulfills the secure ranked search functionality with little relevance score information leakage against keyword privacy.

2. Ranked searchable symmetric encryption scheme provides as-strong-as-possible security guarantee compared to previous Searchable symmetric encryption schemes.

### 2.8.3 Multi-Keyword Search

Multi-keyword search is a popular searching scheme in cloud applications, and it allows Users to search with multiple keywords in the query instead of only one keyword. With these multiple keywords, the efficiency of search, and accurate results can be achieved.

Xia et al. (2015) [67] presented a secure multi-keyword ranked search scheme over encrypted cloud data, which simultaneously supports dynamic update operations like deletion and insertion of documents. Specifically, the vector space model and the widely used TF*IDF weightage model are combined in the index construction and query generation. They construct a particular tree-based index structure and propose a "Greedy Depth-first Search" algorithm to provide an efficient multi-keyword ranked search. The secure KNN algorithm is utilized to encrypt the index and query vectors and ensure accurate relevance score calculation between the encrypted index and query vectors.

Xu et al. (2019) [68] proposed an authorized searchable encryption scheme under a multi-authority setting. The presented scheme leverages the RSA function to enable each authority to limit the search capability of different clients based on clients' privileges. After the server implements a multi-keyword search on all ciphertexts, it returns the results to the receiver. The receiver performs a more accurate keyword search on these results. To improve scalability, they utilized multi-authority attribute-based encryption to allow the authorization process to be performed only once, even over policies from multiple authorities. However, it fails to support the data sharing of encrypted data.

In [69], Zhang et al. proposed a scheme to deal with secure multi-keyword ranked search in a multi-owner model. Different data owners use different secret keys to encrypt their documents and keywords in this scheme. In contrast, authorized data users can query without knowing the keys of these different data owners. The authors proposed an "Additive Order Preserving Function" to retrieve the most relevant search results. However, these works do not support dynamic operations.

### 2.8.4 Ranked Multi-Keyword Search

Ranked search greatly enhances system usability by returning the matching files in a ranked order regarding certain relevance criteria (e.g., keyword frequency), thus making one step closer to the practical deployment of privacy-preserving data hosting services in Cloud Computing which is shown in figure 2.16[21].

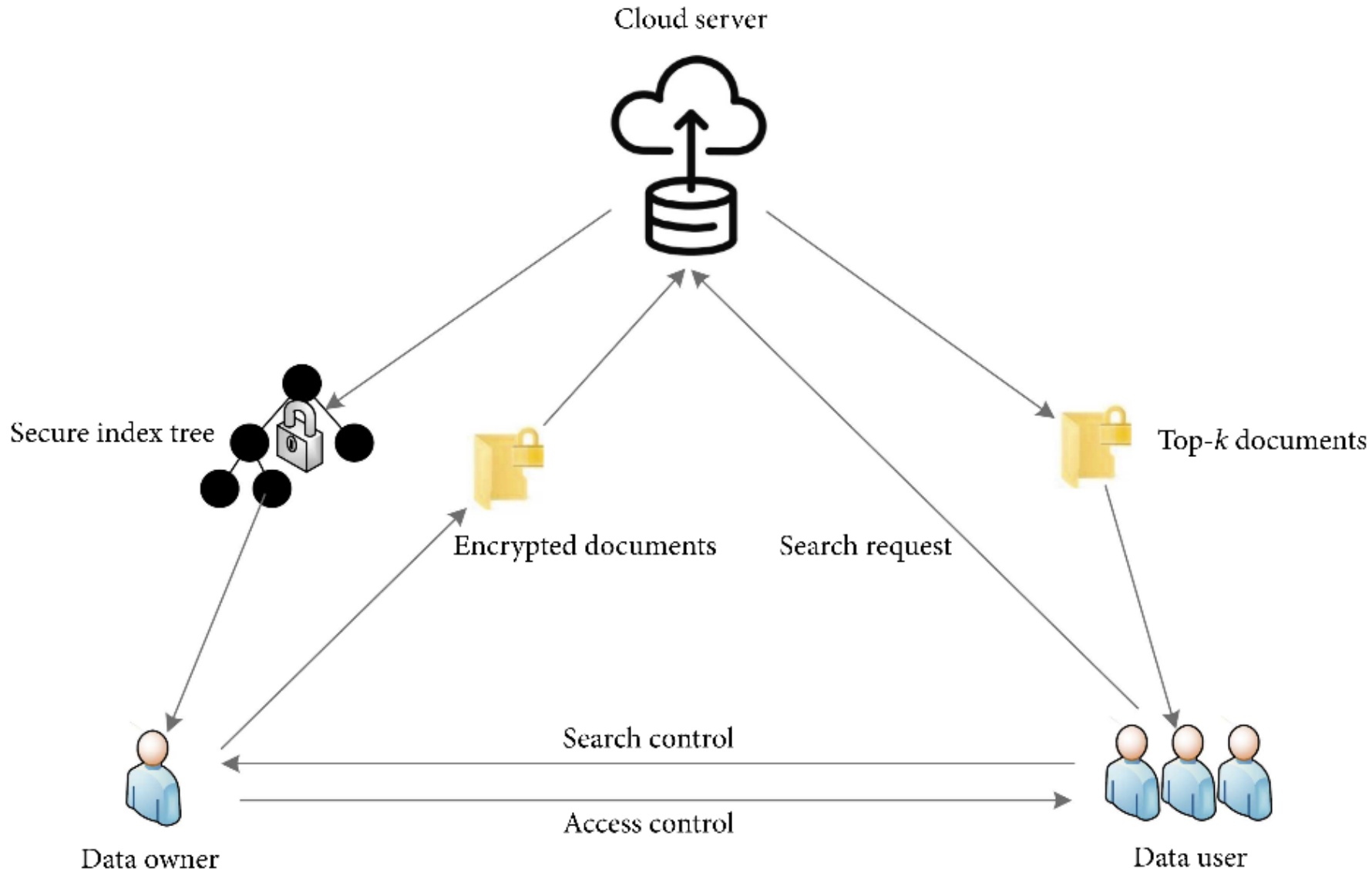

Figure 2.16 Ranked Multi-Keyword Search Scheme

As multi-keyword search can provide more accurate search results, some works are concentrated on the issue of multi-keyword encryption search in the symmetric setting. Cao et al. (2013) [70] defined and solved the challenging problem of the privacy-preserving multi-keyword ranked search over encrypted data (MRSE) in cloud computing. They established strict privacy requirements for such a secure cloud data utilization system. Among various multi-keyword semantics, they have chosen the efficient similarity measure of "coordinate matching," i.e., as many matches as possible, to capture the relevance of data documents to the search query. They further use "inner product similarity" to evaluate such similarity measures quantitatively. They first proposed a basic idea for the MRSE based on secure inner product computation. They then significantly improved MRSE schemes to achieve various stringent privacy requirements in different threat models. However, the scheme does not consider the keyword weight within the document, which makes the search result not accurate enough.

Li et al. (2020) [71] proposed an efficient and privacy-preserving Multi-keyword Ranked Search scheme with Fine-grained access control (MRSF). MRSF can realize highly accurate ciphertext retrieval by combining coordinate matching with Term Frequency-Inverse Document Frequency (TF-IDF) and improving the secure KNN method. Besides, it can effectively refine users' search privileges by utilizing the polynomial-based access strategy. Formal security analysis shows that MRSF is secure regarding confidentiality of outsourced data and the privacy of index and tokens.

[21] Dai, H., Dai, X., Li, X., Yi, X., Xiao, F. and Yang, G., 2020. A Multibranch Search Tree-Based Multi-Keyword Ranked Search Scheme over Encrypted Cloud Data. *Security and Communication Networks*, *2020*.

In this keyword search, the documents are retrieved by considering the weights of the keywords. Further, these weights are assigned by using the TF ∗ IDF ranking model. Here, TF is a term frequency in a specific document, and IDF is an inverse document frequency. IDF can be calculated as:

$$\mathrm{IDF}_t = \log \frac{N}{\mathrm{DF}_t}$$

Where N is the number of documents and $\mathrm{DF}_t$ is the number of documents having the term t. The similarity score between document $D_j$ and query $Q$ can be calculated as:

$$\mathrm{Score}(D_j, Q) = \frac{1}{|D_j|} \sum_{w_i \in Q} \mathrm{TF}_{i,j} . \mathrm{IDF}_i$$

$$|D_j| = \sqrt{\sum_{w_i \in D_j} (\mathrm{TF}_{i,j})^2}$$

Where $|D_j|$ denotes the Euclidean length of the document $D_j$, and $w_i$ is the keyword in document $D_j$.

Ding et al. (2017) [72] investigated the multi-keyword top-*k* search problem for big data encryption against privacy breaches and attempted to identify an efficient and secure solution to this problem. Specifically, for the privacy concern of query data, they construct a special tree-based index structure and design a random traversal algorithm, which makes even the same query produce different visiting paths on the index and maintain the accuracy of queries unchanged under stronger privacy. To improve the query efficiency, they proposed a group multi-keyword top-*k* search scheme based on the idea of partition, where a group of tree-based indexes is constructed for all documents. Finally, they combine these methods into an efficient and secure approach to address our proposed top-*k* similarity search. Note that multi-keyword is potentially the future mainstream encrypted search scheme with higher searching accuracy. However, current ongoing research cannot provide an authoritative method.

### 2.8.5 Conjunctive Keyword Search

Conjunctive Keyword search (CKS) allows users to search with different keywords to produce individual search results and then intersects all individual results to generate a final result. This final result contains documents related to the user search. Conjunctive keyword search returns "all-or-nothing," which means it only returns those documents in which all the keywords specified by the search query appear.

With a conjunctive keyword search, allow a client to find documents containing all of several keywords in a single query (i.e., a single run over the encrypted data). Building a conjunctive

keyword search scheme from a single keyword scheme in a naive way provides the server with a trapdoor for each keyword [73]. The server searches for each of the keywords separately and returns the intersection of all results. This approach leaks which documents contain each keyword and may allow the server to run statistical analysis to deduce information about the documents and keywords.

The first scheme for conjunctive keyword-searchable encryption was introduced by Golle et al. in [29]. The authors provided two constructions in a symmetric key setting. Their first construction is based on the Decisional Diffie-Hellman (DDH) assumption. It has proven secure in the random oracle (RO) model. Unfortunately, the drawback of the scheme is that the size of trapdoors is linear in the total number of data stored on the server. This makes the use of their scheme impractical in many settings.

Their idea for conjunctive searches is to assume particular keyword fields are associated with each document. Emails, for example, could have the keyword fields: “From,” “To,” “Date,” and “Subject.” Using keyword fields, the user must know in advance where (in which keyword field) the match has to occur. The communication and storage cost linearly depend on the database's number of stored data items (e.g., emails).

### 2.8.6 Fuzzy Keyword Search

This was introduced over encrypted data in the cloud to make the search process more user interactive. Fuzzy keyword search allows users to search with minor mistakes in the keyword. This search scheme returns the documents of matched keywords. If the keyword has some minor typos, it returns the closest documents to the keyword by using keyword similarity semantics.

Initially, the concept of searching using fuzzy keywords in encrypted cloud data was given by J. Li. et al. (2010) [74]. They stated that the search system used in this method could give an accurate result even if the user slightly misspells the keyword. This technique attempted to make the search procedure user interactive. On the other hand, in traditional techniques, no result is found when there are minor errors in the spelling of keywords entered. Hence, it makes the user’s task very complicated. To handle this problem, they implemented fuzzy keyword searching. It also focused on preserving the privacy of keywords. If the user spells incorrectly, integrate edit distance with a wildcard-based technique to build fuzzy keyword sets to address minor misspelling issues and format inconsistency. Using this method calculates the closest matching keyword.

Fuzzy keyword search greatly enhances system usability by returning the matching files when users’ searching inputs match the predefined keywords or the closest possible matching files based on keyword similarity semantics when the exact match fails.

### 2.8.7 Semantics Keyword Search

The search results are purely based on user authentication, and only authorized users can input the search request. It increases the flexibility of search when the user forgets the exact keyword. The user can search by using some of its synonyms. Zhangjie Fu et al. (2013) [75] suggested a synonym-based multi-keyword search system in encrypted data in the cloud. This is the first method suggested based on semantics.

Apart from other methods, the main difference in this method is that here, the keyword set is extended by adding some of its synonyms. Also, that keyword needs to be extracted from the file collection before outsourcing it to the cloud. Here uses a better text feature weighting method, which adds a new component module to indicate the distinguishability of the term based on the original TF-IDF (term frequency-inverse document frequency) method. The new element *Cd* has been added to the equation of TF-IDF,

$$\text{Weighting factor} = TF * IDF * Cd$$

The keyword set should be extended by adding synonyms to accomplish an efficient, meaningful search for outsourced data. If a keyword has more than two synonyms, then all synonyms are added to the keyword set. So, using this improved method will extract keywords from outsourced text files. All keywords separated from a single text form a keyword set at the end. The redundant keywords are deleted to reduce the burden of storage.

## 2.9 Security Requirements for Searchable Encryption

There are three main research directions in Searchable Encryption:

- Improve the efficiency
- The security, and
- The query expressiveness.

Efficiency is measured by the computational and communication complexity of the scheme. A variety of different security models have been proposed to define the security of a scheme formally. Since security is never free, there is always a tradeoff between security on the one hand and efficiency and query expressiveness on the other. Searchable encryption schemes that use a security model with a more powerful adversary are likely to have a higher complexity. In a searchable encryption scheme, the security of the documents and keywords stored on the server should be guaranteed. In addition, the security of query keywords should also be assured.

As an encryption system, a searchable encryption system must meet some security properties. It is an encryption system; hence it should guarantee data privacy, i.e., ciphertext doesn’t reveal any private information of the data plaintext. Besides this, there are other security requirements as follows:

- **Search pattern**

Search pattern refers to the information that can be derived in the following sense: given that two searches return the same results, determine whether the two searches use the same keyword or predicate. A search query induces it. The search pattern of a query for keyword $W$ is defined as whether a data file contains this keyword. This definition is only about the relation of inclusion but is irrelevant to the information of keyword and index. Using deterministic trapdoors directly leaks the search pattern. Accessing the search pattern allows the server to use statistics.

- **Access pattern**

The access pattern is defined as the information extracted from knowledge of a sequence of search results that contains the keyword. For example, one query can return a document *x,* while the other could return x and ten documents. This implies that the predicate used in the first query is more restrictive than in the second query.) It is defined as a sequence of search results $(D(w_1),\ldots, D(w_n))$, where $D(w_i)$ is the search results of $w_i$. In other words, $D(w_i)$ is a collection of documents in D that contains the keyword *w*. Here, the adversary could further deduce the private information of the index and the trapdoor from the access pattern. To avoid such leakage, the search results of queries must be indistinguishable.

- **Keyword Privacy**

In a searchable encryption system, the privacy of the keyword is a fundamental requirement, so Keywords in the index are encrypted as ciphertext. This property guarantees that an attacker cannot guess which keyword the user is searching. Similar to a cryptosystem, privacy can resist Chosen Keyword attack (CKA).

- **Secure Channel**

A searchable encryption scheme needs a secure channel to guarantee security. The main concept is to use a secure channel between the data owner and to cloud service provider while uploading the encrypted document. The key distribution method between the Data Owner and the Data User (in the PESK scheme) also should be protected by sharing through a secure channel.

- **Key revocation**

Key revocation is a key feature which associates with security. Key revocation is an important security property for managing users' authorities. In an information system, users usually join or leave a user group that has some permission in the access control system. Suppose a user has permission to search on the encrypted files stored in a storage system; his authority should be canceled in the system when he leaves this system for some reason; that is to say, his private key should be revoked. Then he has no longer the permission to search. Key revocation can revoke users' search authority when these users are no longer permitted to execute keyword searches. This is an effective measure to control users' authority.

- **Index Privacy**

The index privacy is twofold. First, the cloud server should not learn the content of the index since the content of the index directly reflects the content of the documents. Second, the cloud server should deduce no information about the document through analyzing the encrypted index. Such information includes 1) whether a document contains a particular keyword(s) and 2) whether different documents contain a common keyword. Concerning index privacy, if the cloud server deduces any association between keywords and encrypted documents from the index, it may learn the significant subject of a document, even the content of a short document [76]. Therefore, the searchable index should be constructed to prevent the cloud server from performing such kind of association attack.

- **Keyword privacy**

Users usually prefer to keep their search from being exposed to others like the cloud server. The most critical concern is to hide what they are searching for, i.e., the keywords indicated by the corresponding trapdoor. Although the trapdoor can be generated cryptographically to protect the query keywords, the cloud server could do some statistical analysis of the search result to make an estimate. As a kind of statistical information, document frequency (i.e., the number of documents containing the keyword) is sufficient to identify the keyword with high probability. When the cloud server knows some background information about the data set, this keyword-specific information may be utilized to reverse engineer the keyword.

- **Trapdoor Privacy**

A trapdoor is generated for each query request to allow the cloud server to search over the encrypted index. Intuitively, the trapdoor contains the query information but in an encrypted form. Given a trapdoor, the cloud server should learn nothing about the user's query from it.

- **Trapdoor Unlinkability**

The trapdoor generation function should be a randomized one instead of being deterministic. In particular, the cloud server should not be able to deduce the relationship of any given trapdoors, for example, to determine whether the same search request forms the two trapdoors. Otherwise, the deterministic trapdoor generation would give the cloud server advantage of accumulating frequencies of different search requests regarding different keyword(s), which may further violate the keyword mentioned above privacy requirement. So. the fundamental protection for trapdoor unlinkability is to introduce sufficient non-determinacy into the trapdoor generation procedure.

## 2.10 Performance

The objective of the searchable encryption would be to develop a solution that minimizes the computation on both server and user ends and that reduces communication overhead when designing a solution for the problem of searching for encrypted data. One of the primary goals is to lower the computation at the user end since he/she is paying for the services of the server. Here, the server would take most of the work in computation and perform the search. The searchable encryption performance depends on many criteria, which are discussed below.

- **Decrease Encrypted File Size**

For better performance, the scheme should follow the goal is to reducing the encrypted file size on the server. However, it should be aware of the fact that as the encrypted file size decreases, the number of output bits from the cryptographic encryption scheme reduces, therefore lowering the randomness and the security of the scheme. Moreover, an elaborate and full search capability might require word-by-word encryption, resulting in increased memory usage.

- **Index creation time**

Index plays a crucial role in searchable encryption as the search operation is performed on the index rather than on document contents. (58… Index creation time is based on the number of keywords from all the documents. Efficient index construction guarantees that the search process takes less amount of time. There are different ways to create an index with the keywords of documents. An inverted index is a popular technique to construct the index. The vector space model is a widely used technique in multi-keyword search scenarios. Tree-based index construction improves efficiency in the search process.

- **Search time**

Searching plays a significant role in a cloud environment as the latency with which the documents have retrieved the usability of the cloud, i.e., more is the time required, more is the financial burden on the user, and less will be the cloud usage. Search time depends on search functionality. The efficiency of any Searchable Encryption scheme depends on search time; thus, this computational measurement becomes an important metric. The schemes with the inverted index require $O(r)$ time, where r is the number of documents matched with the search keyword. On the other hand, schemes with a tree-based index require $O(\log n)$ time, where n is the total number of documents. Hence, search time should be efficient when designing SE schemes.

- **Query accuracy**

The size of the query depends on the way the index is generated, and the search is performed. The query size is proportional to the number of keywords in the dictionary, i.e., $O(M)$. Thus, the time to generate a query is approximately $O(1)$. On the other hand, the query can be generated at a fixed-size length, i.e., $O(r)$. Thus it is important to generate the query with much accuracy so that the search result would be more accurate and efficient.

- **Low query latency**

Although searching encrypted data is fundamentally more complicated than searching public data, queries against large data sets should take on the order of seconds, not minutes or hours.

- **High server throughput**

The system should be capable of processing many queries concurrently without significantly increasing the number of search servers.

- **Support for Boolean queries**

The query language must be expressive enough to allow users to issue precise queries. We dismiss regular expression searches and focus on more straightforward boolean queries as with most search engines. In practice, we focus on the operator; or and not are relatively trivial.

- **Competitive search accuracy**

Information retrieval systems employ a wide variety of heuristics based on query proximity and frequency within documents, overall keyword frequency, and other metrics to identify the most relevant matches. For a large data set, the searchable encryption scheme must be capable of supporting enough to produce accurate search results.

## 2.11 Applications of Searchable Encryption

This section presents real-world situations where the Searchable Encryption schemes can be applied. Searchable encryption has great potential for use in many applications where confidential data is outsourced to the third-party service provider without affecting the usage of the data stored in the cloud [30]. Hence, searchable encryption is an important domain with great potential for applicability in many applications where confidential data are outsourced to third-party servers and require retrieval later. Apart from applying searchable encryption to secure the outsourced personal documents, some of the applications of SE are discussed below in Figure 2.17[22].

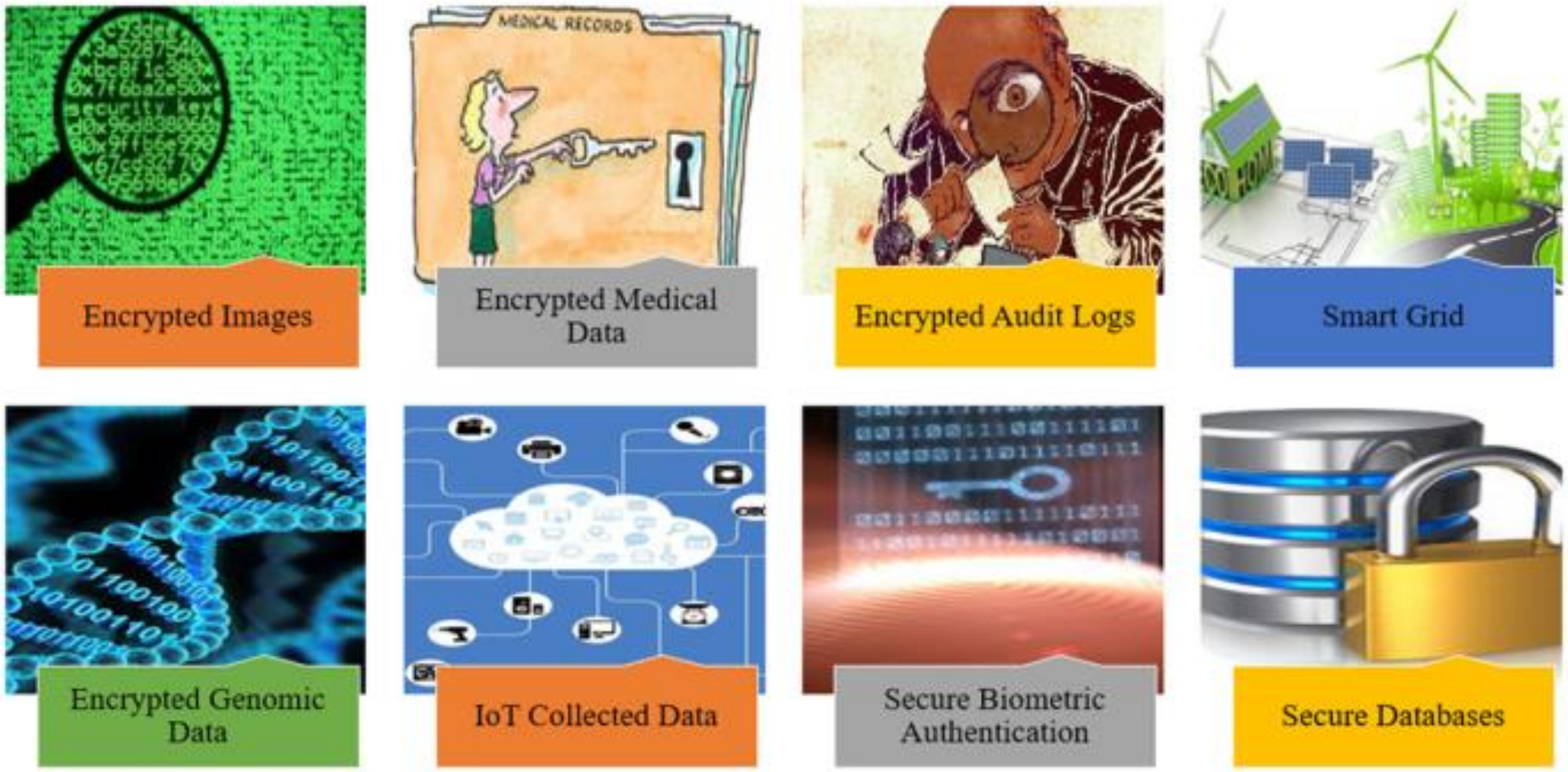

Figure 2.17 Application of Searchable Encryption

### Encrypted images

Searching for an image of interest from extensive data collection is a resource-intensive job that is used in online shopping, criminal/suspect identification, disease detection, medical images of patients, etc. Outsourcing images directly over the public cloud is a security challenge and hinder images' privacy. Thus, similar to text/documents, images are also encrypted before outsourcing. This leads to the extension of SE schemes to encrypted images, which various researchers' study. In most schemes, image descriptors are generated using bag-of-features representation, Fisher vector, or enhanced Fisher vector. The searchable index so generated is encrypted. The search capability is provided using various schemes, such as OPSE and Min-Hash algorithm, homomorphic encryption, and inner product similarity.

---

[22] Handa, R., Krishna, C.R. and Aggarwal, N., 2019. Searchable encryption: a survey on privacy-preserving search schemes on encrypted outsourced data. *Concurrency and Computation: Practice and Experience*, *31*(17), p.e5201.

### Encrypted audit log

Audit logs can be a critical entry point for an adversary to extract confidential information stored in the database/server. For example, an audit log containing information about the actions performed by the users can violate the user's privacy. In contrast, access to results returned to a user by audit records can violate access control policies. Thus, securing audit logs is also required for which encryption is used. Waters et al. [77] studied the problem of searching encrypted audit logs based on identity-based encryption. The server, which builds the log records, encrypts the entries using the public key. In contrast, the corresponding private key is generated using the master secret key. Furthermore, Schneier and Kelsey [78] proposed a scheme to avoid deletion or update audit logs using check-pointing, where the keys are chained. The new key generated to encrypt an audit record is based on the previous key. Hence, only a trusted party with the previous authentication keys can authenticate the records.

### Encrypted smart grid data

With advancements in the Internet of Things (IoT), smart meters are deployed in households, periodically collecting information about the usage of electricity and other information and outsourcing it to the cloud server. Maintaining the security and privacy of these data is of importance. By utilizing the consumption information, it is possible for an adversary to find the appliances used in a household or determine if a person is at home. The data being outsourced to the cloud are encrypted, making SE applicable to smart grid data to provide privacy and security; thus, privacy-preserving search ability using encryption extended for smart grid data, and various studies evolved.

### Encrypted genomic data

To perform analytics on genomic data, i.e., to find patterns and correlations, researchers require volunteers to share their genome information, which exposes them to privacy risks. The data were earlier anonymized, but there are still chances of re-identifying individuals to overcome the privacy risk. Hence, an alternative is to utilize encryption and then use searchable encryption to extract the information of interest. Various researchers have proposed schemes where the encrypted genomic data can be searched, and computations can be performed efficiently. Çetin et al. [79] proposed a scheme to perform a search over a homomorphically encrypted genomic sequence to identify the presence of a disease. Zhao et al. [80] proposed the concept of side-wise encryption to allow the searching on encrypted genomic data. Jha et al. [81] utilized edit distance and computed a Smith-Waterman similarity score between two sequences to perform computations on genomic data.

### Encrypted health care data

Telemedicine refers to the delivery of health care services through information and communication technologies. The services offered under telemedicine include consultancy to patients through video conferencing, storing medical records online, e-health services, patient monitoring, wireless applications, and nursing call centers. In EHR systems, the details of a patient are entered, which are utilized by the doctors to diagnose the disease. These EHRs can be shared with other doctors for consultancy. The users (i.e., patients) realize that the control over the medical data is lost as soon as these data leave the premises. Thus, encryption is preferred to maintain medical data privacy, which introduces barriers to searching the records and sharing them with authorized users. To overcome this, various researchers have worked in this field and proposed schemes to provide searching and sharing capabilities of medical data; for instance, Li et al. [82] utilized attribute-based encryption (ABE) to provide searchability for personal health records.

### IoT collected data

Various *m*-Health monitoring systems are developed (such as MediNet) to monitor the health of the person using wireless sensor networks. The data collected by these devices include blood pressure, breathing rate, ECG, blood glucose, etc. These data are forwarded to the central server. Various analyses are performed to get insights into the data, such as sleep pattern analysis, physical exercise assistants, etc. Various insurance companies have started offering insurance discounts based on the physical exercise of the individual. These companies utilize different wristbands' data captures and offer discounts and loyalty points. In all these applications, data privacy can be breached during storage, transmission, diagnosis, computing, etc. There is a possibility that many companies may share the data with insurance companies, research institutions, and government agencies for financial gains. Encryption is preferred, which motivates researchers to propose efficient SE schemes to address these issues.

### Secure biometric authentication

Searchable Encryption finds its application in secure biometric authentication. Storing biometrics in clear form can lead to privacy issues as the biometrics cannot be changed like passwords in the case of biometric theft. To handle this, the biometric details are encrypted and then stored. Encryption is a possible method for maintaining privacy. The biometric details are encrypted and then stored to handle this. Encryption is a possible method for maintaining privacy. However, there should be an efficient method to recover the biometric for comparison and provide authentication. Moreover, the biometrics scanned are not the same every time due to added noise. Hence, the hash value for slightly different biometrics will be different. Thus, handling biometric authentication is a challenge that various researchers deal with. Cheon et al. [83] proposed a scheme for biometric authentication using homomorphic encryption and message authentication code. Zhang et al. [84] proposed a scheme based on PEKS for biometric authentication.

### Secure databases

One method to provide security of the data stored in databases is by utilizing the cryptographic support available in the database systems. This can provide security from outsiders as the encryption and decryption of the data are handled by DBMS. However, to provide data security from insiders, encryption is required where the owner of the data independently encrypts the data before outsourcing. This prevents unnecessary overheads of encryption and decryption by the DBMS. Thus, one domain where searchable encryption can be extended is secure databases. It involves the execution of queries on remote databases such that the privacy of the data is retained along with secure data-sharing capabilities. It can be considered an extension of searchable encryption. Secure databases permit the parties to securely search and retrieve the relevant data from the database without revealing the remaining data. An example of such a scenario is retrieving data from census data.

Pappas et al. [85] proposed Blind Seer (BLoom filter INDex SEarch of Encrypted Results) to provide secure searches (such as range queries, Boolean queries, negations, etc.) along with database updates using the MySQL database. They utilized search trees to store the keywords using Bloom filters, where leaf nodes refer to the records stored in the database. An exciting area of research within a secure database is providing data confidentiality from malicious administrators.

An interesting area of research within a secure database is providing data confidentiality from malicious administrators. A solution in this direction is presented in the CryptDB system [86]. However, it utilizes a fully trusted proxy for query encryption, which is a matter of concern as the existence of such a trusted proxy in a real-world scenario is questionable.

# 3. CHAPTER THREE: LITERATURE REVIEW

Searchable encryption schemes enable users to store the encrypted data in the cloud and execute keyword searches over the ciphertext domain. Over the past recent years, encrypted search has evolved toward the ability of data sharing with the protection of users' privacies. So far, much work has been proposed under different threat models to achieve various encryption techniques and search functionalities, such as single keyword search, index-based search, attribute-based encryption, identity-based encryption, public-key encryption, etc.

This chapter is discussed several papers to discuss the background and the related works, including searchable encryption, methods, schemas, and performance over the cloud storage system. Since then, there has been much work conducted in this field. We selected and explained simply various techniques focused on secure keyword search in cloud environments. This chapter has studied some recent search techniques on large-scale encrypted data in the cloud. We also identify their strength and weakness. The rest of the report demonstrates how our scheme overcomes those weaknesses.

## 3.1 Secured File Storing and Sharing System

Most cloud service providers keep multiple copies of files to ensure reliability to their users. Nevertheless, this becomes a severe privacy issue when the replicated copies remain hidden in the server even after the owner has deleted the original file. So before searching for encrypted data, we must ensure to store and share the files securely over the cloud. A Prototype of a Secured File Storing and Sharing System for Cloud Storage Infrastructure titled "Crypted Cloud" was proposed by Patwary and Palit (2019) [19]. They have designed a system using the existing cryptographic approaches, which ensures assured deletion of files, resolving all the limitations that remained unpatched by the previous models File Assured Deletion (FADE), Simplified File Assured Deletion (SFADE), and SFADE+. The primary focus of their proposed model was to ensure user privacy and security through a user-friendly system that is less computationally intensive and does not rely upon any trusted entity. They have implemented a prototype named CryptedCloud as an overlay system on top of Google Drive, as shown in figure 3.1. It factually exhibits that CryptedCloud assures file deletion without relying upon any trusted body with no compromise to the performance and user-friendliness.

Their model consists of two distinct components; a client application and a server. The client-side application will handle all the complex cryptographic operations and network communication with both server and cloud storage services. On the other hand, the server is connected to a database that stores and retrieves information. By completing the sign pp process with email, passphrase, and recovery information, the user needs to sign in to use the service. After signing in, the user can perform operations such as upload, download, share, and delete files from cloud storage. Here, we discussed only the upload and share process as these are similar to our proposed model.

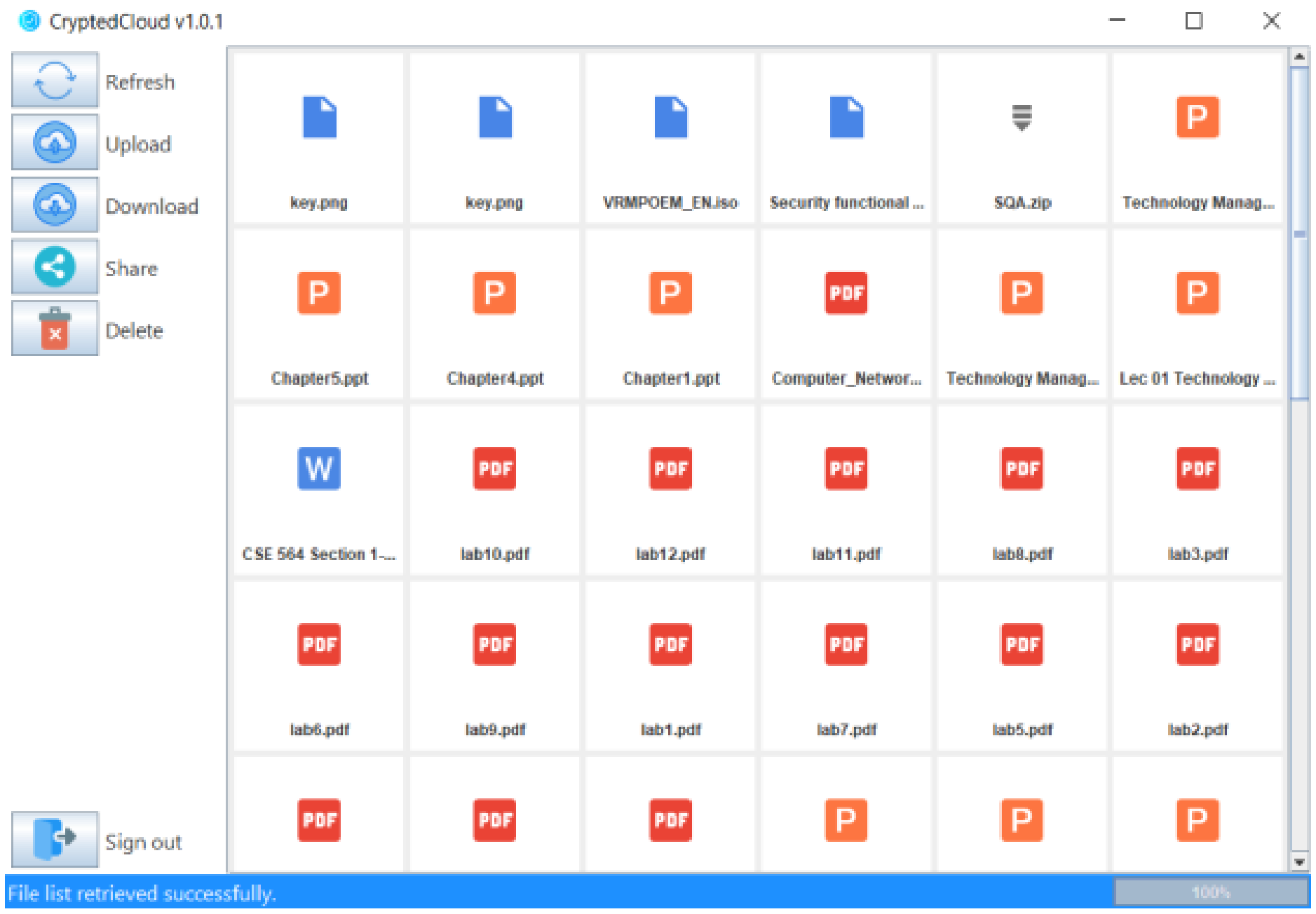

Figure 3.1 Basic User Interface of Crypted Cloud Client Model

## Upload

They have added a layer of encryption before uploading the file to the server. This will ensure that nobody but the uploader can access the file. The upload operation is described below:

1. When the user requests to upload a file, the client application generates a random key (R). The random key is way too vital to be guessed. It is a variable-length key of 256 bit variant of the AES algorithm.
2. The random key (R) is used to encrypt the selected file (F). To achieve stronger encryption, we have decided to use.
3. The encrypted file (F') is uploaded to the cloud storage.
4. After receiving the encrypted file (F') successfully, the cloud service provider will send a unique id (Cf) for that file as a response.
5. The random key (R) is encrypted by the user's passphrase (Ph).
6. Lastly, the unique id provided by the cloud service (Cf), the user's email (E), the encrypted random key (R') and one small piece of information sent by the client application, user's role (Ro) are sent to the server. Thus, the upload process is completed.

The entire file upload process is described using a flow diagram in Figure 3.2.

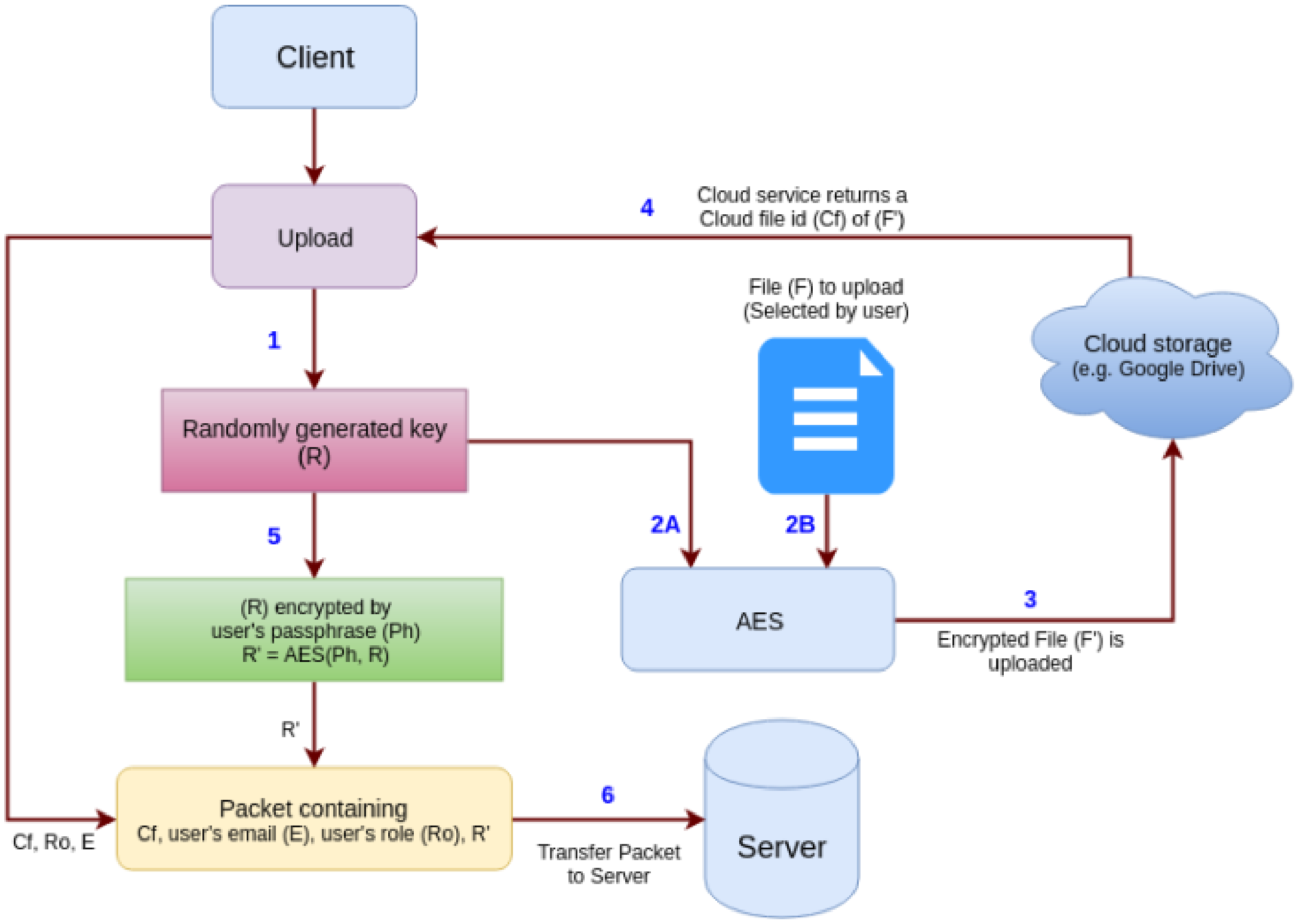

Figure 3.2 Upload process flow diagram

**Share**

We are discussed the entire file sharing process flow step by step. Here, we assume that the Data Owner is User 1 and the Data User is User 2.

1. At the beginning of the file sharing process, to share a file with User 2, User 1 needs to provide User 2's email ($E_2$) and access role ($RO_2$).

2. The client application receives the user's input and takes the User's email address ($E_2$) into account.

3. The client application will communicate with the server to check if the provided email address ($E_2$) is associated with any account.

4. Server will check if User 2 is registered.

5. If no user account is associated with ($E_2$), an email will be sent to the email address asking to register in our service to receive the file. Nevertheless, if User 2 is registered in the service, the client application will request the server for encrypted file information (Fi') using the cloud file id (Cf).

6. Server will provide the encrypted file information (Fi') to the client application.

7. The client application will then check if User 1 is the file's owner (F').

8. If User 1 is not the owner, sharing will fail. Otherwise, the client application will decrypt the encrypted random key (R') using user 1's passphrase (Ph) to retrieve the plain random key (R).

9. Plain random key (R) will then be re-encrypted by user 2's public key (Pu2)

10. In the end, a packet is formed combining the email address of user 2 (E2), re-encrypted random key (R''), cloud file id (Cf), and user 2's access role ($RO_2$) and sent to the server. The server stores this packet in its database so that it can be used by user 2's client application while downloading the file (F').

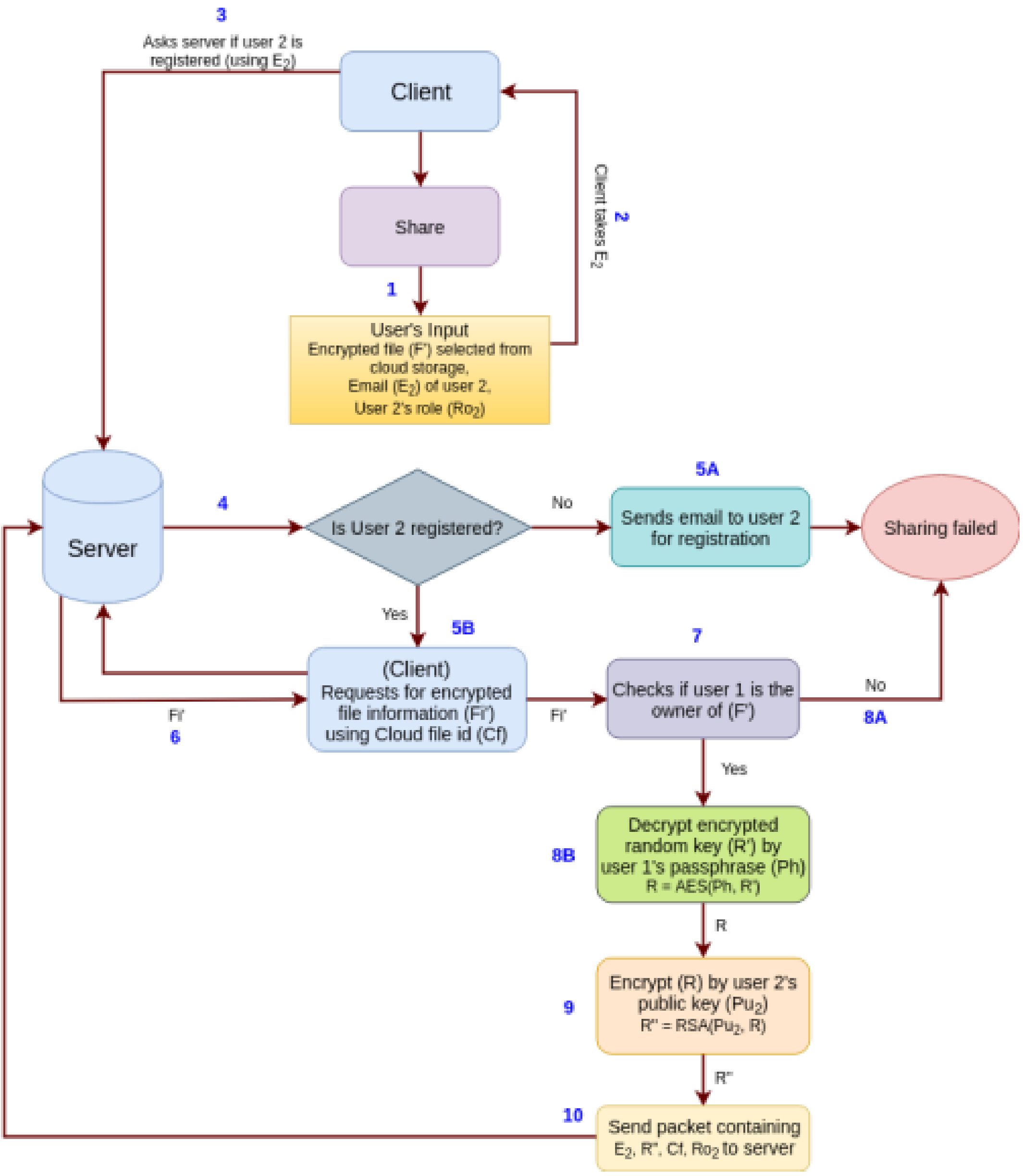

Figure 3.3 The sharing process flow model

According to the efficient data storing and sharing process, there are some limitations of their prototype. The model does not guarantee data integrity. If somehow data gets changed in cloud storage, there is no way to know if the data was modified in the cloud storage. As the model added an additional layer on top of the cloud storage service, it adds a risk of becoming the second point of failure and reduced performance due to cryptographic operations. Although the server does not require to be a trusted entity, the contents stored in the database may be altered by the server admin or an intruder through direct access to the database. There will not be any security breach, but the user will surely lose access to the file.

## 3.2 Searchable Symmetric Key Encryption (SSKE)

To our best knowledge, the first practical solution to the problem of searching on encrypted data was introduced by Song et al. in [21]. Their solution has two approaches: performing a sequential scan without an index (non-keyword-based) and performing a search with an encrypted index (keyword-based).

First, in the non-keyword-based approach, they considered a document a list of words of the same length and used a specially designed stream cipher to encrypt the document. Thus, it is allowed to search on the ciphertext directly. Search is performed by sequentially scanning the entire ciphertext. The underlying idea of this scheme is that the ciphertext is obtained by XORing each of the keywords in the plaintext with a sequence of pseudorandom bits. They introduced a basic scheme, which provides provable secrecy. Another name for provable secrecy is “reductionist security,” meaning that one provides a reduction of the security of the scheme to the security of some underlying primitive [87]. This means that as long as the underlying primitives are secure, the scheme composed of these primitives will be secure. Then, three configurations of this scheme are provided to achieve the basic control searching, support for hidden searches, and final query isolation, respectively. The schemes and the properties that they achieve are discussed below.

**Scheme I: The Basic Scheme**

The data owner is considered encrypting a document containing the sequence of words $W_1, \ldots, W_n$. Intuitively, the scheme worked by computing the bitwise exclusive or (XOR) of the clear-text with a sequence of pseudorandom bits with a special structure. This structure allowed to search on the data without revealing anything else about the clear text. More specifically, the data owner has generated a sequence of pseudorandom values $S_1, \ldots, S_n$ using some stream cipher (namely, the pseudorandom generator $G$), where each $S_i$ is n – m bits long. To encrypt an n-bit word $Wi$ that appears in position $i$, the data owner has taken the pseudorandom bits $S_i$, sets $T_i = < S_i, F_{ki}(S_i)>$ , and outputs the ciphertext $C_i = W_i \oplus T_i$. Note that only the data owner can generate the pseudorandom stream $T_1, \ldots, T_n$. so no one else can decrypt. There was some flexibility in how the keys $K_i$ may be chosen.

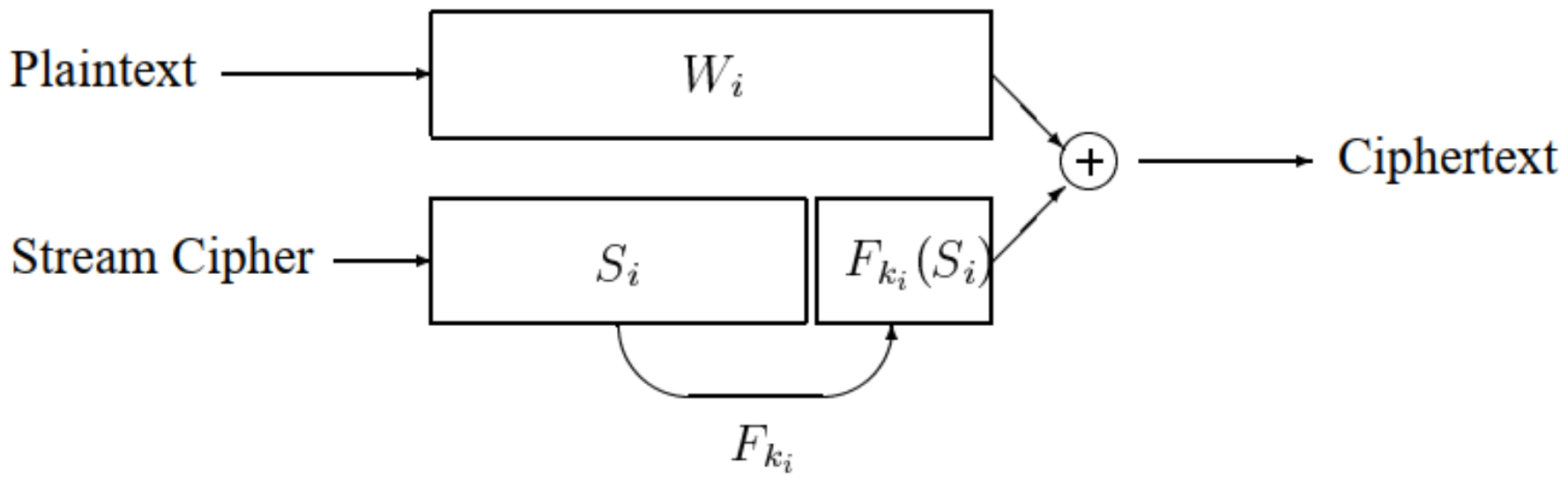

Figure 3.4    The Basic Scheme of SSKE

One possibility was to use the same key $K$ at every position in the document. Another alternative was to choose a new key, $K_i$, for each position independent of all other keys.  The basic scheme provides provable secrecy if the pseudorandom function F and the pseudorandom generator G are secure. By this, they meant that, at each position where $K_i$ was unknown, the values $T_i$ are indistinguishable from genuinely random bits for any computationally-bounded adversary.

**Scheme II: Support for Hidden Searches**

If the data owner is wanted to search for a word $W$ to the server without revealing the word, the hidden search scheme can be applied. They proposed a simple extension to the basic scheme to support this goal.

Suppose the data owner does not want to reveal the word $W$ while searching for a cloud server. In that case, he can simply pre-encrypt each word $W$ in the document using a deterministic encryption algorithm $E_{k''}$, i.e., a block cipher. It is important to note that $E_{k''}(x)$ depends only on $x$; it does not depend on the position of x in the document. So, if we let $X_i = E_{k''}(W_i)$, the data owner will get the sequence $X_1, \ldots, X_L$. Then, he can get the ciphertext by $C_i = X_i \oplus T_i$, where $X_i = E_{k''}(W_i)$ and $T_i = < S_i, F_{ki}(X_i) >$. Thus, to search for a word $W$, the data owner needs to calculate $X = E_{k''}(W)$ and $k = f_{k'}(X)$ and sends $<X, k>$ to Bob. This approach satisfies the hidden search feature if the encryption algorithm E is secure.

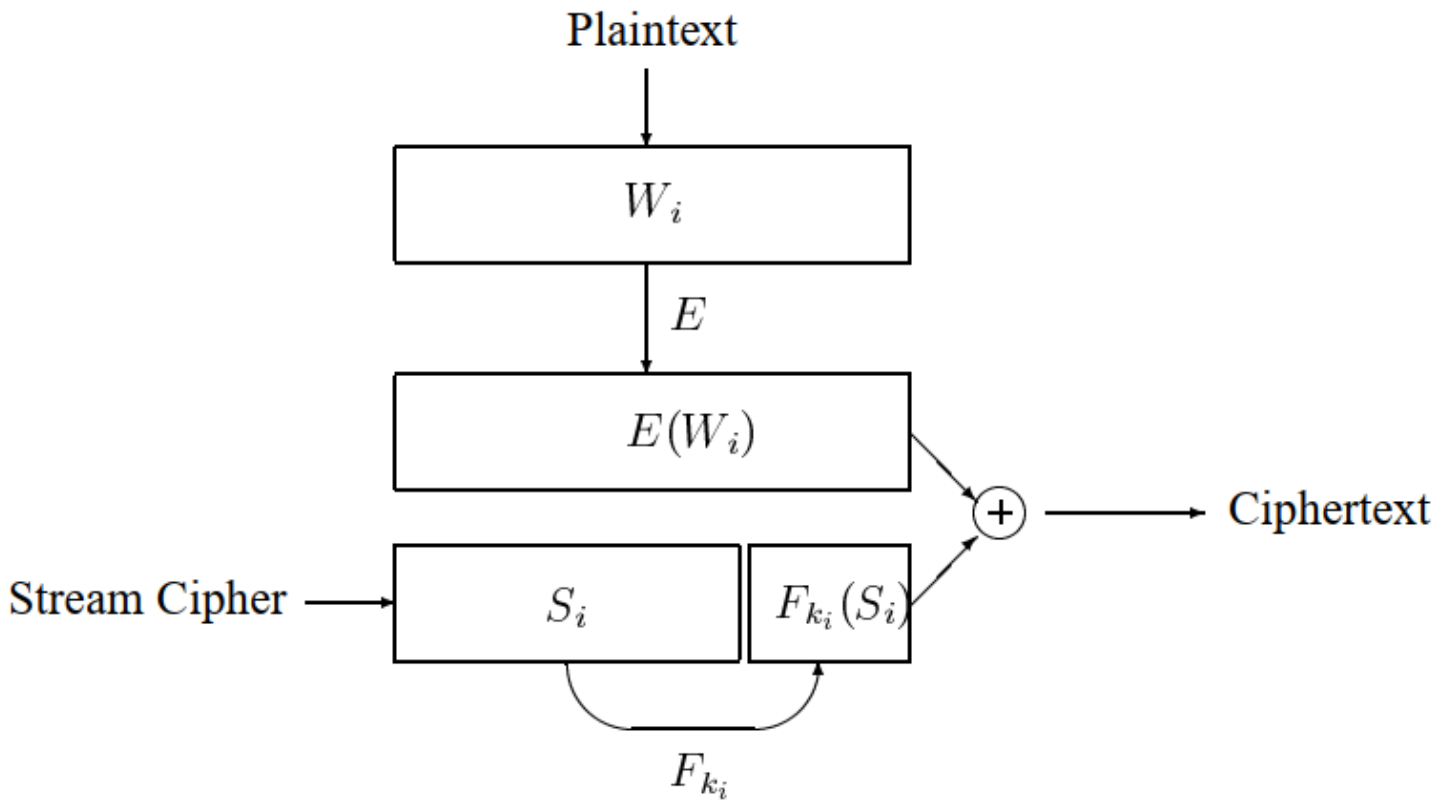

Figure 3.5    The scheme for the hidden search of the SSKE

**Scheme III: The Final Scheme**

The general idea of the scheme is to divide a message into a list of words $W_i$. Each word is then encrypted using a symmetric encryption scheme and key $k_i$, resulting in $E(k_i, W_i)$. Then, $E(k_i, W_i)$ is divided into a left half $L_i$ and a right half $R_i$. A pseudo-random bit stream $S_i$, of size $L_i$, is generated using a stream cipher (which can be constructed from a PRF). The scheme generates $F_{ki}(S_i)$, where $F$ is a keyed PRF, $k_i = F_{ki}(L_i)$, and $k_i$ is a secret key chosen by the user to create a searchable encrypted word. The encrypted word is created as $C_i = (L_i, R_i) \oplus (S_i, F_{ki}(S_i))$, where the purpose of XORing $(S_i, F_{ki}(S_i))$ is to randomize the encrypted word $(L_i, R_i)$. This hides the keyword distribution before the query since identical words in different locations of a message and other messages encrypt to identical ciphertexts. The encrypted word is divided into two halves $(L_i, R_i)$, so the data owner can decrypt a message without storing all encrypted words of a message, which would have defeated the purpose of the scheme. Specifically, if $k_i$ is generated directly as $F_k(E(k_i, W_i))$ instead of $F_{ki}(L_i)$, then to decrypt, the user must know all encrypted words $E(k_i, W_i)$ if a message.

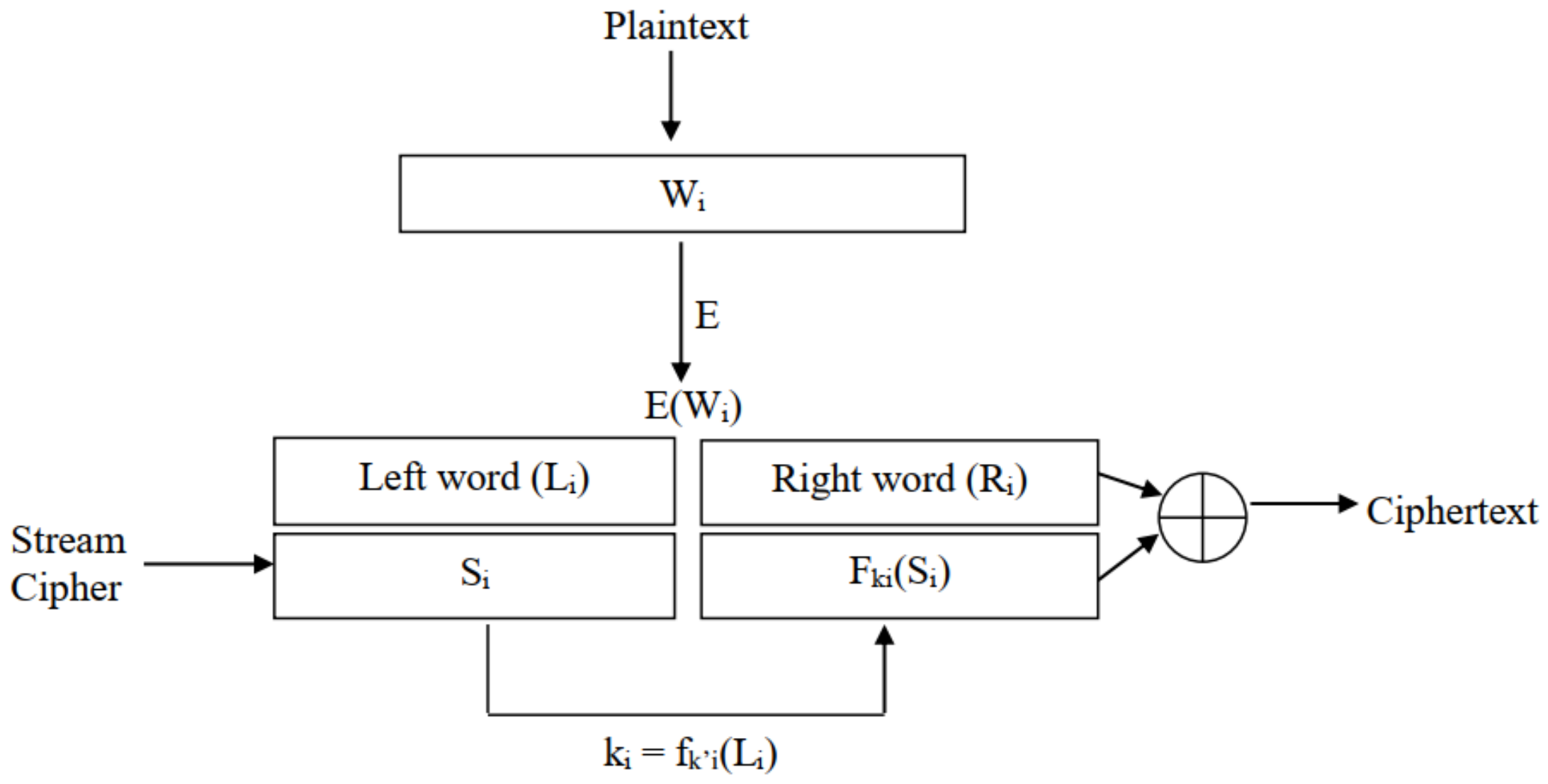

Figure 3.6 The final scheme of the SSKE

Second, in the keyword-based approach, they are considered the searching with an encrypted index. An index consists of a list of keywords along with pointers that point to documents where the keyword appears. The keywords that are in the index are words that the data owner may search for in the future. An easy but time-consuming way to search with an index is encrypting both the keywords and the pointers in the index. So, the cloud server can search for *E(W)* and return the encrypted list to the data owner, and can decrypt this list to see the document entries. Then, the data owner has to send the cloud server a request to retrieve the documents that he needs. As we can see, this approach needs two roundtrips between the user and the server. Another approach to using an encrypted index is encrypting the list of pointers with a key $k_W$ where kw=fk''(E(W)). So to search for W, the data owner sends <E(W), $k_W$> to a cloud server. The cloud server can do statistical analysis if the list of pointers is not a fixed size. Thus, to prevent statistical attacks from the cloud server, the data owner should keep the list of pointers at a fixed size.

However, this scheme requires the server to traverse each document word by word, which leads to a search complexity linear to the document size. This technique resulted in higher costs as the word-by-word scanning of the documents is required. It has a low length of document collection because the server needs to scan the entire ciphertext of a document when it determines whether a particular keyword is contained in the document. In addition, plaintext is vulnerable to statistical attack according to the frequency of the query keyword occurring in the document. In the scheme, it was considered encrypted indexes. However, their solution is incomplete because their index updates are insecure with low efficiency.

Furthermore, they suggested a sequential scan that could be executed with or without an index. The scheme without an index usually requires specially developed encryption algorithms, making it more complex in many specific systems. When the dataset documents are large, the index-based scheme is preferred since it gives faster search results. Nevertheless, this system causes trouble in a situation where storage and updating of records are needed.

## 3.3 Index-Based Search

Utilizing database and text retrieval techniques such as secure indexing to store selected data per document in a separate data structure from the files makes the search operation generally quicker and well adapted to big data scenarios. Secure indexes are helpful in searching for encrypted data only in multi-user settings. The encrypted documents and their indexes on the remote server are frequently updated. This secure index will only allow the user to search for the word W in the index if the user has a valid trapdoor for W. The index does not reveal any information about its contents without a valid trapdoor. There will be an index generated for each document on the remote server. Goh et al. (2003) [15] defined secure indexes and formulated security schemes to achieve index security. They then introduced a secure index construction scheme called Z_IDX based on bloom filters and pseudo-random functions. Bloom filters are memory-efficient data structures, and pseudo-random functions are deterministic and efficient functions that produce pseudo-random values indistinguishable from random sequences

The author addressed some of the limitations (e.g., use of fixed-size words, special document encryption) of the SWP scheme (discussed in 3.2) by adding an index for each document independent of the underlying encryption algorithm. The idea is to use a Bloom filter (BF) as a per-document index.

A Bloom Filter [88] is a data structure that is used to answer set membership queries. It is represented as an array of b bits that are initially set to 0. In general, the filter uses r independent hash functions $h_t$, where $h_h : \{0, 1\}* \rightarrow [1, b]$ for $t \in [1,r]$, each of which maps a set element to one of the b array positions. For each element e (e.g., keywords) in the set $S = \{e_1, . . . e_m\}$, the bits at positions $h_1(e_i), . . . , h_r(e_i)$ are set to 1. To check whether an element x belongs to the set S, check if the bits at positions $h_1(x), \ldots, h_r(x)$ are set to 1. If so, x is considered a member of set S. Using one BF per document, the search time becomes linear in the number of documents. The proposed scheme uses Bloom Filter. Each different word in a document is processed by a pseudo-random function twice and then inserted into the BF.

The index generation has to generate one BF per document. Thus, the algorithm is linear in the number of distinct words per document. The BF lookup is a constant time operation and has to be done per document. Thus, the time for a search is proportional to the number of documents, in contrast to the number of words in the SWP scheme.

Goh's scheme uses the existing cryptographic schemes to generate the algorithm. Also, this approach can be used with compression schemes. Additionally, it is flexible to be modified for occurrence searches and conjunctive searches. It can perform a search for Boolean queries "AND" and "OR" involving multiple words. This scheme is useful, and it is better to be used in multiple user settings where the indexes and documents are frequently updated on the server. If just one user uses the indexes or if they are not updated regularly, the user should use a local index. Z_IDX was tested to implement searches over encrypted data. It is an efficient method when considering index updates but not efficient in search time at the cloud server. It is guaranteed index security but not trapdoors. Goh's scheme has a linear search time and suffers from false positives due to the choice of data structure. Its search time is linear with the number of documents.

In this scheme, the construction of the index works as follows: Given a security parameter *s*, $K_{priv}$ is generated using a pseudorandom function f where $K_{priv} = (k_1,\ldots,k_r) \in \{0,1\}$, $T_w$ is generated for the word w building the index using $K_{priv}$ and D, and finally searching the index with $T_w$ and ID. Goh's index scheme consists of four algorithms. These are:

- Keygen (*s*): Input is a security parameter, s, and output is the master private key $K_{priv}$.
- Trapdoor ($K_{priv}$, *w*): Takes $K_{priv}$, and the word *w*as input and outputs the trapdoor $T_w$.
- BuildIndex (D, $K_{priv}$): Inputs are document D and $K_{priv}$, and output is the index ID.
- SearchIndex ($T_w$, ID): Given $T_w$ and *ID*, outputs 1 if *w* is in the document D and 0 otherwise.

Curtmola et al. [26] proposed the first inverted index-based encrypted searchable index in the index-based searchable approaches. They presented an indexing scheme that achieves the highest time-efficient search function using a uniquely designed linked list data structure. The document list for each keyword is encrypted and obfuscated into an array. They worked off of the previous approach, keeping a single hash table index for all documents and getting rid of false positives introduced by bloom filters. The hash table index for all documents contained entries where a trapdoor of a word that appeared in the document collection is mapped to a set of file identifiers for the documents it appeared. A single encrypted hash table index is built for the entire file collection. Each entry consists of the trapdoor of a keyword and an encrypted set of related file identifiers.

However, according to the design, the position and the content of the inverted list will be disclosed to the cloud server once the keyword is searched. As a result, one keyword can only be searched once before re-generating the index for the keyword. Also, an inverted index construction scheme has an inherent problem: directly updating is difficult.

Subsequently, as an improvement, Kamara and Papamanthou (2013) [89] proposed a new search scheme based on a tree-based index, which can handle dynamic updates on document data stored in leaf nodes. However, their scheme is designed only for single-keyword Boolean search.

In the dynamic index-based searchable approaches, Zhang et al. (2016) [90] proposed an efficient private keyword search (EPKS) scheme that supports binary search and extends to dynamic settings (DEPKS) for the inverted index–based encrypted data. They designed an EPKS scheme whose complexity is logarithmic in the number of keywords. The scheme is based on the groups of prime order and enjoys strong notions of security, namely statistical plaintext privacy and statistical predicate privacy. Also, they have extended the EPKS scheme to the DEPKS scheme to support dynamic keyword and document updates. The extended scheme maintains the properties of logarithmic-time search efficiency, plaintext privacy, and predicate privacy. It has low communication and computation cost for updates compared to existing dynamic search encryption schemes.

It was considered a cloud data storage service in which a data owner has a set of documents $D$ to be outsourced to the cloud server in an encrypted form. To enable efficient query over encrypted documents, we consider the inverted index-based data structure for storing the outsourced files. Specifically, the data owner builds an encrypted searchable inverted index set $C$ with keywords $w_1, w_2, \ldots, w_n$ from $D$, then both the encrypted index set $C$ and the encrypted document set $E(D)$ is outsourced to the cloud server. For every query of a keyword $w_i$, a data user computes a search token $TK$ and sends it to the cloud server. On receiving $TK$ from the data user, the cloud server queries over the encrypted index set $C$ and return the candidate encrypted documents.

Finally, the data user decrypts the candidate documents and verifies each document by checking the existence of the keyword. The Keyword structure of the efficient private keyword search (EPKS) scheme is shown in figure 3.7.

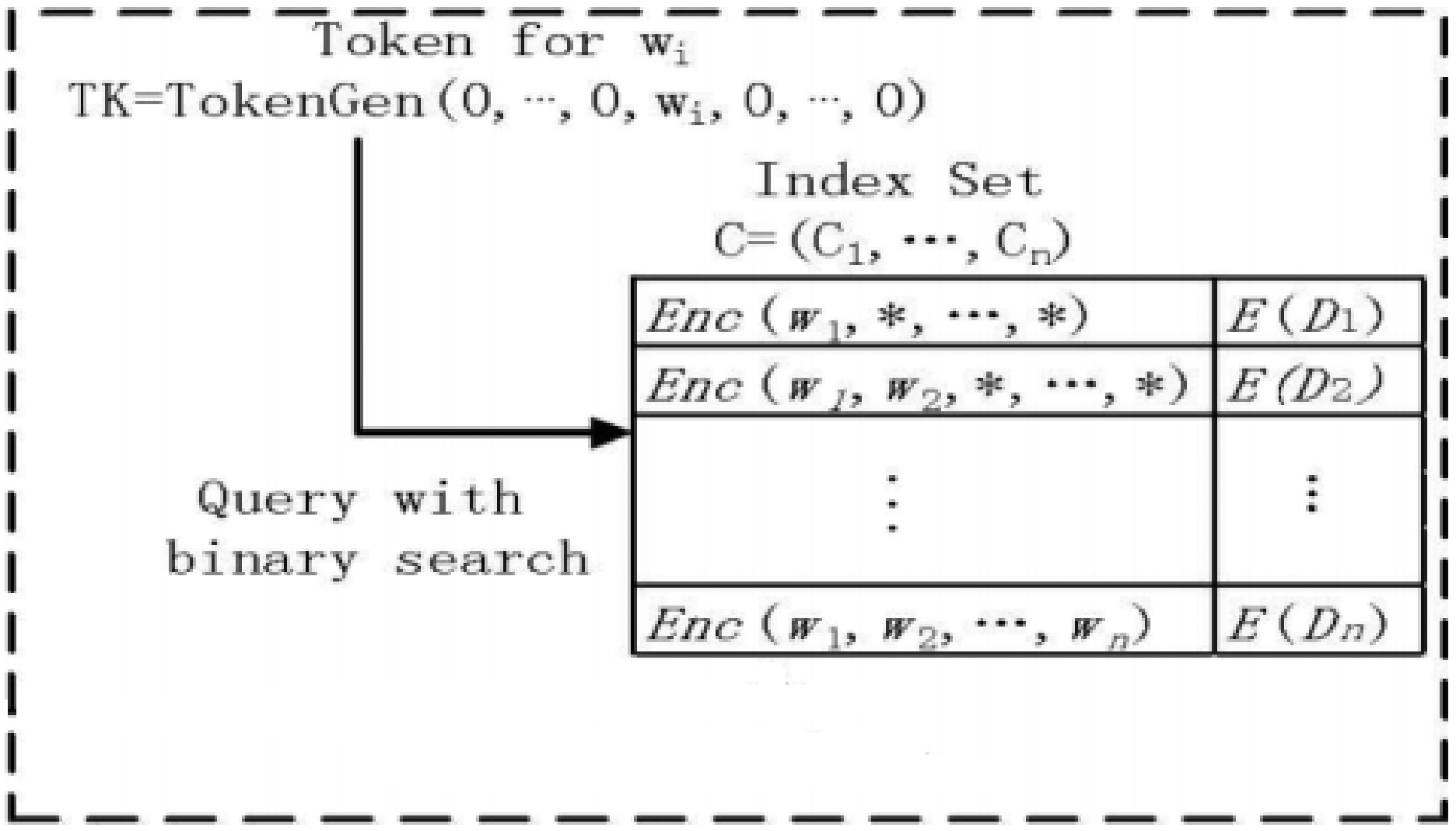

Figure 3.7 Keyword structure of efficient private keyword search (EPKS) scheme

The EPKS scheme consists of five algorithms as follows:

- Setup ($1^{\lambda}$) → (*pp, sk*): On input the security parameter $1^{\lambda}$, output public parameters *pp* and a secret key *sk.*
- Enc (*W, sk, pp*) → *C*: On input the keywords set $W = \{w_1, \ldots, w_n\} \subseteq \mathrm{W}^{\lambda}$, the symmetric key *sk*, and public parameters *pp*, output searchable encrypted index $C = (C_1, \ldots, C_n)$.
- TokenGen ($w_i$, *i, sk, pp*) → *TK*: On input the keyword $w_i \in W$, its sequence number *i*, the secret key *sk*, and public parameters *pp*, output a search token *TK.*
- Test (*TK*, $C_i$, *pp*) → {0, 1}: On input of a search token *TK*, each encrypted index item $C_i$, and public parameters *pp*, output a bit indicating whether the item matches with the queried keyword corresponding to the search token.
- Query(*TK, C, pp*)→E($D_i$) or ⊥: On input a token *TK*, the searchable encrypted index $C = \{C_1, \ldots, C_n\}$, and public parameters *pp*, perform a binary search by running the Test algorithm, output the candidate set of encrypted documents E($D_i$) or ⊥.

They are adding new keywords into the index after they have been generated. Let us be a predefined public parameter, the maximum dimension of vectors used in the algorithms. In other words, as long as the current number of keywords is less than, a new keyword can be easily added to the index without setting up the whole system again.

- Update Token ($w_i$, *i, sk, pp*) → $C_i$: On input of the updated keyword $w_i$, its sequence number $i \in [n + 1]$, the symmetric key *sk*, and public parameters *pp*, output an updated ciphertext $C_i$.

- Update ($C_i$, E($D_i$), *i*, $\tau u$) → (*C, n*): On input of the updated ciphertext $C_i$, the corresponding encrypted document set E($D_i$), the sequence number $i \in [n+1]$, and the update type $\tau u \in$ {*update, add, delete*}, output a new encrypted index *C* and a new counter *n*, which indicates the number of keywords contained in the underlying plaintext of current ciphertexts.

The EPKS and DEPKS schemes are significantly more efficient in terms of keyword search complexity and communication complexity than existing randomized SSE schemes. It is noted that our proposed scheme is the most closely with the EPKS scheme.)

## 3.4 Privacy-Preserving Searchable Encryption

Chaudhari and Das (2019) [23] presented a single keyword-based searchable encryption scheme for the applications where multiple data owners upload their data. Then multiple users can access the data with fine-grained access control. The scheme uses attribute-based encryption that allows users to access the selected subset of data from the cloud without revealing his/her access rights to the cloud server. They proposed a privacy-preserving single keyword-based searchable encryption scheme (PSE) with fine-grained access control. The proposed PSE scheme provides a keyword-based search facility over attribute-based encrypted data with hidden access policy. The scheme is applicable in a scenario with multiple data owners and multiple data receivers is shown in Figure 3.8.

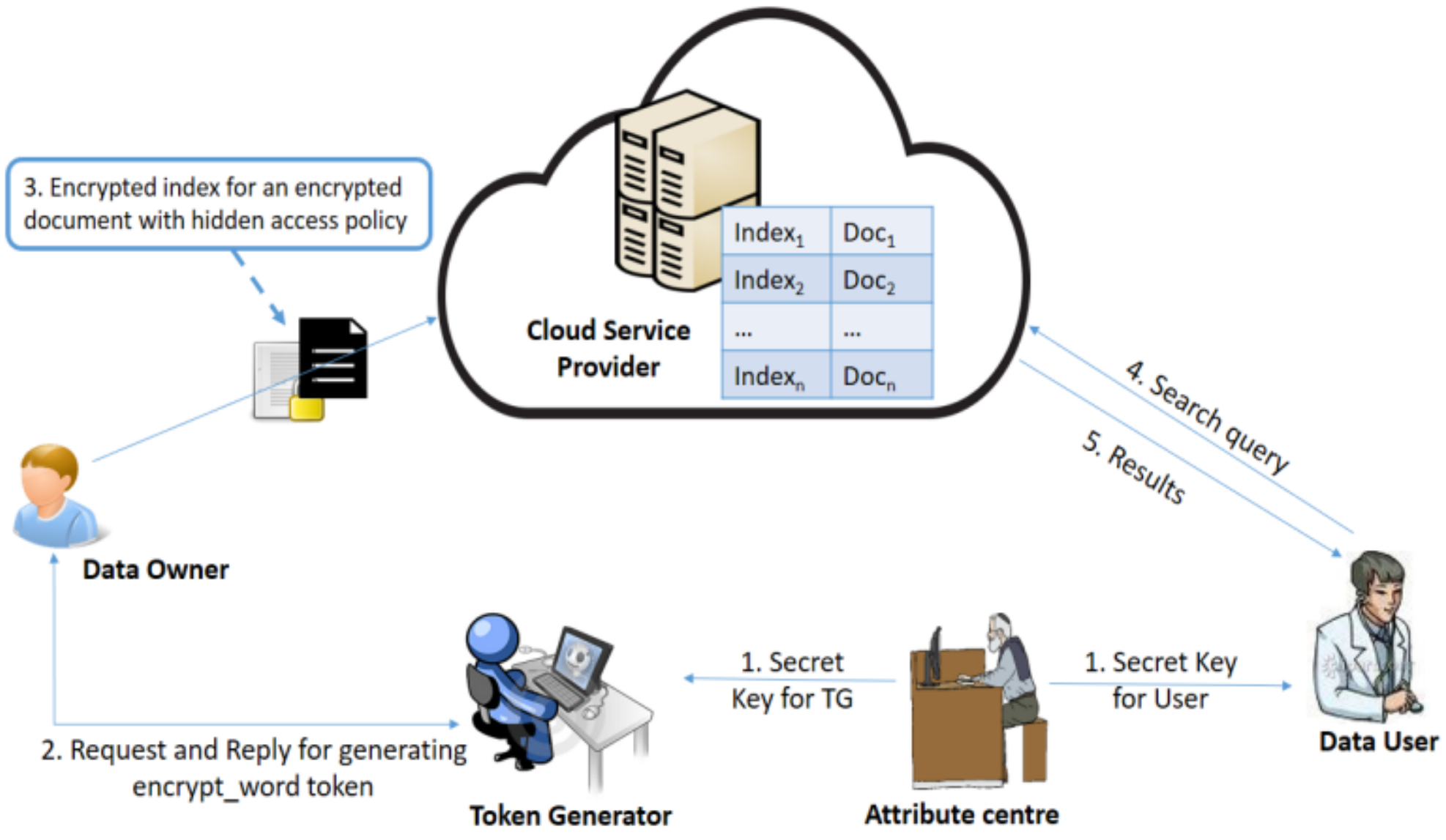

Figure 3.8 System Model of PSE Scheme

The scheme allows each user in the system with a set of attribute values, where a trusted authority verifies the user's attributes and assigns him a secret key. The system model of the PSE scheme comprises the entities of Attribute Centre, Token Generator, Data Owner, Cloud Service Provider, and Receiver (Data) User. They have implemented the scheme on Google cloud instance, and the scheme's performance was found to be practical in real-world applications.

They have presented attribute-based searchable encryption with an access policy enforced by the data owner and hidden inside the ciphertext. The scheme is designed with a single-sender and multi-receiver setup, aimed at facilitating a data owner (sender) to encrypt the index of keywords related to his document and upload it along with the access policy and the encrypted document on cloud storage, where the access policy is decided by the data owner and kept hidden inside the ciphertext. The user (receiver) sends his search query in the form of a trapdoor to the cloud storage server. The cloud server uses this trapdoor to search over all encrypted indexes uploaded to the cloud storage. The documents corresponding to the indexes for which the search operation returns true are sent back to the user as the result of his query

The PSE scheme's system model comprises the following entities.

**Attribute Center:** The Attribute Centre (*AC*) is a trusted third party of the system. *AC* is responsible for generating system parameters and issuing keys to users of the system (Step 1).

**Token Generator:** The Token Generator (*TG*) is a trusted third party of the system which assists a data owner in generating an encrypted index (Step 2). *TG* is involved in the process of generating an encrypted index. They have chosen to place the *TG* on the data owner side instead of the data receiver side, which facilitates a user to generate a trapdoor from his secret key autonomously without waiting for the token from any trusted authority, which ultimately reduces the per-query interaction with the trusted third party and helps to gain efficient response time for search procedure.

**Data Owner:** Data owner encrypts and stores the data on a cloud storage server (Step 3). The encrypted data consists of (i) the index of encrypted keywords and (ii) the encrypted document.

**Cloud Service Provider:** Cloud Service Provider (CSP) provides storage and computation services for the system's entities.

**Receiver (Data) User:** The receiver user generates and submits a trapdoor to CSP. The CSP searches the encrypted indexes using this trapdoor (Step 4). The documents corresponding to the indexes for which the search operation returns true are returned to the user. Finally, the user decrypts the resultant documents (Step 5).

The scope of the PSE scheme is to generate the index of encrypted keywords for a document and perform a privacy-preserving search over the encrypted index. The PSE scheme is a 5-tuple defined as follows:

– **Setup** ($1^l$): The Setup($1^l$) is run by the AC. It takes as input parameter a security parameter $l$ and outputs the master private key $MK$, $TG$'s secret key $T\,SK$, and public key $PK$.

– **KeyGen** ($MK$, $L$): The KeyGen($MK$, $L$) is a randomized algorithm, and the AC runs it. The algorithm takes the master key $MK$ and a set of attributes $L$ of a user as input. It outputs secret key $SK_s$ for that user. The key $SK$ is used to generate a trapdoor for performing search operations.

– **Encrypt Index** ($PK$, $W$, $T$, $TSK$): This is a protocol run between the data owner and $TG$, where $W$ is the set of keywords associated with document $M$. The data owner starts the computation to generate the encryption for each keyword $w$ included in the keyword set $W$. The data owner gets an encrypted token for each $w$ from $TG$ to perform the encryption of the keyword. To generate encrypted word tokens, $TG$ uses his secret key $TSK$. Finally, the data owner outputs a set of encrypted words $CT_w$, known as an encrypted index for document $M$.

– **Trapdoor** ($P\,K$, $SK$, $w$): The receiver user invokes this randomized algorithm to make a trapdoor for retrieving the documents from CSP whose associated index contains an encrypted entry for the word $w$ and for which he possesses sufficient access rights. The algorithm outputs a trapdoor $t_w$ generated for $w$.

– **Search** ($t_w$, $CT_w$ ): This is a deterministic algorithm run by the CSP. The algorithm inputs the trapdoor $t_w$ sent by the user and encrypted index $CT_w$. The algorithm returns true if the word in $t_w$ matches with any of the keywords included in $CT_w$ and the user's key satisfies the access policy of $CT_w$.

One of the key features of the PSE scheme is that once the secret key is obtained, the user can generate the search query himself in the form of a trapdoor using the secret key assigned to him. In the PSE scheme, the trapdoor generated by the user neither reveals the keyword used for the search nor the user's attributes. Each data owner encrypts the index with the help of a trusted authority. The trusted authority makes the index secure by using the master secret key elements inside the index. The inclusion of the master secret key elements in the index prevents adversaries, who can adaptively generate search queries for chosen keywords, not to learning the keywords from the index. After encryption, the index is uploaded along with the encrypted document to the cloud. When a user sends the trapdoor, the cloud server performs the search operation with the input of the trapdoor and encrypted index. The search process is repeated for each encrypted index related to each separate document. The search operation returns true if (1) the keyword inside the trapdoor is included in the index, and (2) the access policy of the encrypted index is fulfilled with the user's attributes.

However, the major limitation of the scheme is that the user has to acquire the search token from the Attribute Centre, which increases the per query interaction overhead for search operation on the user side. Although the scheme can work in a single-sender multi-receiver setup, it cannot perform well in a multi-sender multi-receiver scenario because each data sender has to securely communicate with each data receiver to issue the secret key or search token, which will cost a large communication overhead. It can leak the information that an adversary can obtain from a search.

## 3.5 Keyword Search with Access Control (KSAC)

The system can support multi-keyword search in place of a single keyword search with access control to make the searching system more practical. To support multiple keywords search, Shen et al. (2016) [25] considered conjunctive keywords with access control search over encrypted data. They first proposed a scalable framework where the user can use his attribute values and a search query to derive a search capability locally. A file can be retrieved only when its keywords match the query, and the user's attribute values can pass the policy check. Using the framework, they proposed a novel scheme called KSAC, which enables Keyword Search with Access Control over encrypted data. KSAC utilized a recent cryptographic primitive called HPE (Hierarchical Predicate Encryption) to enforce fine-grained access control and perform multi-field query search. Meanwhile, it also supported the search capability deviation and achieved efficient access policy updates and keyword updates without compromising data privacy. KSAC also provided noises in the query to hide users' access privileges to enhance privacy.

In this paper, they systematically studied the issue of Keyword Search with Access Control (KSAC) over encrypted cloud data. Their main contributions are summarized as follows. First, they proposed a scalable framework, shown in Figure 3.9, that integrates multi-field keyword search with fine-grained access control.

In the framework, every user authenticated by an authority obtains a set of keys called credentials to represent his attribute values. Each file stored in the cloud is attached with an encrypted index to label the keywords and specify the access policy. Every user can use his credential and a search query to locally generate a search capability and submit it to the cloud server, performing search and access control. Finally, a user receives the data files matching his search query and allows access.

Second, to enable such a framework, they have made novel use of Hierarchical Predicate Encryption (HPE) [35] to realize the derivation of search capability. Based on HPE, they proposed the scheme named KSAC. It enables the service of both the keyword search and access control over multiple fields. It supports the efficient update of access policy and keywords. KSAC also introduces some random values to enhance the protection of users' access privacy.

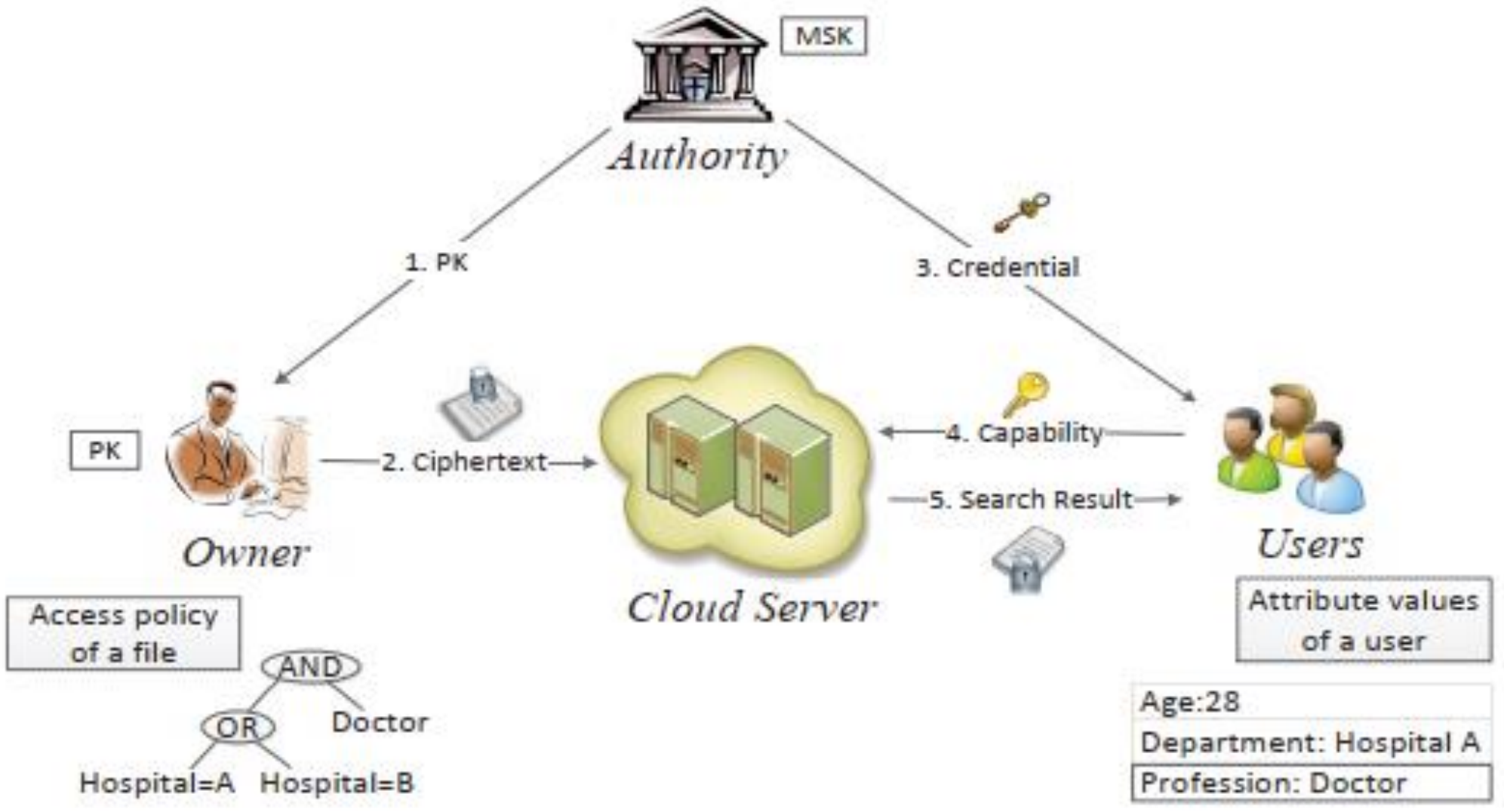

Figure 3.9 The framework of Keyword Search with Access Control

They considered a cloud-data-sharing system consisting of four entities, i.e., authority, data owners, data users, and cloud server (shown in Figure 3.9). Each entity's responsibility is discussed below.

**Authority:** Authority is responsible for authenticating the user's identity. It issues a set of keys (step 1) as a credential to represent the user's attribute values (step 3).

**Data Owners:** The Data Owner creates data files, designs the encrypted indices containing keywords and access policy for each file, and uploads the encrypted files along with the indices to the cloud server (step 2).

**Data Users:** The Data User generates a search capability according to his credential and a search query and submits it to the cloud server for file retrieval (step 4).

**Cloud server**: Cloud Server stores the encrypted data and performs search when receiving search capabilities from users (step 5).

To better understand here, we have discussed some basic system elements. Such as:

**Attribute:** The "*attribute*" is used to characterize the identity of a user. A user's attribute values are denoted as $A := (A_1 = a_1; \ldots; A_e = a_e)$, where $A_i$ is the $i$-th attribute field and *Ai's attribute value*. $e$ denotes the number of attribute fields. For example, if a user's attribute values are "Age=37, Profession=Professor", then "37" (i.e., $a_i$) is an attribute value of attribute "Age" (i.e., $A_i$).

**Keyword and Search Query:** The keywords refer to the specific terms that characterize file content. For a file, its keyword set can be expressed as $W$:= ($W_1$ = $w1$; ...; $W_d$ = $w_d$), where $W_i$ is the $i$-th keyword field, and $w_i$ is the keyword of $W_i$. For example, the keywords of an Electronic Health Records (EHR) may be "Name=Alice, Age=30, illness =headache". Then "illness" is a keyword filed (i.e., $W_i$), and "headache" is a keyword (i.e., $w_i$) over this field.

**Search Capability:** A search capability includes the information of the search query and the user's attribute values. For instance, if a user's attribute values are "Age=37, Profession=professor," and the query is "Date=2013/1/1 ^ Topic=meeting", then the search capability includes the logical information of "Age=27, Profession=professor, Date=2013/1/1 ^ Topic=meeting".

**Index Format:** According to the property of HPE, they gave a novel design of the encrypted index, as shown in Figure 3.10. The index information includes the specified access policy, the representative keywords, and the symmetric keys to encrypt the file content. To build an encrypted index, the data owner first produces the "*body*" component to lock the symmetric key *EK* by utilizing the access policy vector (APV) and a random vector (RV). *EK* is used as the symmetric key to encrypt and decrypt the file content. This body component ensures that only the users whose attribute values satisfy the access policy can recover *EK* for file decryption. APV will reject the recovery of EK performed by the unauthorized user, and the cloud server's attempt to obtain *EK* by stealthily using the authorized user's search capability will be refused by RV. The data owner further encrypts *M*, the representative keywords, and the access policy and produces the "*head*" component. This design ensures that the file can be retrieved only when the keywords match the query and the user's attribute values satisfy the access policy.

A detailed procedure is given about constructing the head component when given the keywords and the access policy. For the keywords $W = (W_1 = w_1; \cdots; W_d = w_d)$, where $d$ is the number of keyword fields, it can convert into the keyword vector $W$ according to the transformation of the access policy vector (APV) from $S = (s_1; \ldots; s_{n1})$ where $n_1$ is the length of *APV*. Suppose the access policy is $S = (A_1 = a_{1,1} V \ \ldots . V a_{1,p1}) \wedge \ldots \wedge (A_e = a_{e,1} V \ \ldots . V a_{e,pe})$, where $a_{i,j}$ $(1 \leq i \leq e, 1 \leq j \leq p_i)$ denotes the $j$-th required attribute value in the field $A_i$, $e$ is the number of attribute fields, and $p_j$ represents the total number of required attribute values over the $j$-th attribute field. Finally, it calls HPE to encrypt $S \,||\, W$ and $M$ and produce the head component. The construction of the body component is similar.

Figure 3.10 shows that the encrypted index in KSAC contains the keywords and the access policies. When performing a search operation, the cloud server first decrypts the head component to see if the decrypted result is *M*. If it is, the file matches the capability. After obtaining the matching files, the user can decrypt the body component to free *EK*, which will then be used to decrypt the text information of the matching files.

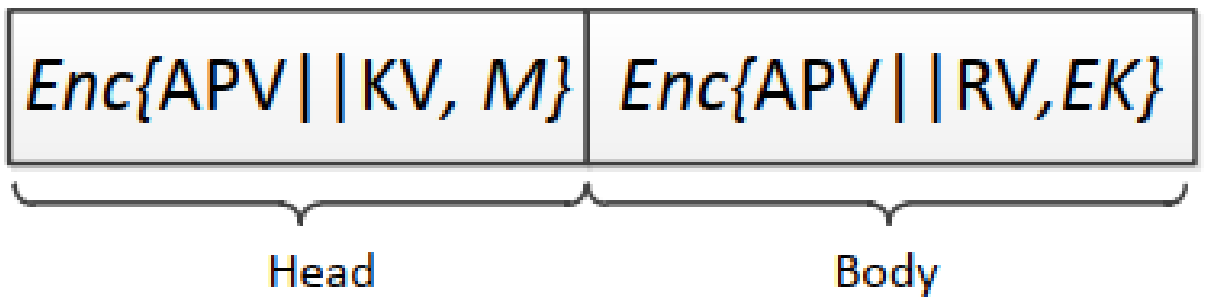

Figure 3.10 The format of the KSAC scheme encrypted index

**An Example:** Alice (i.e., data owner) wishes to share her health records with her personal doctor Bob (i.e., data user) in hospital *A*, then she can combine the access policy *S* "Hospital=A ^ Name=Bob" with keywords of her records before uploading them to the cloud. Bob, who keeps the credential standing for the attribute values "Hospital=A, Name=Bob" distributed by the hospital *A* (i.e., authority), can generate his query like "Name=Alice ^ Date=4/20/2022" to timely learn Alice's body status. Suppose Carl is a new doctor for Alice in hospital A, to grant him access permission. In that case, Alice can change the access policy to "Hospital=A ^ (Name=Carl _ Name=Bob)."

Most existing SE schemes assume that every user can access all the shared files. Such assumption does not hold in the cloud environment where users are granted different access permissions according to the access-control policy determined by data owners. Therefore, it is important to enforce the access policy when searching over encrypted data efficiently. Therefore, the conventional way to perform an encrypted search with access control is to conduct the search operations at the cloud server to take advantage of its large computation power and leave access control enforcement at users' machines to keep their access keys from being disclosed. But this separation of search and access control enforcement could lead to performance degradation, especially when users are assigned different access permissions to search different encrypted cloud data. For example, in conventional practice, a cloud server may perform to search and transfer all the matching files to the users for them to decrypt the files. However, a user may not be allowed to access all the files, and many of the transferred files have to be discarded, which leads to wasted network bandwidth and reduced service efficiency

## 3.6 Identity Based Encryption

Liu et al. (2021) [24] introduced a novel framework called identity certifying authority-aided identity-based searchable encryption (ICA-IBSE), which has the advantage of reducing storage space while maintaining efficiency and security. They formally defined the system model and desired security requirements to represent attacks in a real scenario. In addition, they proposed a provably secure scheme based on the gap bilinear Diffie–Hellman assumption and experimentally evaluated their scheme in terms of its performance and theoretical features.

The identity certifying authority-aided identity-based searchable encryption (ICA-IBSE) comprises five entities, namely the Identity certifying Authority (ICA), Key Generation Center (KGC), Cloud Server (CS), Data Owner (DO), and Data User (DU), which are displayed in Fig. 3.11. The responsibility of each entity is discussed below.

**1) Identity Certifying Authority (ICA):** This authority is responsible for validating the DO's and DU's identities and issuing trapdoor information and an identity certificate. The ICA is generated the public and secret key of ICA ($PK_{ICA}$, $SK_{ICA}$) by taking the system parameter (keywords ($W$) and identity ($ID$) of the Data Owner and Data User) and then is executed the Certificate ID ($Cert_{ID}$) and Trapdoor Information ($TF_{ID}$) to every User ID.

**2) Key Generation Center (KGC):** By validating the correctness of the identity certificate, the KGC generates the DO's and DU's partial secret keys without knowing any information about their identities. The KGC is generated the public and secret key of KGC ($PK_{KGC}$, $SK_{KGC}$) by using system parameter and then provide the partial secret key to the Data Owner ($SK_{DO}$) and Data User ($SK_{DU}$).

**3) Data Owner (DO):** The DO first requests his certificate from the ICA by using his ID and further generates his partial secret key ($SK_{DO}$) by interacting with the KGC. After that, The DO can generate massive quantities of encrypted data ($CT_w$) along with the corresponding encrypted keywords ($W$) and Data User's Identity ($DU_{ID}$), which are uploaded to the Cloud Server.

**4) Data User (DU):** The DU can get the partial secret key ($SK_{DU}$) as the Data Owner does. Subsequently, the DU can generate a trapdoor ($TD_w$) using Data Owner Identity ($DO_{ID}$), Partial Secret key ($SK_{DU}$), and Keyword ($w$) and send the request to the Cloud Server to retrieve the encrypted data associated with the specified keyword.

**5) Cloud Server (CS):** The CS has sufficient storage and computing capacity and is mainly responsible for storing encrypted data and the corresponding encrypted keywords. The CS can search over the encrypted data ($CT_w$) by matching the trapdoor ($TD_w$) the searching for data as the inputs, and 1 is the output if $CT_w$ is matched with $TD_w$; otherwise, 0 is the output.

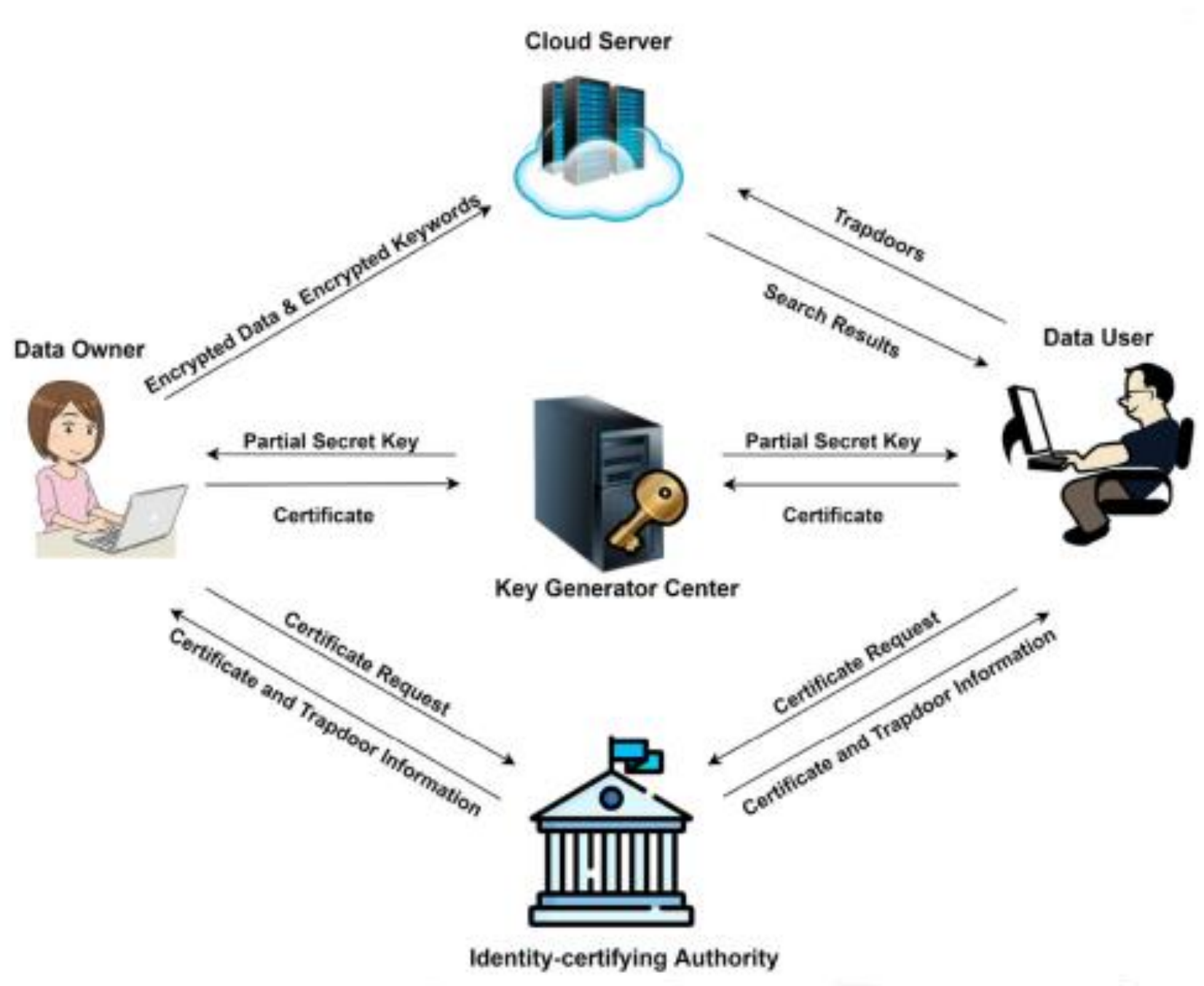

Figure 3.11 Identity Certifying Authority-Aided Identity-based Searchable Encryption Framework

They assumed that the DO and DU are fully trusted, so the DO and DU cannot collude with other parties to reveal their secret keys or the keyword other parties can directly compromise privacy. But the ICA and KGC cannot collude because KGC cannot issue any secret key to a user whose identity cannot be directly or indirectly identified. Therefore, to enable KGC to generate secret keys, there must be some party (e.g., ICA) that can privately authenticate the user's identity. Then, the user can indirectly validate his identity to the KGC. However, if ICA and KGC collude, they can adaptively generate any user's secret keys.

The KGC and CS are honest but curious, which means that they will attempt to retrieve the sensitive information of the keywords from encrypted keywords and trapdoors. The ICA is malicious. In addition to attempting to obtain keyword information as the KGC and CS do, the ICA can generate a potentially malicious ICA key pair.

The communication channels between the DO and DU and the cloud server are insecure. All transmitted information is eavesdropped upon by any one party (e.g., ICA, KGC, and a malicious outsider). However, the communication channels between the DO and DU and the ICA and KGC are secure. Also, because ICA-IBSE schemes are based on secret key settings and do not provide implicit authentication, they require a trusted public key infrastructure KGC to bind public keys with the respective identities of entities through issuing certificates; nevertheless, this process incurs additional overhead. Specifically, the KGC can access any user's secret key; key escrow problems occur when the KGC is malicious. In addition, the users in these schemes require an additional generated certificate or a public–secret value pair to encrypt keywords or generate trapdoors. Consequently, users require more storage space to store certificates and keys. It becomes less convenient for the users in this scheme because they have to use additional information instead of merely using their identities to encrypt keywords and generate trapdoors.

## 3.7 Other Existing Work

Symmetric Searchable Encryption (SSE) is well suited to cloud computing, and various SSE approaches have been proposed [91], [92], [93], [94], [95]. However, SSE is restricted to the key sharing problem of symmetric cryptosystems. Specifically, the Data Owner and Data User must agree on a shared key before encrypting keywords and generating trapdoors.

To further increase the range of applications and reduce the communication overhead of negotiating keys, Boneh *et al.* [96] introduced a searchable encryption method in a public-key setting, called public-key encryption with keyword search (PEKS). Instead of using a shared key as done in SSE, in PEKS, the DO encrypts keywords by using the DU's public key, and the DU generates corresponding trapdoors by using his secret key. PEKS immediately caught the attention of researchers, and many studies have applied PEKS to various applications [97], [98], [99], [100]. However, in 2006, Byun et al. [101] observed that because the entropy of keywords is low, any malicious party, through a so-called keyword guessing attacks, can randomly select keywords to generate the ciphertext and test whether the ciphertext is passable; thus, the malicious party can obtain the information associated with the keywords in the trapdoor.

From the literature, inverted index [102], [103], [104] is a widely used data structure in single-keyword search techniques, and the vector space model achieved many advantages by using the TF*IDF factor to rank the keyword frequency in [105], [106], and [107]. Likewise, in later times, binary search trees decrease the computational overhead with the time complexity $O(\log n)$, where n is the number of documents. The inverted index and vector space model still serve as core indexing techniques. However, the tree and graph structures can be used together to explore new properties and improve the efficiency of building an index.

Many schemes for keyword-based searching over attribute-based encrypted data have been proposed to provide a single keyword search capability over attribute-based encrypted data. Wang et al.'s scheme [108] provides single keyword-based searching along with a partial decryption task delegated to the cloud service provider. The scheme in [109] provides the verifiability of search results in addition to searching over encrypted data. The scheme in [110] provides verifiability of keyword search over key policy attribute-based encryption. The authors of [111] provided a solution that addresses the issue of data sharing, keyword update, and keyword-based search operation over encrypted data. The scheme in [112] provides the disjunctive multi-keyword search facility. Dong et al. [113] presented a scheme that provides an efficient keyword-based searching operation over ABE via an online-offline approach. The scheme in [114] includes proxy re-encryption (PRE) and a secret sharing scheme (SSS) into ABKS. However, in all these schemes, the access control policies are in clear form. To preserve the data confidentiality and the service's purpose, protecting user privacy is a vital requirement [115], [116].

In Identity-based encryption with keyword search existing work, the IBAEKS scheme [117], the user can apply to a trusted key generation center (KGC) for the secret key of a specific identity; he/she can then use the identity as their public key. Thus, in IBAEKS, no additional steps are required to prove the validity of the public key. In Certificate-based authenticated encryption with a keyword search (CBAEKS) scheme [117], the user can apply to a trusted certificate authority for the certificate of a pair comprising a specific identity and public key. Unlike the certificates used in PAEKS schemes, the certificates in CBAEKS not only implicitly authenticate the validity of the identity and public key and act as partial secret keys.

# 4. CHAPTER FOUR: DESIGN OF THE PROPOSED MODEL

Many schemes have been proposed to search on encrypted data with many different schemas in the literature. Most of the proposed schemes built their algorithm on a cloud service provider. The cloud server can search the index first and then return the encrypted documents to the user. But in a real scenario, if we consider most used cloud service providers such as Google Drive, Dropbox, Mega, OneDrive, etc., none of them can give this facility to search on the index first and then return encrypted documents. Though we can search in the cloud storage by plaintext, there is no option to index search first.

In this study, we have proposed a scheme for a cloud-based encrypted data sharing scheme that supports single and multiple keyword searches. We have extended the work of [19], where a prototype was presented and implemented to secure file storage and share with users. We are adding searchable functionality with significant improvement to the system to achieve secure storage, privacy preservation, search and secure sharing of encrypted data.

The scheme supports keyword search, enabling the server to return only the encrypted data that satisfies an encrypted query without decrypting it. The proposed scheme is an efficient scheme for single and multiple keywords searchable encryption, which allows a user with an encrypted text to efficiently search on encrypted data containing the specific keywords over the cloud server.

Also, we have implemented our proposed scheme as an overlay system that works on top of the Google Drive and inverted index search as it is the most popular cloud storage provider and uses by many people, organizations, etc. Our scheme provides a search mechanism for encrypted data both for the data owner himself and data users in a real-life scenario.

We use Google Drive, which stores data on the internet and manages and operates data storage as a service. Google Drive works as cloud storage where we can safely keep various files and any data. Some of the files that we can store here include videos, PDFs, Google docs, photos, etc. It uses a technology known as cloud computing, which means we can access the data anywhere. It can also be managed jointly or individually, making our work easy. Overall, this cloud storage gives us agility, global scale, and durability, with "anytime, anywhere" data access.

For example, a data owner has a set of documents $D$ to be outsourced to the cloud data storage service (Google Drive) in an encrypted form. To enable efficient query over encrypted documents, we consider the inverted index-based data structure for storing the outsourced files. Specifically, the data owner builds an encrypted searchable inverted index set $I$ with keywords $w_1, w_2, \ldots, w_m$ from $D$. The encrypted index set I and the encrypted document set E(D) are outsourced to the cloud server. For every query of a keyword $w$, a data user computes a searchable encrypted text with the help of a trusted third party and sends it to the cloud server. On receiving $TK$ from the data user, the cloud server queries over the encrypted index set $C$ and return the candidate encrypted documents. Finally, the data user decrypts the candidate documents and verifies each document by checking the existence of the keyword.

## 4.1 System Model

The system model of the scheme comprises the following entities.

- Data Owner (DO)
- Data User (DU)
- Cloud Service Provider (CSP)
- Trusted Third Party (TTP)

The responsibilities of each entity are described below.

### 4.1.1 Data Owner (DO)

The Data Owner outsources his documents to the cloud storage to provide convenient and reliable data access to the corresponding data users. He also generates some keywords for each outsourced document to improve search efficiency. The corresponding index is then generated according to the keywords using the secret key SK. After that, and the data owner sends the encrypted documents and the corresponding indexes to the cloud storage. The encrypted data consists of two parts:

i) The index of encrypted keywords
ii) The encrypted documents

The Data Owner must encrypt the documents and keyword collection before outsourcing into a cloud server so that the encrypted keyword set is searchable. However, the number of data owners is independent of the SE scheme.

### 4.1.2 Data User (DU)

The data user will send a plaintext keyword to the trusted third party. The trusted third party will return an encrypted keyword to the data user. After that, the data user will download the encrypted inverted index from the cloud service provider and search locally with the encrypted keyword with the encrypted inverted index. The web application will search over the encrypted indexes using the data user's encrypted keyword. After searching, the web application will return the corresponding FileID based on the keyword search. Then, the data user will send a request to the cloud server to download the encrypted documents. The cloud service provider will return the encrypted documents to the data user based on the FileID by user request.

After that, the data user will send a File ID to a trusted third party along with RSA generated public key. The TTP will verify the FileID with the stored FileID (After documents are uploaded, the data owner would send the FileID of documents to the TTP. So, each document’s FileID is already stored in the TTP). After checking, the TTP will send the decryption key by encrypting it with the public key of the data user. The data user first decrypts the document’s decryption key by using his private key, then decrypt the document.

### 4.1.3 Cloud Service Provider (CSP)

Cloud Service Provider (CSP) provides storage and computation services for the system's entities. It is an intermediary entity that stores the encrypted documents and the corresponding indexes received from the data owner (DO) and then provides data access and searches services to the data user (DU). When the data user sends a request to the cloud service provider by unique file id (FileID), it will return a matching document. The file ID is unique for each document. So it will return a specific document based on data users request to download. We believe that the CS is honest but curious. The CS will follow the protocols, but it may evaluate the data or search query patterns.

### 4.1.4 Trusted Third Party (TTP)

The trusted third party (TTP) is a trusted authority of the system. TTP is responsible for generating system parameters and issuing keys to the system's data owner and data users. TTP will assist the data owner in generating an inverted index. Also, TTP will be responsible for constructing encrypted text that the data user will request. Also, after submitting the file id (FileID) by the data user, TTP will verify the stored FileID of the document. After getting the FileID result of the encrypted file, TTP will generate RSA random public and private keys for the data user. It will be considered a secure channel from TTP to data users. The TTP will send the decryption key of the document through a secure channel to the data user. The decryption key of the document will get by TTP through reference number

One more responsibility of TTP is to control the web server. We are providing the responsibility to handle the web server to the TTP to make the scheme less complex. Otherwise, it would be challenging to control every entity at a time.

## 4.2 Security Model

In our security model, we consider the data owner trustworthy. We will correctly outsource the encrypted data to the cloud server. For the cloud server, we assume it to be honest but curious. Specifically, it will honestly store the encrypted data outsourced by the Data Owner and offer the search function to search on encrypted data both for the data owner and data users. The encrypted keyword preserves the confidentiality of the keyword. It also preserves the receiver's anonymity. However, it may be curious about some private information, such as the plaintext of the data stored in the cloud. Besides, we assume there is no collusion between the cloud server and the data users in our model. Also, any user in the system cannot modify the key allocated to him or generate a new key himself.

This section provides a detailed discussion of the security model for index-based data storage structure. The cloud server searches over a set of searchable indexes instead of searching on encrypted data directly. Our scheme aims to protect privacy associated with the single and multi-keyword search operation, which consists of different types of security listed below.

**Privacy Preservation:** The basic requirement of our scheme is privacy preservation. The encrypted data stored in the cloud should be kept secret from the cloud server. The search and search results should also be kept private from the cloud server and other query users. Our scheme should protect the privacy of indexes and queries simultaneously.

**Data Privacy:** Data privacy is a basic requirement that the outsourced data (or documents) should not be revealed to unauthorized parties, including cloud service providers. Data privacy can be easily achieved by encrypting the documents using a block cipher, such as AES, before outsourcing them to the cloud server. Typically, it can be guaranteed by symmetric encryption algorithms. The user who has the secret key can effectively decrypt the encoded data after retrieving them from the cloud server. The cloud server can only access the encrypted document set and the corresponding secure index that the data owner outsources.

**Index privacy**: Index privacy indicates that the server should not be aware of the keywords embedded in the index. Since the secure index can represent the encrypted documents, any further information (e.g., keywords) should not be deduced from the index by the cloud server. In general, index privacy refers to the information of keywords, document identifiers, and keyword locations. The plaintext information of encrypted searchable index and encrypted keywords should be protected from the cloud server.

**Keyword privacy:** The cloud server could not identify the specific keyword in query, index, or document collection by analyzing the statistical information like term frequency. First, a ciphertext without its corresponding encrypted keyword should not disclose any information about the plaintext keyword it contains to the cloud server and outsiders. Second, an encrypted text should not leak information on keyword values to any outside attackers without the private key that a trusted third party handles. The cloud server cannot identify whether a particular keyword is contained in a query by analyzing indexes or search results. The cloud server should not be able to learn the keywords in search or authentication tags generated by the user.

**Locally Index Search**: To search on the encrypted documents, the data owner and data users need to download the encrypted inverted index from the cloud server first. Instead of searching on an inverted index on the cloud, the scheme should be searching locally so that the cloud server can't learn anything about the keyword and document information, which makes the security and privacy of the data more legitimate.

**Secure Search:** When the data owner gives access to the data user, the data user can generate an encrypted text to search the encrypted data with the help of a trusted third party. Only the authorized data user can generate the encrypted text during the process. Moreover, the identity of the user is used in the search process, so the eavesdropper cannot deduce the real identity of the user

## 4.3 Design Goals

In this proposed model, our goal is to design an efficient and privacy-preserving keyword search scheme under the abovementioned system model and security model. In order to realize secure and efficient dynamic searchable encryption schemes over encrypted data, the following goals should be achieved. The details are listed below:

**Assumptions**: It is assumed that Trusted Third Party does not collude with any adversary. We consider the public cloud storage server semi-trusted, honest in performing the work assigned to him, but curious to learn the information from the data stored in it.

**Functional Goals**: The keywords extracted from a document are encrypted with a secret key. After getting the encrypted text generated by a trusted third party, the Data Owner/User should be able to search independently. On the other hand, the cloud server should be able to perform and fulfill the keyword search operation efficiently.

**Accurate Single-Keyword Search:** Accurate similarity results are retrieved by effective utilization of outsourced cloud data instead of getting unrelated results to design a search scheme for single-keyword searches [28].

**Multi-keyword Search:** To design a searchable encryption scheme that enables the cloud server to support multi-keyword search over encrypted data, it should meet the requirements of search patterns and achieve better search efficiency.

**Efficiency:** In order to achieve the privacy requirement, additional computational costs will be inevitably incurred, i.e., processing the keyword search over the encrypted data will undoubtedly increase the computational cost compared with those doing over the plaintext sets. Therefore, the proposed scheme should be adequately efficient in computation, communication, and storage for practical applications. Our scheme is very efficient in index construction, encrypted text generation, and search processing. The scheme aims to achieve search efficiency by exploring a secured inverted index and an efficient search algorithm.

**Non-impersonation:** The encrypted data belongs to the data owners or users who can control data access. Access control prevents unauthorized data access from the users who are not allowed to access the data. In other words, only an authorized user should access the secure documents, i.e., no adversary should be able to behave as an authorized user. The search capabilities are provided to the end-users after proper authentication and authorization. Simultaneously, data access activities are always be carried out with the participation and monitoring of a trusted third party.

**Dynamic Index Update**. The proposed scheme provides single or multi-keyword queries, accurate results, and dynamic updates on document collections. It may do more frequent updating operations than general searchable encryption schemes. However, all indexes are encrypted and outsourced to the cloud server. Considering the cloud environment, the number of indexes may be huge.

**Secure Communication:** In our system, we assume that the trusted third party will never reveal the secret key and random encrypted keys to unauthorized data owners, data users, and even the server; furthermore, the clients' private keys should be kept secret and cannot be stolen by attackers. In addition, we assume that the communication channels in this system are all encrypted and authenticated against eavesdropping.

## 4.4 Notations

For simplicity and readability, Table 4.1 describes the notations used throughout the report.

| **Notations** | **Description** |
|---|---|
| F | The plaintext document collection, denoted as a collection of n documents $F = \{F_1, F_2, \ldots, F_n\}$. Each document F in the collection can be considered as a sequence of keywords where $F_i$ is the ID of the *i*-th document. |
| n | The total number of documents in F. |
| C | The encrypted document collection is stored in the cloud server, denoted as $C = \{C_1, C_2, \ldots\ldots, C_n\}$. |
| D | A number of collections, initialized as {1,2,3,…,d}, where D is set as the possible maximum size of the outsourced document collection. |
| W | The dictionary, namely, the set of keywords denoted as $W = \{w_1, w_2, \ldots, w_m\}$ from F, which we denote as a dictionary. |
| m | The total number of keywords in W. |
| k(F) | The set of all keyword variants contained in the collection of documents F. |
| $F_w$ | The set of all files containing the keywords w. |
| FID | The file id returned by the cloud server |
| I | The unencrypted form of searchable inverted index $I = (I_{w1}, I_{w2}, \ldots, I_{wm})$ for the whole document collection F. Each $I_{wi}$ is a list that contains $F = \{ F_1, F_2, \ldots, F_n \} \subset F$ where $w_i \in F_j$. |
| $\bar{I}$ | The searchable encrypted inverted index generated from I. |
| $W_q$ | The subset of W represents the keyword in the query. |
| Q | The query vector for the keyword set $W_q$ |
| T | The encrypted keyword text is the encrypted form of Query Q. |
| $\bar{F}_w$ / Re | The returned name list of valid search results from the cloud server is denoted as $Re = \{R_1, R_2, \ldots R_k\}$ |

Table 4.1 Notations of the proposed model

## 4.5 System Definition

The scope of the proposed scheme is to generate the index of encrypted keywords for a document and perform a privacy-preserving search over the encrypted index. It includes the encryption and decryption of a document. Our proposed encrypted search scheme is shown in Algorithm 1, containing the following polynomial-time algorithms.

– **Setup**($1^l$): The Setup($1^l$) is run by a trusted third party. It takes as input parameter a security parameter $l$ and outputs the secret key *SK*, randomly generated 256-bits AES keys $R_{1,....,}$ $R_n$.

– **Encrypt Data** ($F$, $R_n$): The encryption is done by the Data Owner (DO) to encrypt each document (F) with the random key $R_n$ along with a reference number and upload the document to the cloud. The cloud service provider will return the File ID (FID) of the document.

– **Generate Index** (*SK*, *I*): The protocol is run by the Data Owner (DO), where *w* is the set of keywords associated with document D. The data owner starts the computation to generate the encryption for each keyword *w* included in the keyword set *W* to make an inverted index. It takes as input the secret key *SK* and an inverted index *I* for the document File ID (FID), which was got from the document set D. It outputs the encrypted searchable index $\bar{I}$.

– **Store** ($R_n$, FID): This protocol is the responsibility of the Data Owner (DO). The Data Owner stores the Random AES keys for each document and File ID (FID) to the Trusted Third Party.

– **Generate Encrypted Keyword** (*SK*, *w*): The Data User (DU) invokes this randomized algorithm to make an encrypted text for retrieving the documents from Cloud Server whose associated index contains an encrypted entry for the word *w*. The algorithm outputs an encrypted keyword $W_q$ generated for *w*.

– **Search** ($\bar{I}$, $W_q$): This is a deterministic algorithm run by the Web Application. The algorithm takes as input the encrypted keyword $W_q$ search by the Data User and encrypted Index $\bar{I}$, The algorithm returns true if the word in $W_q$ matches with any of the keywords included in the encrypted Index $\bar{I}$.

– **Decrypt Data (**C, $R_n$): The encryption is done by the Data Owner (DO)/Data User to decrypt the encrypted document (C) with decryption key $R_n$.

**Keys**

For our scheme, we have used three kinds of Encryption Keys.

**1. Secret Key:** Secret keys are like a hash function. A secret key is used to encrypt only the keywords in the Inverted Index.

Here, r = $\{0,1\}^n$ is for randomness.

So, Secret Key = Hash function (r + master secret label for identifying + Client Randomness + Server Randomness). So the secret key is a total of 384 bits which are explained in table 4.2

| | | |
|---|---|---|
| 128 bits | 128 bits | 128 bits |
| K for symmetric (i.e. RC4) | Initialization Vector (CBC) | Key for hash |

Table 4.2 Secret Key format size

**2. 256-bit AES Keys:** These keys will be used to encrypt each document. Also, with each random key like $R_1 \ldots R_n$, we need to generate the reference keys to determine which random keys are used for which document. Mainly reference numbers will help recognize the encryption/decryption keys. Note that, Encryption and Decryption keys will be the same random key.

**3. RSA Keys:** This RSA algorithm will generate Public (PK) and secret (SK) keys to give the decryption key (R) to the data used by a secure channel. By secure channel, we mean these RSA keys.

## 4.6 Scheme Scenarios

In our design model, we are proving three scenarios where the data owner, data user, and trusted third party interact to get access and search for encrypted data over the cloud. The scenario schemes with their mechanism are listed below:

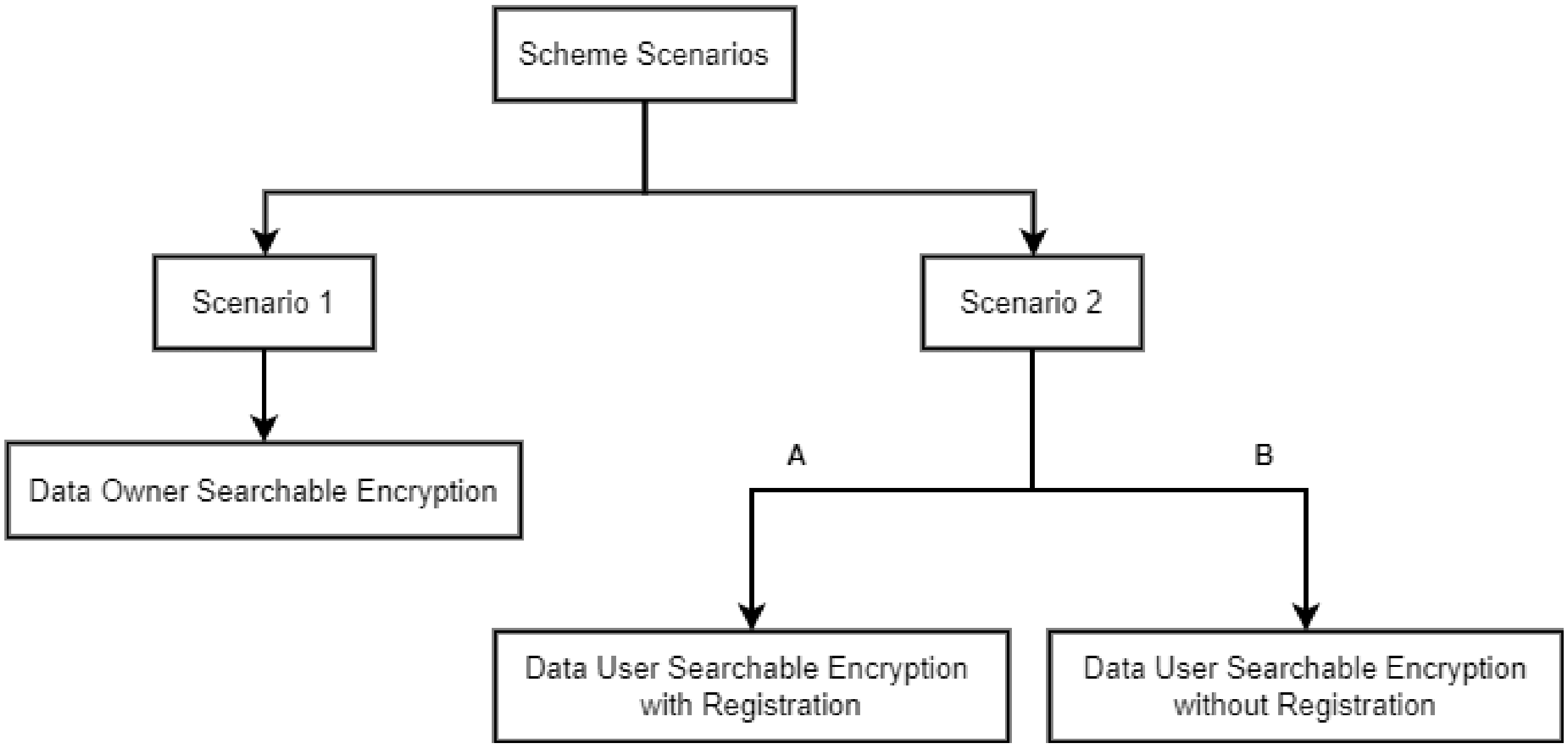

Figure 4.1 Scheme scenarios of the proposed model

**Scenario 1 – Data Owner Searchable Encryption:** The Data Owner will outsource the encrypted documents to the cloud storage and search by keyword on those encrypted documents by himself and get the result.

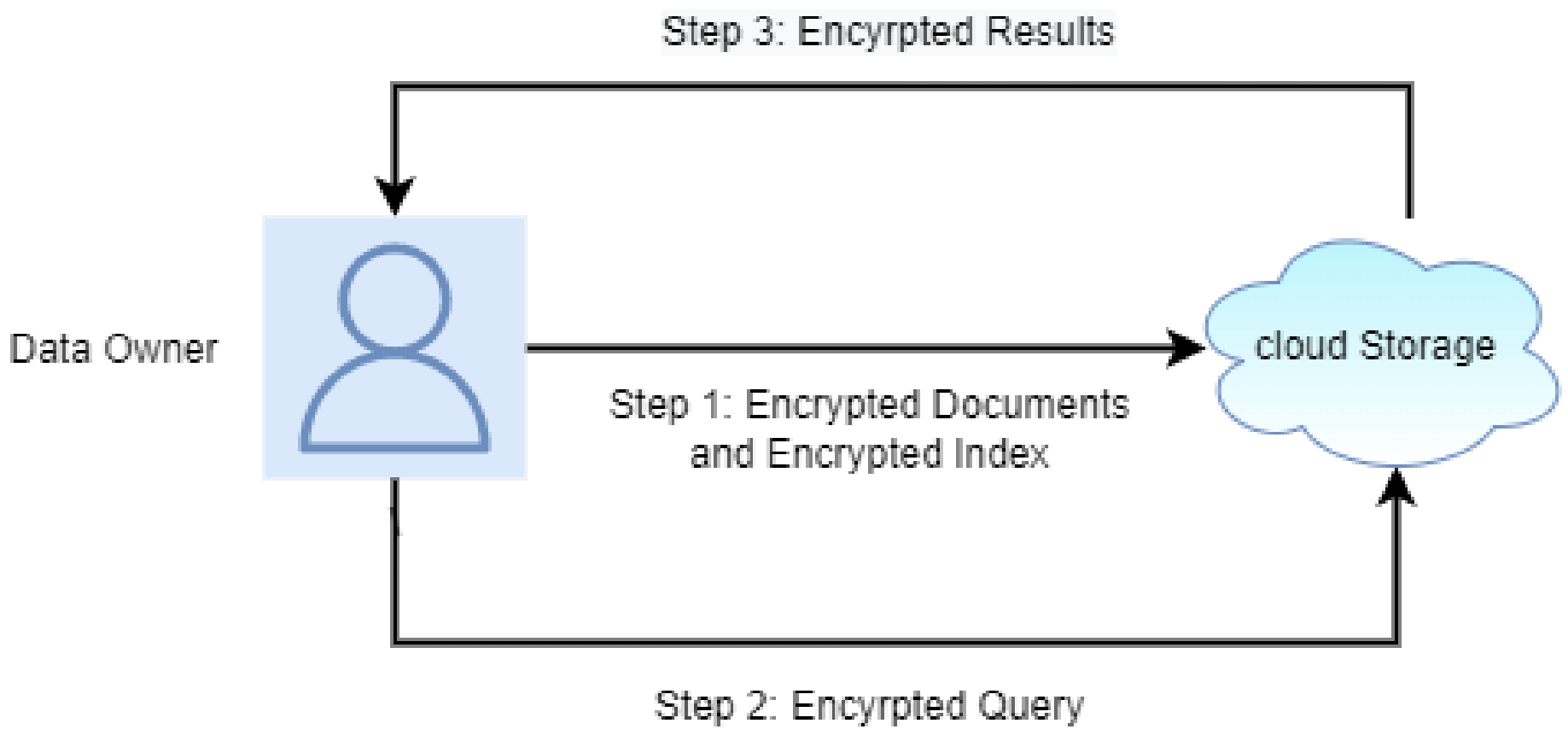

Figure 4.2 Data Owner Searchable Encryption Scenario

**Scenario 2(A) – Data User Searchable Encryption with Registration:** The Data Owner will outsource the encrypted documents and share those documents with the data user. Assume that the Data User is already registered in the web application and enlisted with Trusted Third Party. So that, The Data User can search by keyword on those encrypted documents and decrypt them with help from the Trusted Third Party.

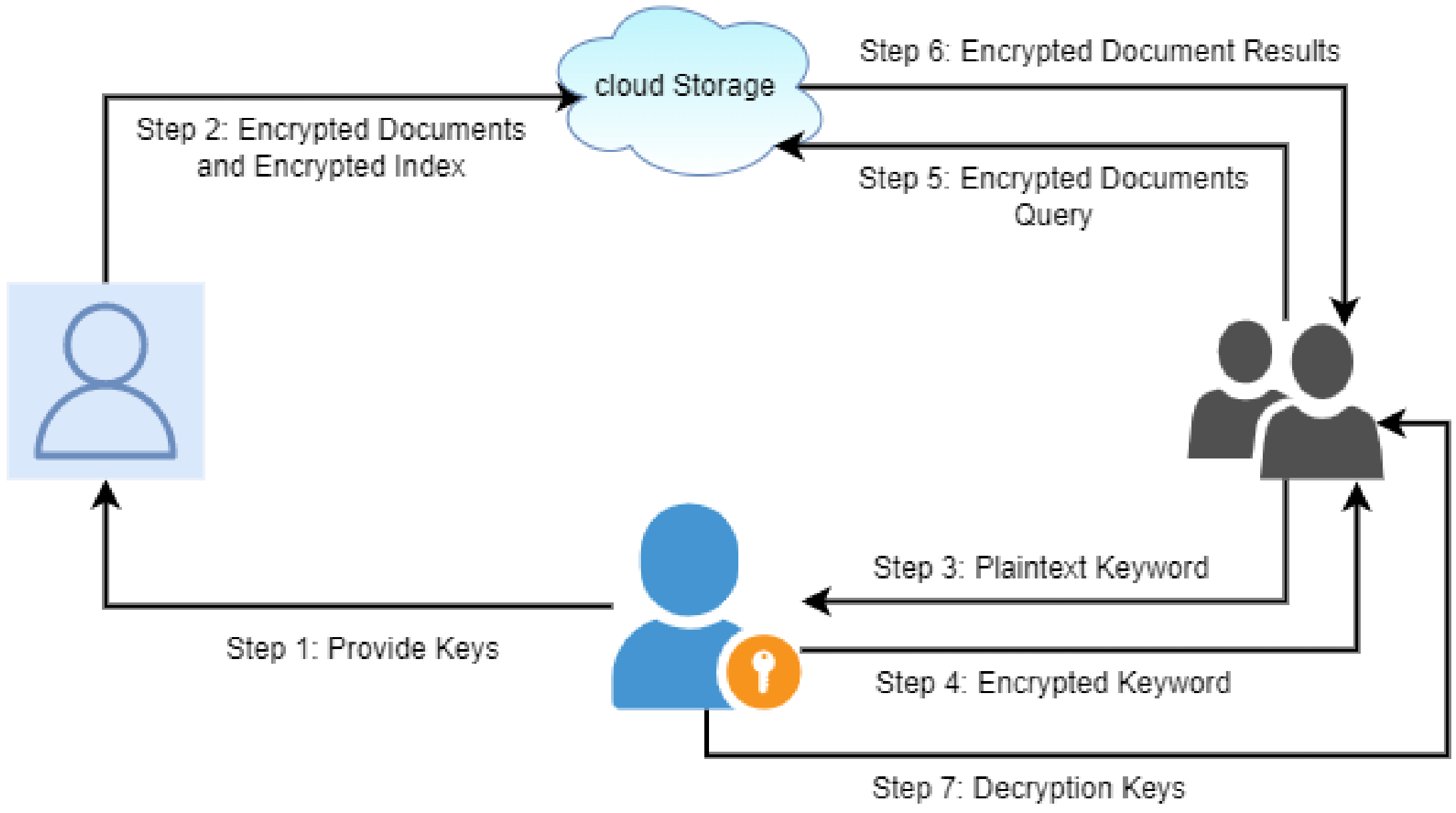

Figure 4.3 Data User Searchable Encryption with Registration Scenario

**Scenario 2(B) – Data User Searchable Encryption Without Registration:** The Data Owner will outsource the encrypted documents and share those documents with the data user. Assume that the Data User is not registered in the web application and then get an email notification that a folder has been shared with him to get access and download his desired documents; he needs to register in the web application first and needs to enlist with Trusted Third Party by Identity Based Encryption (IBE). After that, The Data User can search by keyword on those encrypted documents and decrypt them with help from the Trusted Third Party.

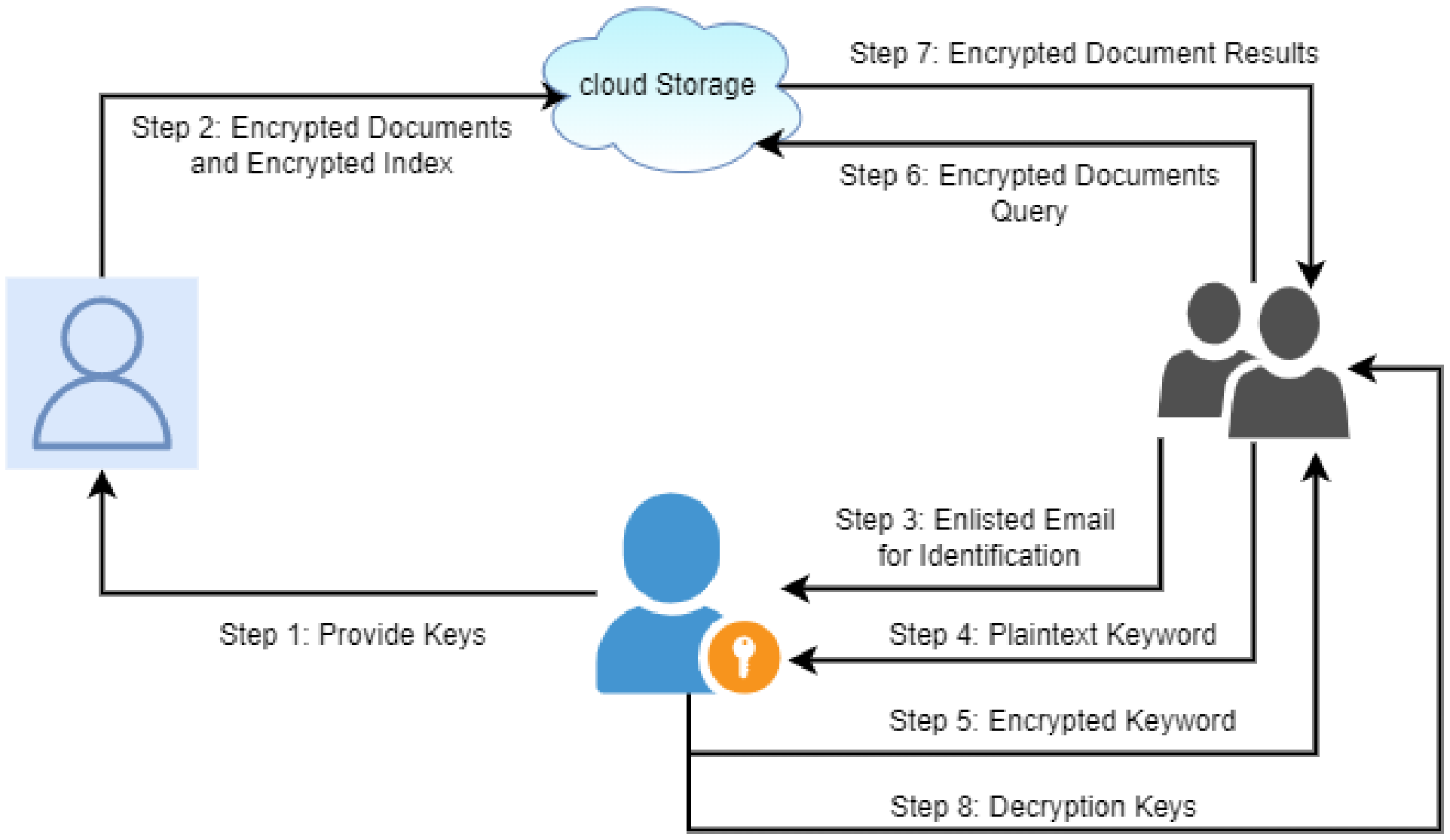

Figure 4.4 Data User Searchable Encryption without Registration Scenario

The details of each scenario are listed below step by step.

### 4.6.1 Scenario 1 – Data Owner Searchable Encryption

In scenario 2, we assume that the Data Owner is already registered in the web application. The following procedures with every step are discussed below:

1. To start the process of encrypting the documents and inverted index, the Data Owner needs to collect the secret and random keys. For that, he will send a request to the Trusted Third Party (TTP) for keys.

2. The Trusted Third Party will run the Setup algorithm to get the parameter. Then he will generate the secret and random keys with reference numbers

3. After generating, The Trusted Third Party will send those keys to the Data Owner.

4. The Data Owner (DO) will encrypt the documents by the random keys based on the reference numbers and outsources the encrypted documents to the cloud storage.

5. After receiving the encrypted file successfully, the Cloud Service Provider (CSP) will send a File ID for each document.

5. The data owner (DO) will generate the inverted index based on keywords from each document and File ID.

6. Then, he will encrypt the inverted index with a secret key and upload the index to cloud Storage.

7. After uploading encrypted documents and inverted index, the Data Owner will send the folder ID with a secret key and File ID with the reference number of random keys to the Trusted Third Party for later use in the decryption process.

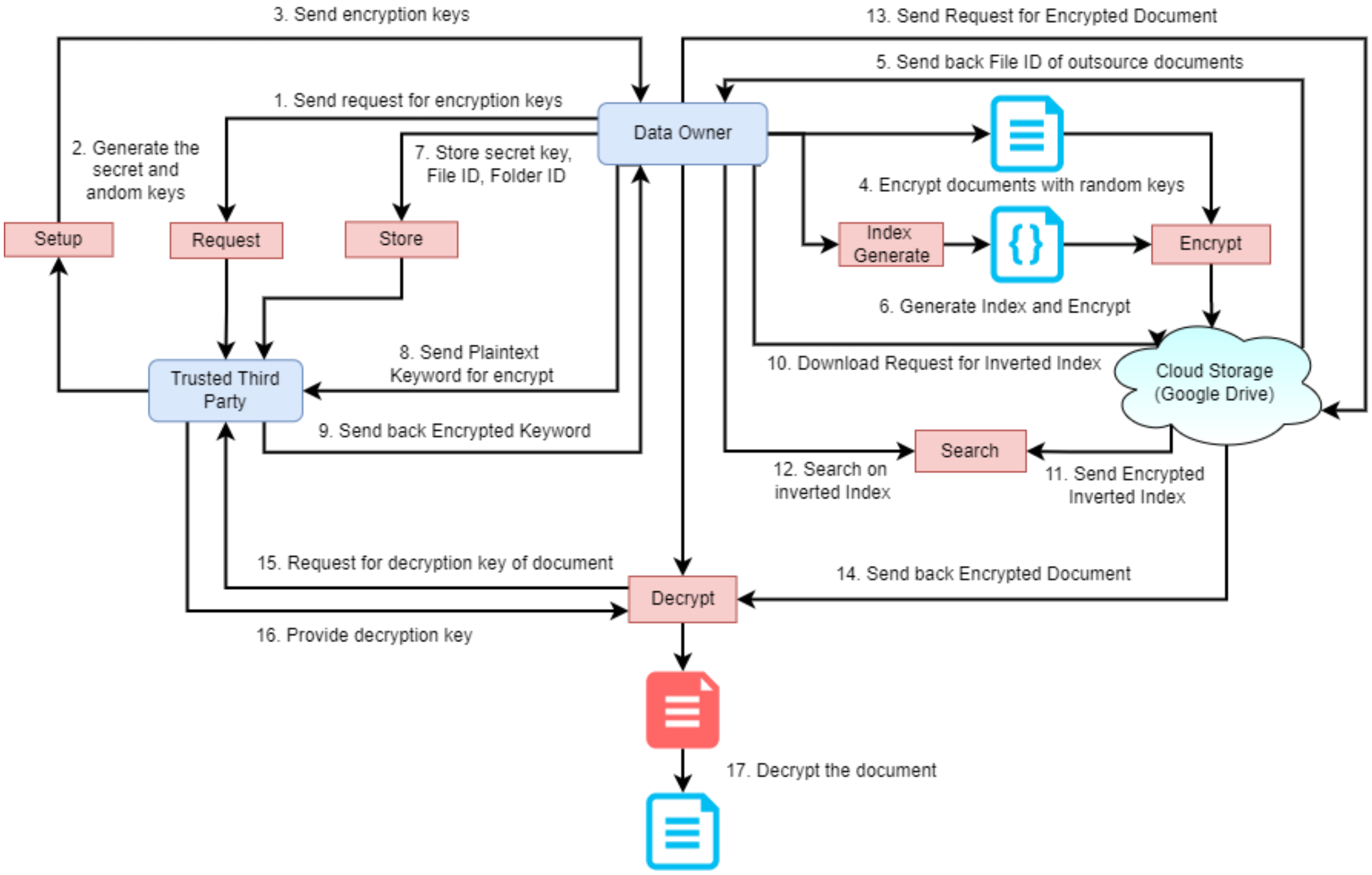

Figure 4.5 Flow diagram of Data Owner Searchable Encryption

8. The Data Owner will send the plaintext keyword to the trusted third party (TTP) for encrypted text. He can search by any keyword which is relevant to the document.

9. The Trusted Third Party will generate an encrypted keyword by combining the secret key and plaintext keyword. He will send the encrypted keyword to the Data Owner.

10. After getting the encrypted keyword for search, the Data Owner will download the encrypted inverted index from the cloud server. The index will store on the web server.

11. After getting a request from the data user, Cloud Service Provider will send the encrypted inverted index list to the Data Owner.

12. In this stage, the data user will search by his encrypted keyword from the Trusted Third Party (in stage 9) to the encrypted inverted index to get the File ID of the documents.

13. After successfully searching, the Data Owner will get the document's file ID and send a request to the cloud service provider for that encrypted document.

14. The cloud Service Provider will send back the encrypted document to the Data Owner.

15. Now, the Data Owner has the encrypted document. So, he needs a decryption key to decrypt the document. He/she will send a request for the decryption key to the Trusted Third Party.

16. The Trusted Third Party will check the decryption key of the document. He already has those decryption keys listed with reference numbers from the Data Owner. (Step 7).

17. Finally, the Data Owner will decrypt the encrypted document with a decryption key.

### 4.6.2 Scenario 2(A) – Data User Searchable Encryption with Registration

In scenario 2(A), we assume that the Data Owner and the Data User are already registered in the web application. It is assumed that the Data Owner has shared the whole document folder with the Data User already. The following procedures with every step are discussed below:

1. To start encrypting the documents and inverted index, the Data Owner needs to collect the secret and random keys. S(he) will send a request to the Trusted Third Party (TTP) for keys.

2. The Trusted Third Party (TTP) will run the Setup algorithm to get the parameter. Then he will generate the secret and random keys with reference numbers and send those keys to the Data Owner (DO).

3. The Data Owner (DO) will encrypt the documents by the random keys based on the reference numbers and outsources the encrypted documents to the cloud storage.

4. After receiving the encrypted file successfully, the cloud service provider (CSP) will send a File ID for that file as response.

5. The data owner (DO) will generate the inverted index based on keywords from each document and File ID.

6. Then, he will encrypt the inverted index with a secret key and upload the index to cloud Storage.

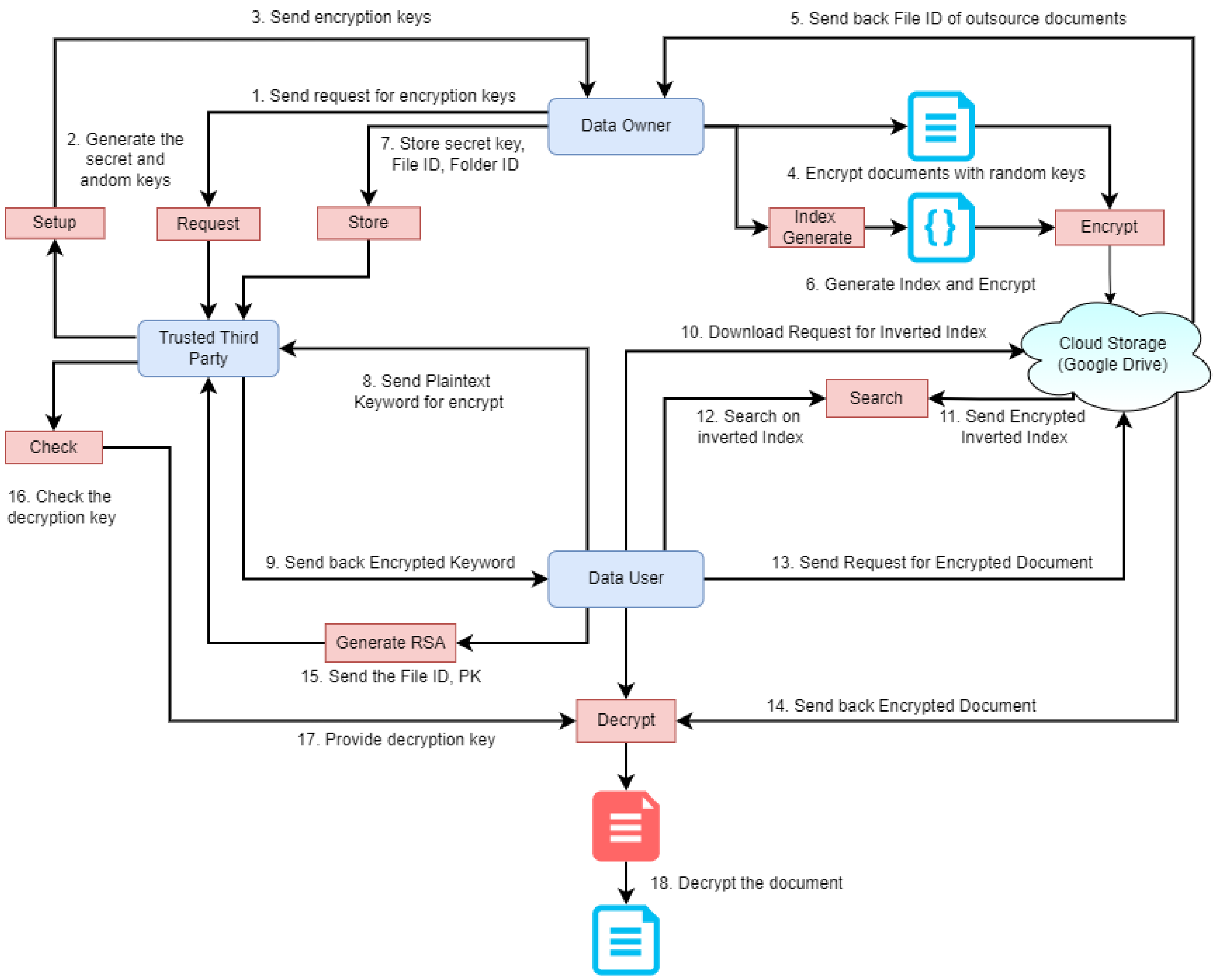

Figure 4.6 Flow diagram of Data User Searchable Encryption with Registration

7. After uploading encrypted documents and inverted index, the Data Owner will send the folder ID with the secret key and File ID with the reference number of random keys to the trusted third party (TTP) for later use in the decryption process.

8. For the search process, the Data User will search with a single keyword of document. He can search by any keyword which is relevant to the document. The data user will send the plaintext keyword to the trusted third party (TTP) for encrypted text.

9. The Trusted Third Party will verify the data user and generate an encrypted keyword by combining the secret key and plaintext keyword. He will send the encrypted keyword to the data user.

10. After getting the encrypted keyword, the data user will download the whole encrypted inverted index. The index will store on the web server.

11. So, he needs to download the inverted index first for searching. After getting a request from the data user, Cloud Service Provider will send the encrypted inverted index list to the data user.

12. In this stage, the data user will search by his encrypted keyword from the Trusted Third Party (in stage 9) to the encrypted inverted index to get the File ID of the documents.

13. After successfully searching, the Data User will get the File ID of the documents and send a request to the cloud service provider for those encrypted documents.

14. The cloud Service Provider will send back those encrypted documents to the Data User.

15. Now, the Data User has encrypted documents. So, he needs decryption keys to decrypt those documents. For that, s(he) will randomly generate RSA public key (PK) and private key (PR) and send the File ID, Public key (PK) to the Trusted Third Party.

16. The Trusted Third Party will check the decryption keys of the documents. He already has those decryption keys listed with reference numbers from the Data Owner (Step 7).

17. After that, the Trusted Third Party will encrypt the decryption keys of the documents by the Data User's Public Key (PK) and send them to the Data User.

18. Finally, the data user will decrypt the document's encryption keys using his Private Key (PR) and decrypt the encrypted documents by a decryption key.

### 4.6.3 Scenario 2 (B) – Data User Searchable Encryption without Registration

1. At the beginning process, the Data Owner needs to provide the email address (E) of the Data User with whom the file needs to be shared.

2. The web application receives the Data Owner's input and takes Data User's email address (E) into account.

3. The web application will communicate with the Trusted Third Party to check if the provided email address (E) is associated with any account.

4. The Trusted Third Party will check if the Data User is registered already.

5. If no user account is associated with (E), an email will be sent to the email address asking to register in the web service to receive the file. However, if the Data User is registered in a web service, the Data User can access the shared folder.

6. The Data User will send a plaintext keyword for search to the Trusted Third Party.

7. The Trusted Third Party will verify the data user and generate an encrypted keyword by combining the secret key and plaintext keyword. He will send the encrypted keyword to the data user.

8. After getting the encrypted keyword, the data user will download the whole encrypted inverted index.

9. Cloud Service Provider will send the encrypted inverted index list to the data user. The index will store on the web server. The Data User will search by his encrypted keyword from the Trusted Third Party (in stage 7) to the encrypted inverted index to get the File ID of the documents.

10. After successfully searching, the Data User will get the File ID of the documents and send a request to the cloud service provider for those encrypted documents.

11. The cloud Service Provider will send back those encrypted documents to the Data User.

12. Now, the Data User has encrypted documents. So, he needs decryption keys to decrypt those documents. For that, he will randomly generate RSA public key (PK) and private key (PR) and send the File ID, Public key (PK) to the Trusted Third Party.

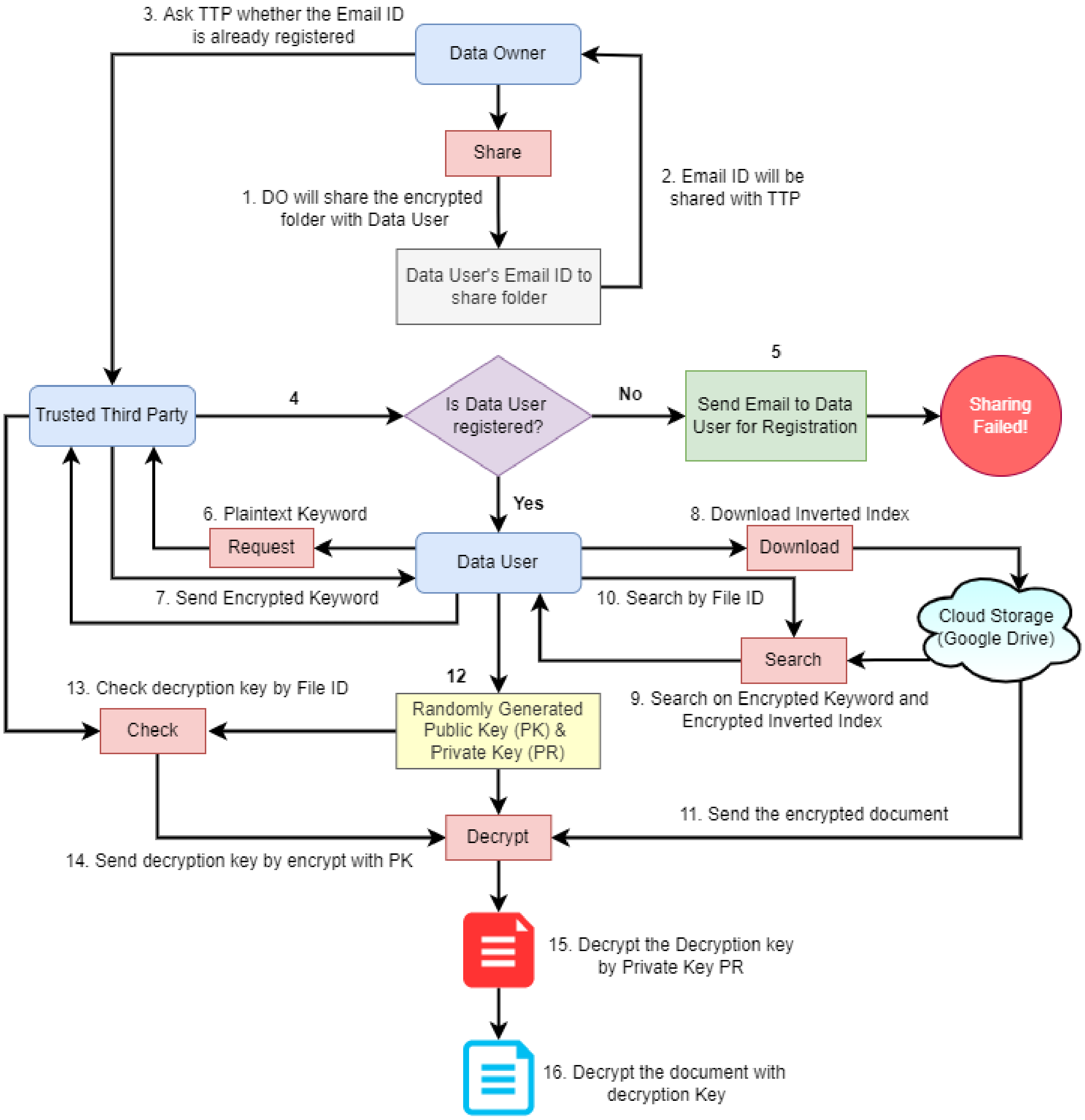

Figure 4.7 Flow diagram of Data User Searchable Encryption without Registration

13. The Trusted Third Party will check the decryption keys of the documents.

14. After that, the Trusted Third Party will encrypt the decryption key of the document by the Data User's Public Key (PK) and send it to the Data User.

15. The data user will decrypt the document encryption key using his Private Key (PR).

16. Finally, he will decrypt the encrypted document with a decryption key.

## 4.7 Details of Every step

In the above three scenarios, some steps such as Sign Up, sign in, setup, request, generate index, search, download, encrypt, decrypt, etc., are used. The details of those steps are explained below with the figure as follows.

### 4.7.1 Sign Up

1. During the Sign Up process, the Data Owner/User needs to provide his Email (E) and Passphrase (Ph) as input.
2. Then, Ph will be hashed using the Bcrypt algorithm as H(Ph). Bcrypt is a password-hashing function designed by Niels Provos and David Mazières, based on the Blowfish cipher.
3. Email (E) and Hashed Passphrase are sent to the server. (TTP will be responsible for storing the information of Sign Up of the Data Owner/Data User) by the web application.

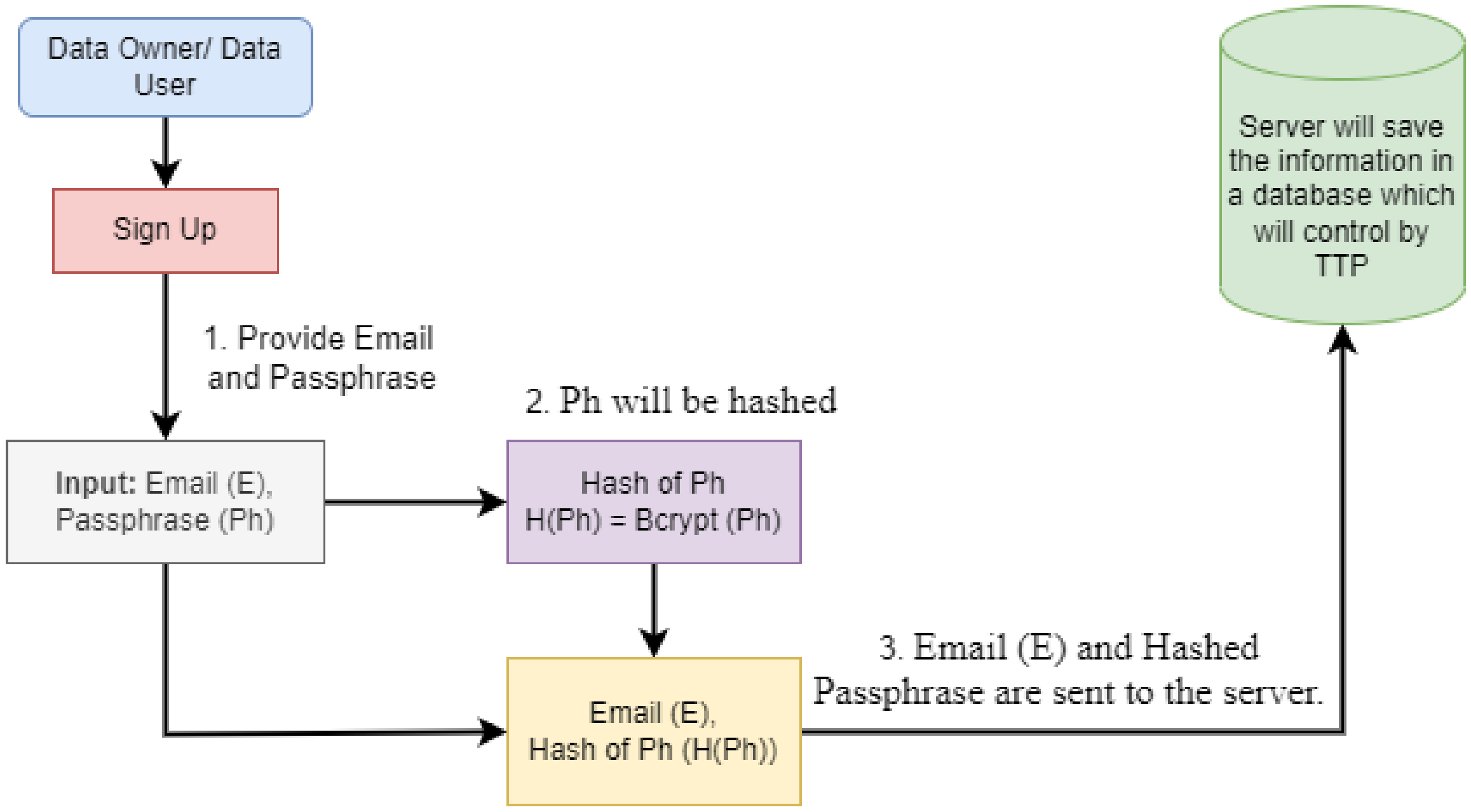

Figure 4.8 Sign up process flow diagram

### 4.7.2 Sign In

1. At first, the user needs to provide his/her credentials (Email, E and Passphrase, Ph) to the web application.
2. The web application receives the user's input and takes the email address (E) into account.
3. The web application then requests the server for user information (Ui) corresponding to the provided email address.
4. The server returns the appropriate user information (Ui) associated with the email address (E). User information (Ui) contains the previously stored hash value of the user's passphrase, H(Ph).
5. The web application then checks whether the previously stored hash of the user's passphrase H(Ph) matches the passphrase provided in step 1.
6. If both the hash values match, the data owner/user is signed in successfully, and he will be able to access all the web application functionalities. But if the hash values do not match, the data owner/user will not be allowed to use functionalities such as; generating the inverted index, encrypting the document, uploading, searching keywords, etc.

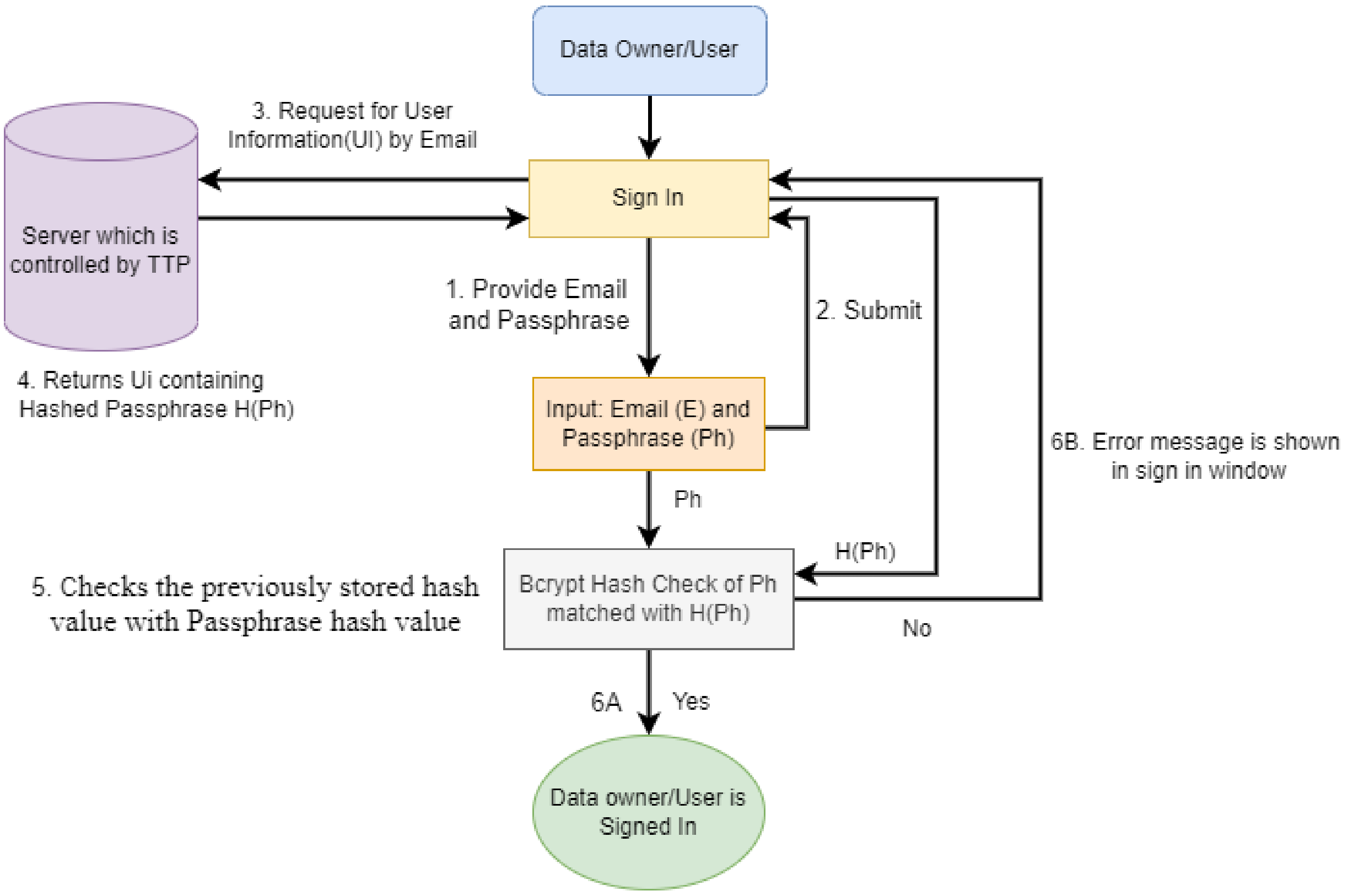

Figure 4.9 Sign In process flow diagram

### 4.7.3 Setup and Key Generate

1. Data owner will send a request to the trusted third party that he needs secret and random keys to encrypt the documents. He will mention how many random keys he will need to encrypt the documents.

2. After getting the request, the trusted third party will take a security parameter for input.

3. After taking the security parameter, the secret key SK for the inverted index will be generated.

4. Random keys $R_{1...}R_n$ for each document to encrypt. Here the random keys will be generated by 256-bit AES with the reference number. The reference number will help recognize which key will be used for encryption and decryption.

5. After generating those keys, the trusted third party will send those keys to the data owner.

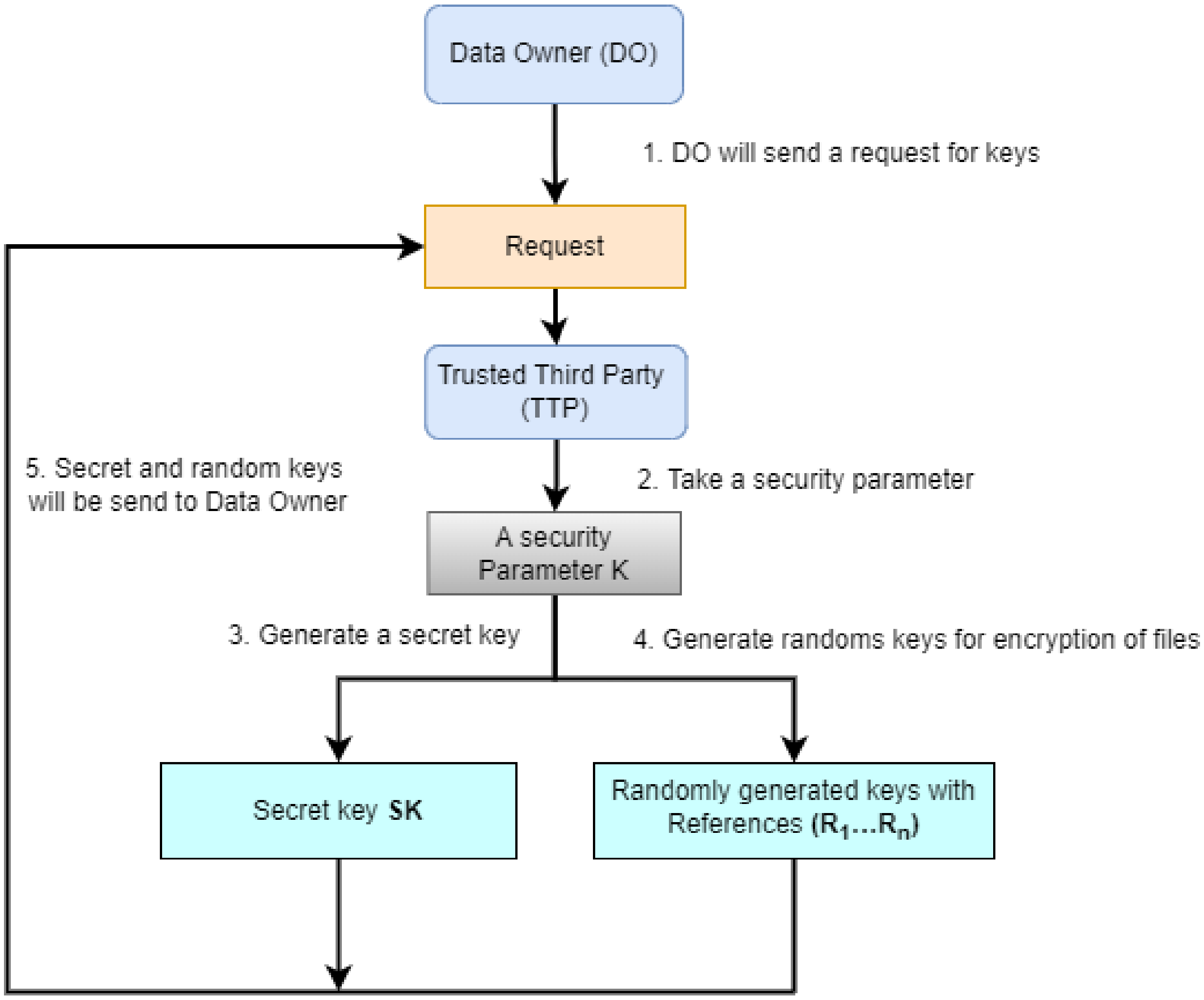

Figure 4.10 Setup and key generate flow diagram

### 4.7.4 Encrypt and Upload

1. Data owner will encrypt those documents using random keys $R_1 \ldots R_n$ respectively.

2. He will upload those documents to Google Drive.

3. Google drive will send the File ID of each document.

4. After getting the File ID, he will generate the Inverted Index. Here, the Inverted Index has two columns: one is for a keyword, and another is for File ID. In Keyword Column, one single key will be saved. In particular, with that keyword, one or more documents' File ID will be saved as many documents may have the same keyword. Also, the data owner will do this step one by one. For example, Google Drive will send a File ID after uploading the document. Then he will include the keyword and File ID of that document in the inverted Index list.

5. After completing the build of Inverted Index, the Data owner will encrypt the Keyword column using Secret Key SK and upload the index to the google drive.

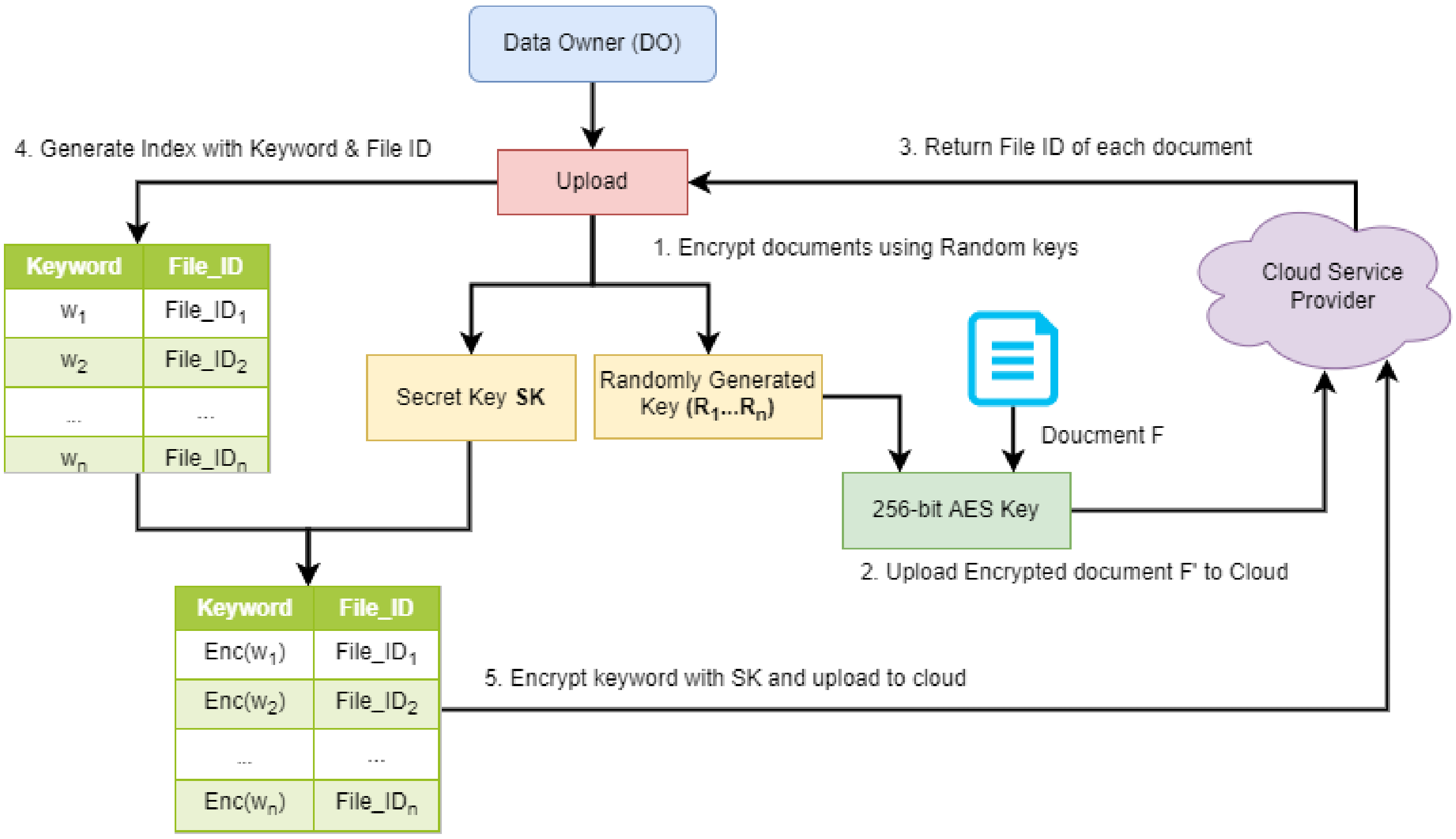

Figure 4.11 Encrypt and Upload flow diagram

### 4.7.5 Store and Share

We assume that the data owner will upload multiple documents in the same folder and share the folder with other data users. The data owner will follow the below steps to store and share documents.

1. After uploading the documents and inverted index to the cloud, the data owner needs to store the Folder ID along with the secret key, reference number of random keys, and file ID. It is necessary to sort out by the trusted third party which documents were encrypted by random keys and which folder ID is related to which inverted index to generate the encrypted text later. Here folder ID will need to share to recognize by the trusted third party by which the secret key was used to generate the encrypted inverted index. Thus, the TTP can use the same secret key to generate encrypted text requested by the data user.

2. TTP will save the above information for later use.

3. After completing the storing documents, index, and keys, the data owner will share that folder with the data user.

4. An email will be sent to the data user that the data owner's email id shares a folder with documents.

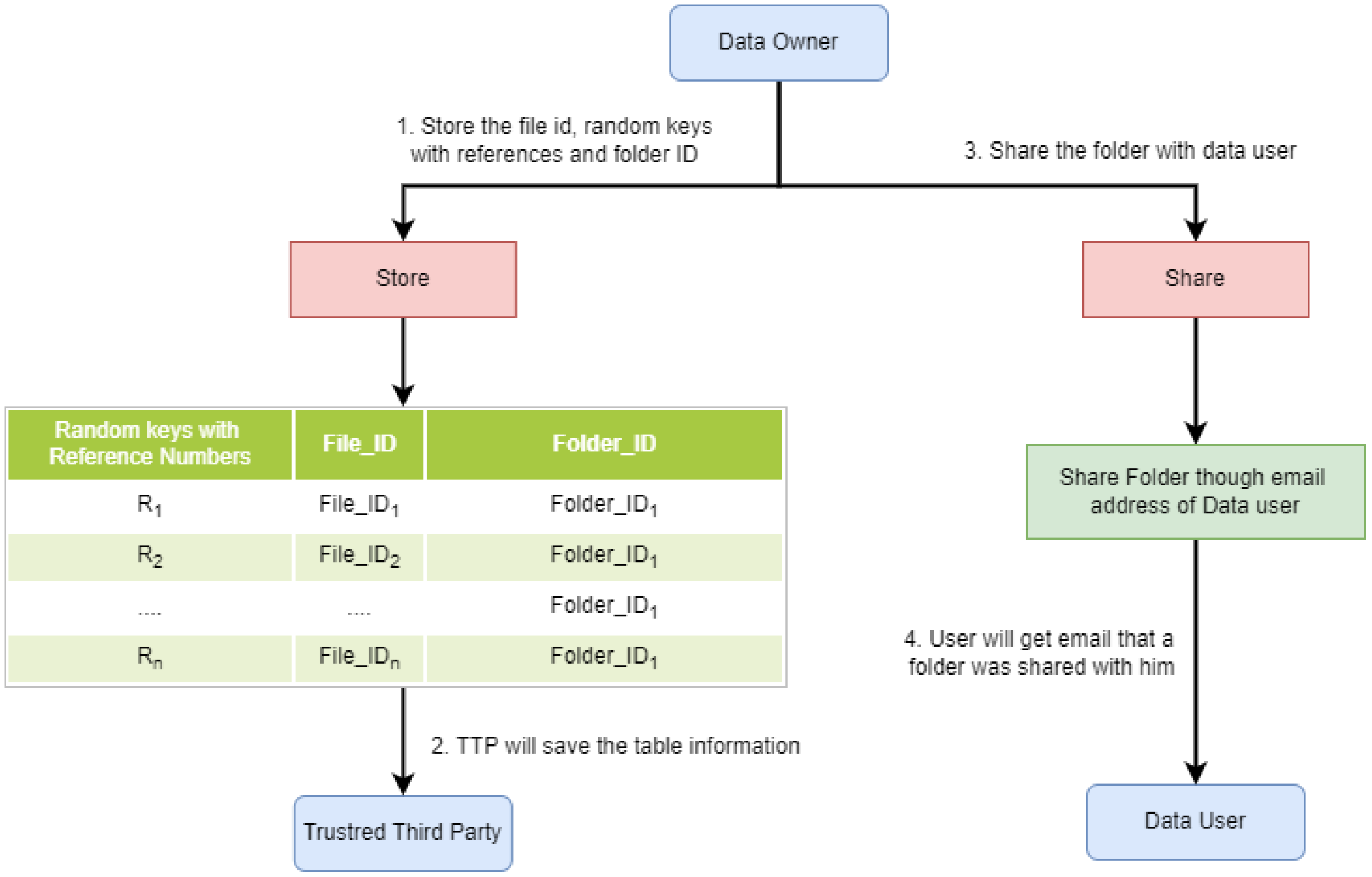

Figure 4.12 Store and Share flow diagram

### 4.7.6 Download Index and Search

1. Firstly, the data user will download the encrypted inverted index from google drive. It will be stored in the same folder which the data owner shared with the data user.

2. Google Drive will return the encrypted inverted index to the data user. As Google Drive has no function to search on the index directly, so we will download it and search on it locally. After downloading the inverted index, it will save to the web application server so that the data user can search on it quickly.

3. After successfully storing the inverted index, the data user will send a plaintext keyword and folder ID to the trusted third party. He will only send the plaintext keyword because we do not want to share the secret keys and cryptographic complexities with the data user.

4. After getting the plaintext, the trusted third will check the folder ID and generate an encrypted text with the secret key SK.

5. Trusted Third Party will send that encrypted text to the data user.

6. The data user can search with that encrypted text to the encrypted inverted index.

7. After searching locally on the web application, it will show whether the file ID was found or not.

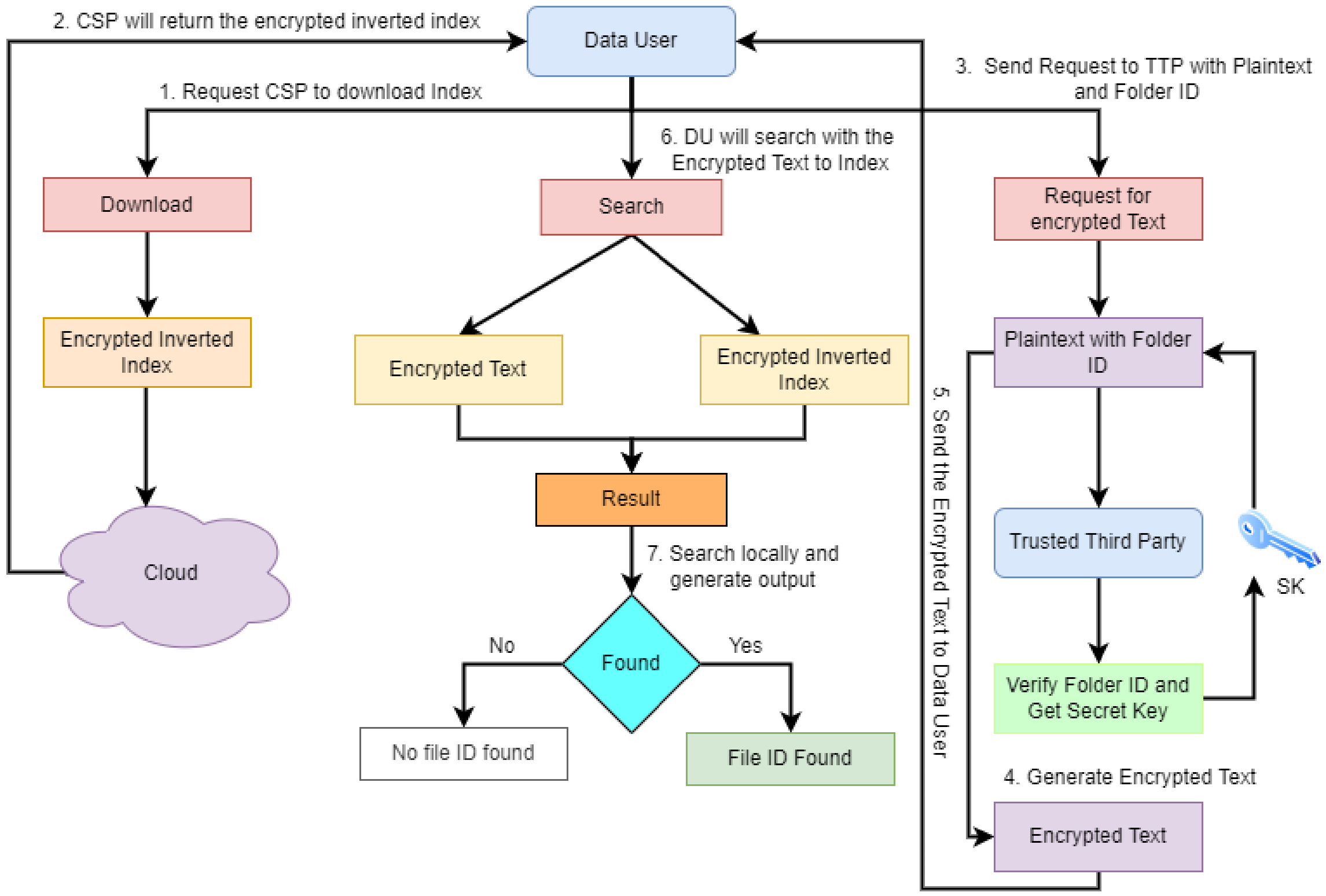

Figure 4.13 Download Index and Search flow diagram

### 4.7.7 Decrypt

1. After getting the file ID, the Data user can download that encrypted document by the API of google drive.

2. Google drive will send the that encrypted document to the data user.

3. Now, the data user has the encrypted document, but he needs to get the decryption key. He will send a request to the trusted third party by that encrypted document's File ID.

4. The Data User will generate a pair of random RSA public (Pu) and private key (Pr). RSA keys are required for the encryption of the decrypted key. A public key (Pu) is used for encryption, and a private key (Pr) is used for decryption. The Data User will send the Public Key (Pu) to the Trusted Third Party.

5. Trusted Third Party will check the decryption key that he got from the data owner.

6. The decryption key of the document will be encrypted by Public Key (Pu) and sent to the data user.

7. The Data user will first decrypt the decryption key of the document by the private key (Pr)

8. Lastly, The Data user will decrypt the document with the decryption key.

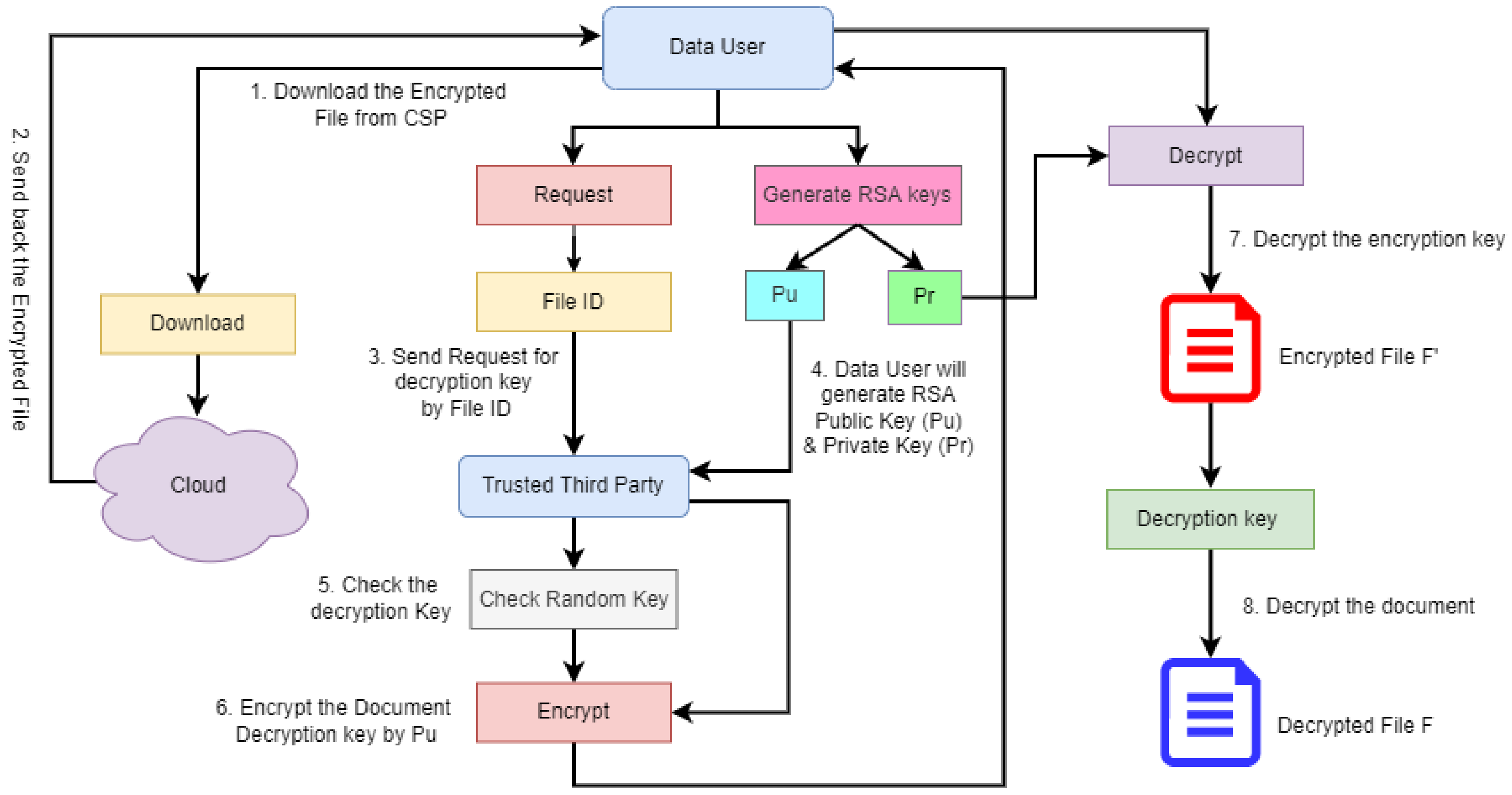

Figure 4.14 Decrypt process flow diagram

# 5 CHAPTER FIVE: IMPLEMENTATION DETAILS

Without practical implementation, theoretical models do not convey much value to the real world. For that, we are wanted to implement our proposed model so that it can be used in day-to-day life. We have implemented a web application called "CryptoSearch," where our proposed model is applied. During implementation, our main focus was to make it as simple as possible so that anyone with very little technical knowledge could operate the web application. At the same time, we have focused on not compromising users' privacy and security in the cloud.

In this chapter, we are focused on the implementation details of our proposed model. We describe how our model is implemented, which technologies, framework and database we have used. We also discuss about the schema diagram, and the design principles we have followed during our implementation. Our model comprises two components; a Frontend side application and a Backend side application. So, we have a separate discussion regarding the frontend side application and the backend side application.

## 5.1 Frontend Side Application

We are called the frontend side application of our implementation, "CryptoSearch Frontend". We have decided to make our frontend application cross-platform as a web application. It would be helped us to run the same application across all the devices regardless of hardware capabilities or operating systems they are running (e.g., PC, Mobile devices, etc.). This would allow us to run the application on the user's device, and we could also optimize the application's performance. Though web applications are run on servers, serious privacy leaks would cause. This means that all the sensitive information (e.g., passphrases, random keys, etc.) would be sent to the server in plaintext. However, we have used cryptography in every process so that it will not violate our proposed model's purpose.

### 5.1.1 Technologies Used

We had to decide which cloud service provider we would be working on. There were several options, e.g., Google Drive, Dropbox, Amazon AWS, etc. We chose to go with Google Drive as it is one of the free solutions available. It is a Google product, so it works seamlessly with Google's vast ecosystem, including the Android operating system, Gmail, Google Home, and, most importantly, the Google Workspace suite of office apps. Using Google Drive's integration with Workspace lets easily collaborate on cloud-based documents with co-workers of a company. For interacting with Google Drive from our client application, we used Google Drive API v3. For Google Drive authentication, we have used OAuth v2.0. We have used ReactJS with Node JS (version 16.15.0), NPM (Node Package Manager) (version 8.5.5), Redux, and Redux Persist for front-end design. Our frontend and backend server communication are done via REST API using JSON (JavaScript Object Notation) format.

### 5.1.2 Design Principles

During the design phase of our Crypto Search Web Application, we have focused on both external and internal design. We have followed some design principles that would make our web application more modular with zero compromises to the performance.

The exterior of the user interface (UI) is designed with user-friendliness in mind. Most of the complex operations are performed in the background without the requirement of user interaction. The basic user interface of the CryptoSearch Application is shown in Figure 5.1.

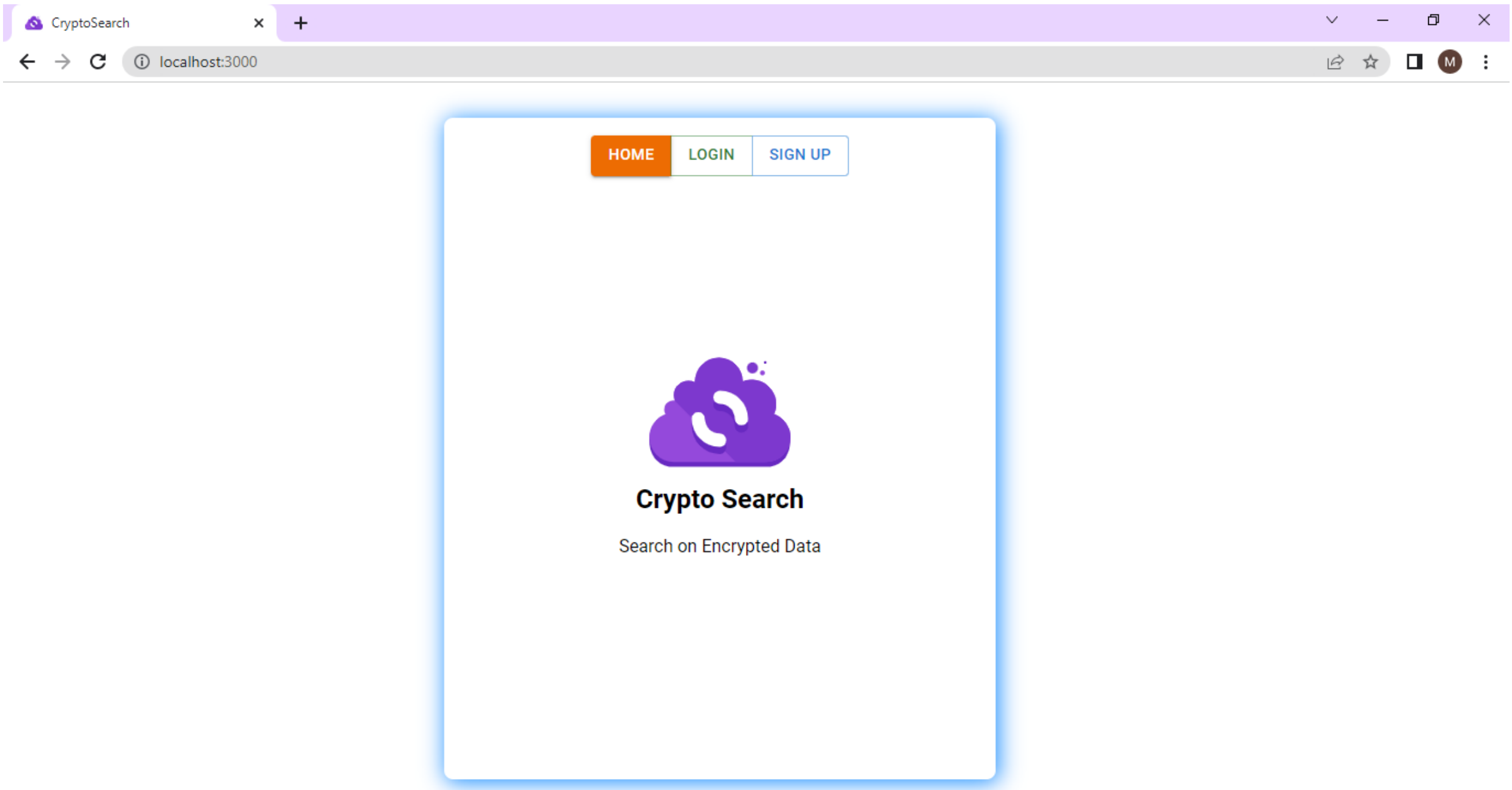

Figure 5.1 Basic User Interface of CryptoSearch Web Application

Only a single portion of the UI is shown in this figure. We will cover the rest of the parts in Section 5.1.3, Functional specification, later in this chapter, which will detail each operation that can be performed through the UI.

The internal architecture is designed in a very modular manner. Two separate threads are working concurrently. One is the main thread, or in our case, it is the UI thread. The other is the background thread. UI thread handles all the user interactions that take place in the UI. On the other hand, the background thread performs all the operations, e.g., communicating to the server, storing data, searching and retrieving data from cloud storage, and performing all the cryptographic operations like encryption/decryption, hashing, etc.

These two threads are separated for users to have a lag-free experience. Otherwise, UI would remain stuck until that operation is finished, which a time-consuming operation is performed. e.g., the user is uploading or downloading a file of size 1 GB. Downloading and decrypting such a big file would take a noticeable amount of time. It would have been stuck until 1 GB was utterly downloaded and decrypted if it were a single-threaded application.

To make this application more modular, we are explicitly taking one measures. The read constant values from a configuration file. It will allow us to change the values anytime without changing the source code. The configuration file is written in JSON format. The configuration can be easily modified as JSON data is easier to read is shown in figure 5.2.

```
cryptography: {
  bcryptSaltGenerationRounds: 1,
  aesAlgorithm: 'aes-256-ctr',
  aesInitializationVector: [32, 27, 169, 241, 200, 189, 141, 73, 29, 56, 165, 241, 66, 53, 42, 108],
  rsa: {
    algorithm: 'rsa',
    modulusLength: 4096,
    publicKeyEncoding: {
      type: 'spki',
      format: 'pem'
    },
    privateKeyEncoding: {
      type: 'pkcs8',
      format: 'pem',
      cipher: 'aes-256-cbc',
      passphrase: 'b8ab#4zf9hc44f2@2b573e7324142fe6b0d!dbe$'
    },
  },
},
randomKeyLength: 64,
secretKeyLength: 64,
googleDrive: {
  scopes: [
    'https://www.googleapis.com/auth/drive'
  ],
  credentialsPath: './google-api-credentials.json',
},
```

Figure 5.2 Configuration file containing JSON data

### 5.1.3 Data Selection

We have selected the electronic health record dataset for testing our implementation result. An electronic health record (EHR) is a digital version of a patient’s paper chart. EHRs are real-time, patient-centered records that make information available instantly and securely to authorized users. While an EHR does contain patients' medical and treatment histories, an EHR system is built to go beyond standard clinical data collected in a provider’s office and can be inclusive of a broader view of a patient’s care. EHRs are a vital part of health IT and can:

- Contain a patient’s medical history, diagnoses, medications, treatment plans, immunization dates, allergies, radiology images, and laboratory and test results.
- Allow access to evidence-based tools that providers can use to make decisions about a patient’s care.
- Automate and streamline provider workflow.

One of the key features of an EHR is that health information can be created and managed by authorized providers in a digital format capable of being shared with other providers across more than one health care organization. EHRs are built on sharing information with other health care providers and organizations - such as laboratories, specialists, medical imaging facilities, pharmacies, emergency facilities, and school and workplace clinics. So, they contain information from all clinicians involved in a patient's care.

EHR technology has become widespread and has attracted attention in healthcare institutions and research [119]. Cloud services build efficient EHR systems and obtain the most excellent benefits of EHR implementation. Many issues have recently surfaced in building an ideal EHR system in the cloud, especially the trade-off between flexibility and security. The security of patient records in cloud platforms is still a point of contention. With cybersecurity concerns rising, encryption can go a long way to protect patients' records [120]. First, it allows only authorized users to access sensitive data. Second, it protects against data breaches, whether the data is in transit or at rest. EHR solutions can code the information so that authorized users and programs can only read it. This makes transferring sensitive patient data such as test results or the transfer of medical histories to referrals much more secure than archaic paper records.

Encryption also reduces potential damage in case data gets stolen. It also enables role-based access control - this way, only employees with authorization can view the decrypted data. Also, with the digitization of traditional medical records, medical institutions encounter difficult problems, such as electronic health record storage and sharing [121]. Patients and doctors spend considerable time querying the required data when accessing electronic health records. However, the obtained data are not necessarily correct, and access is sometimes restricted. So, it's important to search on the encrypted patient's data without violating the security issues. We have chosen a 100,000-patient database that contains in total 100,000 patient's information, 361,760 admissions, and 107,535,387 lab observations [122]. Note that, as the patient's information is susceptible to data sets, all of the information in data sets is virtually created.

### 5.1.4 Functional Specification

In software development, functional specification specifies the functions that a system or component must perform. This documentation usually describes what is required to be done by the user, and at the same time, it also specifies what should be the inputs and outputs of an operation. This section will discuss how our CryptoSearch Web Application works. This will also cover all the user interface parts (UI). We have used the Google Chrome browser (Version 101.0.4951.67 (Official Build) (64-bit) to demonstrate the implementation. The process of each step is discussed below.

### Sign Up

At the first launch of the CryptoSearch Web Application, the user will need to sign up to use the application. The user needs to provide his/her email address and passphrase for signing up. At this stage, we are assuming that a user does not have a CryptosSearch account yet. So, we will first discuss the Sign-up page. A new account can be created by clicking the "Sign Up." Landing on the account creation page, the user will find the following information already placed in the specified fields. e.g., email and passphrase. The user will need to provide an email and passphrase for his/her account. The passphrase should be at least 8 characters in length, including upper, lower, numeric, and special characters. Otherwise, Sign Up process will be failed.

Suppose erroneous input is detected in the fields during account creation. In that case, an error message will be shown in the body area. e.g., suppose the passphrase provided is less than 8 characters. In that case, the front-end application will not validate the input and generate an error. If a user is already registered and attempts to re-register, it would also generate an error and show a notification that the email id already exists. We have kept the sign-up process as simple as possible. Note that the user can be the data owner or the data user. The sign-up procedure is the same for all the users who will use the application. The sign-up page is shown in figure 5.3.

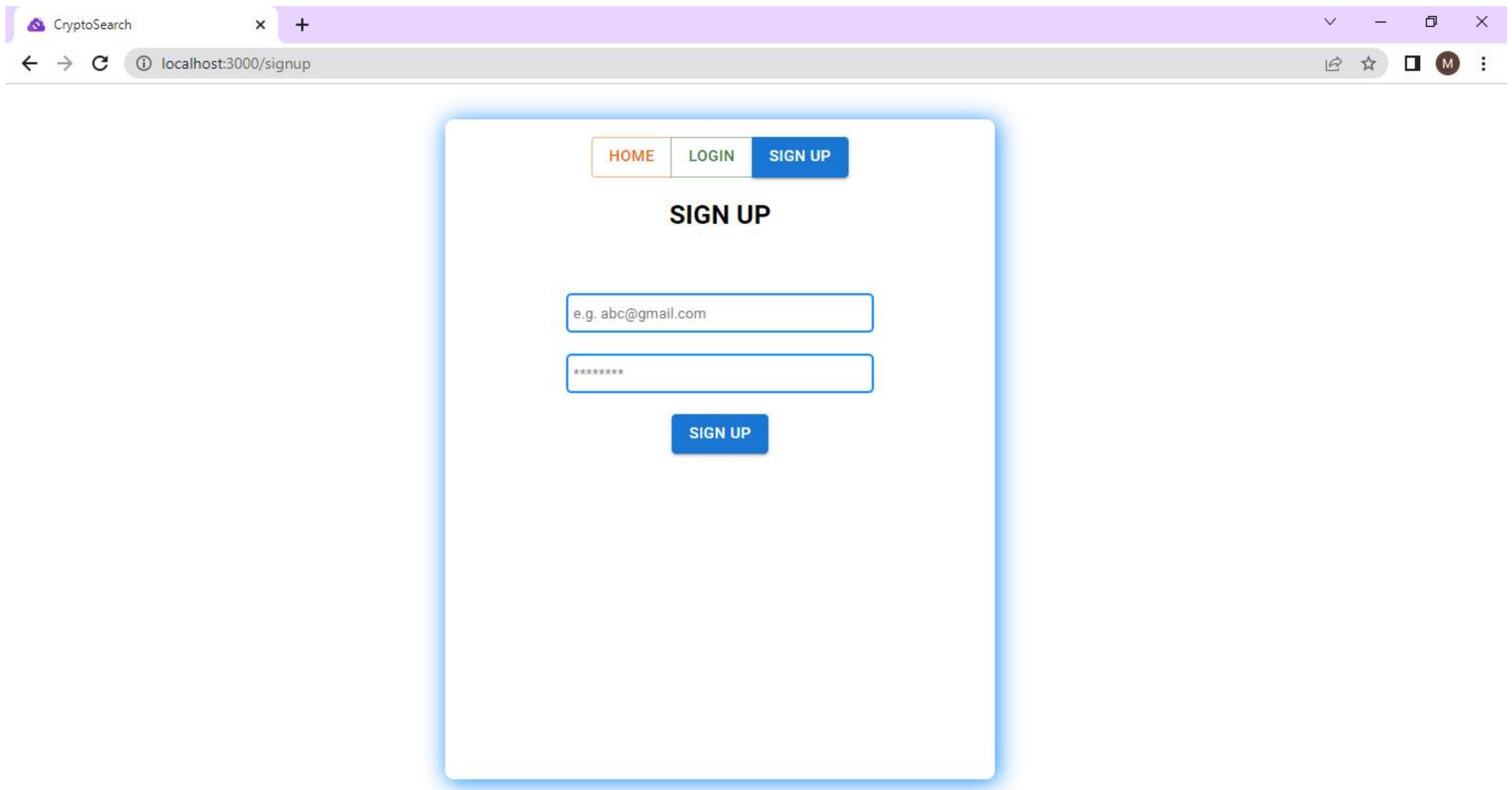

Figure 5.3 CryptoSearch Sign Up Page

### Google Drive Access

After completing the sign-up process, the "Crypto Search" web application will ask to access Google Drive, as shown in figure 5.4. This process is needed to access the folder and files of the user from Google Drive to upload, share and search functionalities.

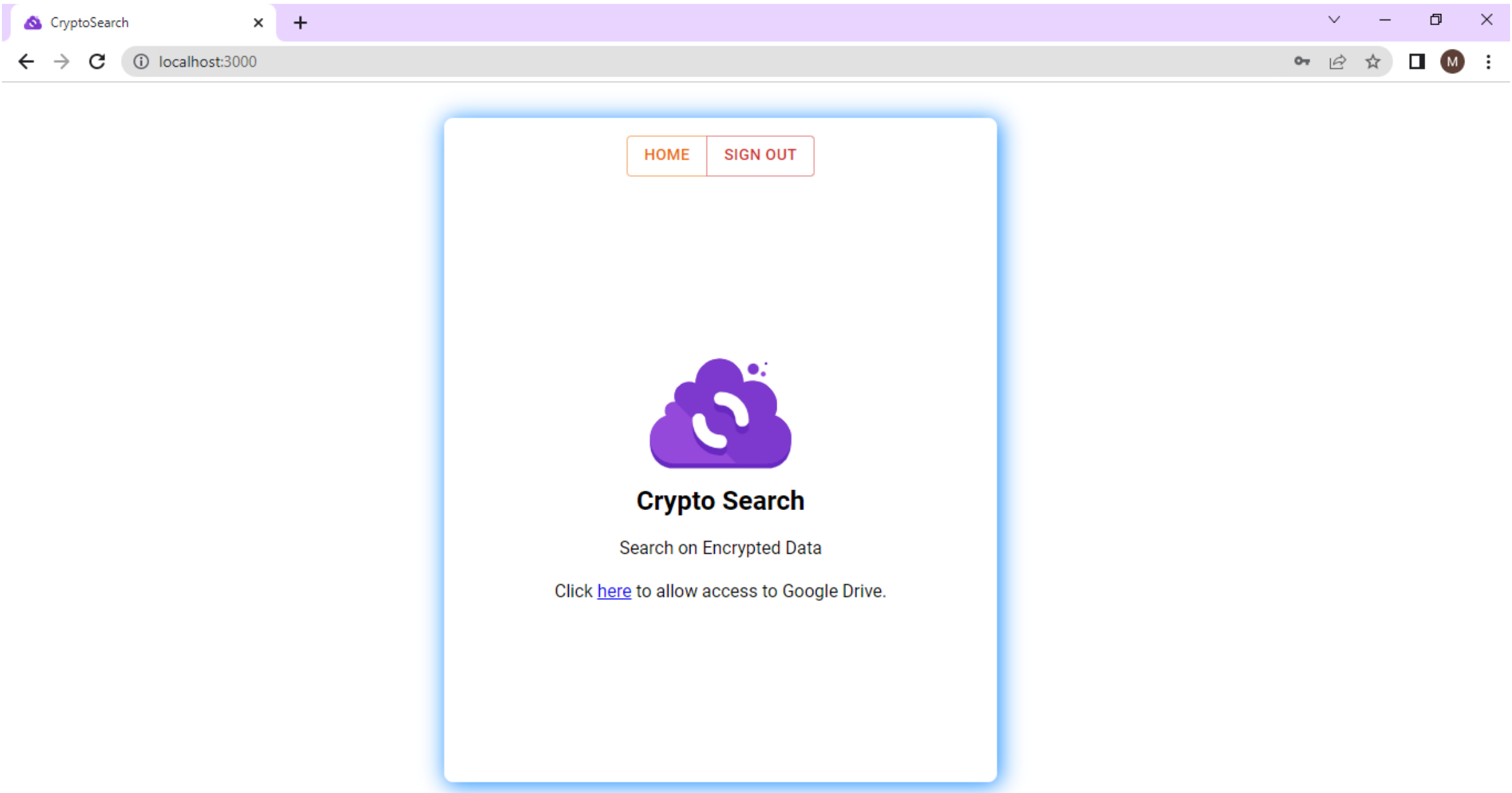

Figure 5.4    Required Google Drive Access

After clicking the allow access to Google Drive, the web application will automatically open a new tab in the browser to sign in to continue the web application “CryptoSearch.” The user needs to enter his/her email address associated with the Google Drive account to continue. Google’s sign-in page is shown in Figure 5.5. Depending on the user's region, the language of this page might be different. However, the information required will remain the same.

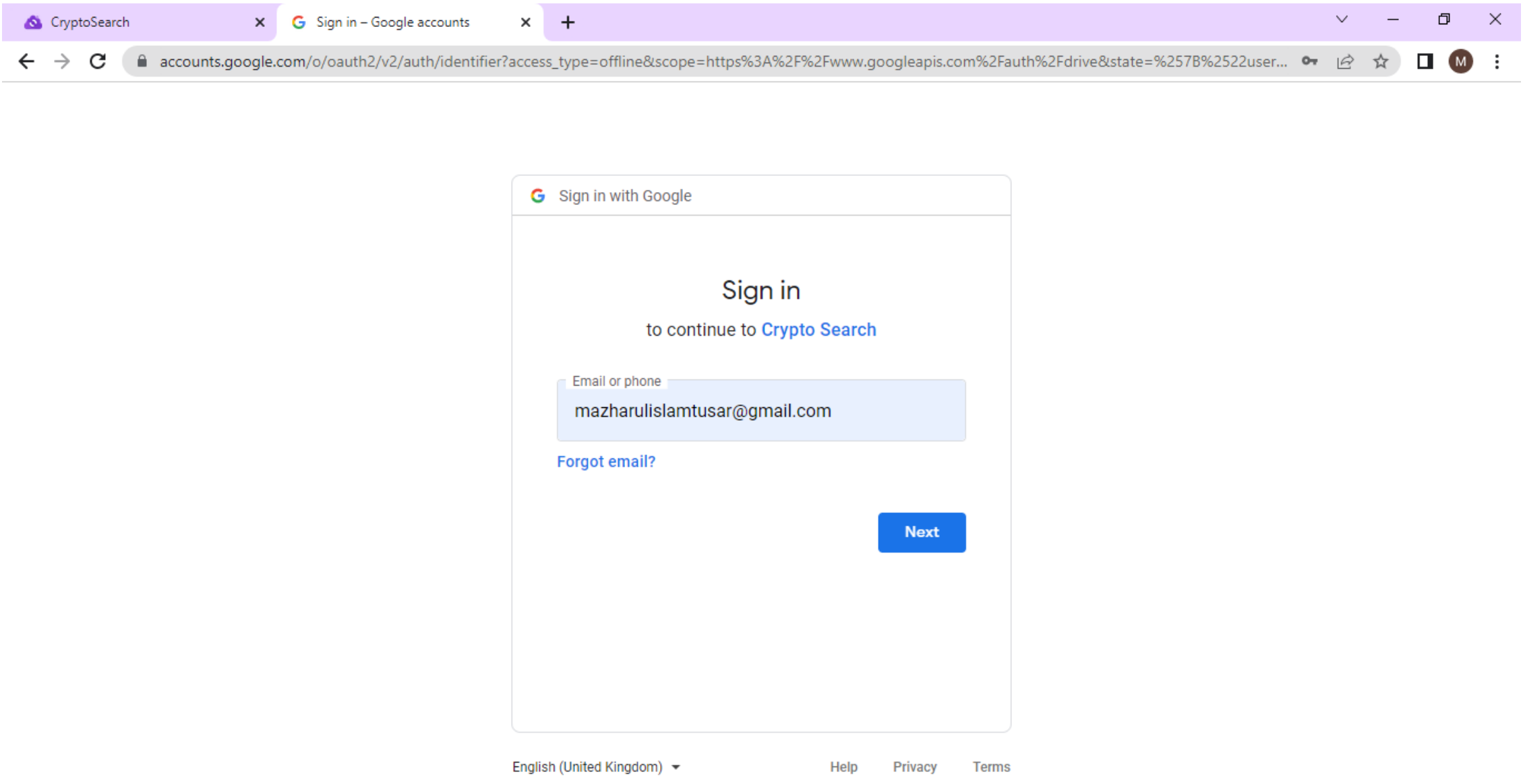

Figure 5.5    Google's Sign-in page

After entering the email address, the user must click on “Next.” This will bring up the next page, which will ask the user to enter the password. There is no chance of a password hack as the password validation will take place on Google's end, and it will not be and cannot be traced by our application. The password prompt for sign-in is shown in Figure 5.6.

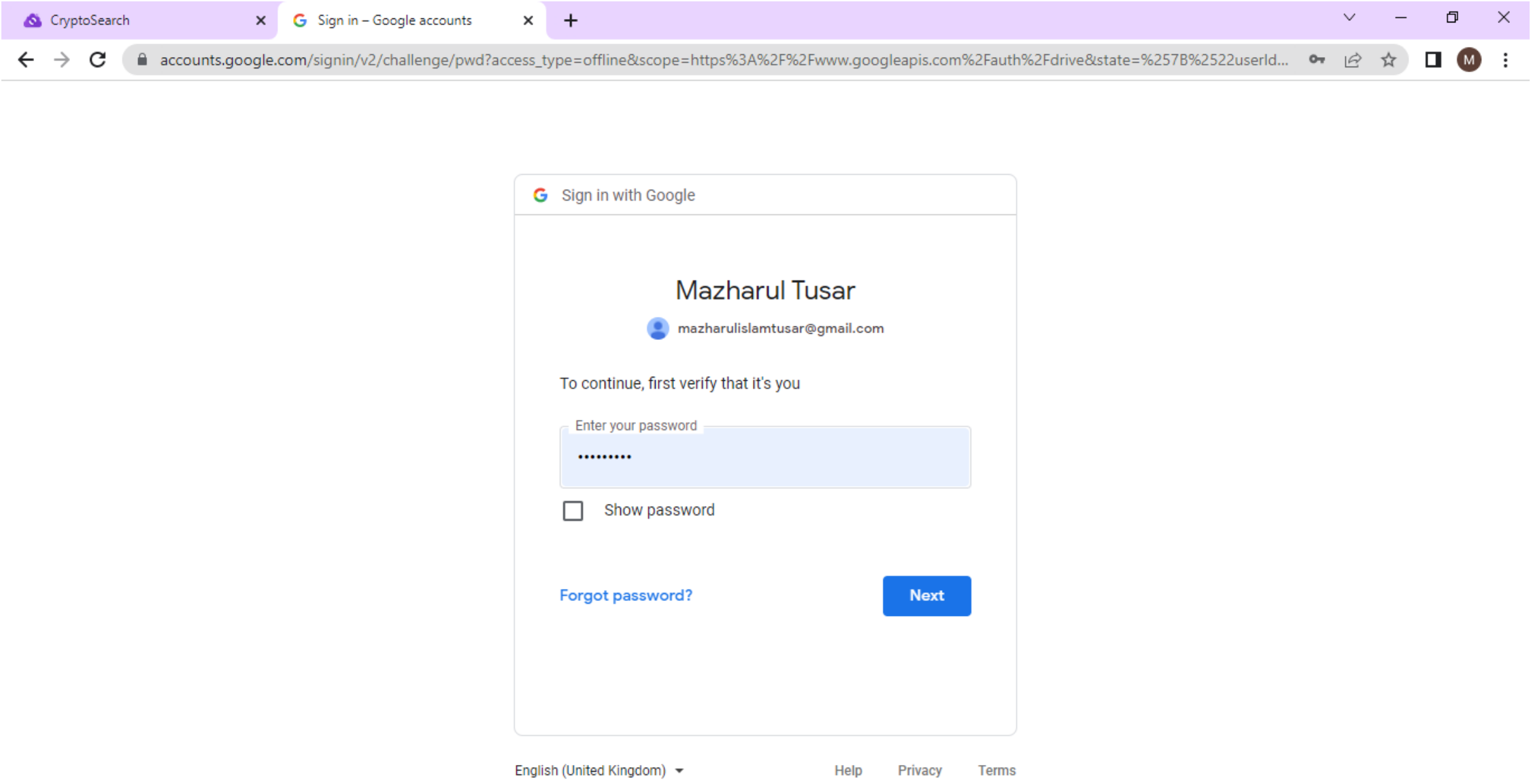

Figure 5.6 Password prompt for signing in

If the user has two-step authentications enabled in the Google account, s(he) will be prompted to enter the secret code sent to him/her via email, SMS or the Google Authenticator app (shown in Figure 5.7). This will also be performed by Google and is not related to CryptoSearch.

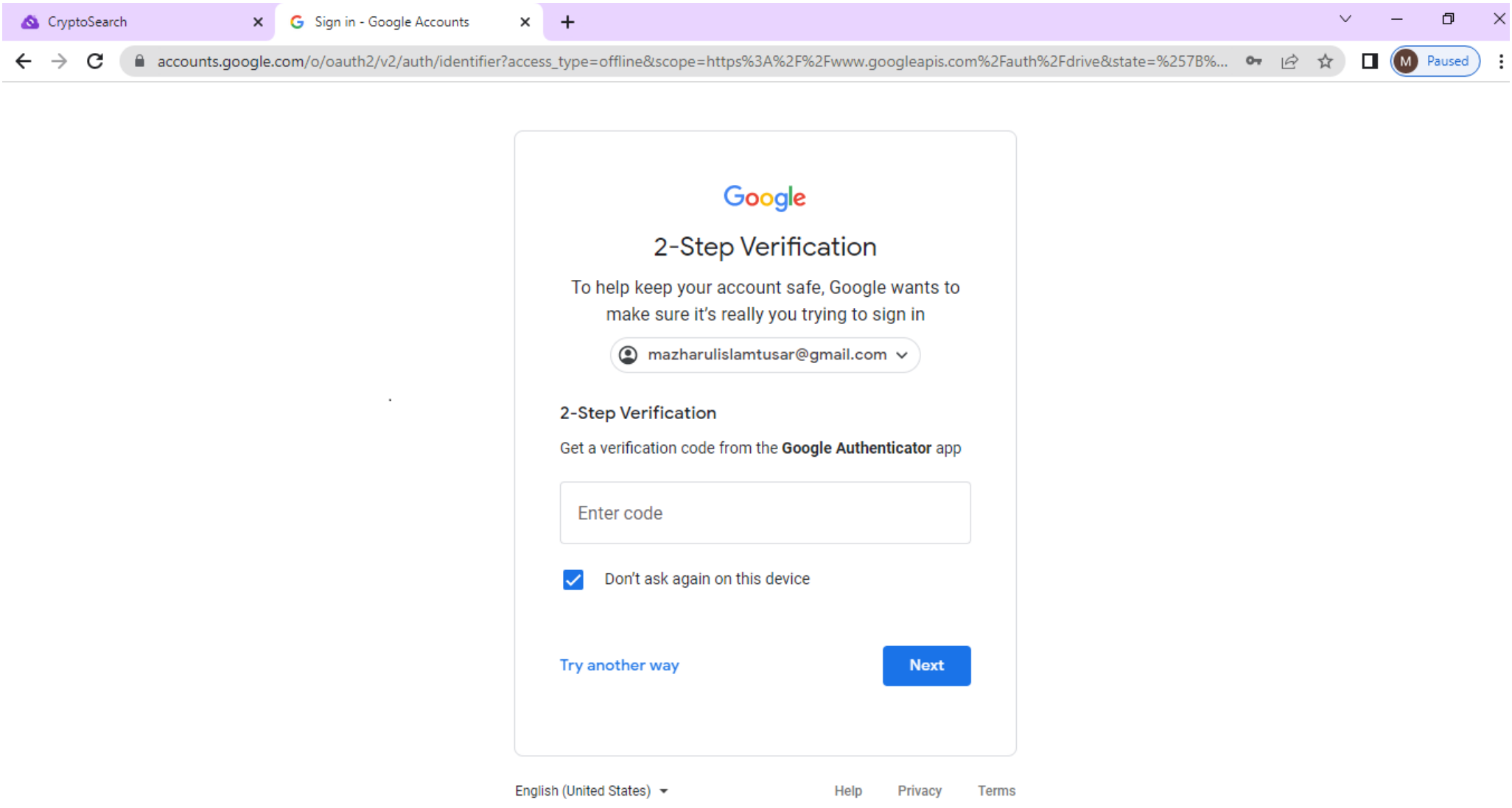

Figure 5.7 Two-step verification performed by Google

After the password is entered, the user needs to click on "Next". It will ask the user to grant/allow Crypto Search's access to his/her Google Drive cloud storage (shown in Figure 5.8). The client application requires access to the user's cloud storage for performing an upload, download, share and delete operations. None of these operations will be performed by the Front-end application without the user's consent. In order to grant the access, the user needs to make sure that the "See, edit, create and delete all of your Google Drive files" is checked and click "Continue".

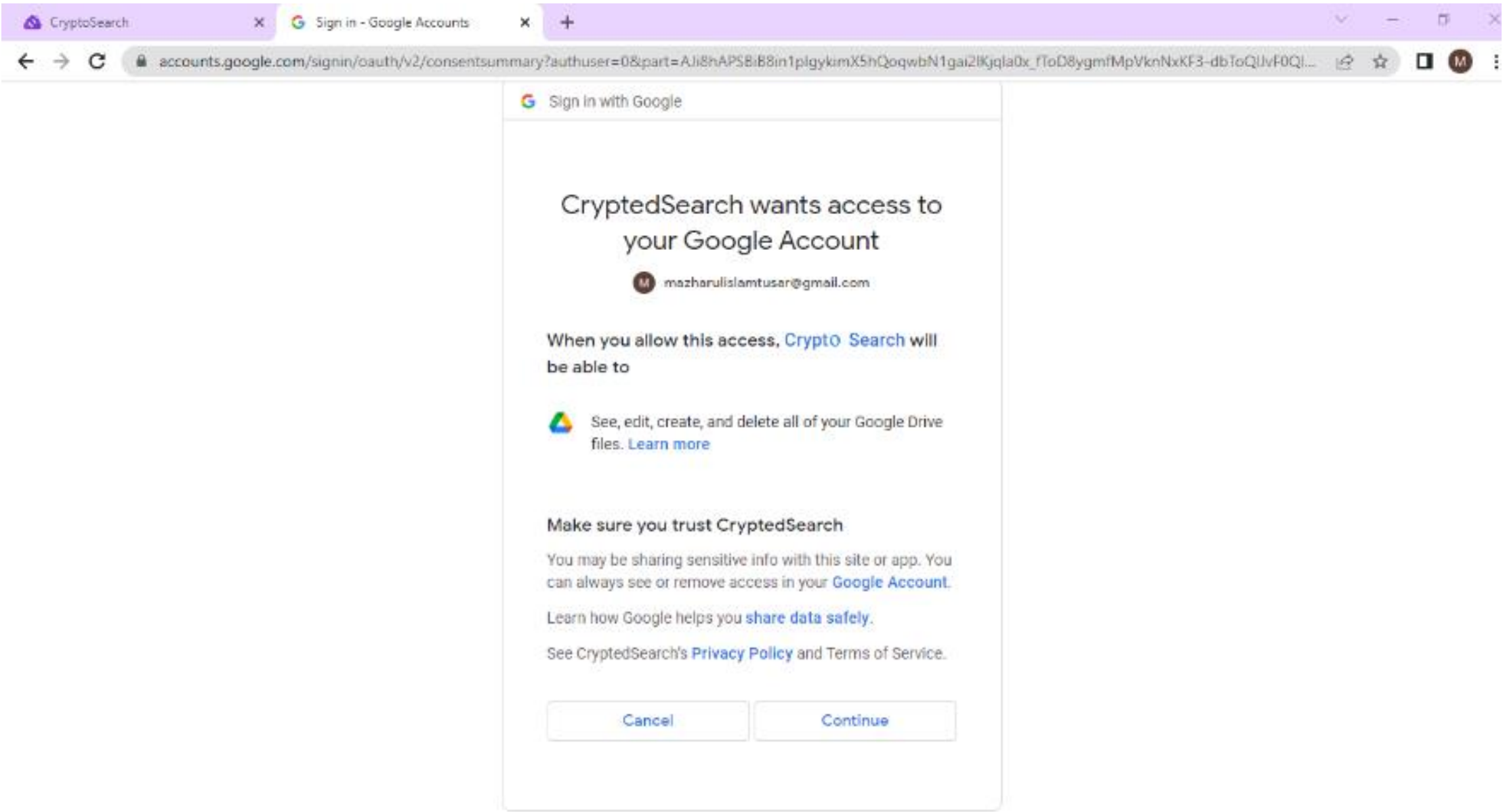

Figure 5.8 Grant CryptoSearch access to Google Drive

The user will have to go through this Google Drive authentication only once. For later use, the Google Drive credentials will be stored safely (by Google Drive API v3). It is mentioned that only the client application can access the user's cloud storage, not the server-side application.

The message will be shown in the prompt about the successful authorization of Google Drive. Again, the user will need to Crypto Search homepage tab and reload the page to use the service. The successful message notice is shown in figure 5.9.

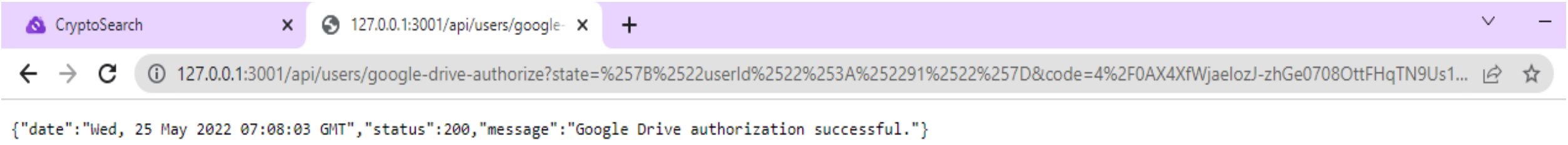

Figure 5.9 Google Drive Authorization Success Notice

## Login

After successful Google Drive authorization, the user will need to select the login option (shown in figure 5.10). On the login page, the user will provide his/her email and passphrase, which he has used during the sign-up process. The web application will verify the input email and passphrase and give access to use the application.

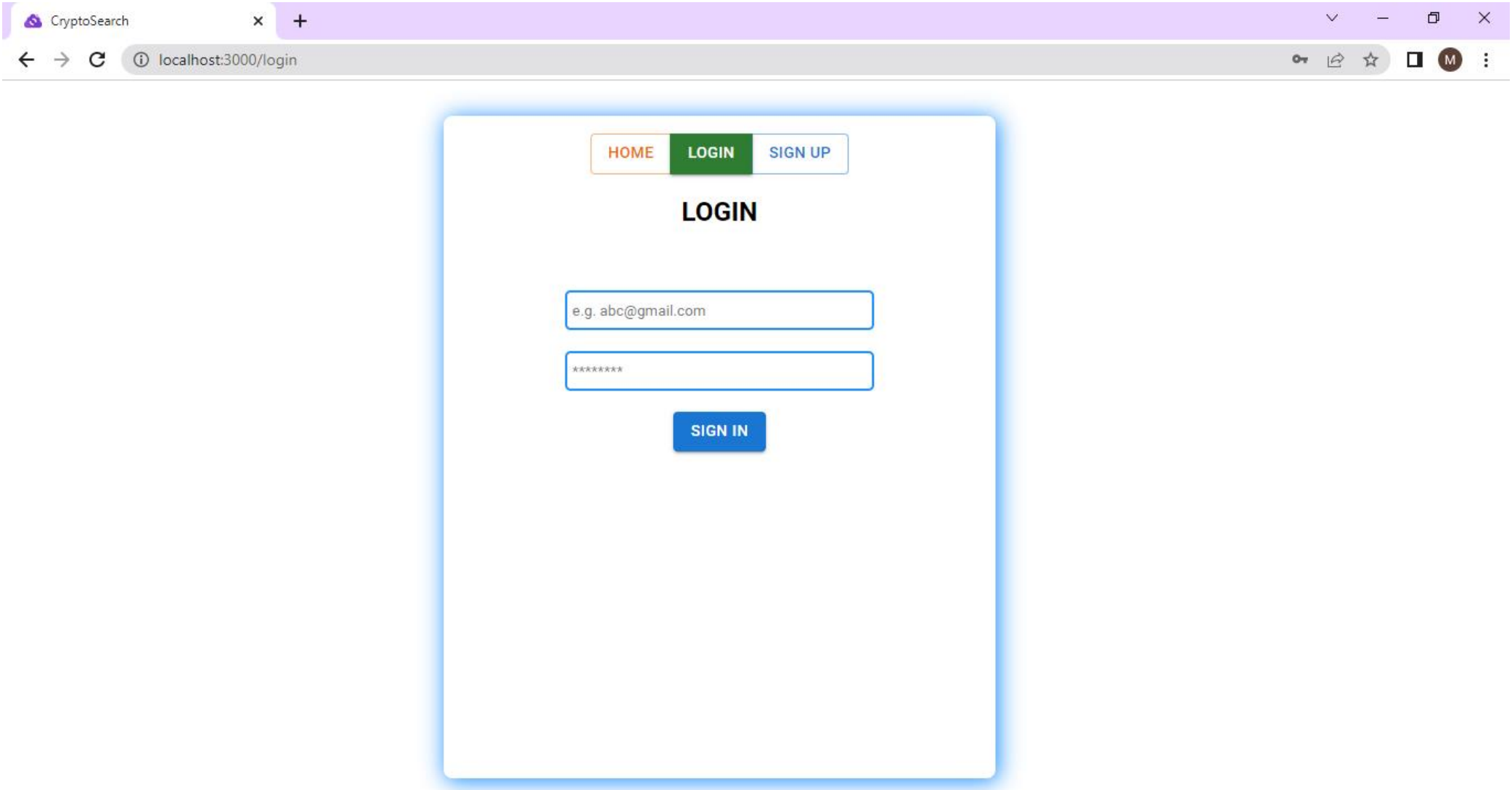

Figure 5.10 CryptoSearch Login Page

Suppose the user provides the wrong passphrase during the login. In that case, the web application will be shown an error message that "Incorrect Passphrase Provided," which is shown in figure 5.11. So, the user will need to provide a valid email and passphrase in the login phase, which he was used during the sign-up phase.

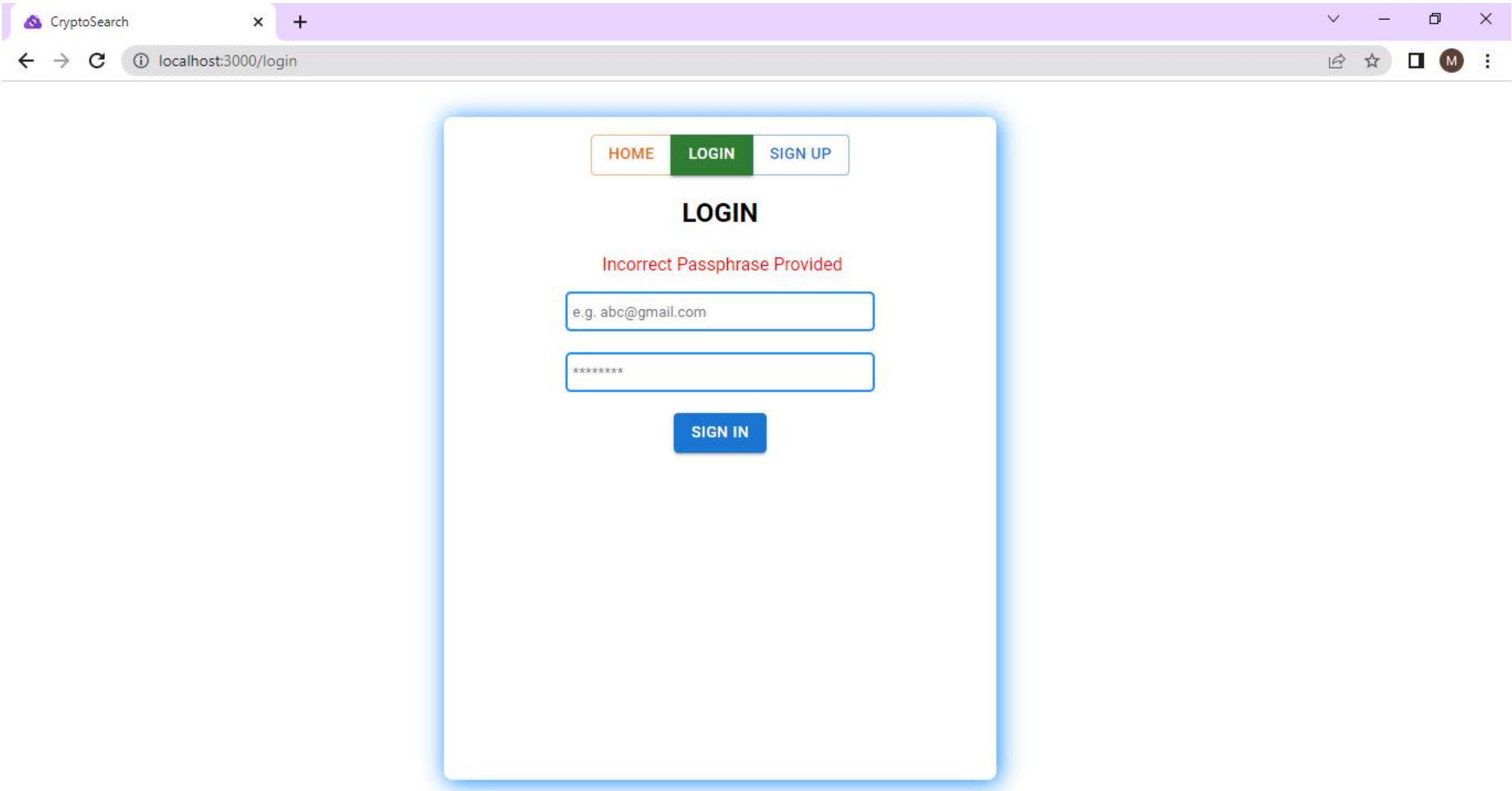

Figure 5.11 Error message for incorrect passphrase

After login into the web application and authorization of Google Drive successfully, the data owner can use the CryptoSearch securely to upload the document and search for it. The Homepage of the Crypto Search Web Application after login is shown in Figure 5.12.

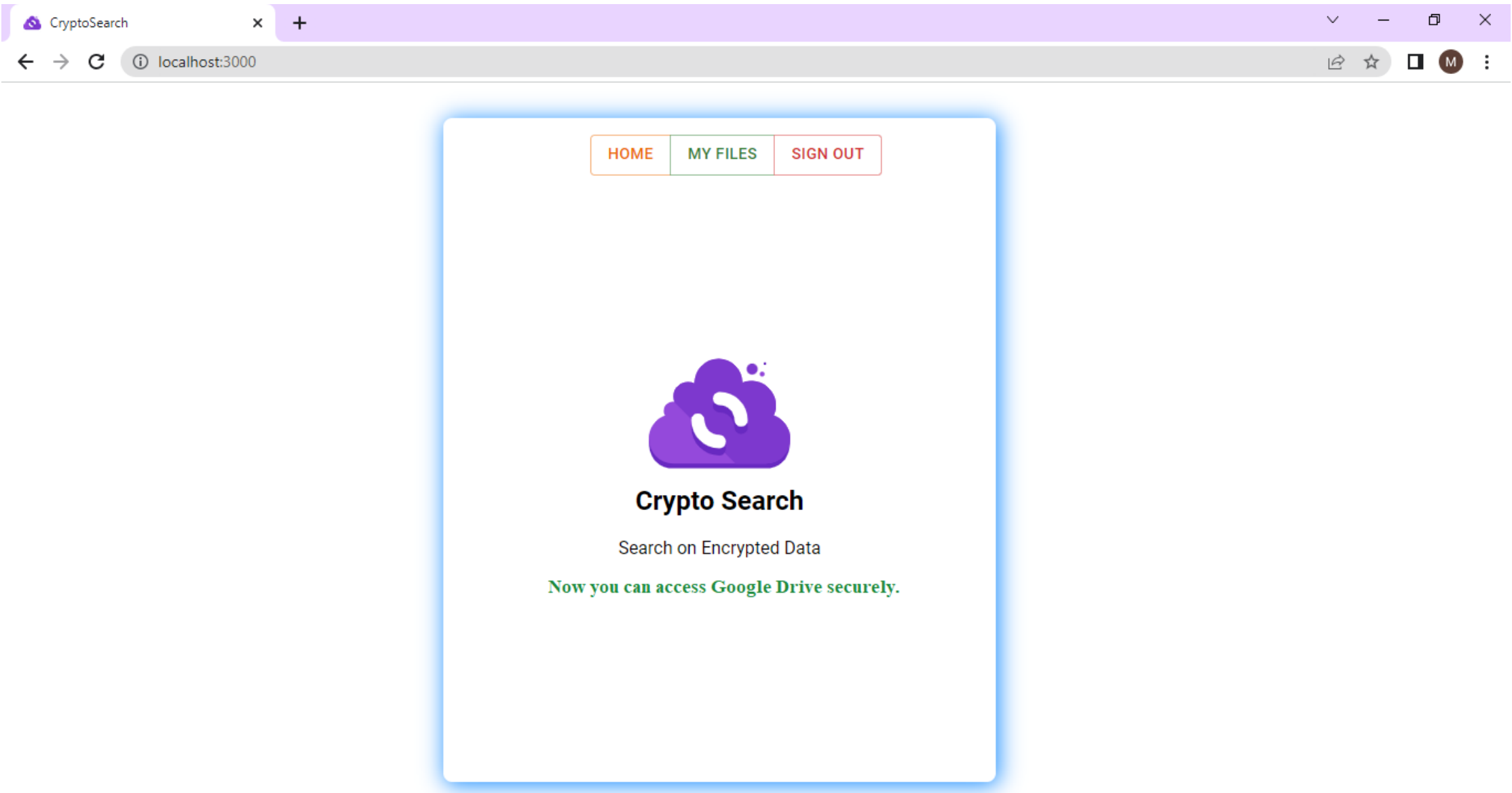

Figure 5.12 Homepage of CryptoSearch Web Application

## Folder List

We can consider the medical authority as a data owner who has access to all medical records documents of the patients. So, after successful login, the data owner can see all of the datasets folders in my files option. Each folder has three functions: share, search, and upload correspondingly. Also, the folder ID has shown below each folder name. In the right-sidebar, we have kept each function's notification panel, which will show what functions are working in the background. The notification bar "files retrieved successfully" means that all folder lists are retrieved from Google Drive and are shown here by following our designed user interface. Clicking on the "Refresh" button will retrieve the latest changes added/deleted files from Google Drive. The Google drive folder lists are shown in figure 5.13, and the user interface of folder lists is shown in figure 5.14 respectively.

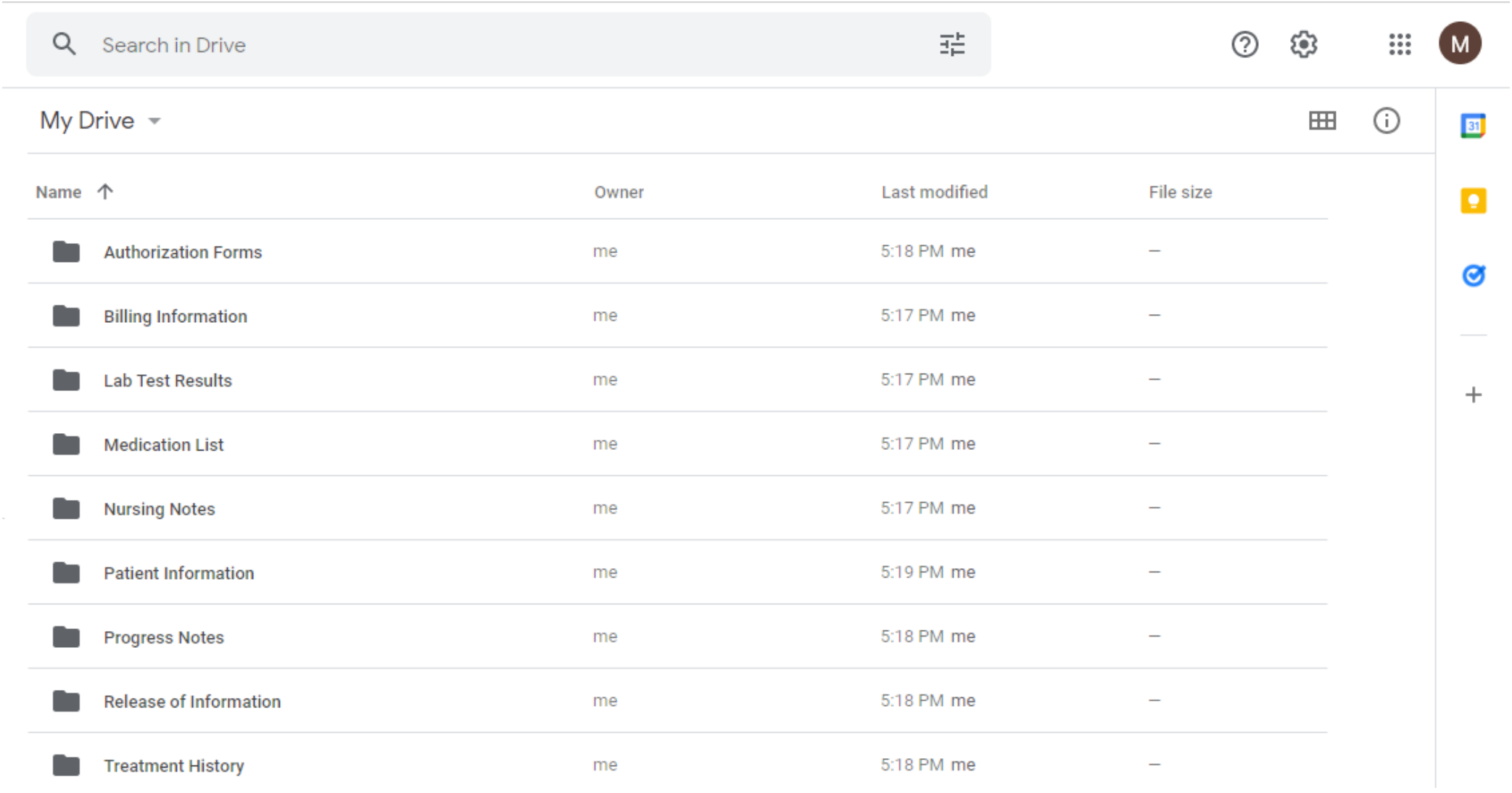

Figure 5.13 Folder list of the documents in Google Drive

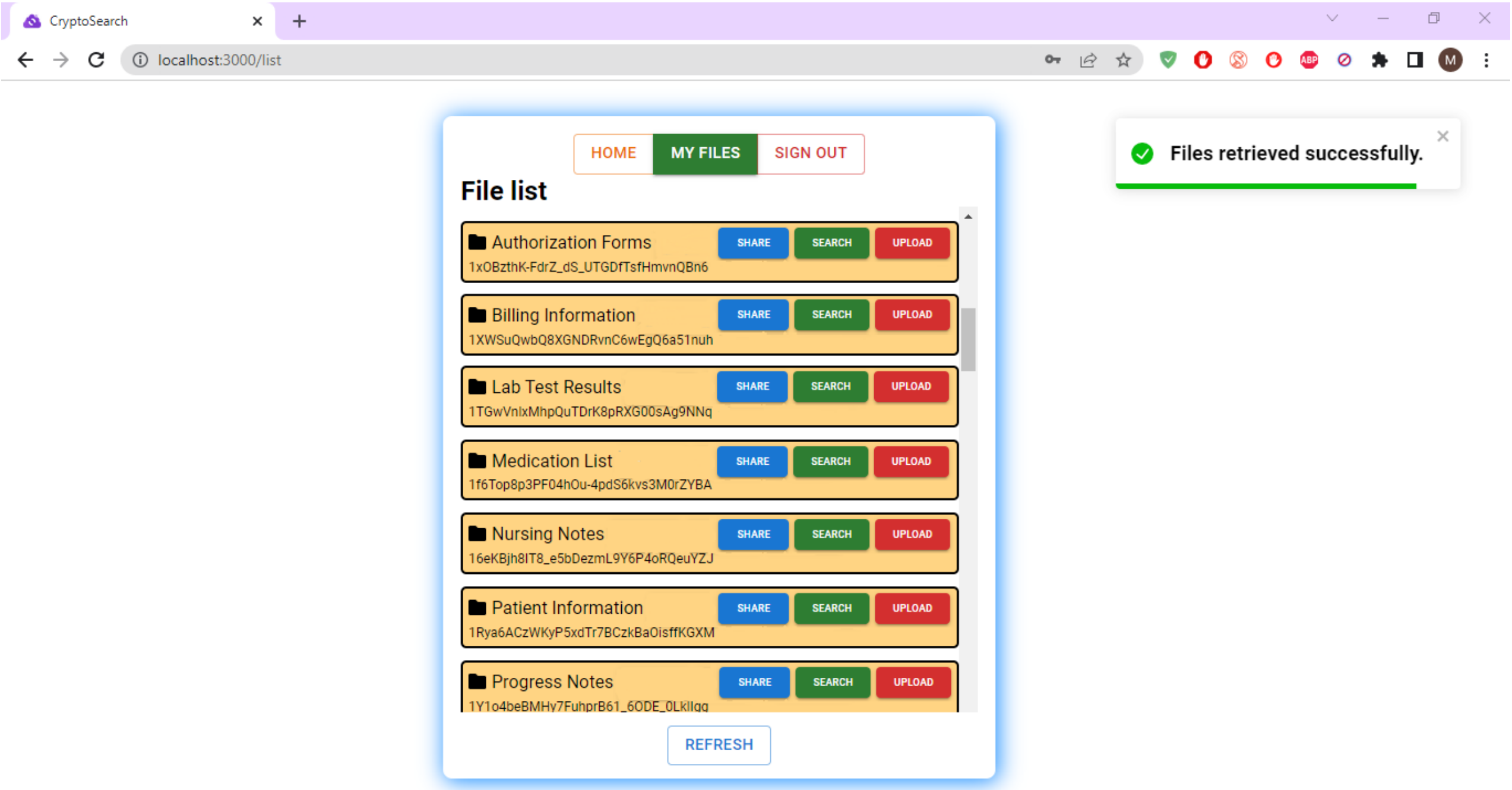

Figure 5.14 Folder List of the Documents in CryptoSearch

### Main User Interface

Upon successful login, the user will be able to access all the functionalities of CryptoSearch web application. From my files, the user can choose whichever file he/she wants to upload, share, and search in a specific folder. The user will find three different functions with three different colors to perform individual operations of each folder. Those are-

1. Upload
2. Share
3. Search

We will describe the use cases of all these functions and how these functions work one by one.

### 1. Upload

As we have mentioned, we are using EHR datasets in section 5.1.3. We have selected a specific folder, “Patient Information,” to show how these functions are working. There is a total of 100,000 patients' list of individual records in the dataset. A sample patient details fill-up form is shown in figure 5.15, which is in PDF format titled “patient 1”. Is it mentioned that we are using only 5 documents, “Patient 1”, “Patient 2”, “Patient 3”, “Patient 4” and “Patient 5” to explain how our proposed model is worked both theoretically and practically.

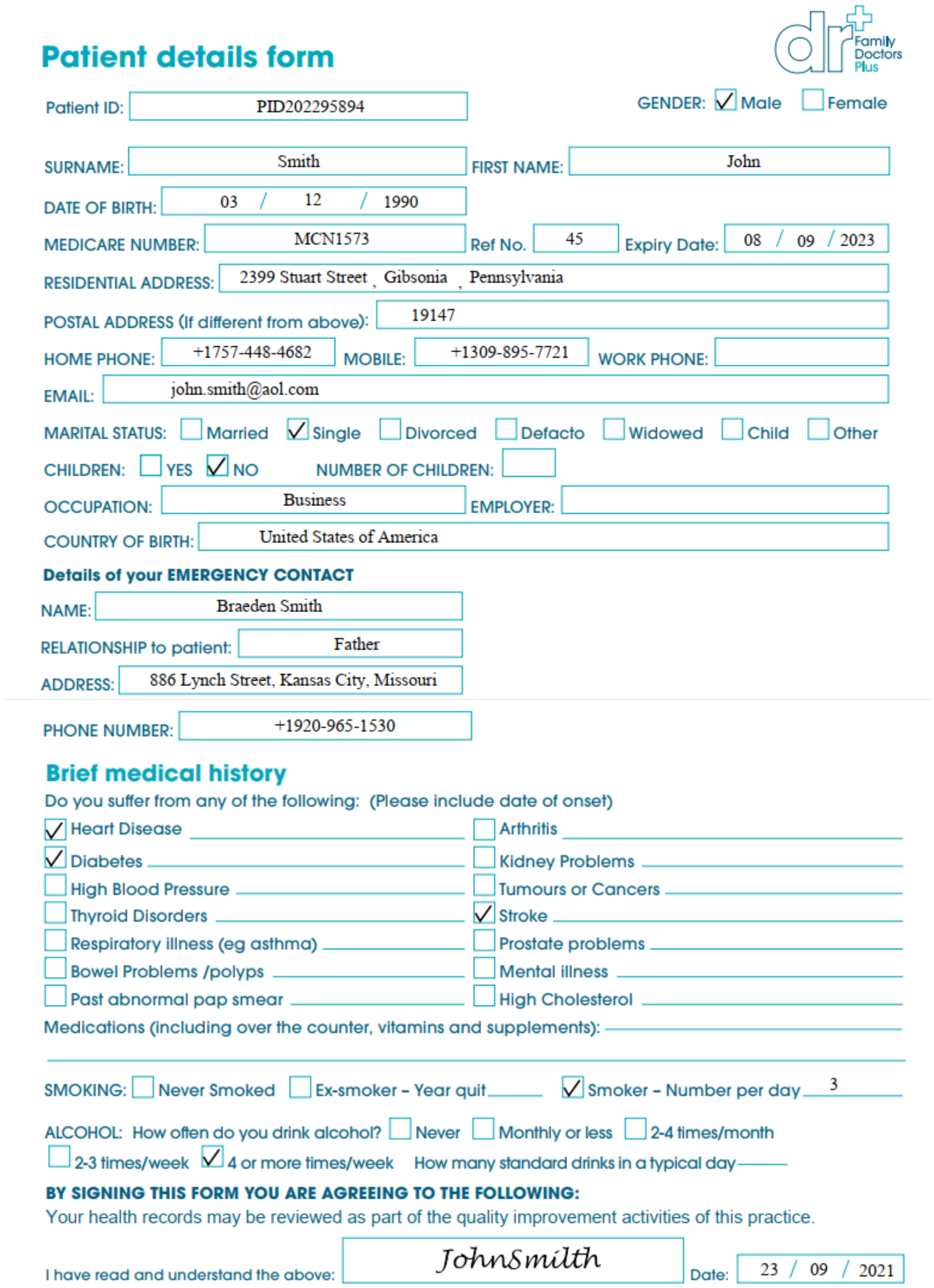

## Patient details form

Family Doctors Plus

Patient ID: PID202295894

GENDER: ☑ Male ☐ Female

SURNAME: Smith FIRST NAME: John

DATE OF BIRTH: 03 / 12 / 1990

MEDICARE NUMBER: MCN1573 Ref No. 45 Expiry Date: 08 / 09 / 2023

RESIDENTIAL ADDRESS: 2399 Stuart Street , Gibsonia , Pennsylvania

POSTAL ADDRESS (If different from above): 19147

HOME PHONE: +1757-448-4682 MOBILE: +1309-895-7721 WORK PHONE:

EMAIL: john.smith@aol.com

MARITAL STATUS: ☐ Married ☑ Single ☐ Divorced ☐ Defacto ☐ Widowed ☐ Child ☐ Other

CHILDREN: ☐ YES ☑ NO NUMBER OF CHILDREN:

OCCUPATION: Business EMPLOYER:

COUNTRY OF BIRTH: United States of America

**Details of your EMERGENCY CONTACT**

NAME: Braeden Smith

RELATIONSHIP to patient: Father

ADDRESS: 886 Lynch Street, Kansas City, Missouri

PHONE NUMBER: +1920-965-1530

### Brief medical history

Do you suffer from any of the following: (Please include date of onset)

☑ Heart Disease ☐ Arthritis

☑ Diabetes ☐ Kidney Problems

☐ High Blood Pressure ☐ Tumours or Cancers

☐ Thyroid Disorders ☑ Stroke

☐ Respiratory illness (eg asthma) ☐ Prostate problems

☐ Bowel Problems /polyps ☐ Mental illness

☐ Past abnormal pap smear ☐ High Cholesterol

Medications (including over the counter, vitamins and supplements):

SMOKING: ☐ Never Smoked ☐ Ex-smoker - Year quit ☑ Smoker - Number per day 3

ALCOHOL: How often do you drink alcohol? ☐ Never ☐ Monthly or less ☐ 2-4 times/month

☐ 2-3 times/week ☑ 4 or more times/week How many standard drinks in a typical day

**BY SIGNING THIS FORM YOU ARE AGREEING TO THE FOLLOWING:**

Your health records may be reviewed as part of the quality improvement activities of this practice.

I have read and understand the above: JohnSmilth Date: 23 / 09 / 2021

Figure 5.15 Patients details form

The user needs to select the "upload" function of a specific folder to upload the document. As the data owner are willing to upload the patients document in the "Patient information" folder, s(he) choose the upload option and move to a new page which is shown in figure 5.16.

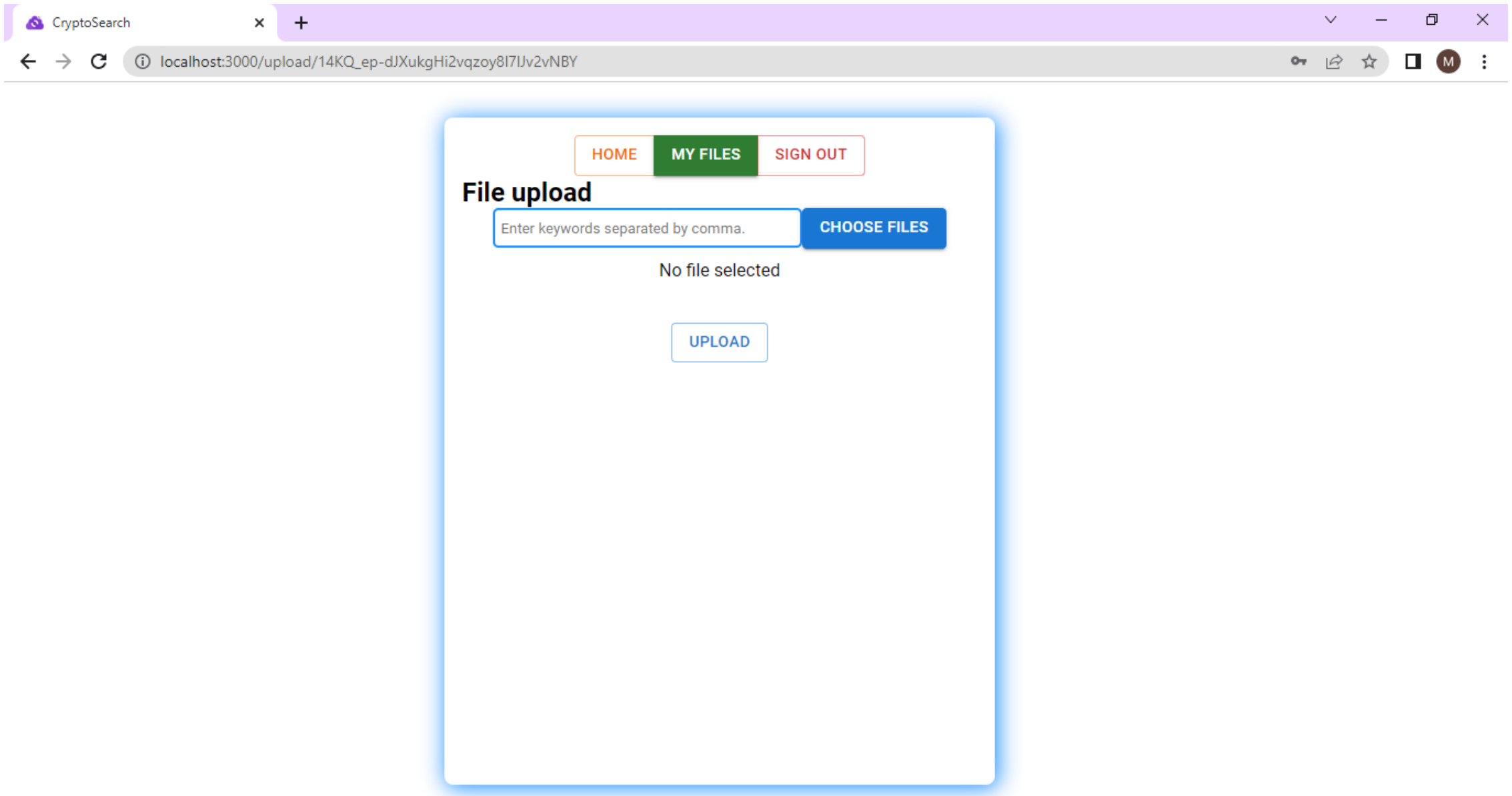

Figure 5.16 File upload page of the CryptoSearch

Now, suppose the data owner wants to upload the document "patient 1" in the patient information folder. In that case, s(he) will need to click on the choose files option. A new window will appear to upload the document, which is shown in figure 5.17.

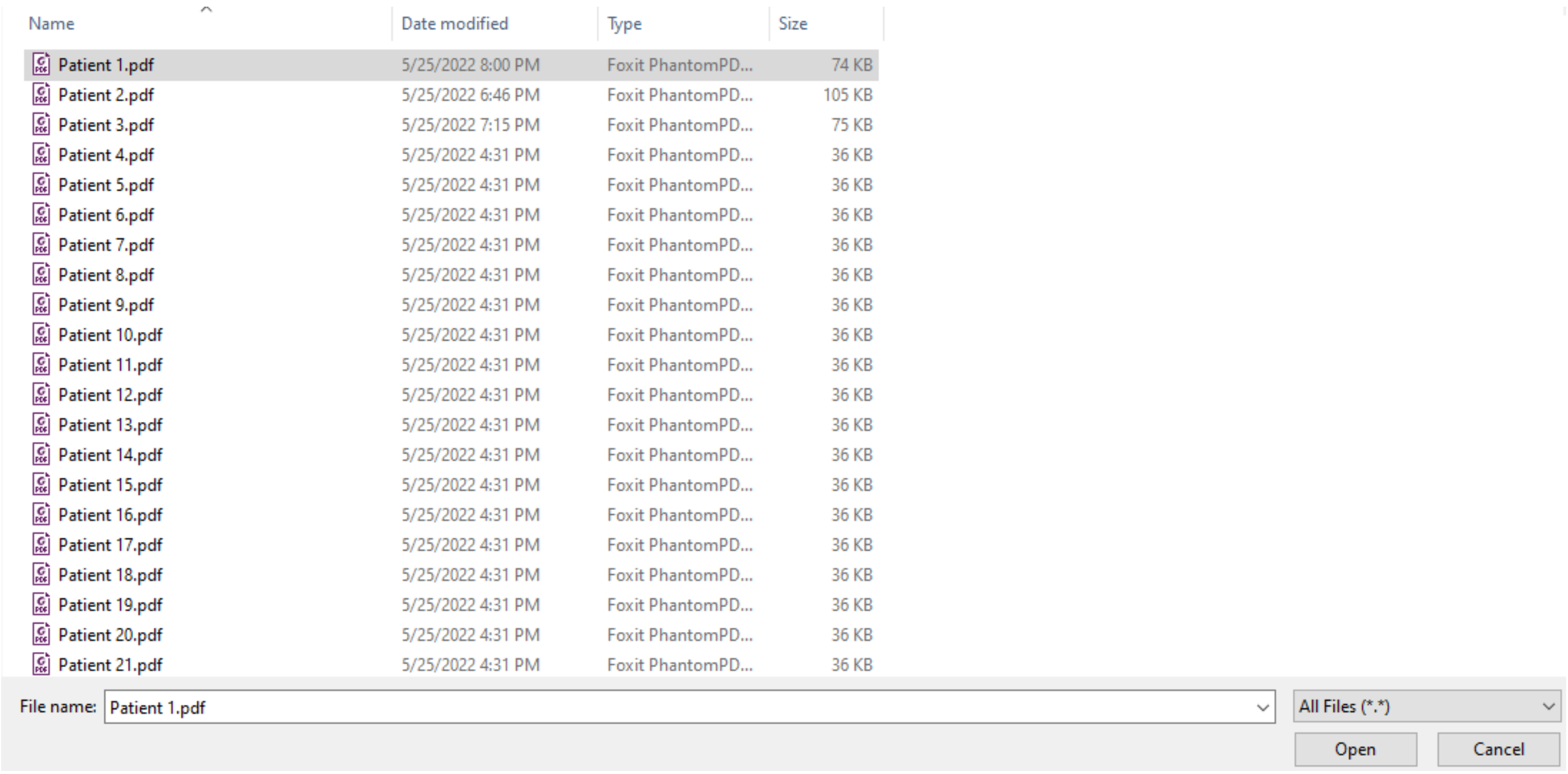

Figure 5.17 Document Upload Process using CryptoSearch Frontend

The CryptoSearch web application supports single keyword, multiple keyword, multiple documents upload respectively which are discussed below.

**I. Keyword Set (Single keyword)**

After selecting the document of "Patient 1", the data owner needs to provide the keyword based on the document shown in figure 5.18. The data owner can choose any single keyword regarding the document, i.e., Patient ID, Patient Name, Medicare Number, Medical History, etc. The keyword can contain both the uppercase, lowercase alphabet and numeric numbers.

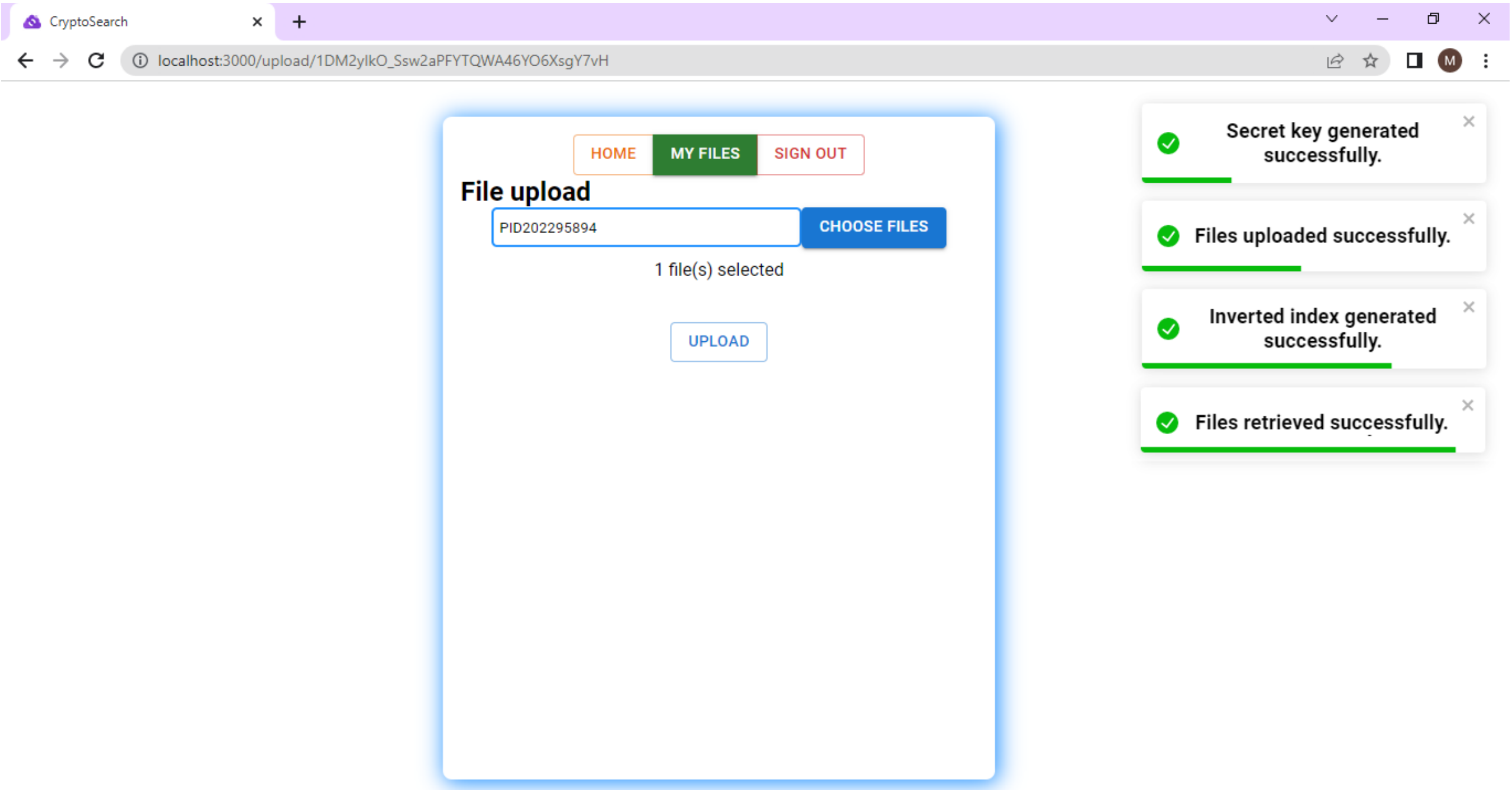

Figure 5.18 Patient 1 document is uploaded with single keyword

After choosing the keywords, the data owner needs to click the "upload" option to upload the document "Patient 1" in Google Drive. During the upload process, on the right side, the notification panel will show four notification bars about the upload process of a document. Each notification bar is shown each individual works about the backend process during the upload process. Note that each notification bar will stay remain 6 second when click on the "Upload" button and after that, it will vanish from the notification bar. During this time, the notification bar will show a green bar that the document is uploaded successfully with maintain each steps of our design. the Each of them is explained below briefly.

- First Notification Bar - Secret Key Generation

It will show that the secret key was generated successfully, which means it is observed from the Trusted Third Party and used to encrypt of the choosing keywords.

- Second Notification Bar – File Upload

It will show that the selected file is uploaded in the Google Drive successfully. Google Drive will generate the File ID of that document and return it.

- Third Notification Bar – Inverted Index Generation

The secret key and File ID of the document is generated by inverted index.

- Fourth Notification Bar – Files Retrieve

After the uploading process of the document, the system needs to refresh to get the updated list. The newly added document will retrieve to show in the file list of the folder.

After successful upload process of a document, the CryptoSearch will take to the file list page and in the notification bar it will show a notification that files retrieved successfully which is shown in figure 5.19.

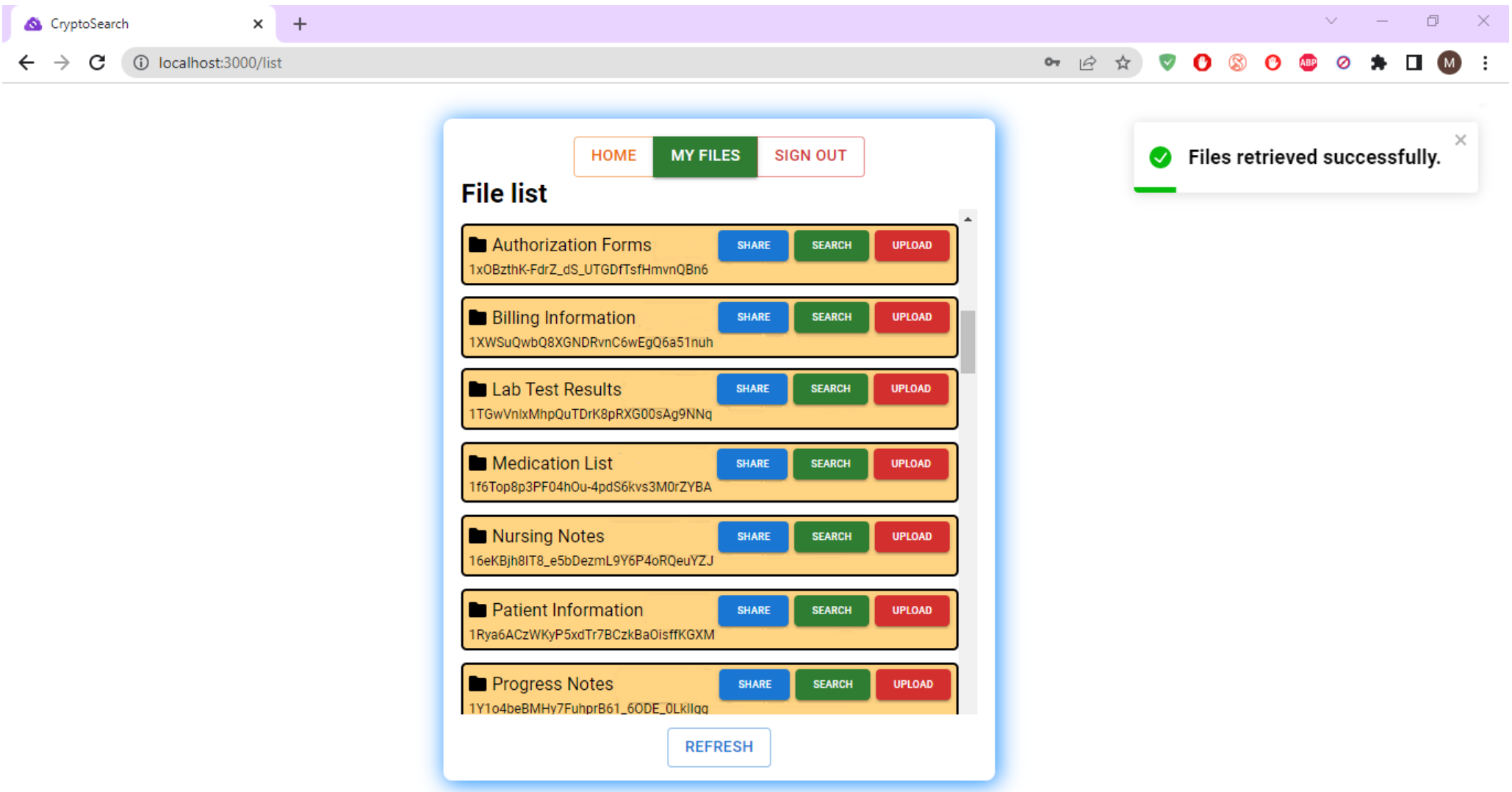

Figure 5.19 The notification panel after upload process of the CryptoSearch frontend

Another document titled “Patient 3” is uploaded with the single keyword “Diabetes,” shown in figure 5.20.

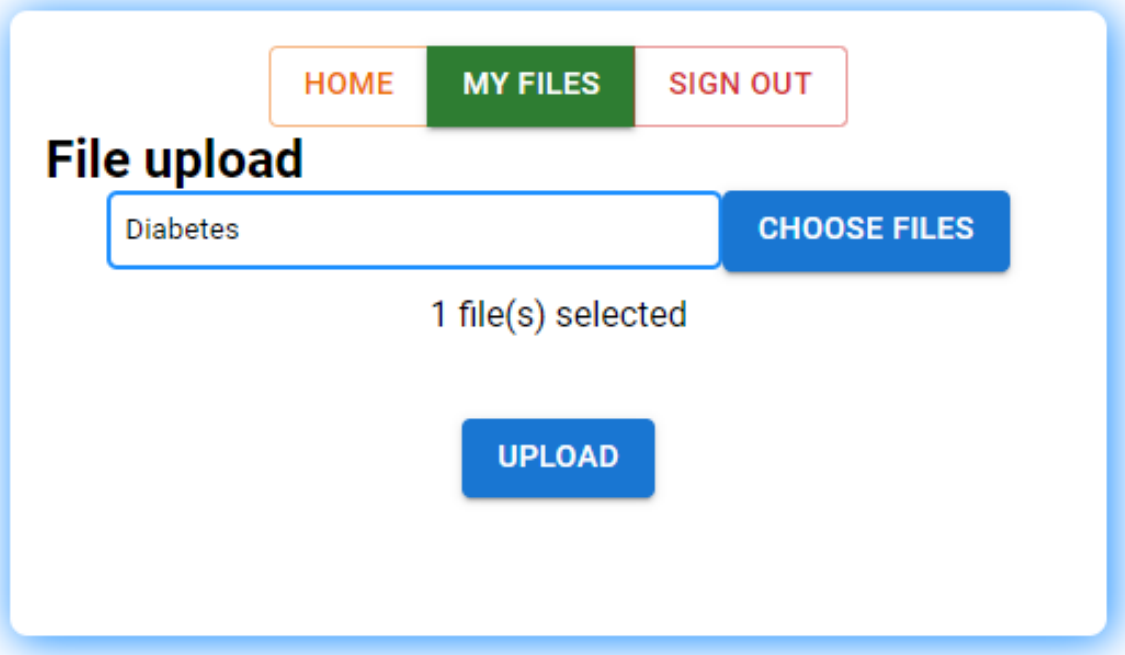

Figure 5.20 Patient 3 document is uploaded with single keyword

### II. Keyword Set (Multiple Single keyword)

The data owner can select multiple single keywords for a specific document so that it might be efficient and easy for the data owner/user to search on the document during the searching process. The multiple single keyword searches can be chosen by a comma (,) after each keyword. For the patient 1 document, we choose the Patient ID, Medicare Number, and Medical History keyword, shown in figure 5.21. Note that patient 1 and patient 3 have the same "Diabetes" disease.

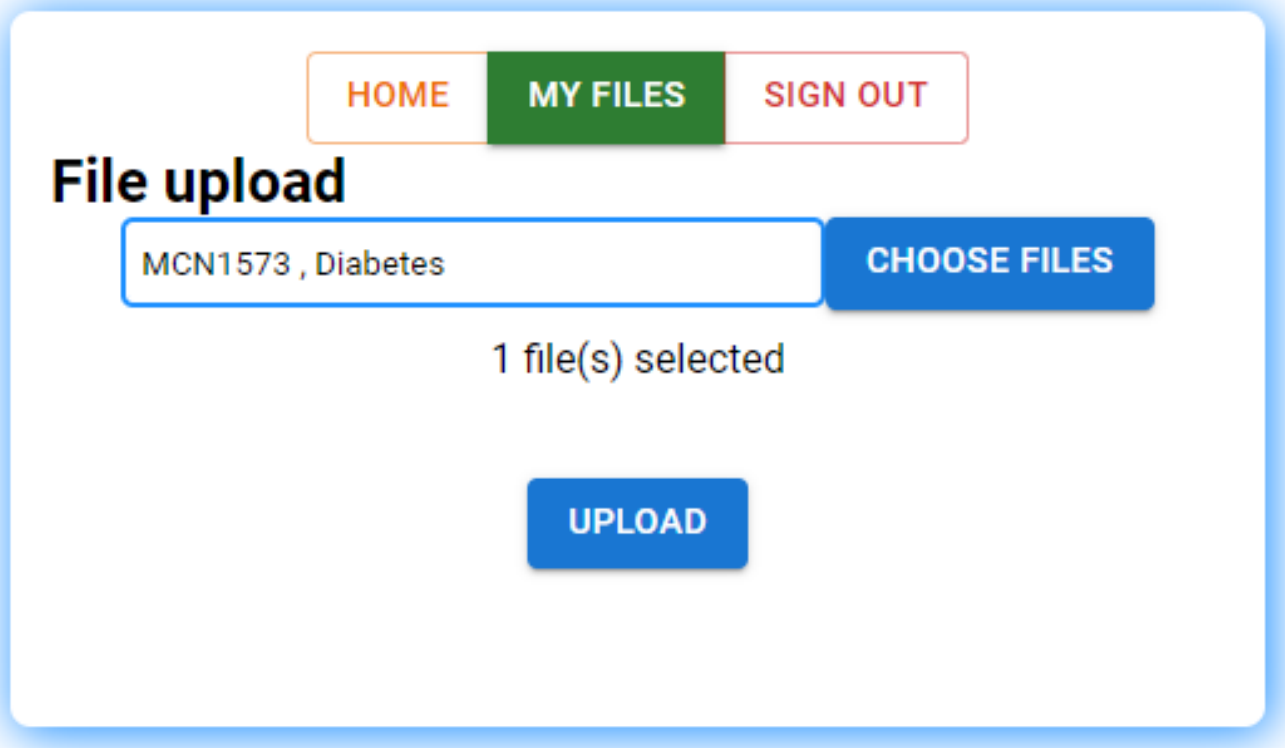

Figure 5.21  Patient 1 document is uploaded with multiple single keyword

### III. Keyword Set (Multiple keywords)

Our implementation system also supports multiple keywords within a document. Sometimes it's necessary to search with multiple keywords to get an accurate result for the documents. So, the document titled "Patient 2" is uploaded to the cloud server with multiple keywords of the patient name "Aliana Lucy", and "High Blood Pressure," which is shown in figure 5.22.

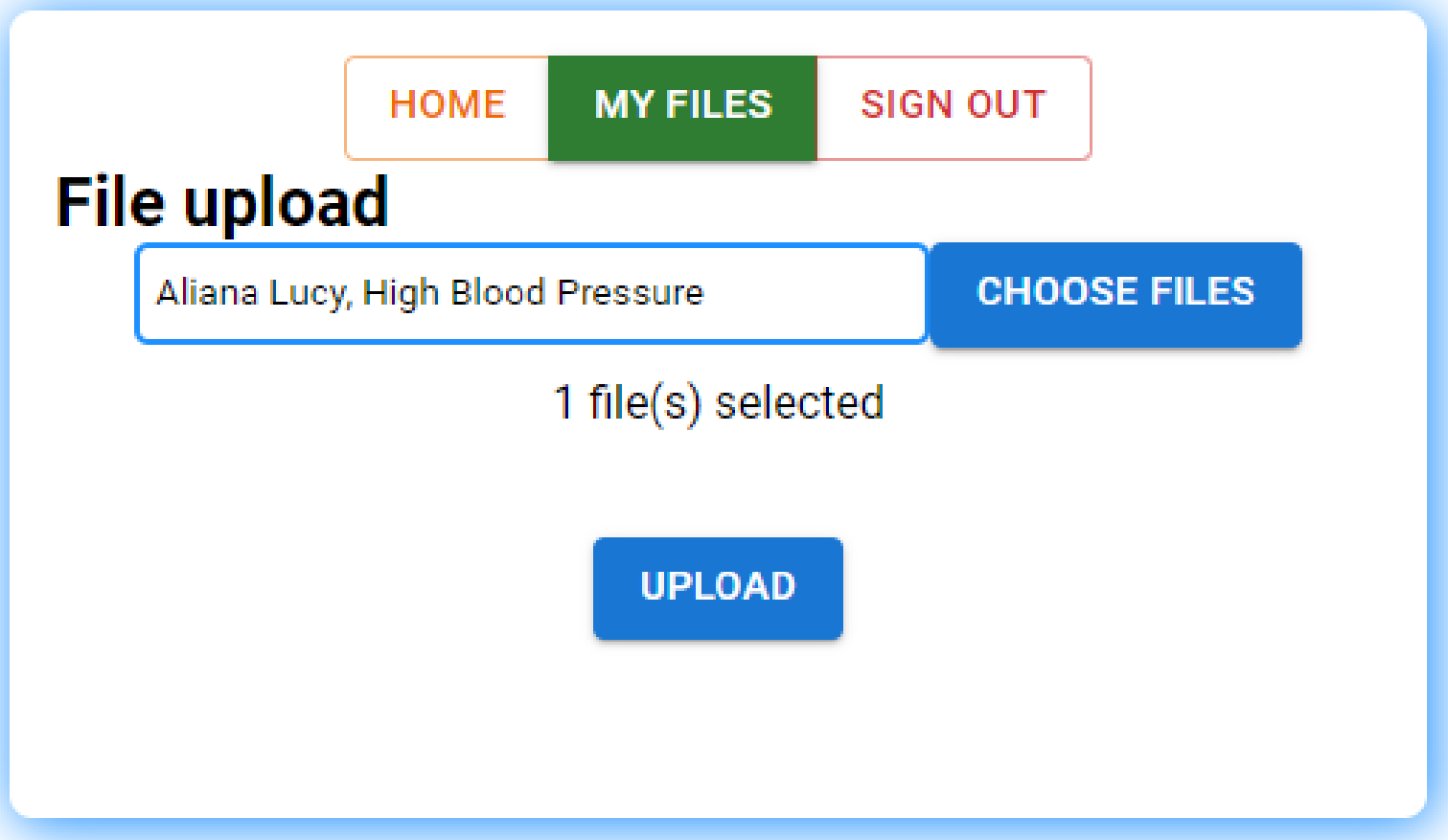

Figure 5.22  Patient 2 document is uploaded with multiple keywords

### IV. Upload Multiple Documents

Our proposed model also supports uploading multiple documents with the same keyword simultaneously. We have uploaded the “patient 4” and “patient 5” document, which has the same medical history as “Stroke,” disease which is shown in figure 5.23. So, the data owner can upload multiple documents at a time with a common set of keywords.

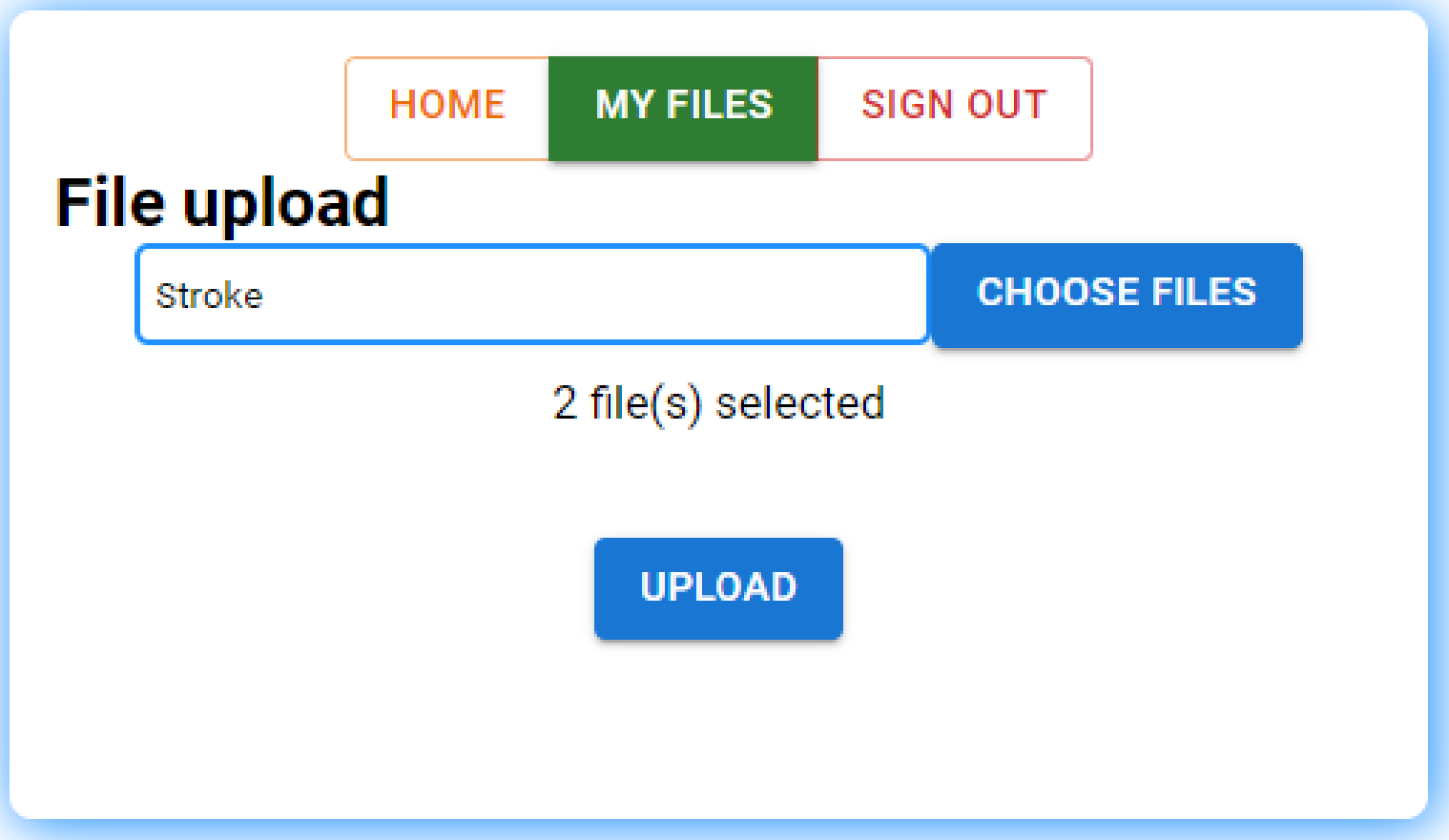

Figure 5.23 Multiple documents is uploaded upload with same keyword

We have uploaded all the patient’s documents; however, we have selected only 5 documents for the explanation. After successfully document upload with keyword, the folder with document lists is shown in figure 5.24.

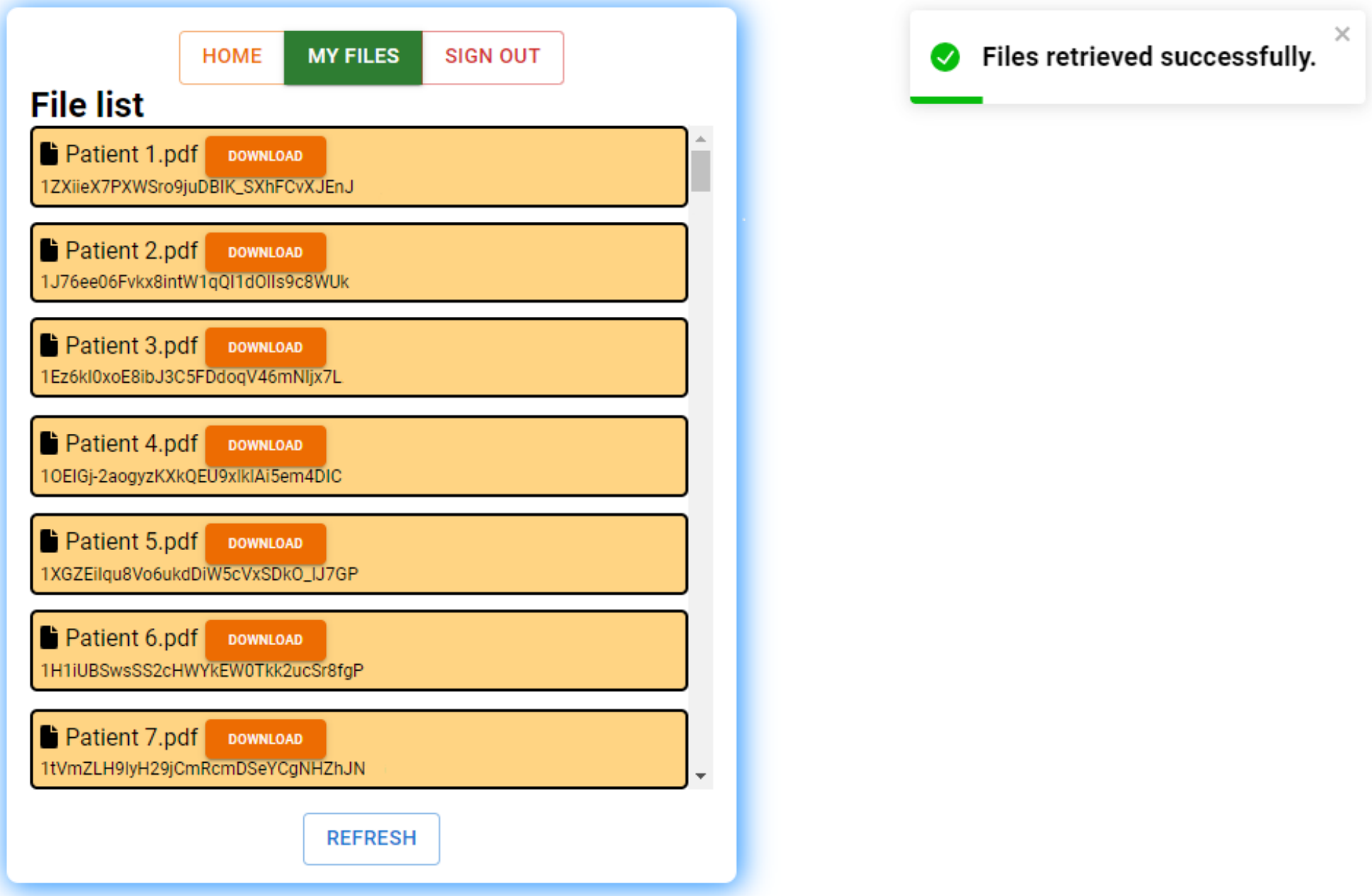

Figure 5.24 Document lists of the Patient information folder in CryptoSearch

The document lists of the “Patient Information” folder in the Google Drive is shown in the figure 5.25. Note that the folders and document list of the folder is shown as same from Google Drive to CryptoSearch. The web application is checked simultaneously if there any modification, add, delete of the data and then files retrieved successfully.

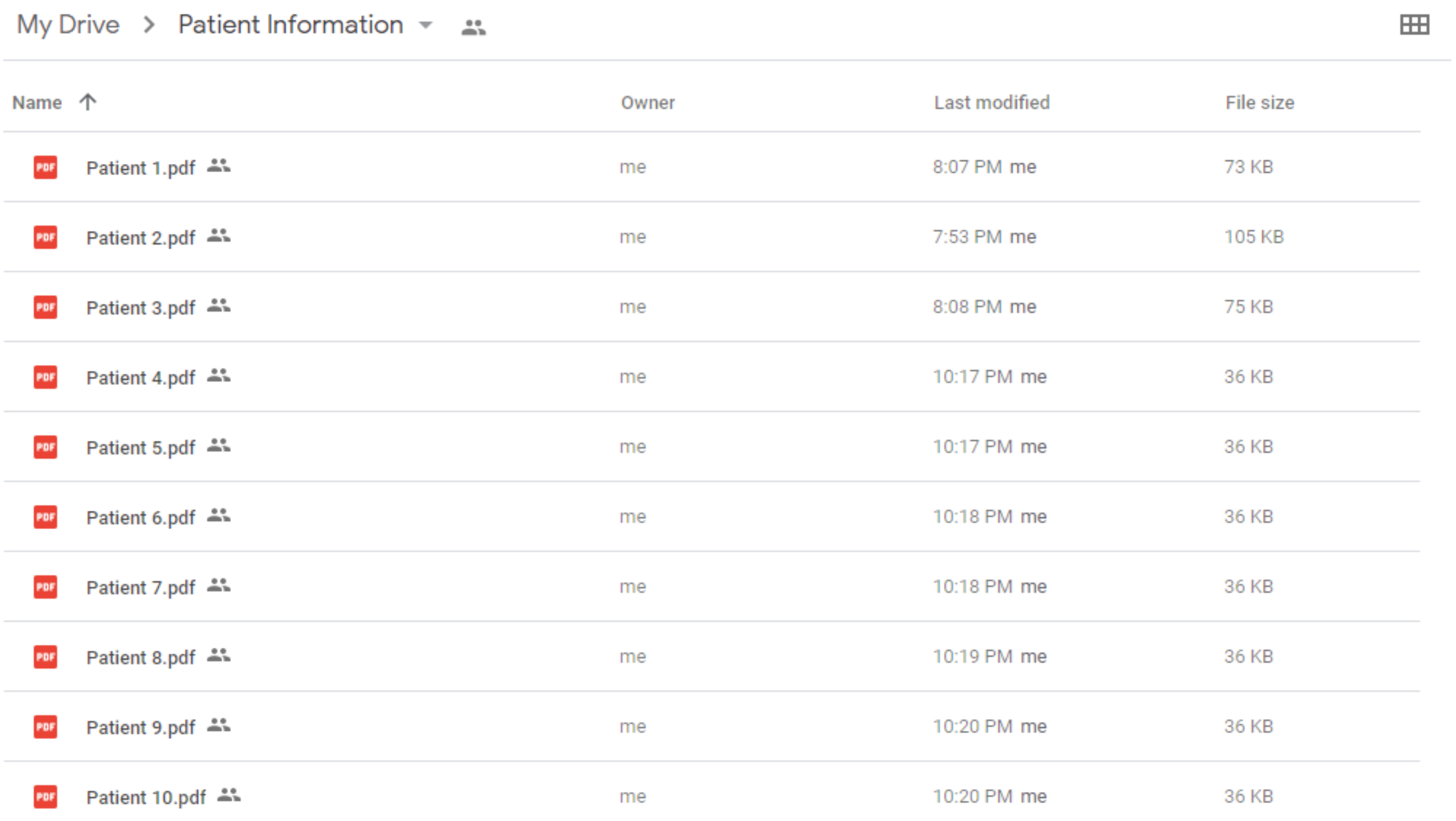

Figure 5.25 Document lists of the Patient information folder in Google Drive

### Encrypted Document

After successful upload process, if the data owner wants to check the document which he has uploaded, he needs to click the “download” button from the file list to download the document. note that, the document was encrypted by the 256-bit AES key. To verify if the encryption process is worked or not we have uploaded and checked the text, PDF, doc, image file which are discussed below.

We have uploaded a text file titled “Electronic Health Record.txt,” in the “Patient Information” folder. After downloading the “Electronic Health Record.txt” text document, we can see that the document is in encrypted form, which is shown in figure 5.26. We can check that on the left, there is the plaintext format of the document, and on the right, there’s the encrypted format of the document, which means the encryption process has been worked successfully.

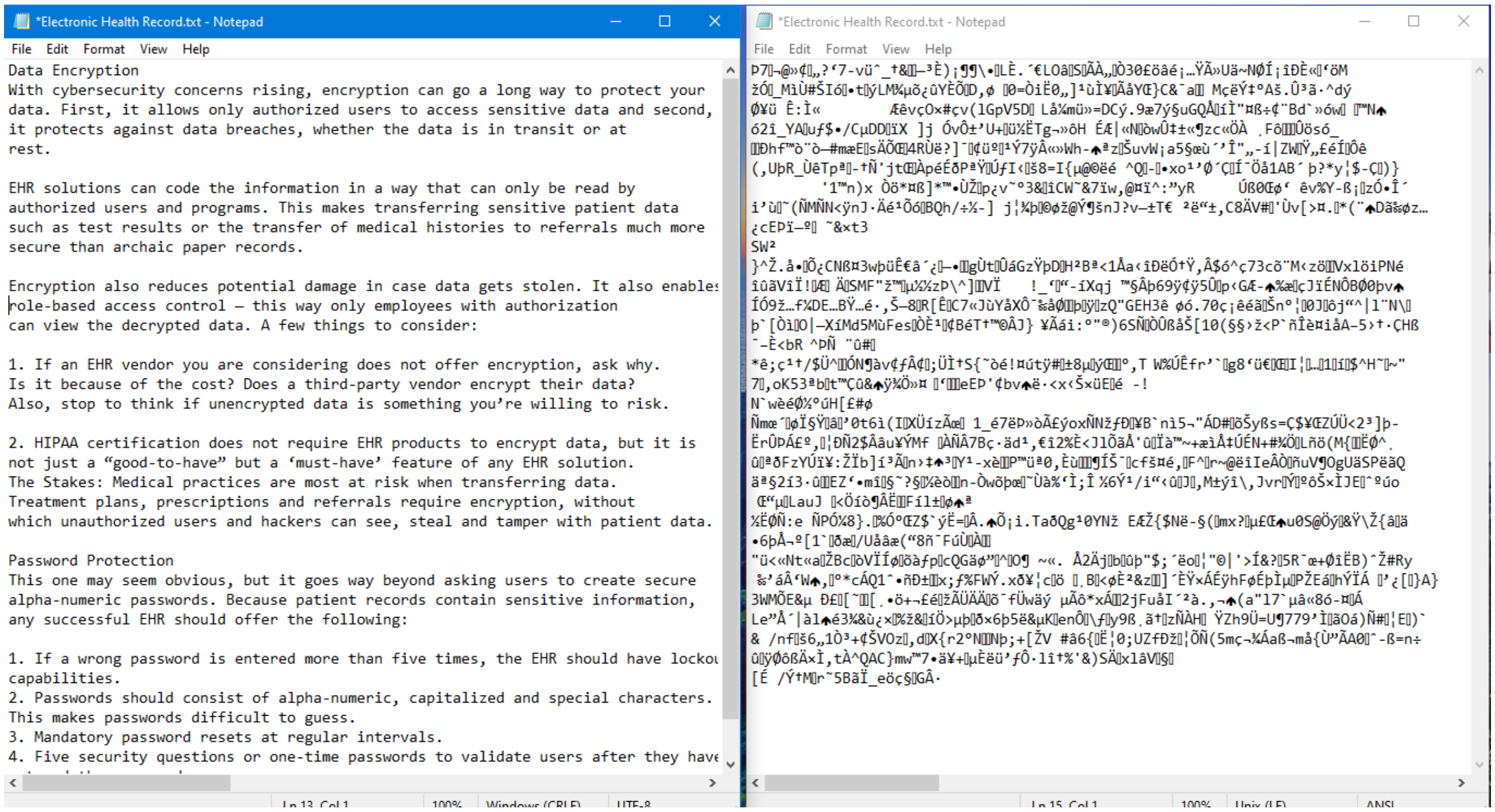

Figure 5.26 Plaintext and Encrypted format of a text file

We have uploaded all the patients document in PDF form. We have checked the "Patient 1" document, in the PDF (Portable Document Format) which is shown in figure 5.27.

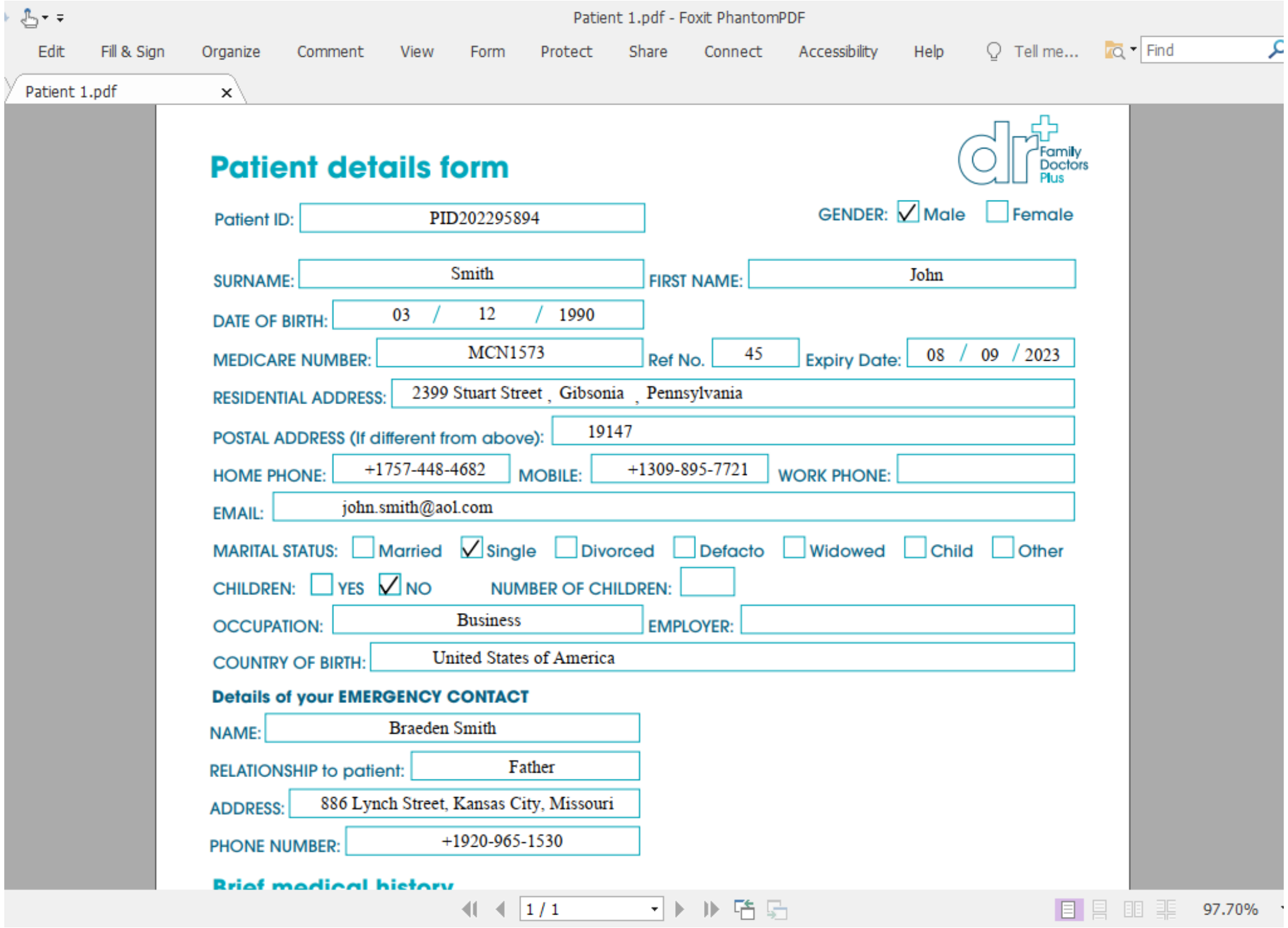

Figure 5.27 Patient 1 document in the PDF format

After download, we have tried to open the document using Foxit Phantom (PDF Version: 10.0.1.35811) software and were unable to open the encrypted document, which is shown in figure 5.28.

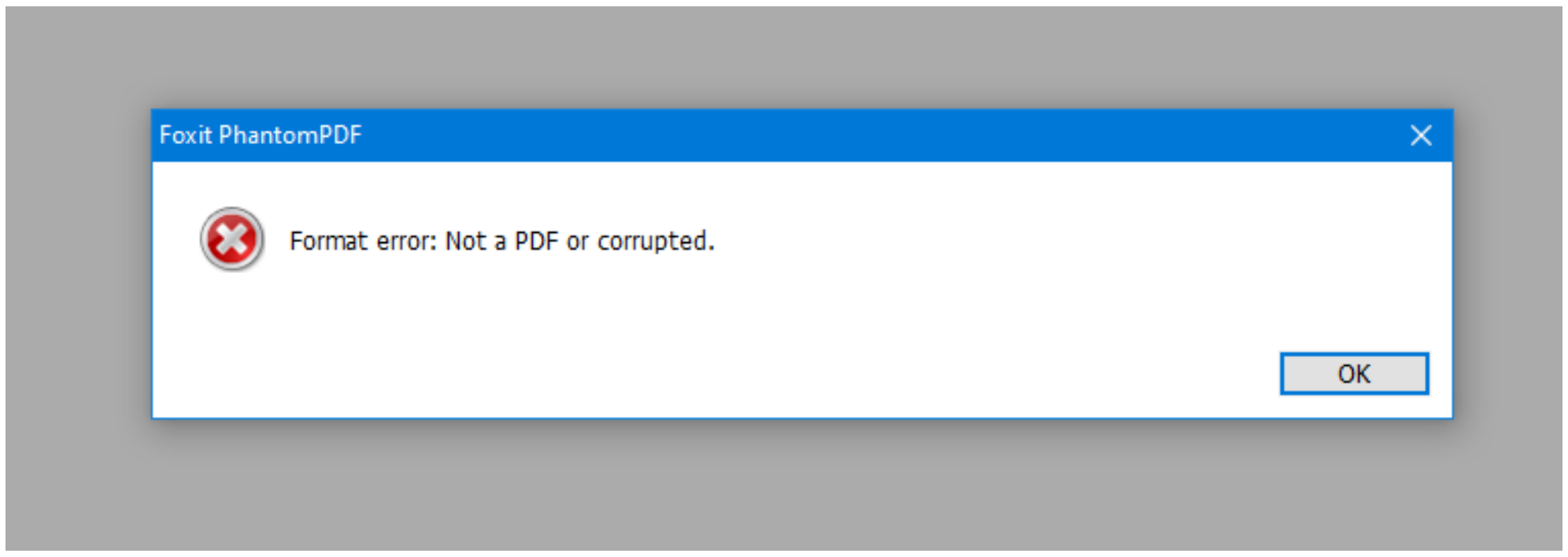

Figure 5.28 Patient 1 document in the encrypted PDF format

We have checked by uploading the patient form in doc file which is shown in figure 5.29.

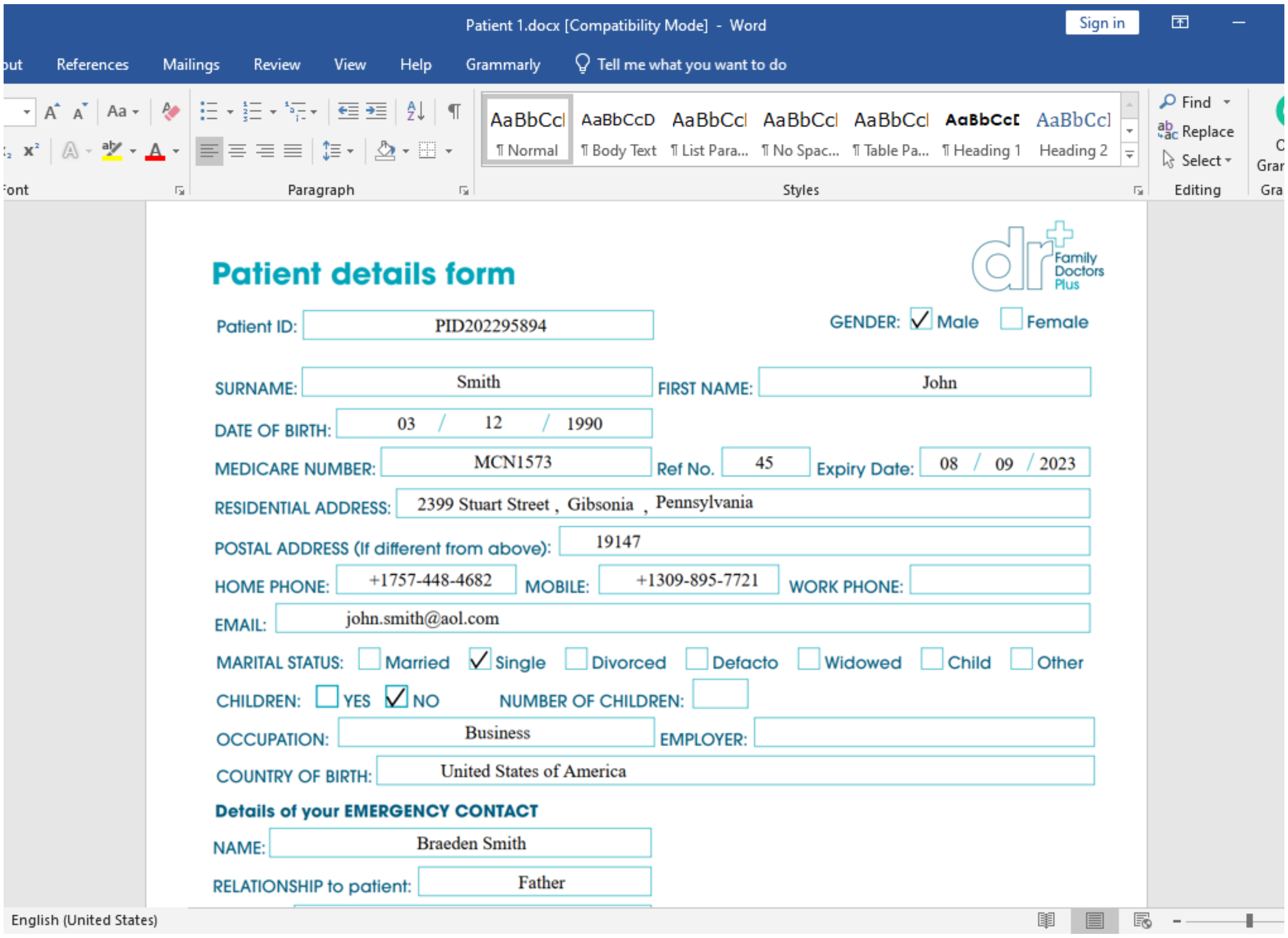

Figure 5.29 Patient 1 document in the docx format

After uploading the doc file, we also have tried to open the "Patient 1.docx" document, to check if it is working to encrypt or not. To open the "Patient 1.docx" document, we tried to open the document using "Microsoft Office Professional Plus 2019" software and showed an error shown in figure 5.30 as it was in encrypted form and successfully worked for the Docx format file also.

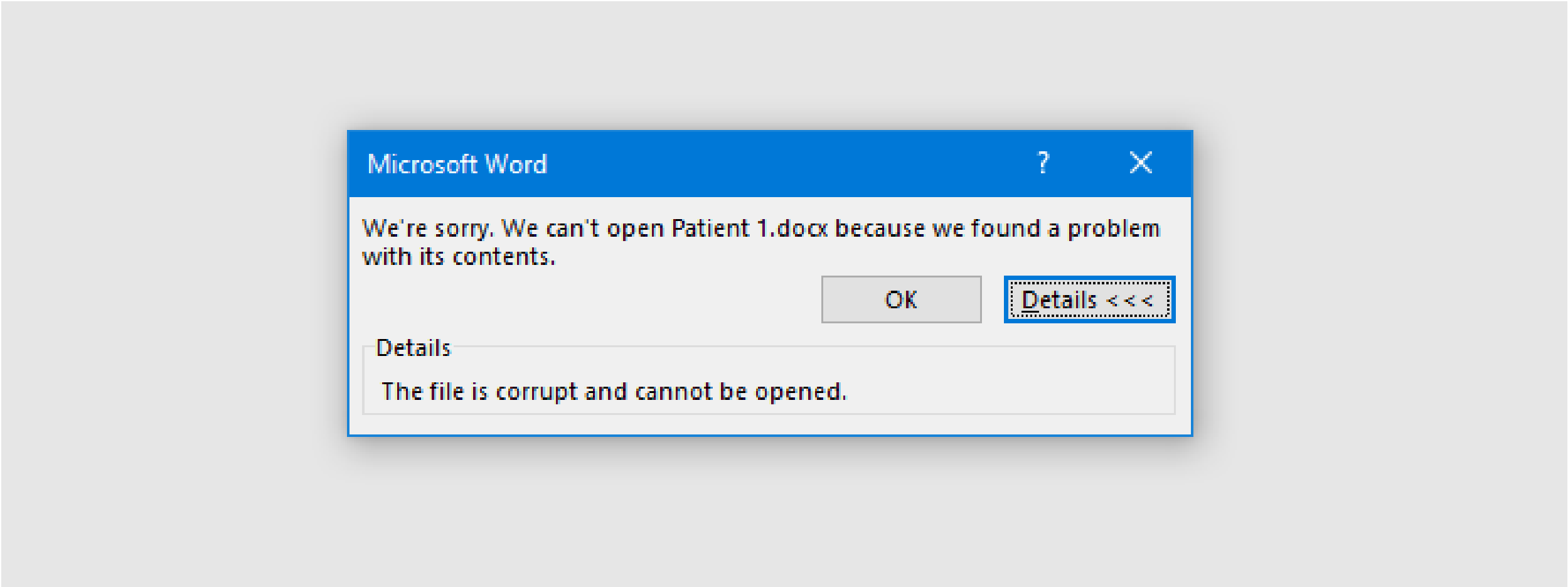

Figure 5.30 Patient 1 document in the encrypted docx format

Lastly, we have uploaded the image file which contents the features of Electronic Health Records titled "EHR.PNG" format which is shown in figure 5.31.

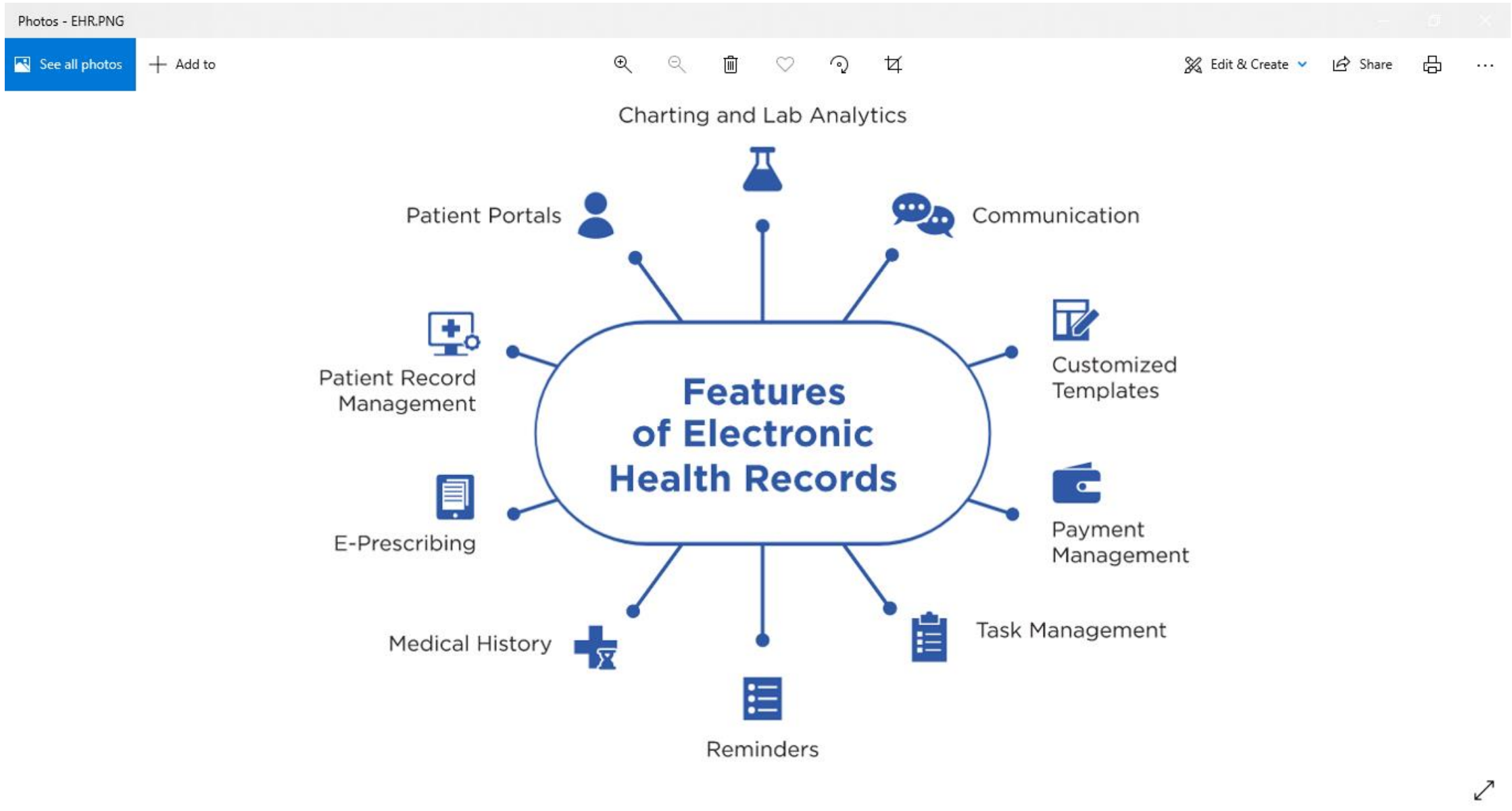

Figure 5.31 EHR image in PNG extension

After uploaded the image which is in PNG format, we have downloaded the imagine file and tried to open the file by using windows 10 default photos software. The file is showed an error that the file format is not supported as it is encrypted which is shown in figure 5.32.

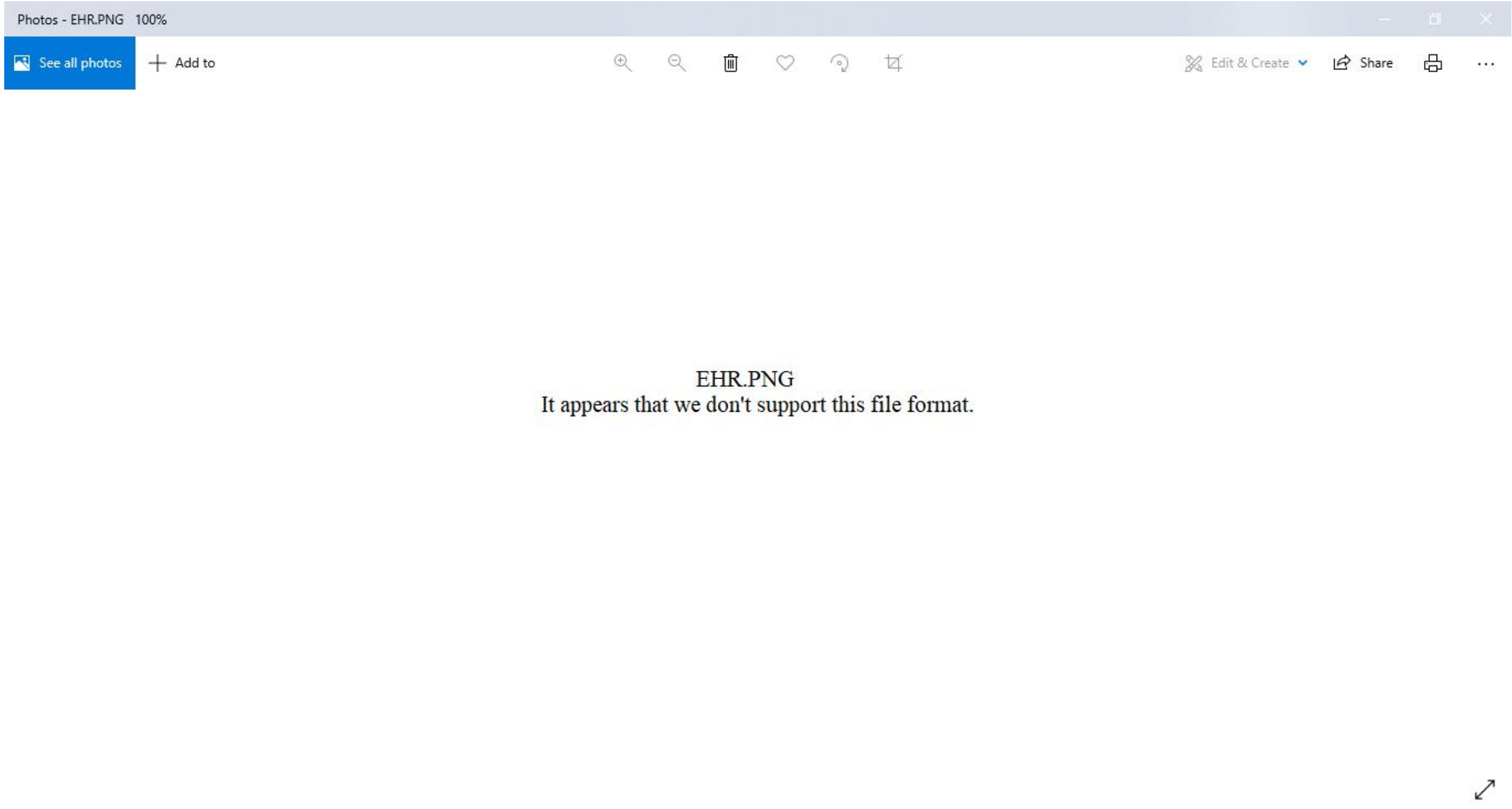

Figure 5.32 EHR image in encrypted format

**Index Generation**

In our proposed scheme, we have used an inverted index to set the keyword and document in the way of form that the secret key is used to encrypt the keyword. The random keys are used to encrypt the document. As we have implemented the proposed scheme over Google Drive, it is mentioned that the google drive is provided a unique ID (File ID) for each document. We have used the JSON format {“key”: “value”} to generate each encrypted keyword and File ID, which is in the form normally,

{ “key” : [“File ID”] }

Where,

Key = The encrypted form of the keyword, encrypted by Secret key, SK and

File ID = The document’s unique ID

The index of the documents uploaded in the “Patient Information” folder is shown in figure 5.33.

Searchabel Inverted Index.json

```json
{
  "2a3a24b1ac8b90f3d8c7a928f0a53010129a4c": [
    "1tq9Nm07OsnaRT4w8xyBiOOLjUKJ54qYs"
  ],
  "3a3020a4a9d981f5d8d7bc6df4ba": [
    "11QZ0EU3svpQmwhsKHJK0K9ebjIGWO7_U"
  ],
  "283e24b5e59893e9d8c7bd69f5e92c1f07c94ccec7bf52": [
    "1E1L6x6kdw_81637SlNxbsK10R67xUTTu"
  ],
  "392d23a9b79085eec4": [
    "17pek60jXU_aiQpiP_0XAOcH5Wa4coFm"
  ],
  "2c2a3aaeb08b82a7d8c7f06bf8a73f1b059a": [
    "1sLDT96dS9wxmAwfyRn9j1zJFPh6ubL8K"
  ],
  "282d38b2b19885e297c5a267fba5391304": [
    "13WiozkG1bhxuMvYwuY9oTs_ujGZy9Y-e"
  ],
  "353a39b5a495d1eedbd9be6deaba": [
    "1HsW6h48-_40fnelxfAr-1zFCebbsJJ5Y"
  ],
  "3c3636a3a08d94f4": [
    "17EKMJcf0-NF8D05WZrsYy82Ee6IYzeKy",
    "1xsjqCncks6d74ZFv3F47EGGlZpCQ0llX"
  ],
  "353c39f0f0cec2": [
    "1_mrSKwmsOU1UCtjpAHua85YteTGCxZmP",
    "1atP5wnlnt-b5RsoRHw-GHi5x_Ei-azcl",
    "1UdA1ZPjCsGJ8ZXEac4ZU4L_lU_y35Cqf"
  ],
}
```

Figure 5.33 Generated Inverted Index

## 2. Share

The data owner would like to share the folder with the data user after the uploading process successfully. Sharing is the important mechanism in our proposed model to get access to the data. In this case, if the data owner is the medical authority, then they might share those patients' information with the specialists or doctors for an appointment, receptionists, etc. The data owner can share the encrypted document folder with the data used by selecting the "Share" option from the web application. A new page will appear from where the data owner needs to put the email address of the data user to whom he wants to share the folder, which is shown in figure 5.34. By clicking the share button, the sharing process will be completed, and in the right notification panel, a message will be shown that "File Shared Successfully."

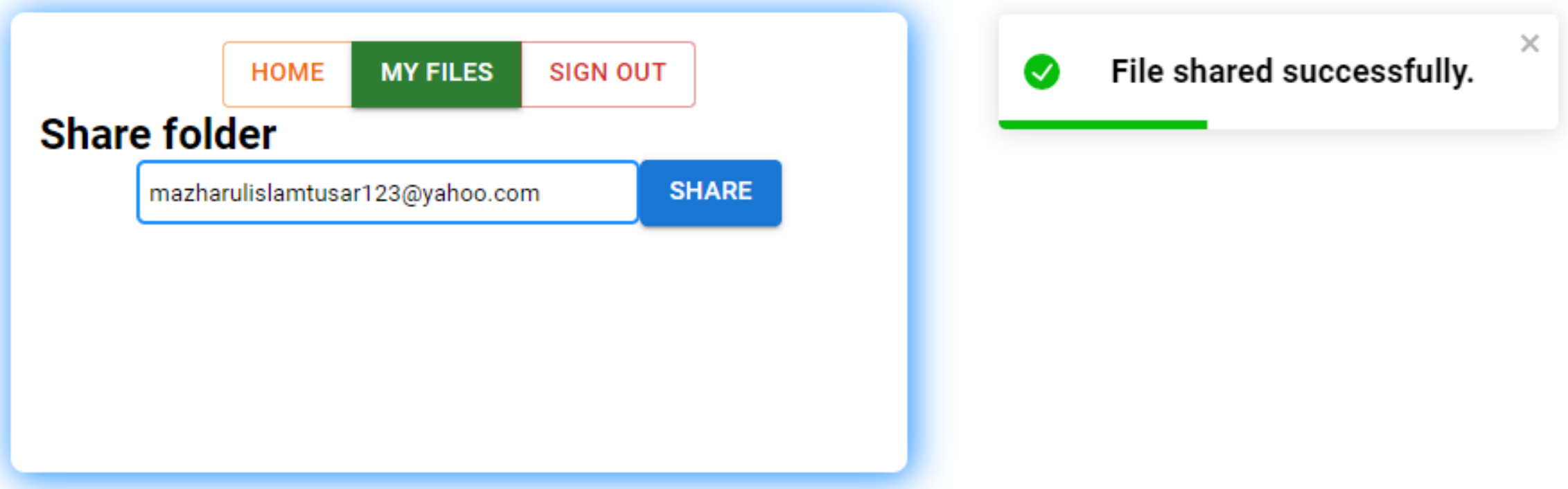

Figure 5.34 Share folder process with the data user

Normally when we use Google Drive and want to share documents of a folder with someone else or groups, we need to choose the “Share” option and insert the email address of the user with whom we want to share the folder. Google Drive generates the folder ID of a shared folder and creates a link to that shared folder which is shown in Figure 5.35. There is an option on whether the shared person can “edit”, “commentor”, or “view” the document with some restrictions. But in our implementation, we have kept only the share option as simple as that so that the data user can access the folder easily.

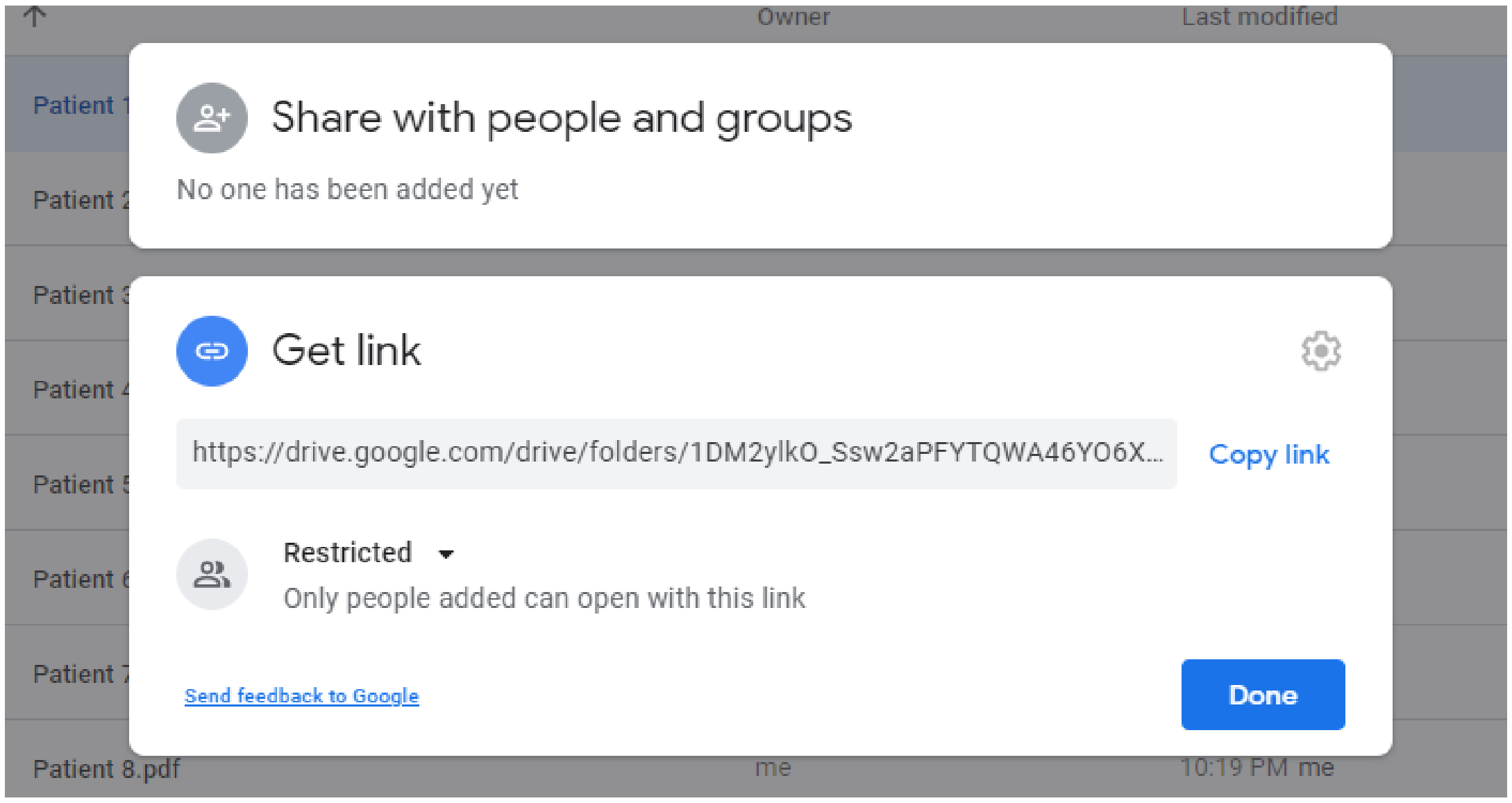

Figure 5.35 Share Option in the Google Drive

After a successful sharing process, the data user will get an email that the data owner has shared a folder with him/her. To get access to the folder, he/she needs to sign up on the “Crypto Search” web application, which is shown in Figure 5.36. The registration link will be given directly in the email

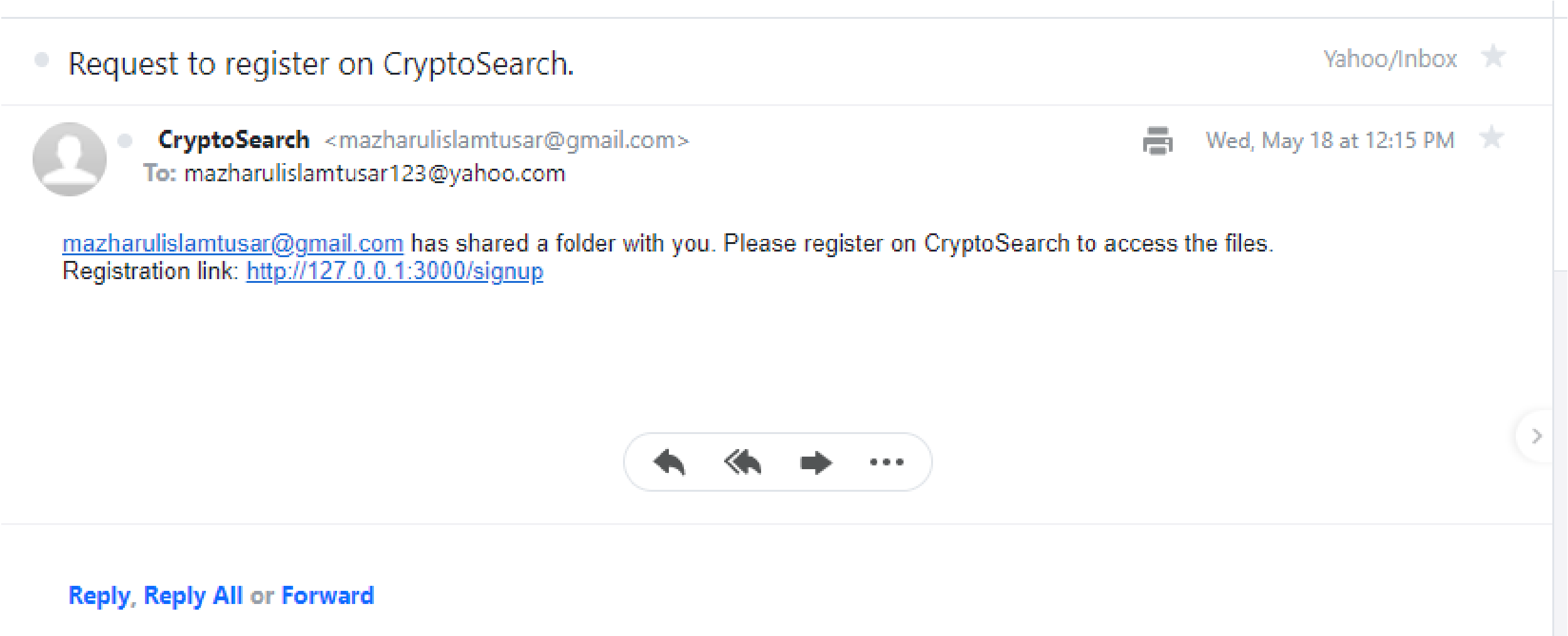

Figure 5.36 Email message to sign up on Crypto Search

The data user needs to sign up in the "Crypto Search" to get access to the shared folder. In this process, the proposed scheme will follow scenario no. 3, which is discussed in section 4.6.3, followed by Identity Based Encryption (IBE). Moreover, we have kept the sharing process as much as possible so that the data owner can understand and the data user gets the email immediately.

### 3. Search

We have used 5 patient documents to explain the implementation details discussed in the upload process before. We set keywords corresponding to the patient's documents listed below in table 5.1.

| Keywords | Documents ID |
|---|---|
| PID202295894, MCN1573, Diabetes | Patient 1 |
| Aliana Lucy, High Blood Pressure | Patient 2 |
| Diabetes | Patient 3 |
| Stroke | Patient 4, Patient 5 |

Table 5.1 Keywords and document ID lists

#### I. Single Keyword Search

The search process can be done both by the data owner and data user, which is described in section 4.6.1, and by the data user who got access to the shared folder, which is described in sections 4.6.2 and 4.6.3. We assume that the data owner is the hospital authority and that the specialists or doctors are data users. Now, if we consider the "patient 1" document, which was set by the 3 keywords "PID202295894" which is the Patient ID number, "MCN1573" which is the Patient Medicare number, and "Diabetes" which is in the patient's medical history. So, suppose the user wants to search by the patient's ID number. In that case, he/she will get the "Patient 1" medical information document and get the notification in the right panel that that Patient ID found the File ID of that document as a keyword. So, the user can download the document in plaintext form by clicking the download button, shown in Figure 5.37.

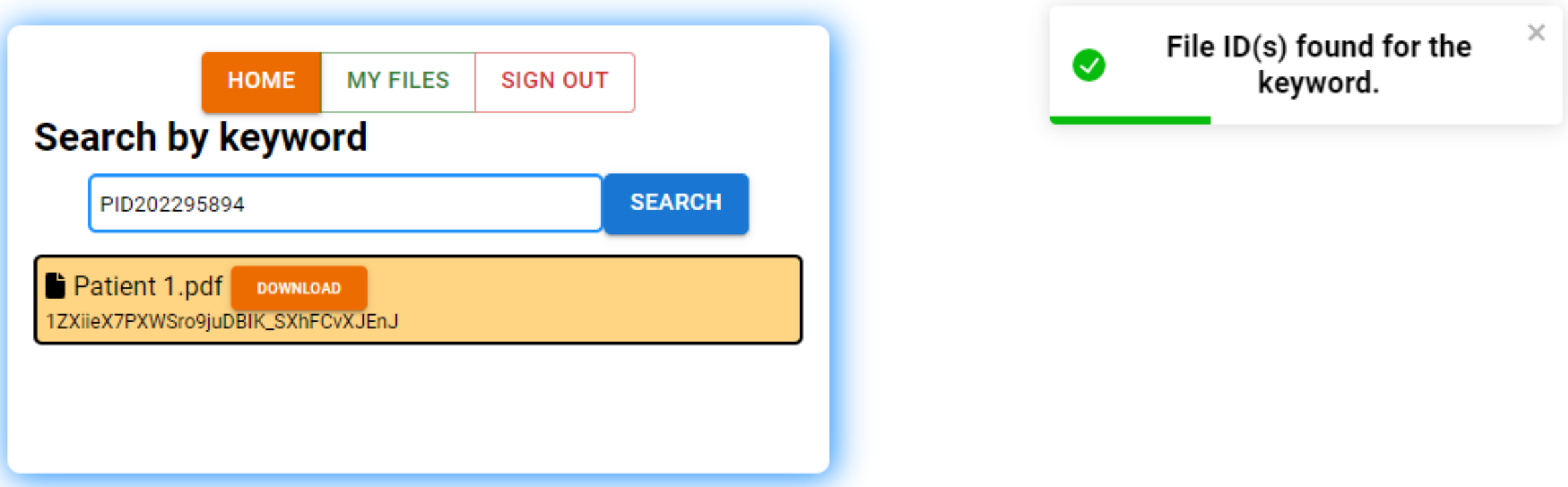

Figure 5.37 Search by keyword result of the document

## II. Alternative Single Keyword Search

The data user can search by the Medicare number of the "Patient 1" document alternatively. This will then return the same document as a result of the "Patient 1" shown in Figure 5.38. From the table, we can see that we have set 3 individual keywords for the "Patient 1" document. So, if the user searches any one of those keywords, he/she will find the same document.

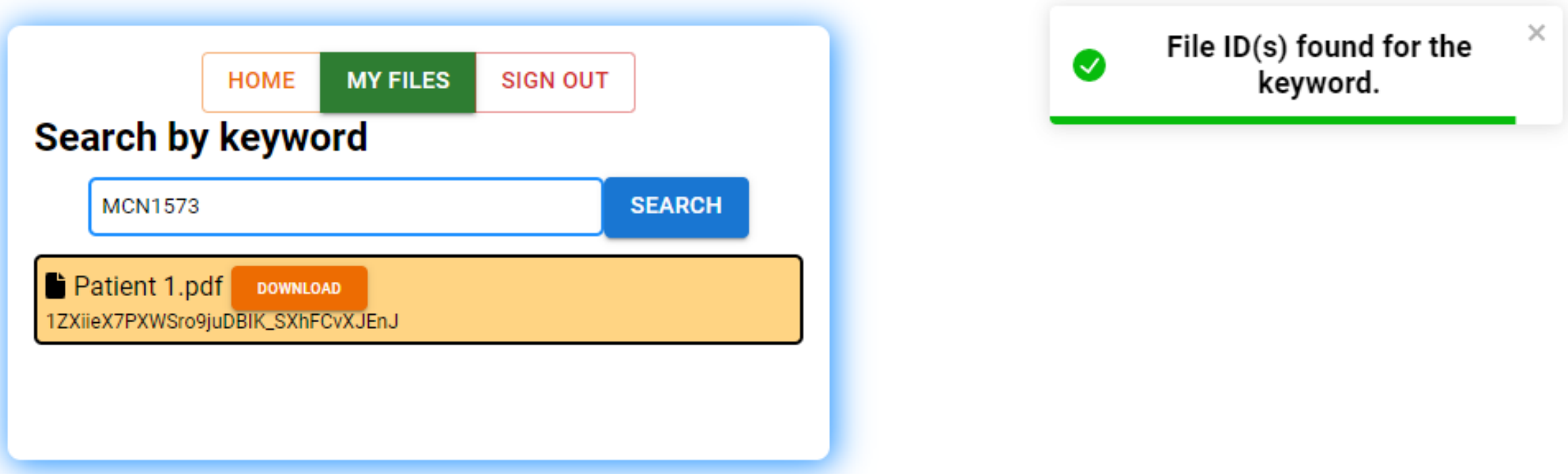

Figure 5.38 Search by alternative keyword result of the document

## III. Multi Keyword Search

The data user can search by multiple keywords at a time to find the specific document. Multiple keywords can help find the document more accurately than a single keyword search. However, both single and multiple keywords are supported in our proposed model and implementation. For example, the data user can search by the patient name "Aliana Lucy" set for the "Patient 2" document and get the document shown in figure 5.39.

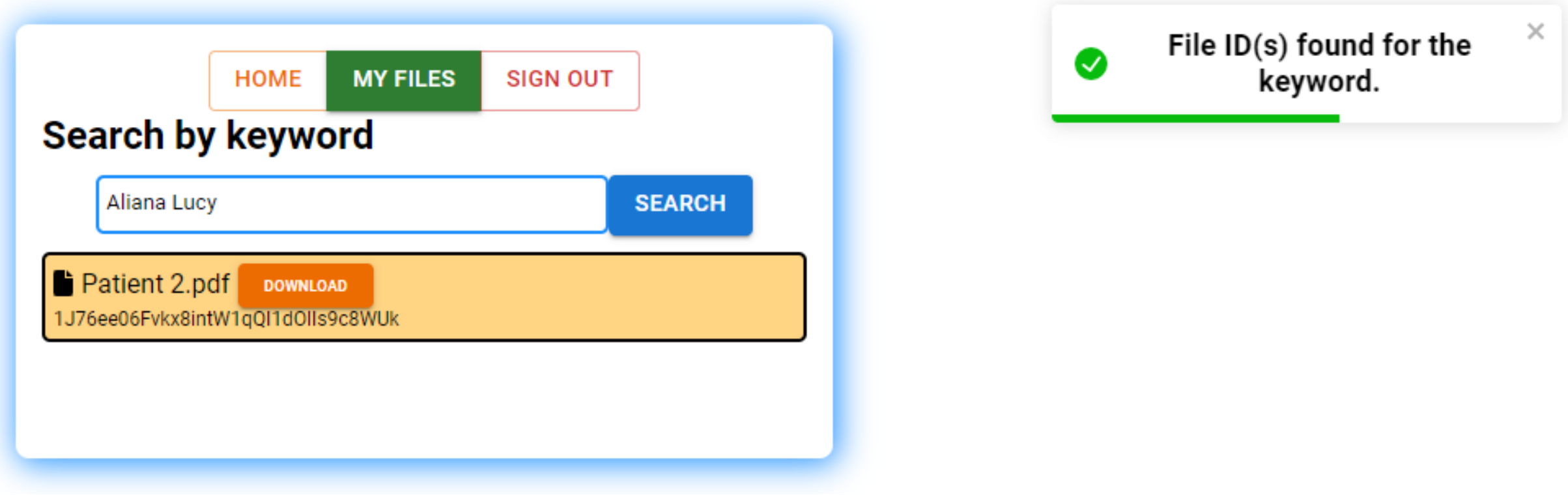

Figure 5.39 Search by multiple keywords result of the document

## IV. Multiple documents retrieval

The data owner can set the exact keywords for multiple documents for similar objectives. In this situation, the data user would like to expect more than one document when he likes to search in the application. For example, we can see that the "Diabetes" keyword was set to the keyword both for "Patient 1" and "Patient 3" documents which are listed in table 5.1. So, if the data user searches by the keyword "Diabetes", he/she will get two documents shown in figure 5.40.

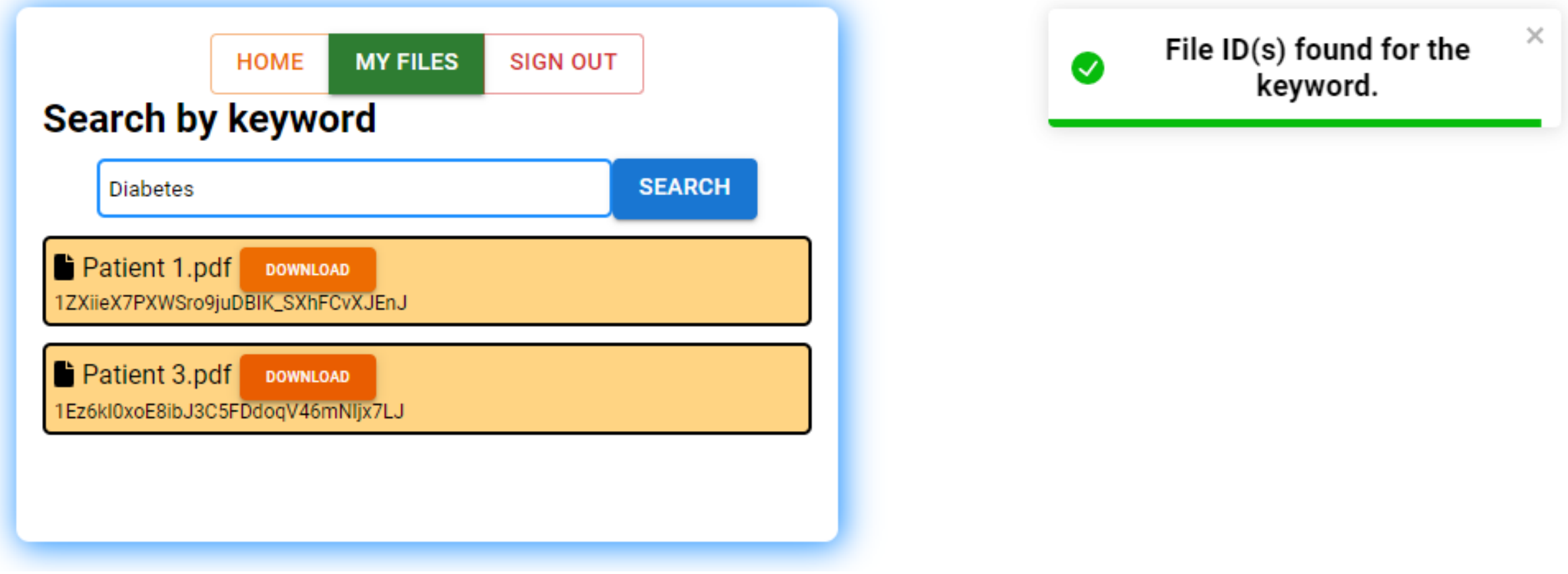

Figure 5.40 Multiple documents retrieval by keyword search

## V. Error Message

Suppose the data user searches by the wrong keyword, which was not set by the data user or which is not the relevant keyword of the documents. In that case, he/she will get an error message that "No file found". For example, we have set the "Stroke" keyword for "Patient 4" and "Patient 5" documents. But if the data user searches by the "Kidney Problems" which is not set in the keyword, he/she won't get any document shown in figure 5.41. Also, In the right notification panel, the message will be shown that no file is found for the specified keyword. So, the data owner/data user always must be searched by the specific keyword of the document.

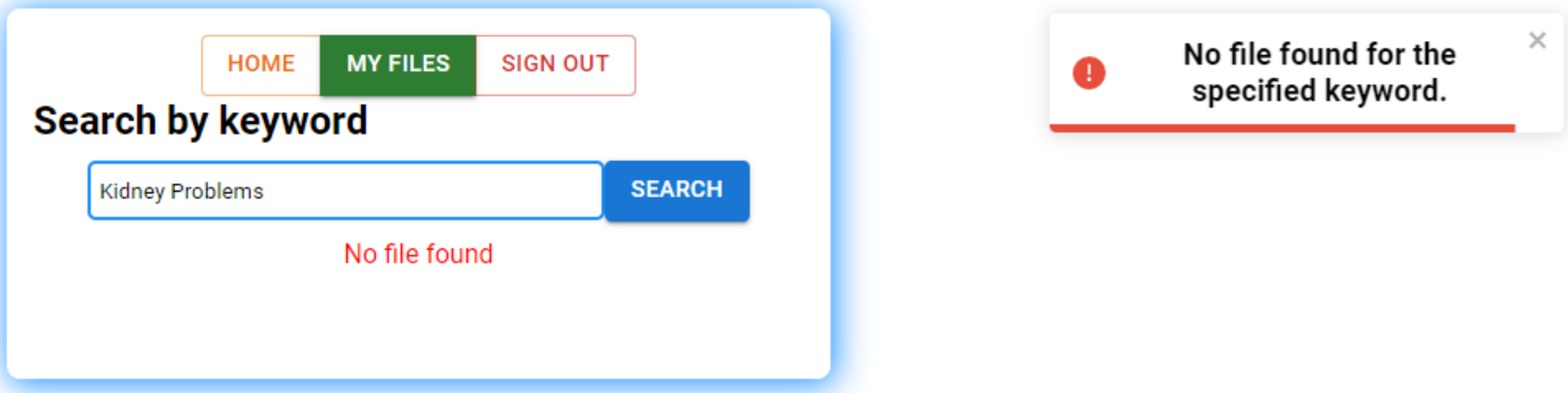

Figure 5.41 No file found by the unspecified keyword search

**Sign Out**

The Data Owner/ Data User can sign out from the web application by clicking the "Sign out" button after uploading or downloading his/her document successfully. After the sign-out process, the "Crypto Search" homepage will return.

Therefore, In the functional specification section, we have shown the proposed model practically with every scenario and step with example. We have kept the functional complexity on the backend side so that anyone can use our system with some little knowledge of the upload, search, and download. Overall, our theoretical and implementation model is suitable for our reality-based scenario both for the data owner who outsources the data and the data users with whom the data is shared.

### 5.1.5 Performance and Analysis

This section mainly discusses the complexity and performance of the implemented proposed scheme of searching on encrypted data over the cloud. We have implemented the proposed scheme through frontend (React JS, Node JS, Express JS) and backend (MongoDB, NoSQL) technology. The experiment is conducted on a laptop with Google Chrome web browser of the Windows 10 operation system, Intel Core i7 4200M CPU, 8GB RAM, 250GB SSD.

We have performed the performance analysis of current searchable encryption schemes based on encryption time, index creation time, and search time. Our results can be summarized as follows:

- Encryption time is linear in document size.
- There is a substantially limited memory overhead.
- Inverted index generation time is O (1) for each encrypted keyword and document.
- The overhead of the secret key generation is fairly low.
- Search time is linear with the set of keywords in the document.

## 5.2 Backend Side Application

The server-side application of our implementation is called “CryptoSearch Backend”. The primary objective of our server-side is to store data and serve them to the user when requested. As there will be no processing on the server side, we wanted to make an API that can be called from the frontend application. We had several options to choose from, e.g., ASP .NET MVC, ASP .NET Core, Spring MVC, Spring Boot, etc. All of them were good options.

### 5.2.1 Technologies Used

We have implemented our front-end application in React JS with Node JS. We preferred NoSQL database for our back-end side application. So, we chose the NoSQL database to store the information of User ID, File ID, etc. NoSQL databases are quite useful for working with large sets of distributed data. NoSQL databases are used in nearly every industry. Use cases range from the highly critical (e.g., storing financial data and healthcare records) to the more fun and frivolous (e.g., storing IoT readings from a smart kitty litter box). We wanted to build our service as a RESTful web service. MongoDB allows us to manage document-oriented information and store or retrieve information. MongoDB is an open-source NoSQL database management program.

Document databases store data in documents similar to JSON (JavaScript Object Notation) objects. Each document contains pairs of fields and key values. The key values can typically be a variety of types, including things like strings, numbers, Booleans, arrays, or objects. Key-value databases are a simpler type of database where each item contains keys and values. In our case, keys are the secret, and random generated keys and values are the File ID of the document.

While a variety of differences exist between relational database management systems (RDBMS) and NoSQL databases, one of the key differences in the way the data is modeled in the database. By deciding which database to use, we have found one or more of the following factors typically lead to selecting a NoSQL database:

- Fast-paced Agile development
- Storage of structured and semi-structured data
- Huge volumes of data
- Requirements for scale-out architecture
- Modern application paradigms like microservices and real-time streaming

### 5.2.2 Design Principles

While designing the Crypto Search Web Application, we had to concentrate on two separate elements, the source code design pattern and the database design. We tried to follow the proper design pattern for our source code. There were several reasons behind that. The source code will be more organized, extensions can be made to the server capabilities in the future, and it will be much easier for developers to contribute. So, we chose to follow the hybrid design pattern. This allowed us to separate business logic from user request handling.

Other than that, another critical task performed by our server is to communicate with the database. The database communication part is separated from JSON to maintain the separation of concern. The database design was another crucial decision we had to make as it has a severe performance impact if mixed designed. We have used 3 database modules, such as File Info, UID, and, shown in figure 5.42.

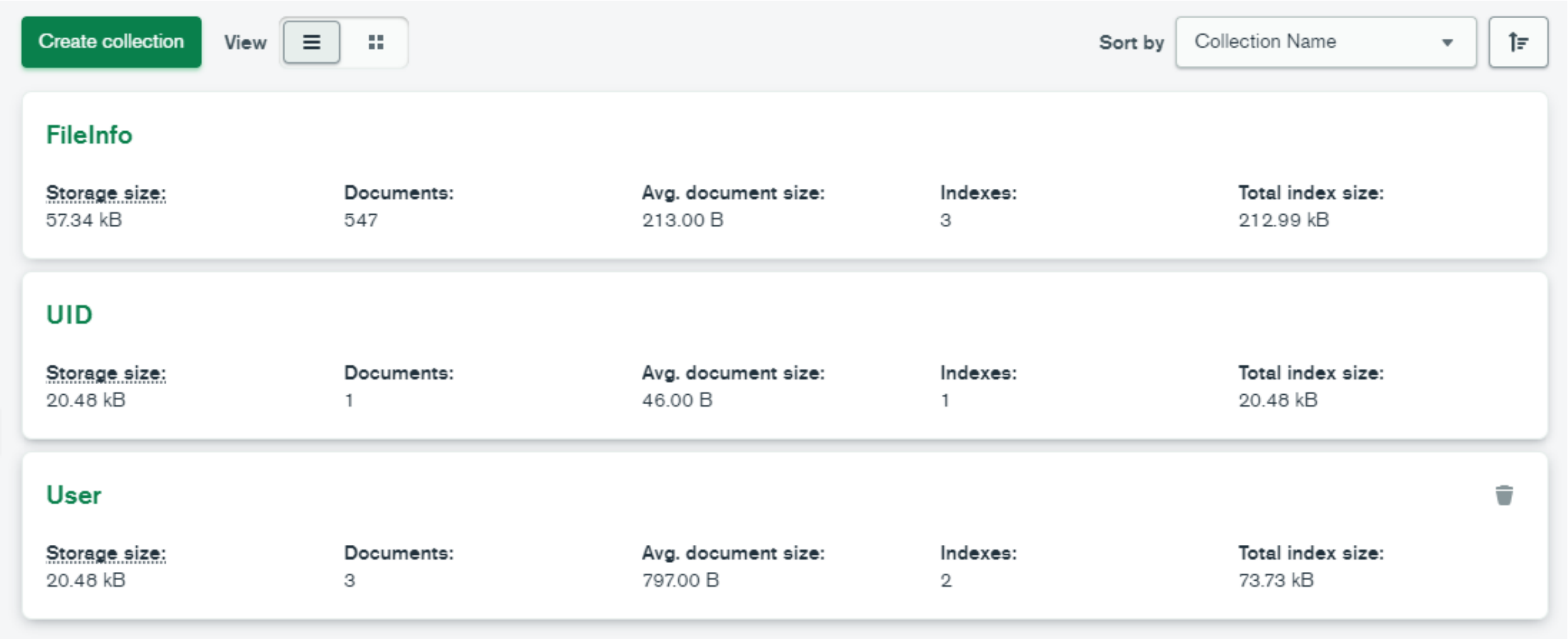

Figure 5.42 Collections of the database lists

Some key elements of each file id consist of the file, the random key generated, file path, folder ID, and file ID. It also shows which file is authorized by which data owner by using a User ID. The file ID database is shown in Figure 5.43.

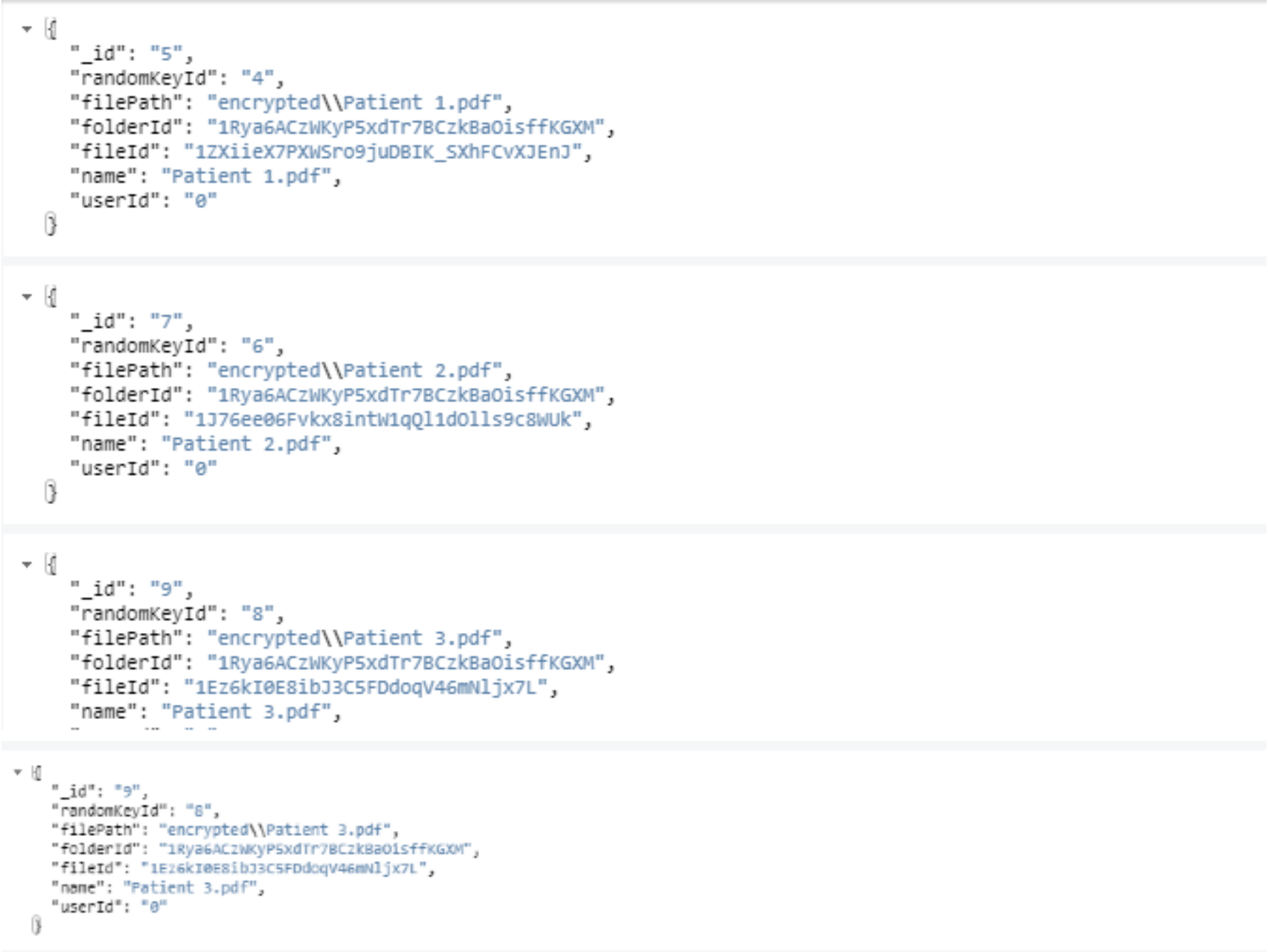

Figure 5.43 File ID database

The user id database contains the id of the data owner/data user, including email, hashedPassword, access token from Google Drive, etc., as shown in figure 5.44.

```
_id: "91"
email: "mazharulislamtusar@gmail.com"
hashedPassword: [illegible]
cloudServiceProvider: "GOOGLE_DRIVE"
credentials: Object
  access_token: "ya29.a0ARrdaM9-_FDCI2a3K4_AT_mVoK7eMpVS5weL7wrr9qxgEfJkfQf2AN9gvMB6oZg…"
  refresh_token: "1//0gZgwymqHv0zhCgYIARAAGBASNwF-L9IrACL3j_5AhbGMhQjb3aL3Cqv_7xeTdqMvkW…"
  scope: "https://www.googleapis.com/auth/drive"
  token_type: "Bearer"
  expiry_date: 1653466082913
googleDriveRefreshToken: "1//0gb5dZE1VBc-tCgYIARAAGBASNwF-L9IrUcCg0gN6eMXgHpCcCx7DWIILurlHet5axJ…"
```

Figure 5.44 User ID database

### 5.2.3 Internals and Code Snippets

The entire source code is organized in two different folders. e.g., code related to core components is placed in a backend folder, code related to UI is kept in the frontend folder, and so on. As the complete source code is too large to be a part of this report, we will provide some of the code's essential pieces.

How the UID of each file is generated is shown in figure 5.45.

uid-generator.js

```javascript
const { MongoDb, } = require('../mongodb');
const { UIDGenerator } = require("../uid-generator");

const uidGenerator = UIDGenerator.create();

module.exports.generateUidAsync = async () => {
  const database = MongoDb.getInstance().getDatabase();
  const uid = database.collection('UID');
  const lastGeneratedUid = await uid.findOne({ _id: 'LAST_GENERATED_ID', });

  lastGeneratedUid && uidGenerator.setLast(lastGeneratedUid.value);

  const generatedUid = uidGenerator.generate();

  await uid.updateOne({ _id: 'LAST_GENERATED_ID', }, {
    $set: {
      value: generatedUid,
    },
  }, { upsert: true, });

  return generatedUid;
};
```

Figure 5.45 UID generator of the file

We have generated the inverted index of each keyword and file id. We also uploaded this inverted index to google drive, which will help find the file by keyword search. In figure 5.46, it is shown a small part of how the inverted index is generated.

```
      // uploads inverted index to google drive...
      fileResponse = await googleDriveService.uploadFileAsync({
        googleDriveClient: request.googleDriveClient,
        folderId: folderId,
        fileInfo: {
          name: configuration.invertedIndexFileName,
          mimeType: configuration.invertedIndexFileMimeType,
          content: fs.createReadStream(invertedIndexFilePath),
        },
      });

      invertedIndexes[folderId] = fileResponse.data.id;

      user.invertedIndexes = invertedIndexes;

      const database = MongoDb.getInstance().getDatabase();
      const users = database.collection('User');

      await users.updateOne({ _id: user._id, }, {
        $set: { invertedIndexes: invertedIndexes, },
        $addToSet: { folders: folderId, },
      });
    }

    return {
      message: 'Inverted index generated successfully.',
      data: {
        invertedIndexId: fileResponse.data.id,
      },
    };
  });
```

Figure 5.46 Inverted Index generate

Other files and folders of the code can be found similarly. As our frontend side is totally detached from the backend side, we do not have any view class on our server side.

# 6. CHAPTER SIX: CONCLUSIONS AND FUTURE WORK

## 6.1 Summary of the work

We have proposed a searchable and sharing scheme with secure storage and sharing based on searchable encryption technology. It solves the security problems of data storage, sharing, and searching on encrypted data over the cloud. We have designed and implemented a web application that uses a cryptographic approach to secure the user data and search on it.

This work aims to make searching and sharing functionality of the encrypted data simultaneously, keeping the security concern intact. We have proposed a practical step from signing up to generating keys, encrypting, uploading, searching, downloading, and sharing encrypted data with other users. We have used the inverted index to generate the searchable keys while the third-party entity is responsible for generating and distributing keys among data owners and users.

We have designed the model in such a way that it has high performant, modular, efficient, and at the same time, user-friendly architecture, which also resolves the issues that were introduced with the other models. So, we have shown a new system that concentrates on key areas such as; secure and faster sharing of encrypted files among users and efficient search on encrypted documents. It can be used without the need for any support or modification from the cloud service provider and keeping the architecture as simple as possible, which makes our system easier to use by data users.

We have implemented the proposed scheme in the practical scenario. Based on this, it is expected that the proposed method can be practically used in the industrial field dealing with personal information data, such as credit information evaluation and personal health, which requires privacy protection. The proposed schemes can achieve better efficiency than the existing works in terms of storage, search, and updating complexity.

Therefore, we conclude that the proposed scheme improves the tradeoff between memory overhead, search capability, and security when storing data in encrypted form.

## 6.2 Limitations

During the design phase of our web application, our primary focus was to design an elegant-looking user interface that looks simple and familiar to use for regular users. We put much effort into improving the user experience. However, there are still some limitations to our proposed models, such as:

1. As our proposed model adds an additional layer on top of cloud storage service, it risks becoming the second point of failure. Suppose a data user uses Google Drive as cloud storage and uses our model on top of it. If the user uploads a file through our service, the file will be uploaded to the cloud in an encrypted form. Suppose he wants to search on the folder to get the file, and for some reason, our server-side crashes. In that case, the user will not be able to use the service for the time being until the server is up and running again.

2. When the data owner has shared the encrypted document folder with the data user, the data user gets an email that he needs to sign up to our web application to access the folder. Without signing up, he cannot access the shared folder. So, the data user must sign up in the system, which is not very important to access the folder. Google Drive gives services to share the data with many users by generating the folder URL directly. Nevertheless, we are using a sign-up process to access the folder making the scheme more complex.

3. We have used a trusted third-party entity in our scheme who would manage the key generation, key distribution to the data owner, store the folder ID along with file encryption AES keys with reference number, generate searchable encrypted text for data user with the user identity, which leads the scheme less secure and vulnerable. Though there is no way of colluding with cloud storage to the trusted third party to access the document of a folder, if the trusted party is honest and curious about the document, the entity can act as an adversary or intruder to a data user. The whole folder document might be vulnerable.

4. Also, the data owner, data user, and the contents stored in the web server may be altered by the trusted third party through direct access to the web server. Suppose the trusted third party changes the encrypted passphrase of a user by accessing the database. There will not be any security breach, but the user will surely lose access to the file.

5. Another shortcoming of using our model has reduced performance due to cryptographic operations. We have optimized our design for better performance. However, this does not eliminate the cost of cryptographic operations that are performed underneath. We have used the RSA cryptosystem in our proposed model, which is well known for heavy resource consumption, especially key pair generation, and decryption.

6. Our proposed scheme permits the Data User to retrieve the encrypted document by searching based on keywords. Our method of keyword searching is limited to the exact keyword search. This requirement heavily burdens data users and significantly compromises the system's scalability.

## 6.3 Future Work

This proposed model has tremendous improvement opportunities in both architectural design and implementation. We can add new features and functions to make our searching and sharing system more versatile by eliminating the limitations. There is still room for improvement in terms of the scalability of the schemes, such as:

1. If we can improve the design of our proposed model so that the file(s) can be searched and shared without the need for the user to Sign Up. It will not only improve the design of our searching system, but it will also be an excellent achievement for entire cryptography.

2. We have used a trusted third-party entity which leads the system less secure. So, for better security concerns, we can remove this entity and do the responsibilities of this entity with the web server. Removing the trusted third party would be a great advantage. In that case, the system will be more secure and trustworthy.

3. The inverted index model still serves as a core indexing technique. However, the tree, graph structures, and vector model can be used together to explore new properties and improve the efficiency of building an index.

4. The incorporation of the keyword list into our scheme limits the capability of the number of searchable words. The next objective can be to provide the offer the freedom to search for any word in the document.

5. As our web application performs all the operations in the background, the user might want to stop or kill a running task. Adding a task manager could improve the user experience to a greater extent. It would allow users to have complete control over the running tasks.